\documentclass[12pt]{article}

\usepackage[hypertex,colorlinks=true,linkcolor=red,citecolor=blue]{hyperref}
\usepackage{graphicx}
\usepackage{amsmath,amssymb}

\newcounter{comment}

{\refstepcounter{comment}%
\begin{quote}
\ttfamily\small$\blacksquare$ \textbf{\underline{Comment} $\sharp$\thecomment:}}%
{\end{quote}}

{
\begin{quote}
\ttfamily\small$\blacktriangleright$ \textbf{\underline{Reply} $\sharp$\thecomment:}}%
{\end{quote}}

\newcommand{\GeV}{{\rm GeV}}

\newcommand{\intsum}{\int\!\!\!\!\!\!\!\!\!\!\sum_{(n-1)}}
\newcommand{\intsumm}{\int\!\!\!\!\!\!\!\!\!\!\sum_{(n)}}
\newcommand{\intk}{\int\!\!\!\!\!\int\!\!d^2{\bf k}_{\perp}}
\newcommand{\intkb}{\int\!\!\!\!\!\int\!\!d^2\bar{\bf k}_{\perp}}
\font\cmss=cmss12 
\def\1{\hbox{{1}\kern-.25em\hbox{l}}}
\def\bfZ{\relax{\hbox{\cmss Z\kern-.4em Z}}}

\def\ru1{\rule[-0.4truecm]{0mm}{1truecm}}

\begin{document}
\numberwithin{equation}{section}

\begin{titlepage}

\centerline{\large \bf  The concept of  phenomenological light-front wave functions}
\centerline{\large \bf -- Regge improved  diquark model predictions --}

\vspace{7mm}

\centerline{\bf D.~M\"uller$^{a}$ and D.~S.~Hwang$^{b}$}

\vspace{7mm}

\vspace{4mm} \centerline{\it $^a$Institut f\"ur Theoretische
Physik II, Ruhr-Universit\"at Bochum} \centerline{\it D-44780
Bochum, Germany}

\vspace{4mm} \centerline{\it $^b$Department of Physics, Sejong
University } \centerline{\it Seoul 143--747, South Korea}

\vspace{20mm}

\centerline{\bf Abstract}

\vspace{10mm}

\noindent
We introduce a classification scheme for  parton distribution models and we model
generalized parton distributions (GPDs), their form factors, and parton distribution functions (PDFs), integrated and  unintegrated ones,
in terms of unintegrated double distributions that are obtained from the  parton number conserved overlap of effective light-front wave functions. For a so-called ``spherical" model we present general expressions for all twist-two related non-perturbative quantities in terms of one effective light-front wave function, including chiral-odd GPDs. We also discuss the Regge improvement of such quark models from the $s$-channel point of view and study the relations between zero-skewness GPDs and unintegrated PDFs on a more general ground.  Finally, we provide a few phenomenological applications that emphasize the role of orbital angular momenta.

\vspace{0.5cm}

\noindent \vspace*{12mm}


\vspace{3mm}


\end{titlepage}

\newpage

\tableofcontents

\newpage

\noindent
\section{Introduction}

During the last two decades we witnessed the development of new phenomenological approaches to study hadronic physics, in particular the proton.
On one hand a renaissance of  transverse momentum  dependent parton distribution functions (PDFs), so-called unintegrated PDFs \cite{Ralston:1979ys,Collins:1981uw}
(uPDFs\footnote{
These functions are often called transverse momentum dependent parton distributions and they have the acronym TMDs.
Since we will also consider other unintegrated quantities such as GPDs and double distributions (DDs) that are related to ``Wigner" or ``phase-space" distributions or to so-called ``mother" functions or generalized transverse momentum dependent parton distributions (GTMDs), we adopt a simple naming scheme by adding the prefix ``u",
staying for unintegrated, to the common acronym for the various parton distributions.}), occurred in the description of semi-inclusive reactions and on the other hand general parton distributions (GPDs) have been proposed to describe hard exclusive processes on amplitude level \cite{Mueller:1998fv,Radyushkin:1996nd,Ji:1996nm,Collins:1996fb}. Certainly, both suggestions are based on the partonic picture and partially on factorization theorems \cite{Collins:1996fb,Collins:1998be} which state that hard physics can be systematically evaluated while soft physics, encoded in parton distributions, is universal, i.e., independent of the considered process.
Unfortunately, such a strict  factorization of hard and soft physics does not hold for the uPDF description of semi-inclusive processes \cite{Collins:1981uk,Collins:1984kg,Ji:2004wu,Col11}.
 Triggered by the promise that one can
address with GPDs and to some extend with uPDFs both the so-called ``nucleon spin puzzle", i.e., the
mismatch of the small quark spin contribution that is measured in polarized deep inelastic scattering with the large quark spin contribution predictions from
constituent quark models \cite{Ashman:1989ig}, and the transverse distribution of partons,
enormous efforts have been made to measure both hard exclusive processes and semi-inclusive processes.

The question arises: What can we learn from such measurements?
The generic answer to this question is that we will get a deeper insight into the nucleon as a composite system of partonic degrees of freedom.
Unfortunately, Quantum Chromodynamics (QCD) remains an unsolved theory and the phenomenological approach is based on factorization theorems, the  parametrization of non-perturbative quantities, and fitting is finally used to access them. It is clear that for the interpretation of such fitting results one needs a model based framework. Early attempts to interpret or predict non-perturbative quantities within pure constituent quark degrees of freedom do not only imply the `nucleon spin puzzle', however, were also not fully successful even in the case of unpolarized PDFs, cf.~Refs.~\cite{Gluck:1977ah} and \cite{Gluck:1994uf}. In this circumstance it should be useful to have a fresh look, and parameterize partonic quantities and observables in a more universal way, which offer the possibility to have a deeper understanding of both the reliability of QCD tools, one is using, and the hadronic structure.

Numerous model studies for uPDFs and GPDs were performed in various quark models, e.g., such as the MIT bag model \cite{Chodos:1974je}, chiral quark soliton model \cite{Diakonov:1987ty}, and various constituent quark models.
Both uPDFs and GPDs are  embedded in a more unifying object, which is defined as the expectation value of a two-parton correlation function, e.g., so-called ``Wigner" or ``phase-space" distribution \cite{Belitsky:2003nz}.  Model studies are utilized to demonstrate that GPDs and uPDFs inherit some common aspects from the embedding object, e.g., the transverse momentum ${\bf k}_\perp$ and momentum transfer square  $t$  dependence might be indirectly linked and the role of quark orbital angular momentum plays a complementary role. Alternatively, in the Hamilton approach to QCD both of them probe different aspects of the nucleon wave function.   An interesting possible attempt for it is to employ light-front wave functions (LFWFs), where it seems to be a rather straightforward task to build  (unintegrated) PDFs, however, it is more intricate for GPDs.
In a series of papers, e.g., Refs.~\cite{BerTer76,BerTer77a,AznBagTer82,Schlumpf:1992ce,Sch92,BolJakKroBerSte95,Diehl:1998kh,ChoJiKis01,TibMil01,
TibMil01a,BofPasTra02,AhmHonLiuTan06,Mukherjee:2002gb,JiMisRad06}, wave function models have been employed to provide information on non-perturbative
quantities.
Moreover, such studies are only a subset of investigations, which are also performed in other frameworks. One might wonder that some uPDF studies, done in different model frameworks, provide rather equivalent results \cite{Avaetal09}. To reveal the reason for such findings, a classification scheme for models is needed.

Unfortunately, since the underlying Poincar{\'e} symmetry is not manifestly implemented,
see, e.g., Ref.~\cite{LeuSte77,Kei93,Jau98,KruTro09}, GPD modeling in terms of LFWFs is not straightforward and
might yield inconsistent results \cite{BofPas07}.
The GPDs are governed by both polynomiality, arising from Poincar{\'e} invariance, and positivity constraints in the so--called outer region in which the modulus of the momentum fraction is larger than the skewness parameter. In this region a GPD may be considered as an overlap of LFWFs which can be interpreted as the exchange of a quark in the $s$-channel, while in the central region we can view it as an exchange of a mesonic like parton state.  Using GPD duality among these two regions, see Ref.~\cite{Kumericki:2008di} and references therein, will open the road for unifying the QCD phenomenologies of hard exclusive and inclusive measurements in terms of phenomenological LFWFs. Working along the line of Ref.~\cite{Hwang:2007tb}, we will follow this idea and choose an effective two-body LFWF as an appropriate tool to parameterize different aspects of the nucleon in a universal manner.

In the following section, we introduce our notation, give the LFWF overlap representation for the eight quark uPDFs that appear in the leading--power description of semi-inclusive deep inelastic scattering. We introduce a classification scheme for uPDFs models and illustrate then that some known results for uPDFs can be represented in terms of one effective LFWF, considered to be unspecified. Therefore, our model represents a certain class of models.  In Sec.~\ref{Sec-GPDs&uPDFs} we introduce the eight GPDs at twist-two level with {\em zero} longitudinal momentum transfer in the $t$-channel and give an analog classification scheme as for uPDFs models. Utilizing the LFWF overlap representation, we discuss
then the correspondence of uPDFs and these GPDs on general ground.
In Sec.~\ref{Sec-GPDs} we then turn to the non-trivial part, namely, the construction of a whole set of twist-two chiral even and odd GPDs from the parton number conserving overlap of LFWFs in the outer region.  We introduce a ${\bf k_\perp}$--dependent double distribution (DD) as a tool that allows us to restore Lorentz covariance.
We note that in a LFWF longitudinal and transversal momenta are tied to each other by the underlying Lorentz symmetry, which is explicitly given by the  quantization procedure, and present a representation that allows to restore the $t$ dependence from its transverse part ${\bf \Delta_\perp}$. However, there will be still left non-trivial constraints on the longitudinal momentum dependence, for which we provide one solution that allows to build up rather flexible LFWFs. In Sec.~\ref{Sec-phenomenology}
we introduce both power-likely and exponentially ${\bf k}_\perp$-dependent LFWF models and confront them extensively in the scalar diquark sector with phenomenological  and lattice estimate findings. We also shortly comment on LFWF modeling in the axial-vector sector and present a ``pomeron'' inspired model for sea quarks,
from which we extract a negative and sizeable $D$-term contribution \cite{Polyakov:1999gs}. Finally, we give the conclusions. Two appendices are devoted to technicalities for the construction of proper two-body LFWFs and  the overlap representation of GPDs.

\section{Preliminaries: PDFs in terms of LFWFs}

According to Dirac one may set up a Hamiltonian approach to relativistic Quantum field theory in three different forms which are characterized  by the
form of the hypersphere at some given ``time" \cite{Dir49}.
Instead of initializing the fields in the three dimensional space at time $t=0$ one can alternatively take the tangential
plane of the light-cone at light-cone time $\tau\equiv t + z/c=0$.  This so-called front form is perhaps more suited for quantifying the partonic picture in which we think of the proton as a bunch of partons that move nearly on the light-cone, e.g.,
specified by $n^\mu=(1,0,0,-1)$ \cite{DreLevYan69,DreYan69,Wes70,BroDre80,BroPauPin97}.  This leads to the concept of LFWFs $\psi^S_m(X_i,{\bf k}_{\perp
i},s_i)$. They are the probability amplitudes for their
corresponding $n$-parton states $|n, p^+_i, {\bf p}_{\perp
i},s_i\rangle$, which build up the proton state with spin projection $S=\{+1/2 (\Rightarrow),-1/2(\Leftarrow)\}$ on the $z$-axis:
\begin{eqnarray}
\label{Def-ProSta} |P,S\rangle = \sum_n
\int\![dX\, d^2{\bf k}_\perp]_n\, \prod_{j=1}^n \frac{1}{\sqrt{X_j}}\; \psi^S_n(X_i,{\bf
k}_{\perp i},s_i)\, |n, X_i
P^+, X_i {\bf P}_\perp + {\bf k}_{\perp i},s_i\rangle\,,
\end{eqnarray}
where we used a shorthand notation for the $n$-parton phase space:
\begin{equation}
\label{[dXd2k]}
 [dX\, d^2{\bf k}_{\perp}]_n=\prod_{i=1}^n
\frac{dX_i d^2 {\bf k}_{\perp i}}{16\pi^3 } 16\pi^3
\delta\!\left(1-\sum_{i=1}^n X_i\right)
\delta^{(2)}\!\left(\sum_{i=1}^n {\bf k}_{\perp i}\right) \,.
\end{equation}
The $n$-parton states are normalized as following
\begin{eqnarray}
\langle  s_i^\prime,  {\bf p}^\prime_{\perp i},p_i^{\prime +},n
|n, p_i^+,  {\bf p}_{\perp i},s_i\rangle = \prod_{i=1}^n 16 \pi^3 p^+_i \delta(p_i^+ - p_i^{\prime +})\delta^{(2)}({\bf p}_{\perp i}-{\bf p}^\prime_{\perp i}) \delta_{s_i,s_i^\prime}\,.
\end{eqnarray}
A transversely polarized proton state with projection $S=\{+1/2 (\Uparrow),-1/2(\Downarrow)\}$ on the $x$-axis is built from the longitudinal ones by the corresponding linear combination, i.e.,
\begin{eqnarray}
\label{p-trapol}
|P,\Uparrow\rangle = \frac{1}{\sqrt{2}}\left[|P,\Rightarrow\rangle + |P,\Leftarrow\rangle\right], \quad
|P,\Downarrow\rangle = \frac{1}{\sqrt{2}}\left[|P,\Rightarrow\rangle - |P,\Leftarrow\rangle\right].
\end{eqnarray}
The LFWFs depend on the longitudinal momentum
fractions $X_i = k^+_i/P^+$ (the plus component of a four-vector
$V^\mu$ is $V^+=V^0+V^{3} = n\cdot V$), the transverse momenta
${\bf k}_{\perp i}$, and the LF-helicities $s_i$. In principle, they are
determined from the eigenvalue problem,
\begin{eqnarray}
P^- |P,S\rangle  = \frac{M^2}{P^+} |P,S\rangle \,,  \quad
\mbox{with}\quad P^-=P^0-P^3\,,\; P^+=P^0+P^3\,,\;  {\bf
P}_\perp=0\,,
\end{eqnarray}
for the LF-Hamiltonian $P^-$ that is given in terms of quark and gluon creation and annihilation operators.

Let us emphasize that in any Hamiltonian approach the time coordinate, in our case the light-cone time $\tau=t+z/c$, is singled out
and so Poincar{\'e} symmetry is not explicitly implemented. In other words some generators of the Poincar{\'e} transformations depend on
interaction terms and so the behavior of the LFWFs under Poincar{\'e} transformations remains unknown as long as the bound state problem
is unsolved.
For us it is important that both the boost along the $z$-axis and the rotation around the $z$-axis is independent on the
interaction%
\footnote{There are two further kinematical Lorentz transformations, namely, the ``Galilean" boost in $x$- and $y$-directions, which arise from
a combined  boost and rotation transformation.}
\cite{Dir49,KogSop70}
and, hence, each LFWF transforms simultaneously, i.e.,
$
\psi^S_n(X_i,{\bf k}_{\perp i},s_i) \to\psi^S_n(X^\prime_i,{\bf k^\prime}_{\perp i},s_i)\,,
$
where the momentum fractions $X^\prime_i=X_i$ remain invariant and ${\bf k^\prime}_{\perp i}$ arise from the rotation of ${\bf k}_{\perp i}$.
This may be used to label the partonic states with respect to the angular momentum projections $L^z_i$ on the $z$-axis.
The phase of the corresponding LFWF is then given by
\begin{eqnarray}
\label{LFWF-phase}
{\rm arg}\, \psi^S_n(X_i,{\bf k}_{\perp i},s_i) = \pm \exp\left\{\sum_{i=1}^{n-1} L^z_i \varphi_i\right\}\,,
\end{eqnarray}
where $\varphi_i$ is the polar angle, appearing in the parametrization of the vector ${\bf k}_{\perp i}$.
Angular momentum conservation tells us that the longitudinal
nucleon spin is decomposed as following \cite{Brodsky:2000ii}
\begin{eqnarray}
\label{angular momentum sum rule}
S= \sum_{i=1}^{n-1} L^z_i +\sum_{i=1}^n s_i\,.
\end{eqnarray}  Obviously, if we boost the proton to the rest frame
these partonic quantum numbers do not alter  and so they are appropriate for the discussion of the nucleon ``spin puzzle".
Let us also add that it might be not so obvious how a LFWF behaves under discrete parity and time-reversal transformations.

However, in practice
the QCD dynamics is not well understood and it remains very
challenging to develop this concept to a stage at which it can be
used for quantitative evaluations of physical observables or
parton distributions \cite{BroPauPin97}. Nevertheless, it is instructive to represent partonic distributions and form factors  by the overlap of LFWFs.
Certainly, since the full restoration of the underlying Lorentz symmetry depends on the interaction, a LFWF model consideration might induces problems for the evaluation of Lorentz covariant quantities.  Let us have in this section a closer look to uPDFs, which are intensively studied in terms of various quark models. Thereby, we restrict ourselves to the eight leading--power uPDFs for which the potential problem of Lorentz symmetry restoration does not show up.
\\

\subsection{LFWF overlap representations for uPDFs}
\label{Sect-uPDF-overlap}

A Field theoretical definition of a uPDF depends on the choice of both the reference frame and the gauge link  \cite{Ralston:1979ys,Collins:1981uw,BelJiYua02}. We will here not enter a discussion about the gauge link, which has a transversal part that induces a nontrivial phase \cite{BelJiYua02}. We generically define a uPDF as
\begin{eqnarray}
\label{uPDF-gen}
{\bf \Phi}(x,{\bf k}_\perp)= \int\!\!\!\int\! \frac{d^2{\bf y}_\perp}{(2\pi)^2}\, e^{-i {\bf y}_\perp \cdot {\bf k}_\perp}
\int\frac{d y^-}{8\pi}\;e^{ix P^+y^-/2}\;
\langle P| {\bf \hat{\phi}}(0) \{\mbox{gauge link}\}  {\bf \hat{\phi}}(y)\big|_{y^+=0}|P \rangle\,.
\end{eqnarray}
Apart from singularities, caused by the choice of the gauge link path \cite{Collins:1981uw,BelJiYua02}, the operator is well defined for  ${\bf y}_\perp \neq 0$.
To avoid any confusion here, we stress  that the
operator in Eq.~(\ref{uPDF-gen}) might be {\em formally} expanded in a series of operators with well defined geometrical twist\footnote{We use the term twist in the original manner, namely, as a  classification scheme of operators with respect to their irreducible representations of the Lorentz group.} \cite{GeyLazRob99}, however, it is expected that such an expansion has mathematical problems that might be overcome with an appropriate renormalization
prescription.
Integrating over ${\bf k}_\perp$ sets then the operator on the light-cone and establishes the formal definition of a PDF:
\begin{eqnarray}
\label{PDF-gen}
{\bf \Phi}(x,\mu^2)=\int\!\!\!\int\! d{\bf k}_\perp^2\, {\bf \Phi}(x,{\bf k}_\perp)=
\int\frac{d y^-}{8\pi}\;e^{ix P^+y^-/2}\;
\langle P| {\bf \hat{\phi}}(0)  {\bf \hat{\phi}}(y)\big|_{y^+=0,{\bf y}_\perp=0}|P \rangle\,,
\end{eqnarray}
where we adopted the light-cone gauge in which the gauge link is equated to one.
The resulting light-ray operator has now well defined twist and possesses singularities that are removed by a renormalization procedure. Intuitively, we might connect the scale $\mu$ with a cut-off for the ${\bf k}_\perp$ integration. This scale dependence can be evaluated perturbatively and is governed by the
Doskshitzer-Gribov-Lipatov-Altarelli-Parisi (DGLAP) evolution equations \cite{Dok77,GriLip72,AltPar77}.

The quark uPDFs that appear, e.g., in the description of semi-inclusive deep inelastic scattering at leading power in the inverse photon virtuality,
are given by Eq.~(\ref{uPDF-gen}) in terms of three quark operators
$$
{\bf \hat{\phi}}(0)  {\bf \hat{\phi}}(y) \to \hat{\bar{\psi}}(0)\Gamma \hat\psi(y)\quad\mbox{with}\quad \Gamma \in\{\gamma^+,\gamma^+\gamma^5,i\sigma^{j+}\gamma^5\}\,, \quad \sigma^{j+} =\frac{i}{2}\left[ \gamma^j \gamma^+ - \gamma^+  \gamma^j \right],
$$
where the  chiral even ones have parity even or odd.
In the following we adopt the common nomenclator, see e.g., Refs.~\cite{Ralston:1979ys,BorMul97},
\begin{eqnarray}
\label{Phi^{[gamma^+]},Phi^{[gamma^+gamma^5]}}
{\bf \Phi}^{[\gamma^+(1\pm \gamma^5)/2]}(x,{\bf k}_\perp)
&\!\!\! = \!\!\! &
\frac{1}{2}f_1(x,{\bf k}_\perp) \pm \frac{1}{2} S_{\rm L}\, g_1(x,{\bf k}_\perp)
\\&&\!\!\! -
\frac{\epsilon^{jl}_\perp {\bf k}^j_\perp  {\bf S}^l_\perp}{2M} f_{1{\rm T}}^\perp(x,{\bf k}_\perp)
\pm \frac{{\bf k}_\perp \cdot {\bf S}_\perp}{2M} g_{1{\rm T}}^\perp(x,{\bf k}_\perp)\,,
\phantom{\Bigg|}
\nonumber\\
\label{[sigma^{i+}gamma^5]}
{\bf \Phi}^{[i\sigma^{j+}\gamma^5]}(x,{\bf k}_\perp)&\!\!\! = \!\!\!&
{\bf S}_\perp^j h_1(x,{\bf k}_\perp)+ S_{\rm L} \frac{{\bf k}^j_\perp}{M} h_{1{\rm L}}^\perp(x,{\bf k}_\perp)
\\
&&\!\!\!+
\frac{2{\bf k}^j_\perp {\bf k}^l_\perp - {\bf k}^2_\perp  \delta^{jl}}{2 M^2} {\bf S}_\perp^l h_{1{\rm T}}^\perp(x,{\bf k}_\perp) +
\frac{\epsilon^{jl}_\perp {\bf k}^l_\perp}{M} h_{1}^\perp(x,{\bf k}_\perp)\,,
\nonumber
\end{eqnarray}
where $S_{\rm L}$ and ${\bf S}_\perp$ are the longitudinal and transverse component of the polarization vector $S^\mu$, respectively, and the sign of the two-dimensional Levi-Civita tensor is fixed by $\epsilon^{12}_\perp=1$. In the chiral even sector (\ref{Phi^{[gamma^+]},Phi^{[gamma^+gamma^5]}})
the Dirac structure $\gamma^+(1\pm \gamma^5)/2$ projects on quark states with spin $\rightarrow$ ($\leftarrow$) for $+$ ($-$).
In the chiral odd quark sector (\ref{[sigma^{i+}gamma^5]}) the transversity combination
$i\left(\sigma^{1+} \pm i \sigma^{2+}\right)\gamma^5/2$ appears,
for precise terminology see, e.g.,  Ref.~\cite{Jaf96}.  Note  that $( \gamma^+ \pm i
\sigma^{1+} \gamma^5 ) / 2$ projects for $+$ ($-$) on a transversely polarized incoming quark
in $x$ ($-x$) direction. The spin density matrix, having elements
$$\widetilde{\bf \Phi}_{a b}(x,{\bf k}_\perp) \quad\mbox{with}\quad
a=\Lambda^\prime\lambda^\prime,b=\Lambda\lambda \in \{\Rightarrow\rightarrow,\Rightarrow\leftarrow,\Leftarrow\rightarrow,\Leftarrow\leftarrow\}\,,
$$
can be straightforwardly calculated from the definitions (\ref{Phi^{[gamma^+]},Phi^{[gamma^+gamma^5]}},\ref{[sigma^{i+}gamma^5]}). Thereby, a transversely  polarized proton $\Uparrow$, aligned to the $x$-axis, may be represented as superposition (\ref{p-trapol}),
where the transverse polarization vector is then given as $\left(1,i\right)/\sqrt{2}$. Parameterizing the transverse momentum in terms of polar coordinates
$$k^1 \equiv {\bf k}^1_\perp = |{\bf k}_\perp| \cos\varphi\,,\quad k^2 \equiv {\bf k}^2_\perp = |{\bf k}_\perp| \sin\varphi\,, $$
one finds the hermitian $4\times 4$ matrix:
\begin{eqnarray}
\label{tPhi}
\widetilde {\bf \Phi}(x,{\bf k}_\perp) = \left(\!\!
\begin{array}{cccc}
\frac{f_1+g_1}{2} & \frac{|{\bf k}_\perp|e^{i\varphi}}{M}\, \frac{h_{1{\rm L}}^\perp-i h_{1}^\perp}{2} &
\frac{|{\bf k}_\perp| e^{-i\varphi}}{M}\,\frac{g^\perp_{1{\rm T}}+i f_{1{\rm T}}^\perp}{2} & h_1
\\
\frac{|{\bf k}_\perp|e^{-i\varphi}}{M}\, \frac{h_{1{\rm L}}^\perp+i h_{1}^\perp}{2} &  \frac{f_1-g_1}{2} &
 \frac{{\bf k}_\perp^2e^{-i2\varphi}}{2M^2}   h_{1{\rm T}}^\perp &
\frac{-|{\bf k}_\perp|e^{-i\varphi}}{M}\, \frac{g^\perp_{1{\rm T}}-i f_{1{\rm T}}^\perp}{2}
\\
\frac{|{\bf k}_\perp| e^{i\varphi}}{M}\, \frac{g^\perp_{1{\rm T}}-i f_{1{\rm T}}^\perp}{2} &
\frac{{\bf k}_\perp^2 e^{i2\varphi}}{2 M^2}   h_{1{\rm T}}^\perp & \frac{f_1-g_1}{2}
&  \frac{-|{\bf k}_\perp|e^{i\varphi}}{M}\, \frac{h_{1{\rm L}}^\perp+i h_{1}^\perp}{2} \\
h_1 & \frac{-|{\bf k}_\perp| e^{i\varphi}}{M}\,\frac{g^\perp_{1{\rm T}}+i f_{1{\rm T}}^\perp}{2}  &
\frac{-|{\bf k}_\perp|e^{-i\varphi}}{M}\, \frac{h_{1{\rm L}}^\perp-i h_{1}^\perp}{2} &  \frac{f_1+g_1}{2} \\
\end{array}
\!\!
\right)(x,{\bf k}_\perp^2).
\nonumber\\
\end{eqnarray}
Since parity and time-reversal invariance is already implemented, it is not surprising that the spin-density matrix
can be decomposed into four $2\times 2$  matrices, where the (off)diagonal blocks are related to each other.    The trace of the spin-density matrix (\ref{tPhi}) is given by $2 f_1$.  As we will see below from its LFWF overlap representation (\ref{tPhi-LFWF}), it is also a semi-positive definite matrix.
Hence, one can straightforwardly derive positivity bounds \cite{BacBogHenMul00}, e.g., the analog of the Soffer bound for PDFs \cite{Sof94} reads:
\begin{eqnarray}
\label{Soffer-bound-uPDFs}
2|h_1(x,{\bf k}_\perp)| \le  f_1(x,{\bf k}_\perp) + g_1(x,{\bf k}_\perp)\,.
\end{eqnarray}

If we utilize the definitions of the matrix elements ${\bf \Phi}^{[\cdots]}$, cf.~Eq.~(\ref{uPDF-gen}), and  neglect the gauge link, we can
easily write down the LFWF overlap representation for the uPDFs.
For instance, there are {\em two} of them that are related to chiral even twist-two PDFs\footnote{We emphasize that our twist definition differs from the
notion ``twist", often used in the literature without quotation mark,  in which people denote all eight uPDFs  (\ref{Phi^{[gamma^+]},Phi^{[gamma^+gamma^5]}},\ref{[sigma^{i+}gamma^5]}) ``twist-two", partially, not realizing that power and twist counting in semi-inclusive
processes do not match.}, namely, the unpolarized ($f_1$) and longitudinal polarized ($g_1$) ones. They are obtained from Eq.~(\ref{Phi^{[gamma^+]},Phi^{[gamma^+gamma^5]}}) and read as
\begin{eqnarray}
\label{OveRep-f1}  f_1(x,{\bf k}_{\perp}) &\!\!\!=\!\!\!& \sum_n\; \intsum
\left[
\psi^{\ast\, \Rightarrow}_{\rightarrow,n}
\psi^{\Rightarrow}_{\rightarrow,n}+
\psi^{\ast\,\Rightarrow}_{\leftarrow,n}\psi^{\Rightarrow}_{\leftarrow,n}
\right](X_i,{\bf k}_{\perp i},s_i) \,, \\
\label{OveRep-g1}
g_1(x,{\bf k}_{\perp}) &\!\!\!=\!\!\!& \sum_n\; \intsum
\left[\psi^{\ast\,\Rightarrow}_{\rightarrow,n}\psi^{\Rightarrow}_{\rightarrow,n}-
\psi^{\ast\,\Rightarrow}_{\leftarrow,n}\psi^{\Rightarrow}_{\leftarrow,n}
\right](X_i,{\bf k}_{\perp i},s_i)  \,.
\end{eqnarray}
Here, we use for the $(n-1)$-particle phase space integration and spin summation  of the spectator system the shorthand
\begin{eqnarray}
\label{intsum}
\intsum \cdots \equiv \sum_{s_2,\dots, s_n} \int\![dX\, d^2{\bf k}_{\perp}]_n\,
\delta(x-X_{1}) \delta^{(2)}({\bf k}_{\perp}-{\bf k}_{\perp 1}) \cdots
\end{eqnarray}
with the phase space (\ref{[dXd2k]}) and
$\psi^{S}_{s_1,n}(X_i,{\bf k}_{\perp i},s_2,\dots, s_n)= \psi^{S}_{n}(X_i,{\bf k}_{\perp i},s_i)$ is an $n$-parton Fock-state LFWF with the struck quark LF-spin projection $s_1\in\{\rightarrow,\leftarrow\}$ on the $z$-axis,
and the momentum fraction and transverse momentum of the struck parton are denoted as
$$x\equiv X_1\quad \mbox{and} \quad {\bf k}_{\perp}\equiv{\bf k}_{1,\perp}=-\sum_{i=2}^n {\bf k}_{i,\perp},$$ respectively.

As one clearly realizes in Eqs.~(\ref{OveRep-f1}) and (\ref{OveRep-g1}), the unpolarized and polarized uPDFs arise from the sum and difference of diagonal LFWF overlaps in which the struck quark spin is longitudinally aligned and opposite to the proton one, respectively. If the quark spin is pointing in the same direction as the proton spin, the phase of the LFWF (\ref{LFWF-phase}) that is related to the struck quark is given by
$e^{i \bar{L}^z \varphi}$, where angular momentum conservation (\ref{angular momentum sum rule}) tells us that the orbital angular momentum of the struck quark is given by the sums  of spin and orbital angular momentum projections of the spectators, i.e., $\bar{L}^z=-\sum_{i=2}^n s_i-\sum_{i=2}^{n-1} L^z_i$.   In the case that the struck quark spin $\leftarrow$ $[\rightarrow]$ is pointing in the opposite direction of
the proton spin $\Rightarrow$ $[\Leftarrow]$, the phase is given by $e^{i (\bar{L}^z +1) \varphi}$ $[e^{i (\bar{L}^z -1) \varphi}]$. It is obvious that the overall phase $e^{i \bar{L}^z \varphi}$  is not accessible in leading--power uPDFs, and so only the difference of  proton spin and struck quark projections,
\begin{eqnarray}
\label{l^z}
l^z \equiv S -s_1 = L^z_1 - \bar{L}^z \in \{-1,0,1\}\,, \quad S \in \{\Rightarrow,\Leftarrow\}\,, \quad s\equiv s_1 \in \{\rightarrow,\leftarrow\}\,,
\end{eqnarray}
is the relevant partonic quantum number.
Within this definition we say that $f_1$ and $g_1$ contain contributions from both diagonal $l^z =0$ and $l^z =1$  LFWF overlaps.  We can certainly evaluate $f_1$ and $g_1$ for proton spin $\Leftarrow$ rather than $\Rightarrow$, as used in Eqs.~(\ref{OveRep-f1}) and (\ref{OveRep-g1}). Although it is not so obvious how a certain $n$-parton LFWF behaves under parity and time-reversal transformations \cite{BroPauPin97}, we can state that for the LFWF overlaps the following identities must be satisfied:
\begin{eqnarray}
\label{LFWF-parity-1}
\intsum
\psi^{\ast\, \Rightarrow}_{\rightarrow,n}(X_i,{\bf k}_{\perp i},s_i)
\psi^{\Rightarrow}_{\rightarrow,n}(X_i,{\bf k}_{\perp i},s_i) =\;
 \intsum
\psi^{\ast\, \Leftarrow}_{\leftarrow,n}(X_i,{\bf k}_{\perp i},s_i)
\psi^{\Leftarrow}_{\leftarrow,n}(X_i,{\bf k}_{\perp i},s_i)\,,
\\
\label{LFWF-parity-2}
\intsum
\psi^{\ast\, \Rightarrow}_{\leftarrow,n}(X_i,{\bf k}_{\perp i},s_i)
\psi^{\Rightarrow}_{\leftarrow,n}(X_i,{\bf k}_{\perp i},s_i) =\;
\intsum
\psi^{\ast\, \Leftarrow}_{\rightarrow,n}(X_i,{\bf k}_{\perp i},s_i)
\psi^{\Leftarrow}_{\rightarrow,n}(X_i,{\bf k}_{\perp i},s_i)\,.
\end{eqnarray}
Formally, integrating (\ref{OveRep-f1}) and (\ref{OveRep-g1}) over  ${\bf k}_{\perp}$ provides the overlap representation for unpolarized and
polarized twist-two PDFs.

The two twist-tree related chiral even uPDFs, the Sivers function ($f_{1\rm T}^\perp$) \cite{Siv90} and transversally polarized uPDF ($g_{1{\rm T}}^\perp$), arise for a transversally polarized proton (\ref{p-trapol}). We take the polarization along the
$x$-axis and  we find then from Eqs.~(\ref{uPDF-gen},\ref{Phi^{[gamma^+]},Phi^{[gamma^+gamma^5]}}) the LFWF overlap representations:
\begin{eqnarray}
\label{f_{1T}^perp-overlap}
f_{1{\rm T}}^\perp(x,{\bf k}_\perp) &\!\!\!=\!\!\!& \frac{M}{2i k^1}\sum_n\;\intsum
\left[
\psi^{\ast\,\Rightarrow}_{\rightarrow,n} \psi^{\Leftarrow}_{\rightarrow,n}-
\psi^{\ast\,\Leftarrow}_{\leftarrow,n} \psi^{\Rightarrow}_{\leftarrow,n} - \mbox{c.c.}
\right](X_i,{\bf k}_{\perp i},s_i)\,,
\\
\label{g_{1T}^perp-overlap}
g_{1{\rm T}}^\perp(x,{\bf k}_\perp) &\!\!\!=\!\!\!& \frac{M}{2k^1} \sum_n\;\intsum
\left[
\psi^{\ast\,\Rightarrow}_{\rightarrow,n} \psi^{\Leftarrow}_{\rightarrow,n}-
\psi^{\ast\,\Leftarrow}_{\leftarrow,n} \psi^{\Rightarrow}_{\leftarrow,n} + \mbox{c.c.}
\right](X_i,{\bf k}_{\perp i},s_i)\,,
\end{eqnarray}
where we also employed the identities (\ref{LFWF-parity-1},\ref{LFWF-parity-2}) to simplify
the overlaps of the  LFWF superpositions
$$
\psi^{\Uparrow}_{(n)}(\cdots) = \frac{1}{\sqrt{2}} \left[\psi^{\Rightarrow}_{(n)}(\cdots) + \psi^{\Leftarrow}_{(n)}(\cdots) \right].
$$
As one sees both twist-three related uPDFs arise from the differences of a $l^z=0$ with $l^z=1$ and $l^z=0$ with $l^z=-1$ LFWF overlaps. We defined them in such a manner that the naive time reversal odd ($T$-odd) Sivers function and $T$-even transversally polarized uPDF is given by the imaginary and real part of these overlaps, respectively. Note that there exist alternative overlap
representations, in which one might define $f_{1{\rm T}}^\perp$ and $g_{1{\rm T}}^\perp$ as the real and imaginary part, respectively:
\begin{eqnarray}
\label{f_{1T}^perp-overlap-2}
f_{1{\rm T}}^\perp(x,{\bf k}_\perp) &\!\!\!=\!\!\!& \frac{M}{2 k^2}\sum_n\;\intsum
\left[
\psi^{\ast\,\Rightarrow}_{\rightarrow,n} \psi^{\Leftarrow}_{\rightarrow,n}+
\psi^{\ast\,\Leftarrow}_{\leftarrow,n} \psi^{\Rightarrow}_{\leftarrow,n} + \mbox{c.c.}
\right](X_i,{\bf k}_{\perp i},s_i)\,,
\\
\label{g_{1T}^perp-overlap-2}
g_{1{\rm T}}^\perp(x,{\bf k}_\perp) &\!\!\!=\!\!\!& -\frac{M}{2i k^2} \sum_n\;\intsum
\left[
\psi^{\ast\,\Rightarrow}_{\rightarrow,n} \psi^{\Leftarrow}_{\rightarrow,n}+
\psi^{\ast\,\Leftarrow}_{\leftarrow,n} \psi^{\Rightarrow}_{\leftarrow,n} - \mbox{c.c.}
\right](X_i,{\bf k}_{\perp i},s_i)\,.
\end{eqnarray}

To derive the overlap representation in the chiral odd sector, we employ again LFWFs that are labeled by longitudinal spin projections.  By means of Eqs.~(\ref{uPDF-gen},\ref{[sigma^{i+}gamma^5]}) we find then  that  for a longitudinal polarized target the LFWFs for a transversally polarized struck quark, projected on the $x$-axis, are given as superpositions
$$
\psi^{S}_{\uparrow,n} = \frac{1}{\sqrt{2}} \left[
\psi^{S}_{\rightarrow,n} + \psi^{S}_{\leftarrow,n}
\right],\quad
\psi^{S}_{\downarrow,n} = \frac{1}{\sqrt{2}} \left[
\psi^{S}_{\rightarrow,n} - \psi^{S}_{\leftarrow,n}
\right],\quad S\in\{\Rightarrow,\Leftarrow\}\,.
$$
Moreover, the twist-three related uPDF
\begin{eqnarray}
\label{h1L^perp-overlap}
h_{1 {\rm L}}^\perp(x,{\bf k}_\perp) &\!\!\!=\!\!\!&  \frac{M}{2k^1}\sum_n\;\intsum
\left[
\psi^{\ast\Rightarrow}_{\rightarrow,n} \psi^\Rightarrow_{\leftarrow,n} -
\psi^{\ast\Leftarrow}_{\leftarrow,n} \psi^\Leftarrow_{\rightarrow,n}+\mbox{c.c.}
\right](X_i, {\bf k}_{\perp i},s_i)
\end{eqnarray}
is given by the real part of the differences of a $l^z=0$ with $l^z=1$ and $l^z=0$ with $l^z=-1$ LFWF overlaps. Analogously, as in the chiral even sector we might for this $T$-even uPDF also use the equivalent representation
\begin{eqnarray}
\label{h1L^perp-overlap-2}
h_{1 {\rm L}}^\perp(x,{\bf k}_\perp) &\!\!\!=\!\!\!&  \frac{M}{2i k^2}\sum_n\;\intsum
\left[
\psi^{\ast\Rightarrow}_{\rightarrow,n} \psi^\Rightarrow_{\leftarrow,n} +
\psi^{\ast\Leftarrow}_{\leftarrow,n} \psi^\Leftarrow_{\rightarrow,n}-\mbox{c.c.}
\right](X_i, {\bf k}_{\perp i},s_i)\,.
\end{eqnarray}

The other three transversity uPDFs are naturally defined for a transversely polarized proton, where, e.g., the spin points in the $x$ direction, in terms of LFWFs with transverse quark spin projection $s_1\in \{\uparrow,\downarrow\}$
on the $x$-axis
\begin{eqnarray}
\label{OveRep-dq}
\delta q(x,{\bf k}_{\perp})  &\!\!\! \equiv \!\!\! &{\bf \Phi}^{[i\sigma^{1+}\gamma^5]}(x,{\bf k}_\perp)\Big|_{ {\bf S}_\perp= (1,0)}
\nonumber
\\  &\!\!\!=\!\!\!& \sum_n\; \intsum
\left[\psi^{\Uparrow \ast}_{\uparrow,n} \psi^{\Uparrow}_{\uparrow,n}-
\psi^{\Uparrow\ast}_{\downarrow,n}\psi^{\Uparrow}_{\downarrow,n}
\right](X_i,{\bf k}_{\perp i},s_i)
\,.
\end{eqnarray}
Representing these LFWFs as superposition of those with longitudinal spin projections,
$$
\psi^{\Uparrow}_{\uparrow,n} = \frac{1}{2} \left[
\psi^{\Rightarrow}_{\rightarrow,n} + \psi^{\Rightarrow}_{\leftarrow,n} +\psi^{\Leftarrow}_{\rightarrow,n}+\psi^{\Leftarrow}_{\leftarrow,n}
\right],\quad
\psi^{\Uparrow}_{\downarrow,n} = \frac{1}{2} \left[
\psi^{\Rightarrow}_{\rightarrow,n} - \psi^{\Rightarrow}_{\leftarrow,n} +\psi^{\Leftarrow}_{\rightarrow,n}-\psi^{\Leftarrow}_{\leftarrow,n}
\right],
$$
we find that
\begin{eqnarray}
\label{deltaq-overlap1}
\delta q(x,{\bf k}_{\perp}) &\!\!\!=\!\!\!&  h_1(x,{\bf k}_{\perp}) +\frac{(k^1)^2-(k^2)^2}{2M^2} h_{1 {\rm T}}^\perp(x,{\bf k}_{\perp})
+ \frac{k^2}{M} h_1^\perp(x,{\bf k}_{\perp})\,,
\end{eqnarray}
where the three scalar functions $h_1, h_{1 {\rm T}}^\perp,$ and $h_1^\perp$ depend on $x$ and ${\bf k}_{\perp}^2$, i.e., are invariant under rotation around the $z$-axis.
As said the transverse polarization vector is here along the $x$-axis; the result for the $y$-direction follows by the interchange of $k^1 \to k^2$ and $k^2 \to -k^1$.
The twist-two related transversity uPDF
\begin{eqnarray}
\label{OveRep-h1}
h_1(x,{\bf k}_{\perp}) &\!\!\!=\!\!\!&  \sum_n\;\intsum
\psi^{*\,\Rightarrow}_{\rightarrow,n}(X_i,{{\bf k}}_{\perp i},s_i) \psi^\Leftarrow_{\leftarrow,n}(X_i,{{\bf k}}_{\perp i},s_i)
\end{eqnarray}
is given as the overlap of $l^z=0$ LFWFs,
the twist-four related ``pretzelosity" distribution
\begin{eqnarray}
\label{h_1T^perp-overlap}
h_{1{\rm T}}^\perp(x,{\bf k}_{\perp})  =  -
\sum_{n}\;\intsum
\frac{M^2}{2i k^1 k^2}  \left[
\psi^{\ast\,\Rightarrow}_{\leftarrow,n}
\psi^{\Leftarrow}_{\rightarrow,n}
-\mbox{c.c.}
\right](X_i, {{\bf k}}_{\perp i},s_i)
\end{eqnarray}
is the only leading--power quark uPDF that arises from off-diagonal $|l^z|=1$ overlaps, and the twist-three related Boer-Mulders function
\begin{eqnarray}
\label{h_1^perp-overlap}
h_1^\perp(x,{\bf k}_{\perp}) =  -\frac{M}{2 i k^1}\sum_{n}\; \intsum
\left[
\psi^{\ast\,\Rightarrow}_{\rightarrow,n}\psi^{\Rightarrow}_{\leftarrow,n}
-
\psi^{\ast\,\Leftarrow}_{\leftarrow,n} \psi^{\Leftarrow}_{\rightarrow,n} - \mbox{c.c.}
\right](X_i,{\bf k}_{\perp i},s_i)\,,
\end{eqnarray}
which is $T$-odd \cite{BorMul97}, is the imaginary part of  differences of $l^z=0$ with $l^z=1$ and $l^z=0$ with $l^z=-1$ LFWF overlaps.  Thereby, we employed as above in Eqs.~(\ref{LFWF-parity-1},\ref{LFWF-parity-2}) a definite behavior under parity transformation and so the contributions
$$ \intsum
\left[\psi^{\ast\,\Rightarrow}_{\rightarrow,n}(\cdots) \psi^{\Leftarrow}_{\leftarrow,n}(\cdots)-
\psi^{\ast\,\Leftarrow }_{\leftarrow,n}({\cdots})\psi^{\Rightarrow}_{\rightarrow,n}(\cdots)\right]$$
were neglected. These identities guarantees that the transversity (\ref{OveRep-h1}) has a second representation in which both the target and struck quark spin projections are flipped. As in the case of the Sivers function (\ref{f_{1T}^perp-overlap},\ref{f_{1T}^perp-overlap-2}), we might also define the Boer-Mulders function as the real part of the sum of the $l^z=0$  with $|l^z|=1$ LFWF overlaps,
\begin{eqnarray}
\label{h_1^perp-overlap-2}
h_1^\perp(x,{\bf k}_{\perp}) = \frac{M}{2 k^2}\sum_{n}\; \intsum
\left[
\psi^{\ast\,\Rightarrow}_{\rightarrow,n}\psi^{\Rightarrow}_{\leftarrow,n}
+
\psi^{\ast\,\Leftarrow}_{\leftarrow,n} \psi^{\Leftarrow}_{\rightarrow,n} + \mbox{c.c.}
\right](X_i,{\bf k}_{\perp i},s_i)\,.
\end{eqnarray}
Finally, we add that also the ``pretzelosity" distribution (\ref{h_1T^perp-overlap}) has an equivalent representation as real part of off-diagonal $|l^z|=1$ LFWF overlaps
\begin{eqnarray}
\label{h_1T^perp-overlap-2}
h_{1{\rm T}}^\perp(x,{\bf k}_{\perp})  =
\sum_{n}\;\intsum
\frac{M^2}{(k^1)^2- (k^2)^2}  \left[
\psi^{\ast\,\Rightarrow}_{\leftarrow,n}
\psi^{\Leftarrow}_{\rightarrow,n}
+\mbox{c.c.}
\right](X_i, {{\bf k}}_{\perp i},s_i)\,.
\end{eqnarray}

Let us shortly comment on the interpretation of the transversity functions.
If a transverse direction is selected, the rotation invariance around the $z$-axis is distorted for the $l^z=\pm 1$ states. Hence, in the $\delta q$ uPDF (\ref{deltaq-overlap1}) an off-diagonal overlap of the two $|l^z|=1$ LFWFs  or a chiral-odd quark spin flip contribution enters, yielding a $(k^1)^2-(k^2)^2$  proportional term, which is the ``pretzelosity" distribution (\ref{h_1T^perp-overlap}).  Certainly, this distortion is in the concept of operator product expansion a geometric twist-four ($=\rm {dimension} - {\rm spin}$) effect, which can be also seen from the fact that if one integrates over ${\bf k}_\perp$, even weighted with ${\bf k}_\perp$, the distortion vanishes. Furthermore, we also have in $\delta q$  a distortion that is pointing in the $k^2$-direction and proportional to the Boer-Mulders function (\ref{h_1^perp-overlap}), which is induced by the overlap of  $l^z=0$ and $|l^z|=1$ LFWFs. If naive time reversal invariance holds true, as it should be in a pure quark model, this twist-three related effect is absent.

To have a more compact overlap representation at hand for all leading--power uPDFs, we introduce a LFWF ``spinor" for each $n$-parton  contribution, which we write as
$$
\mbox{\boldmath $\psi$}_{(n)}(X_i,{\bf k}_{\perp i},s_i)=
\left(
  \begin{array}{c}
    \psi^{\Rightarrow}_{\rightarrow,n} \\
    \psi^{\Rightarrow}_{\leftarrow,n} \\
    \psi^{\Leftarrow}_{\rightarrow,n} \\
    \psi^{\Leftarrow}_{\leftarrow,n} \\
  \end{array}
\right)(X_i,{\bf k}_{\perp i},s_i)\,.
$$
The spin-density matrix
\begin{eqnarray}
\label{tPhi-LFWF-0}
\widetilde {\bf \Phi}(x,{\bf k}_\perp) = \sum_n \widetilde {\bf \Phi}_{(n)}(x,{\bf k}_\perp)\,,
\quad
\widetilde {\bf \Phi}_{(n)}(x,{\bf k}_\perp) =
\mbox{\boldmath $\psi$}^\ast_{(n)}(X_i,{\bf k}_{\perp i},s_i) \stackrel{(n-1)}{\otimes}  \mbox{\boldmath $\psi$}_{(n)}(X_i,{\bf k}_{\perp i},s_i)
\end{eqnarray}
is then defined as the infinite sum of overlap spin-density matrices that are given by the Cartesian  product of the $n$-parton LFWF ``spinors",
where the symbol $\stackrel{(n-1)}{\otimes} $ also stays for the $n-1$-parton phase space integration.
The explicit representation reads
\begin{eqnarray}
\label{tPhi-LFWF}
\widetilde {\bf \Phi}_{(n)}(x,{\bf k}_\perp)=
\intsum \left(
          \begin{array}{cccc}
\psi^{\ast\,\Rightarrow}_{\rightarrow,n}\psi^{\Rightarrow}_{\rightarrow,n} & \psi^{\ast\,\Rightarrow}_{\rightarrow,n} \psi^{\Rightarrow}_{\leftarrow,n}
& \psi^{\ast\,\Rightarrow}_{\rightarrow,n}\psi^{\Leftarrow}_{\rightarrow,n} & \psi^{\ast\,\Rightarrow}_{\rightarrow,n}\psi^{\Leftarrow}_{\leftarrow,n}
\\
\psi^{\ast\,\Rightarrow}_{\leftarrow,n} \psi^{\Rightarrow}_{\rightarrow,n}& \psi^{\ast\,\Rightarrow}_{\leftarrow,n}\psi^{\Rightarrow}_{\leftarrow,n}
& \psi^{\ast\,\Rightarrow}_{\leftarrow,n} \psi^{\Leftarrow}_{\rightarrow,n}& \psi^{\ast\,\Rightarrow}_{\leftarrow,n}\psi^{\Leftarrow}_{\leftarrow,n}
\\
\psi^{\ast\,\Leftarrow}_{\rightarrow,n}  \psi^{\Rightarrow}_{\rightarrow,n}& \psi^{\ast\,\Leftarrow}_{\rightarrow,n} \psi^{\Rightarrow}_{\leftarrow,n}
 & \psi^{\ast\,\Leftarrow}_{\rightarrow,n}\psi^{\Leftarrow}_{\rightarrow,n} & \psi^{\ast\,\Leftarrow}_{\rightarrow,n}\psi^{\Leftarrow}_{\leftarrow,n}
 \\
\psi^{\ast\,\Leftarrow}_{\leftarrow,n} \psi^{\Rightarrow}_{\rightarrow,n} & \psi^{\ast\,\Leftarrow}_{\leftarrow,n} \psi^{\Rightarrow}_{\leftarrow,n}  &
 \psi^{\ast\,\Leftarrow}_{\leftarrow,n} \psi^{\Leftarrow}_{\rightarrow,n} & \psi^{\ast\,\Leftarrow}_{\leftarrow,n}\psi^{\Leftarrow}_{\leftarrow,n} \\
          \end{array}
        \right)(X_i,{\bf k}_{\perp i},s_i)\,.
\end{eqnarray}
It is straightforward to see that all LFWF overlap representations for uPDFs, we have discussed, can be easily obtained by equating this equation (\ref{tPhi-LFWF}) with the spin-density matrix (\ref{tPhi}).

\subsection{Classification of uPDF models}
\label{uPDF-classification}

Since the infinite number of LFWFs in the overlap representations (\ref{OveRep-f1},\ref{OveRep-g1},\ref{f_{1T}^perp-overlap},\ref{g_{1T}^perp-overlap},\ref{h1L^perp-overlap},\ref{OveRep-h1}--\ref{h_1^perp-overlap}) remains unknown, it is rather popular to introduce effective degrees of freedom and employ some model assumptions. This has been done in different representations and with various dynamical assumptions. Thereby, surprisingly, linearly and even quadratically algebraic relations among uPDFs  have been found in rather different quark models \cite{JakMulRod97,MeiMetGoe07,PasCazBof08,AvaEfrSchYua08,EfrSchTerZav09}.  Certainly,  one might wonder how general such relations are.
To understand that these findings are partially trivial and in some cases rather surprising, we should introduce a classification scheme for uPDF models.
To do so, we simply  ask for the spectrum of the spin-density matrix
$$
\widetilde{\bf \Phi}(x,{\bf k}_\perp) = \sum_m \widetilde{\bf \Phi}_{(m)}(x,{\bf k}_\perp)\,,
$$
where $m$ is a {\em generic} label for the LFWF overlaps, given in terms  of some $n-1$ ``parton'' phase space integrals (\ref{tPhi-LFWF}).
Knowing the degree of degeneration, we derive the constraints among uPDFs from the  eigenvalues
\begin{eqnarray}
\label{tPhi-eigenvalues}
\widetilde {\bf \Phi}^{\rm e.v.} = \left\{\!\!
                                     \begin{array}{c}
\frac{f_1-h_1}{2}+\frac{{\bf k}_\perp^2}{4 M^2} h_{1{\rm T}}^\perp
-
\frac{1}{2}\sqrt{\left(g_1-h_1- \frac{{\bf k}^2}{2 M^2} h_{1{\rm T}}^\perp\right)^2+
\frac{{\bf k}_\perp^2}{M^2} \left(g_{1\rm T}^\perp+ h_{1\rm L}^\perp\right)^2+\frac{{\bf k}_\perp^2}{M^2} \left(f_{1\rm T}^\perp- h_1^\perp\right)^2}
\\
\frac{f_1-h_1}{2}+\frac{{\bf k}_\perp^2}{4 M^2}  h_{1{\rm T}}^\perp
+
\frac{1}{2} \sqrt{\left(g_1-h_1- \frac{{\bf k}^2}{2 M^2} h_{1{\rm T}}^\perp\right)^2+
\frac{{\bf k}_\perp^2}{M^2} \left(g_{1\rm T}^\perp+ h_{1\rm L}^\perp\right)^2+\frac{{\bf k}_\perp^2}{M^2} \left(f_{1\rm T}^\perp- h_1^\perp\right)^2}
\\
\frac{f_1+h_1}{2}-\frac{{\bf k}_\perp^2}{4M^2} h^\perp_{1{\rm T}}
-
\frac{1}{2}\sqrt{\left(g_1+h_1+ \frac{{\bf k}^2}{2 M^2} h_{1{\rm T}}^\perp\right)^2+
\frac{{\bf k}_\perp^2}{M^2} \left(g_{1\rm T}^\perp - h_{1\rm L}^\perp\right)^2+\frac{{\bf k}_\perp^2}{M^2} \left(f_{1\rm T}^\perp+h_1^\perp\right)^2}
\\
\frac{f_1+h_1}{2}-\frac{{\bf k}_\perp^2}{4 M^2} h_{1{\rm T}}^\perp
+
\frac{1}{2}\sqrt{\left(g_1+h_1+\frac{{\bf k}^2}{2  M^2} h_{1{\rm T}}^\perp\right)^2+
\frac{{\bf k}_\perp^2}{ M^2} \left(g_{1\rm T}^\perp - h_{1\rm L}^\perp\right)^2+\frac{{\bf k}_\perp^2}{ M^2} \left(f_{1\rm T}^\perp + h_1^\perp\right)^2} \\
                                     \end{array}
                                   \right\}.
\nonumber\\
\end{eqnarray}
of the spin-density matrix (\ref{tPhi}).
Note that as a consequence of parity invariance the upper two eigenvalues are related with the two lower ones by the substitution $h_{\dots}^{\dots} \to - h_{\dots}^{\dots} $.
Models in which the spin-density matrix possesses degenerate triplet and duplet
states are called ``spherical" \footnote{This notion has been introduced without quotation marks in Ref.~\cite{LorPas11}, where the authors wrote the spin-density matrix as a SO(3,1) $\simeq$  SU(2) $\otimes$ SU(2) representation and suggested that the degeneration of the spin-density matrix is connected with a SO(3) symmetry in three dimensional space.
} and ``axial-symmetric", respectively. We also take the rank of the spin-density matrix, i.e., the number of zero modes
is given by $4-{\rm rank}\, \widetilde{\bf \Phi}$,
as a  further model characteristic.
On the other hand, knowing that such model relations are satisfied in a specific model, one can read off to which
class it belongs and one might give an equivalent representation in terms of effective two-body LFWFs.

\subsubsection{``Spherical" models of rank-one and rank-four}
\label{uPDF-classification-spherical}

The simplest quark models consist of a struck quark and one collective spectator. Hence, we have then only one scalar two-body LFWF and the spin-spin coupling of the struck quark and spectator is fixed. It is clear that all uPDFs (or other non-perturbative quantities) are expressed by one  LFWF overlap and we have seven model dependent constraints among the eight quark uPDFs, entering the spin-density matrix (\ref{tPhi}).  However, among these constraints there are those
which do not depend on the specific choice of the spin-spin coupling. To understand the origin of these constraints, we remark first that a {\em two-body} spin-density matrix
\begin{eqnarray}
\label{bfPhi-rank1}
\widetilde{\bf \Phi}^{\rm rank-1}(x,{\bf k}_\perp) = \widetilde{\bf \Phi}_{(1)}(x,{\bf k}_\perp) = \left(
\psi^{\Rightarrow}_{\rightarrow},  \psi^{\Rightarrow}_{\leftarrow},  \psi^{\Leftarrow}_{\rightarrow},  \psi^{\Leftarrow}_{\leftarrow}
\right)^\ast \otimes
\left(
  \begin{array}{c}
    \psi^{\Rightarrow}_{\rightarrow} \\
    \psi^{\Rightarrow}_{\leftarrow} \\
    \psi^{\Leftarrow}_{\rightarrow} \\
    \psi^{\Leftarrow}_{\leftarrow} \\
  \end{array}
\right)(X_i,{\bf k}_{\perp i})\,
\end{eqnarray}
has rank-one, i.e., it contains three zero modes.

Let us first derive the conditions for having (degenerate) triplet states. From the eigenvalues  (\ref{tPhi-eigenvalues}), given in terms of uPDFs,
it is not hard to realize that for three identical eigenvalues one of the two square roots has to vanish, which implies one set of
three linear constraints:
\begin{eqnarray}
\label{constraint-rank1-1}
&&\!\!\!\!\!\!\!\!g_{1\rm T}^\perp \pm  h_{1\rm L}^\perp =0\,,
\quad
f_{1\rm T}^\perp \mp h_1^\perp =0\,,
\quad
g_1\mp h_1 \mp \frac{{\bf k}_\perp^2}{2 M^2} h_{1{\rm T}}^\perp  =0\,,
\end{eqnarray}
where the upper or lower sign has to be taken consistently.
Furthermore, the quadratic constraint
\begin{eqnarray}
\label{constraint-rank1-2}
&&\!\!\!\!\!\!\!\!\left(h^\perp_{1 \rm L}\right)^2 +\left(h^\perp_{1} \right)^2 + 2 h_1 h^\perp_{1 {\rm T}} = 0
\end{eqnarray}
guarantees that a third degenerate eigenvalue exists and that the forth one is already fixed (since $\sum\widetilde {\bf \Phi}^{\rm e.v.}=2 f_1$).
The three degenerate eigenvalues are given by $(f_1 + g_1 \mp  2h_1)/2 $ and they vanish if the Soffer bound (\ref{Soffer-bound-uPDFs}) for uPDFs is saturated.

A simple example for a rank-one model is the scalar diquark model, e.g., obtained from the Yukawa theory in leading order of perturbation
theory. Since, the two-body LFWF
has even parity, the upper sign in the linear constraints (\ref{constraint-rank1-1}) holds true. Moreover, the two $T$-odd functions are identically zero and, since the spin-density matrix has rank-one, the Soffer bound is automatically saturated, i.e., altogether we have
\begin{eqnarray}
\label{constraint-scadiquark-1}
&&\!\!\!\!\!\!\!\!g_{1\rm T}^\perp +  h_{1\rm L}^\perp =0\,,
\quad
g_1 - h_1 - \frac{{\bf k}_\perp^2}{2 M^2} h_{1{\rm T}}^\perp  =0\,,
\quad
\left(h^\perp_{1 \rm L}\right)^2 + 2 h_1 h^\perp_{1 {\rm T}} = 0\,,
\quad h^\perp_{1} = f^\perp_{1}=0\,,
\\
\label{constraint-scadiquark-2}
&&\!\!\!\!\!\!\!\!2h_1 = f_1+g_1\quad \mbox{or equivalently}\quad   f_1 - h_1 + \frac{{\bf k}_\perp^2}{2 M^2} h_{1{\rm T}}^\perp  =0\,,
\end{eqnarray}
where the last equation simply follows from combining linear constraints. As we realize now the transversity related twist-two
and twist-tree  uPDFs,
$$h_1(x,{\bf k}_\perp)=
\left[
\psi^{\ast\Rightarrow}_{\rightarrow,n}\psi^\Leftarrow_{\leftarrow,n} +\mbox{c.c.}
\right](x,{\bf k}_\perp)\,, \qquad
h_{1\rm L}^\perp(x,{\bf k}_\perp)=\frac{M}{k^1}
\left[
\psi^{\ast\Rightarrow}_{\rightarrow,n} \psi^\Rightarrow_{\leftarrow,n} +\mbox{c.c.}
\right](x,{\bf k}_\perp)\,,$$
may be considered as independent functions and the four remaining ones can be uniquely obtained from the constraints (\ref{constraint-scadiquark-1})
and (\ref{constraint-scadiquark-2}). The ratio of these both functions is just determined by the ratio of the $l^z=0$ and $l^z=1$ LFWFs. Specifying it,
means that all uPDFs are fixed in terms of one two-body LFWF.
Let us add that for a non-trivial gauge-link, expanded to leading order or treated in the eikonal approximation,
the constraint for the $T$-odd functions $h_1^\perp=f_{1\rm T}^\perp$ holds true, too, however, the quadratic equation (\ref{constraint-rank1-2}) is only satisfied if
these $T$-odd functions are neglected \cite{MeiMetGoe07,BroHwaSch02,BoeBroHwa02,GamGolOga03,BacConRad08}.

A more non-trivial example for a rank-one model is provided by the $2u/3-d/3$ flavor sector of the three quark LFWF model in Ref.~\cite{PasCazBof08}.
Since this model is based on SU(4) spin-flavor symmetry, the struck quark couples to a scalar diquark. We conclude that in this sector the three quark model can be mapped to an effective two-body  scalar diquark model. Indeed, by inspection of the formulae set (51-56) in Ref.~\cite{PasCazBof08}, we can confirm this conclusion.

A ``spherical" model, which has a spin-density matrix of rank-four, can in the simplest case only arise from the LFWF overlaps of four orthogonal states. Hence, we have the {\em necessary} condition:
$$
\widetilde{\bf \Phi}^{\rm ``spheric"}(x,{\bf k}_\perp) =\sum_{m=1}^{m_{\rm min.}}  \widetilde{\bf \Phi}_{(m)}(x,{\bf k}_\perp)\,,
\quad\mbox{with}\quad m_{\rm min.} \ge 4\,.
$$
A simple, however, highly non-trivial example is the $u/3+4 d/3$ flavor sector of the three quark LFWF model \cite{PasCazBof08}, in which the spin-density matrix is proportional to the identity matrix times a scalar LFWF overlap.  According to SU(4) spin-flavor symmetry in this sector the struck quark couples to an axial-vector and a scalar diquark.
It is rather surprising that if one sums over all angular momentum states, interference effects cancel all off-diagonal spin-spin correlations and one finds a spin-density matrix that is proportional to the identity matrix. For $u$- and $d$-quark the spin-density matrix is then the sum of a rank-one and a diagonal matrix,
\begin{eqnarray}
\label{model-sperical-SU(4)-1}
\widetilde{\bf \Phi}^{u}(x,{\bf k}_\perp) &\!\!\! \stackrel{\rm sph\&SU(4)}{=} \!\!\!& \frac{1}{3} {\bf 1\!\!\!1}_{4\times 4}\,  f_1^{u/3+4d/3}(x,{\bf k}_\perp)+  \frac{4}{3}\widetilde{\bf \Phi}^{2u/3-d/3}(x,{\bf k}_\perp)\,,
\quad {\rm rank}\,\widetilde{\bf \Phi}^{2u/3-d/3}=1\,,
\\
\label{model-sperical-SU(4)-2}
\widetilde{\bf \Phi}^{d}(x,{\bf k}_\perp) &\!\!\! \stackrel{\rm sph\&SU(4)}{=} \!\!\!&   \frac{2}{3} {\bf 1\!\!\!1}_{4\times 4}\,  f_1^{u/3+4d/3}(x,{\bf k}_\perp)-\frac{1}{3}\widetilde{\bf \Phi}^{2u/3-d/3}(x,{\bf k}_\perp) \,,
\end{eqnarray}
where ${\bf 1\!\!\!1}_{4\times 4}$ is the four dimensional identity matrix.
It is obvious that the spin-density matrices (\ref{model-sperical-SU(4)-1},\ref{model-sperical-SU(4)-2}) together with $\widetilde{\bf \Phi}^{2u/3-d/3}$
can be simultaneously diagonalized  and that for the former two
a degenerated triplet state with non-vanishing eigenvalue exist, which are given by the entry of the corresponding
diagonal matrix. Consequently, we have a ``spherical"
model and the constraints (\ref{constraint-scadiquark-1}) for the scalar diquark model hold true, where the Soffer bound for uPDFs is now unsaturated.
We add that the axial-vector diquark model in the version of Ref.~\cite{JakMulRod97}, in which the sum over the diquark polarization vectors provides
$-g_{\mu\nu}+ P_\mu P_\nu/M_A^2$ with the diquark mass $M_A^2$ and proton momentum $P_\mu$, yields an equivalent result, of course, the effective scalar LFWF overlap is different.

Also the bag \cite{AvaEfrSchYua10}, the chiral quark soliton \cite{LorPasVan11}, and the covariant parton \cite{EfrSchTerZav09} model are ``spherical". So far we did not check if all these models have the same  relative $l^z=1$ to $l^z=0$ LFWF coupling and the same flavor-spin content. Nevertheless, we can make the conjecture that this class of models can be represented by a parity even
 ``spinor" times an effective two-body LFWF,
\begin{eqnarray}
\label{LFWF-spinor}
\mbox{\boldmath $\psi$}^{\rm sca}(x,{\bf k}_\perp) = \frac{1}{M} \left(\!\!
               \begin{array}{c}
                 M \\
                -g^{10} |{\bf k}_\perp| e^{i \varphi} \\
                 g^{10} |{\bf k}_\perp| e^{-i \varphi}  \\
                 M \\
               \end{array}\!\!
             \right){\phi}^{\rm sca}(x,{\bf k}_\perp)\,,
\end{eqnarray}
and a  positive definite   uPDF $f_1^{q}$.
Here, $g^{10}$ is the relative strength of the $l^z=1$  LFWF compared to the $l^z=0$ one. This coupling might depend on ${\bf k}_\perp$ and $x$, however, the underlying Lorentz symmetry tells us that this function cannot be ambiguously chosen. We will come back to this point below in Sec.~\ref{uPDF-scaDiq} and Sec.~\ref{sect-LI}.
The spin-density matrix can then be written as
\begin{eqnarray}
\label{spin-matrix-spherical}
\widetilde{\bf \Phi}^{q}(x,{\bf k}_\perp) \stackrel{\rm sph}{=} \frac{1}{2} {\bf 1\!\!\!1}_{4\times 4} f_1^{q}(x,{\bf k}_\perp)+ g^{\rm sca}_q \left[\left(
\mbox{\boldmath $\psi$}^{\ast\ \rm sca}\otimes \mbox{\boldmath $\psi$}^{\rm sca}\right) -  \frac{1}{4} {\bf 1\!\!\!1}_{4\times 4}
{\rm Tr}\, (\mbox{\boldmath $\psi$}^{\ast\ \rm sca}\otimes \mbox{\boldmath $\psi$}^{\rm sca})\right](x,{\bf k}_\perp)\,,
\end{eqnarray}
where $g^{\rm sca}_q$ is the ``coupling" of the quark $q$ to the scalar diquark sector. Note that $g^{\rm sca}_q$ might be negative.
Since the non-trivial part of the matrix is written as a rank-zero matrix and $f_1^{q}$ is a positive function,
the spin-density matrix (\ref{spin-matrix-spherical}) of our  "spherical" model is always positive definite.
We also emphasize that an identity proportional part of the spin-density matrix arises from the overlap contributions of  four LFWF ``spinors" that belong to (rather) different states. To provide a toy example, we might for instance choose the following complete set of ``spinors":
\begin{eqnarray}
\label{LFWF-spinors-Identity}
\frac{1}{2}\left(
\begin{array}{c}
1 \\
-e^{i \varphi} \\
e^{-i \varphi}  \\
1 \\
\end{array}\!\!
\right),\quad
\frac{1}{2}\left(
\begin{array}{c}
1 \\
e^{i \varphi} \\
e^{-i \varphi}  \\
-1 \\
\end{array}\!\!
\right),\quad
\frac{e^{-i \varphi}}{\sqrt{2}}\left(
\begin{array}{c}
0 \\
e^{i \varphi} \\
0  \\
1 \\
\end{array}\!\!
\right),\quad
\frac{e^{i \varphi}}{\sqrt{2}}\left(
\begin{array}{c}
1 \\
0 \\
-e^{-i \varphi}  \\
0 \\
\end{array}\!\!
\right).
\end{eqnarray}
Here, the first ``spinor" mimics a scalar diquark coupling, cf.~Eq.~(\ref{LFWF-spinor}), the second one the coupling to a pseudo-scalar diquark, and the remaining two might be considered as coupling to an axial-vector diquark that has spin projection $+1$ and $-1$, respectively. Of course, the choice of such a basis is not unique rather we can utilize any unitary transformation to obtain another ``spinor" basis. Hence, from this point of view it is not surprising that quark models with different orbital angular momentum content, e.g., with \cite{PasCazBof08} and without \cite{JakMulRod97} $D$-waves, provide equivalent uPDF models.

\subsubsection{Two kinds of ``axial-symmetrical"  models of rank-two}

Diquark models that are based on two orthogonal states are build from the sum,
$$
\widetilde{\bf \Phi}^{\rm rank-2}(x,{\bf k}_\perp) = \sum_{m=1}^2  \widetilde{\bf \Phi}_{(m)}^{\rm rank-1}(x,{\bf k}_\perp)\,,
$$
of two rank-one spin-density matrices. We suppose that the LFWF ``spinors" in these two contributions are linearly independent and so
the spin-density matrix has rank two and possesses two zero modes. There are now two plus four possibilities to distribute these two zero eigenvalues among the four ones (\ref{tPhi-eigenvalues}). More generally, we might ask for  ``axial-symmetric" models, possessing a degenerated duplet state with non-negative eigenvalue.

The first kind of a ``axial-symmetrical"  model is realized if one set of the linear relations (\ref{constraint-rank1-1}) holds true, i.e., the upper or lower two eigenvalues (\ref{tPhi-eigenvalues}) are degenerated. Consequently, we have the degenerated eigenvalues $(f_1 + g_1 \mp  2h_1)/2 $, which collapses to zero for a rank-two model, i.e., the Soffer bound (\ref{Soffer-bound-uPDFs}) is saturated. An example of such a model is the scalar diquark model that contains a gauge link that implies in a perturbative leading order expansion non-vanishing Boer-Mulders and Sivers distributions that are equal \cite{MeiMetGoe07,BacConRad08}. This equality holds also true if the gauge link contributions are summed by means of the eikonal approximation \cite{Gam11pri}.   More generally, if we extend a given ``spherical" model in the $T$-even sector by two $T$-odd functions which are equal to each other (or differ by a sign if the third and forth
eigenvalues (\ref{tPhi-eigenvalues}) are degenerated), we will end up with an ``axial-symmetrical" model in which the linear relations (\ref{constraint-rank1-1}) hold true and the  Soffer bound (\ref{Soffer-bound-uPDFs}) is unsaturated.  It is known that  the condition $f_{1\rm T}^\perp= h_1^\perp$ is violated for a gauged bag model \cite{Yua03a,CheDAlKocMur06,CouScoVen08,CouScoVen09,Cou10}. One might wonder if a gauged ``spherical" axial-vector diquark  or a three quark LFWF model will become an ``axial-symmetrical" model. To our best knowledge such model calculations are not performed yet.

The other four possibilities for the realization of a rank-two model are that one eigenvalue out of the first two eigenvalues (\ref{tPhi-eigenvalues}) coincides with one of the last two eigenvalues (\ref{tPhi-eigenvalues}). We might write this condition for an ``axial-symmetrical" model of the second kind in the following form:
\begin{eqnarray}
\label{constraint-axialsym}
&&g_1^2 - h_1^2 +
 \frac{{\bf k}^2_\perp}{M^2} \left[
 \left(g^\perp_{1\rm T}\right)^2 +\left(h^\perp_{1\rm L}\right)^2 + \left(f^\perp_{1\rm T}\right)^2 + \left(h^\perp_{1}\right)^2 -  \frac{{\bf k}^2_\perp}{4 M^2} \left(h^\perp_{1\rm T}\right)^2 + 3  h_1 h^\perp_{1\rm T}
 \right]
=
\nonumber\\
&&\pm \sqrt{\left(g_1-h_1- \frac{{\bf k}^2}{2 M^2} h_{1{\rm T}}^\perp\right)^2+
\frac{{\bf k}_\perp^2}{M^2} \left(g_{1\rm T}^\perp + h_{1\rm L}^\perp\right)^2+\frac{{\bf k}_\perp^2}{M^2} \left(f_{1\rm T}^\perp-h_1^\perp\right)^2}
\\
&&\times
\sqrt{\left(g_1+h_1+ \frac{{\bf k}^2}{2 M^2} h_{1{\rm T}}^\perp\right)^2+
\frac{{\bf k}_\perp^2}{M^2} \left(g_{1\rm T}^\perp - h_{1\rm L}^\perp\right)^2+\frac{{\bf k}_\perp^2}{M^2} \left(f_{1\rm T}^\perp+h_1^\perp\right)^2}\,.
\nonumber
\end{eqnarray}
In the case that the two eigenvalues vanish both quadratic relations
\begin{eqnarray}
\label{constraint-rank2-1a}
(f_1+g_1- 2 h_1) \left(f_1-g_1+\frac{{\bf k}^2_\perp}{M^2} h_{1 \rm T}^\perp\right)-
\frac{{\bf k}^2_\perp}{M^2}  \left[
\left(g_{1\rm T}^\perp + h_{1\rm L}^\perp\right)^2+ \left(f_{1\rm T}^\perp - h_1^\perp\right)^2
\right] =0\,,
\\
\label{constraint-rank2-1b}
(f_1+g_1 + 2 h_1) \left(f_1-g_1 -\frac{{\bf k}^2_\perp}{M^2} h_{1 \rm T}^\perp\right)-
\frac{{\bf k}^2_\perp}{M^2}  \left[
\left(g_{1\rm T}^\perp - h_{1\rm L}^\perp\right)^2 + \left(f_{1\rm T}^\perp + h_1^\perp\right)^2
\right] =0
\end{eqnarray}
must be satisfied. Combining them, we might also write the two constraints as
\begin{eqnarray}
\label{constraint-rank2-2a}
f_1^2 - g_1^2 -
 \frac{{\bf k}^2_\perp}{M^2} \left[
 \left(g^\perp_{1\rm T}\right)^2 +\left(h^\perp_{1\rm L}\right)^2 +
 \left(f^\perp_{1\rm T}\right)^2 + \left(h^\perp_{1}\right)^2 -  \frac{{\bf k}^2_\perp}{4 M^2} \left(h^\perp_{1\rm T}\right)^2 + 2  h_1 h^\perp_{1\rm T}
 \right]=0\,,
 \\
 \label{constraint-rank2-2b}
(f_1-g_1)h_1 -\frac{{\bf k}^2_\perp}{2 M^2} \left[(f_1+g_1) h^\perp_{1\rm T}-2(g^\perp_{1\rm T} h^\perp_{1\rm L} - f^\perp_{1\rm T} h^\perp_{1}) \right] =0\,.
\end{eqnarray}

An example of such a rank-two model is the diquark model with axial-vector coupling, where only the transverse polarization
of the diquark is taken into account \cite{BacConRad08}. Such a model will break in general
the underlying Lorentz symmetry and, therefore, it is  not applicable for GPD modeling, however, it might be still used for uPDF modeling.
In this model the linear equations (\ref{constraint-rank1-1}) are not satisfied, however, the quadratic ones (\ref{constraint-rank2-2a},\ref{constraint-rank2-2b}) hold true.  Obviously, if we replace the axial-vector coupling by a vector coupling, we will find
rather analogous results. In the case of a gauge field theory, gauge invariance ensures then that the result also respects Lorentz symmetry.  The so-called quark-target model \cite{MeiMetGoe07}, calculated to leading order, belongs to this rank-two model class\footnote{Of course, the proton mass has now to be identified with the target quark mass. Alternatively, if one sets the quark mass to zero and keep the proton mass, the model constraints (\ref{constraint-rank2-2a},\ref{constraint-rank2-2b}) for a rank-two model are still satisfied.
Thereby, the non-vanishing uPDFs are $f_1=g_1$ and $h_1$.}. We are not aware of an ``axial-symmetrical" model of the second kind that has a rank four
spin-density matrix.

\subsubsection{Models of rank-three}

Diquark models which are based on three orthogonal states have the spin-density matrix
$$
\widetilde{\bf \Phi}^{\rm rank-3}(x,{\bf k}_\perp) = \sum_{m=1}^3  \widetilde{\bf \Phi}_{(m)}^{\rm rank-1}(x,{\bf k}_\perp)
$$
and they possess at least one zero mode.  Hence, the determinant of the spin density matrix, given by the product of eigenvalues, vanish and we find two possibilities two satisfy the nonlinear
constraints, which we already wrote down in Eqs.~(\ref{constraint-rank2-1a}) and (\ref{constraint-rank2-1b}).
Thereby, if the spin-density matrix has truly rank-three, only one root of one of these quadratic relations vanishes.

An example of such a model is an axial-vector diquark model of Ref.~\cite{BacConRad08} where the diquark posses two transversal  and one longitudinal polarization, however, the polarization tensor differs from that of the ``spherical" axial-vector diquark model \cite{JakMulRod97}.
Even if the time-like polarization is taken into account, i.e., the polarization tensor is $-g_{\mu \nu}$, one still has a rank-three model and the same constraint
$$
(f_1+g_1+2 h_1) \left(f_1-g_1-\frac{{\bf k}^2_\perp}{M^2} h_{1 \rm T}^\perp\right)-
\frac{{\bf k}^2_\perp}{M^2}
\left(g_{1\rm T}^\perp - h_{1\rm L}^\perp\right)^2 =0
$$
is satisfied.

\subsection{``Spherical" diquark  models in terms of effective  LFWFs}
\label{uPDF-scaDiq}

In the following we will utilize effective two-particle LFWFs of an effective  struck quark with mass $m$ and spin projection $s=\{-1/2,1/2\}$, where the collective spectator system has mass $\lambda$.
Formally, we might introduce a Hilbert space and write the proton wave function in terms of these {\em new} degrees of freedom as
\begin{eqnarray}
\label{Def-ProSta-1} |P,S\rangle = \sum_{s=-1/2}^{1/2} \int\! d\lambda^2
\int\!\frac{[dX\, d^2{\bf k}]_2}{\sqrt{X_1 X_2}}\, \Psi^S_h (X_i,{\bf
k}_{\perp i},s_i|\lambda^2)\, |\lambda^2, X_i
P^+, X_i {\bf P}_\perp + {\bf k}_{\perp i},s\rangle\,.
\end{eqnarray}
The normalization condition for the states is taken to be the following
\begin{eqnarray}
\label{Def-ProSta-2}
&&\langle s^\prime, {\bf P}_\perp + {\bf k}^\prime_{\perp i},X^\prime_i P^+, \lambda^{\prime 2} |\lambda^2, X_i P^+, X_i {\bf P}_\perp + {\bf k}_{\perp i},s\rangle
\\
&&\hspace{5cm}=  \prod_{i=1}^2 16 \pi^3 X_i \delta(X^\prime_i-X_i)
\delta^{(2)}({\bf k}^\prime_{\perp i}-{\bf k}_{\perp i}) \delta(\lambda^{\prime 2}-\lambda^2) \delta_{s^\prime s}\,.
\nonumber
\end{eqnarray}

If we restrict ourselves to a scalar--diquark spectator, we have four LFWFs \cite{Hwang:2007tb},
\begin{eqnarray}
\label{Def-LF-WF1} &&\!\!\!\Psi^{\Rightarrow}_{\rightarrow}(X,{\bf k}_\perp|\lambda^2)=\frac{m+X M}{M \sqrt{1-X}}
\overline{\phi}(X,{\bf k}_\perp|\lambda^2),\quad
\Psi^{\Rightarrow}_{\leftarrow}(X,{\bf k}_\perp|\lambda^2) = \frac{-k^1- i k^2}{M\sqrt{1-X}}
\overline{\phi}(X,{\bf k}_\perp|\lambda^2),
\nonumber\\
 \\
\label{Def-LF-WF2} &&\!\!\!\
\Psi^{\Leftarrow}_{\leftarrow}(X,{\bf k}_\perp|\lambda^2) =\frac{m+X M}{M\sqrt{1-X}}
\overline{\phi}(X,{\bf k}_\perp|\lambda^2), \quad
\Psi^{\Leftarrow}_{\rightarrow}(X,{\bf
k}_\perp|\lambda^2) = \frac{k^1- i k^2}{M\sqrt{1-X}} \overline{\phi}(X,{\bf k}_\perp|\lambda^2),
\nonumber\\
\end{eqnarray}
in terms of one effective LFWF $\overline{\phi}(X,{\bf k}_\perp|\lambda^2)$, which we might also write in terms of a LFWF ``spinor" (\ref{LFWF-spinor}):
\begin{eqnarray}
\label{LFWF-spinor-model}
\mbox{\boldmath $\psi$}^{\rm sca}(X,{\bf k}_\perp) = \frac{1}{M} \left(\!\!
               \begin{array}{c}
                 m +X M \\
                -|{\bf k}_\perp| e^{i \varphi} \\
                 |{\bf k}_\perp| e^{-i \varphi}  \\
                 m +X M \\
               \end{array}\!\!
             \right) \frac{ \overline{\phi}(X,{\bf k}_\perp|\lambda^2)}{\sqrt{1-X}}\,.
\end{eqnarray}
In our  model (\ref{Def-ProSta-1})  the proton is described as a superposition of struck quark states with aligned and opposite spin projection on the $z$-axis. The former and latter have for a longitudinally polarized proton $(\Rightarrow)$ an orbital angular momentum projection $L^z=0$ and $L^z=1$, respectively, where the phase is $\exp\{i L^z \varphi\}$. Since we have a scalar spectator, we can equate the orbital angular momenta projection $L^z=l^z$ with the spin difference projection (\ref{l^z}), which is as discussed above the relevant quantum number for our partonic quantities.  We emphasize that
the relative normalization in (\ref{Def-LF-WF1}--\ref{Def-LF-WF2}) cannot be changed otherwise we will be not able to satisfy the GPD constraints that arise from Lorentz covariance.

To obtain a spectral representation  with respect to the spectator quark mass one might introduce a $\lambda$ dependent coupling between the struck quark and the spectator system, which has to be independent on the spin in our model,
\begin{eqnarray}
\label{rho(lambda)}
\overline{\phi}(X,{\bf k}_\perp|\lambda^2) = \sqrt{\rho(\lambda)}\, \phi(X,{\bf k}_\perp|\lambda^2),
\end{eqnarray}
where $\rho(\lambda)$  is considered as the spectral mass density.
Since in our model the spin-spin coupling is fixed, it is obvious that all non-perturbative quantities are expressed by the overlap of one effective LFWF, which spectral representation follows from the ansatz (\ref{Def-ProSta-1},\ref{Def-ProSta-2},\ref{rho(lambda)}),
\begin{eqnarray}
\label{def-TMD-overlap}
{\bf \Phi}(x,{\bf k}_\perp)=\int\! d\lambda^2 \rho(\lambda^2) \,
 \frac{\big|\phi(X=x,{\bf k}_\perp|\lambda^2)\big|^2}{1-x}\,.
\end{eqnarray}
Note that such a representation will be essential to recover phenomenological PDF parameterizations, which cannot be obtained from `pure' quark models.
The unintegrated parton density, related to the unpolarized twist-two PDF, read then in the quark spin averaged case (\ref{OveRep-f1}) as
following and the $T$-odd Sivers function $f_{1 {\rm T}}^\perp$ vanishes:
\begin{eqnarray}
\label{f1-DQSM}
f_1(x,{\bf k}_{\perp}) &\!\!\!=\!\!\!&
\frac{(m+ x M)^2 + {\bf k}_\perp^2}{M^2}\, {\bf \Phi}(x,{\bf k}_\perp)\,, \quad f_{1 {\rm T}}^\perp(x,{\bf k}_{\perp}) =0\,.
\end{eqnarray}
The overall normalization of the wave functions (\ref{Def-LF-WF1},\ref{Def-LF-WF2}) is fixed for valence like quarks by the quark number $n$ and for sea-quarks
by the  momentum fraction average $\langle x \rangle$:
\begin{eqnarray}
\label{LFWF-normalization}
\int\!  dx\! \int\!  d^2{\bf
k}_{\perp} \left\{1 \atop x \right\}  f_1(x,{\bf k}_{\perp})=
\left\{{n \atop \langle x \rangle}\right\}\,.
\end{eqnarray}
For the spin difference of longitudinally polarized quarks in an longitudinally ($g_1$) and transversally ($g^\perp_{1{\rm T}}$) polarized hadron
we find a twist-two and twist-three related unintegrated parton density, respectively:
\begin{eqnarray}
\label{g1g1T-DQSM}
g_1(x,{\bf k}_{\perp})  &\!\!\!=\!\!\!& \frac{(m+ x M)^2 - {\bf k}_\perp^2}{M^2}\, {\bf \Phi}(x,{\bf k}_\perp)\,, \quad
g^\perp_{1 {\rm T}}(x,{\bf k}_{\perp}^2) =
2 \left(\frac{m}{M}+ x\right){\bf \Phi}(x,{\bf k}_\perp)\,.
\end{eqnarray}
Analogously, if we rotate the quark spin the  transversity for a transversally ($h_1,h_{1 {\rm T}}^\perp$), longitudinally ($h_{1{\rm L}}^\perp)$, and un-($h_{1}^\perp$)  polarized target is related to twist-two, twist-four, and two twist-three  PDFs:
\begin{eqnarray}
\label{h1hTperp-DQSM}
h_1(x,{\bf k}_{\perp}) &\!\!\!=\!\!\!& \left(\frac{m}{M}+x\right)^2 {\bf \Phi}(x,{\bf k}_\perp)  \,,
\qquad\qquad
h_{1 {\rm T}}^\perp(x,{\bf k}_{\perp})  = -2 {\bf \Phi}(x,{\bf k}_\perp)\,,
\\
\label{h1L-DQSM}
h_{1{\rm L}}^\perp(x,{\bf k}_{\perp}) &\!\!\!=\!\!\!& - 2\left(\frac{m}{M}+ x\right)  {\bf \Phi}(x,{\bf k}_\perp)\,,
\qquad\quad h_{1}^\perp(x,{\bf k}_{\perp}) =0\,.
\phantom{\Bigg|}
\end{eqnarray}

As we have shown in Sec.~\ref{uPDF-classification}, the overlap representation in terms of only one LFWF ``spinor" implies that we have a ``spherical" model of rank-one
and so one quadratic and three linear constraints (\ref{constraint-scadiquark-1},\ref{constraint-scadiquark-2}) hold true for the six $T$-even uPDFs, which one might easily verify from the
explicit expressions (\ref{f1-DQSM},\ref{g1g1T-DQSM},\ref{h1hTperp-DQSM},\ref{h1L-DQSM}). The relative strength of the $l^z=1$ and $l^z=0$ LFWFs is fixed by the ratio
$$
\frac{h_{1{\rm L}}^\perp(x,{\bf k}_{\perp})}{2h_{1}(x,{\bf k}_{\perp})} = \frac{-M}{m+x M}
$$
and, hence, the transversity uPDF (\ref{h1hTperp-DQSM}) might be chosen as independent function. Alternatively, we might consider the unpolarized
uPDF as the independent function.  Indeed, it is trivial to realize that the concept of a (scalar) diquark model has immediately the consequence that all uPDFs, not only twist-two related ones, are expressed by the unpolarized PDF:
\begin{eqnarray}
\label{con-TMD-Dq}
g_1(x,{\bf k}_{\perp}) &\!\!\!=\!\!\!&
\frac{(m+ x M)^2 - {\bf k}_\perp^2}{(m+ x M)^2 +{\bf k}_\perp^2}\, f_1(x,{\bf k}_{\perp})\,,\qquad\;
g_{1 {\rm T}}^\perp(x,{\bf k}_{\perp}) =
\frac{2M (m+ x M)}{(m+ x M)^2 +{\bf k}_\perp^2}\, f_1(x,{\bf k}_{\perp})\,,\qquad
\\
\label{con-TMD-dq}
h_1(x,{\bf k}_{\perp}) &\!\!\!=\!\!\!&
\frac{(m+ x M)^2}{(m+ x M)^2 +{\bf k}_\perp^2}\, f_1(x,{\bf k}_{\perp})\,,\qquad\;
h^\perp_{1 {\rm T}}(x,{\bf k}_{\perp}) = \frac{-2M^2}{(m+ x M)^2 +{\bf k}_\perp^2}\, f_1(x,{\bf k}_{\perp})\,,
\\
\label{con-TMD-dqL}
h^\perp_{1 \rm L}(x,{\bf k}_{\perp}) &\!\!\!=\!\!\!&
\frac{-2M (m+ x M)}{(m+ x M)^2 +{\bf k}_\perp^2}\, f_1(x,{\bf k}_{\perp})\,.
\end{eqnarray}

It is also not surprising that within another spin-spin coupling, e.g., in an axial-vector diquark model of Ref.~\cite{BacConRad08}, the set
of such relations is altering or getting smaller. We also emphasize that our chosen spectral representation  of the LFWF overlap (\ref{def-TMD-overlap})
preserves both the constraints (\ref{constraint-scadiquark-1},\ref{constraint-scadiquark-2}) and the specific model relations (\ref{con-TMD-Dq}--\ref{con-TMD-dqL}). In the case we would have also introduced a spectral representation w.r.t. the struck quark mass, the strength of the spin coupling for the $l^z =0$ LFWF, given by $m+x M$, becomes ``dynamical" in the quantities (\ref{f1-DQSM},\ref{g1g1T-DQSM},\ref{h1hTperp-DQSM},\ref{h1L-DQSM}).  Since the model constraints (\ref{constraint-scadiquark-1},\ref{constraint-scadiquark-2}) are independent on the struck quark mass $m$, they would still be satisfied. On the other hand  the algebraic equations (\ref{con-TMD-Dq}--\ref{con-TMD-dqL}) would not hold anymore, giving an example that interference effects will break uPDF model relations. However,  employing the mean value theorem one would recover Eqs.~(\ref{con-TMD-Dq}--\ref{con-TMD-dqL}) within an effective quark mass that becomes a function of both longitudinal and transversal degrees of freedom $x$ and ${\bf k}_{\perp}$.

We may have also evaluated in our model the remaining other unintegrated parton densities, which are of higher twist in the sense of power counting, and then one may also express the findings in terms of the unpolarized one. We will skip this exercise here, since first some clarification about the classification scheme of TMDs is needed \cite{GoeMetSch05}, %
which is far beyond the scope of the present study.

The above set of formulae (\ref{f1-DQSM},\ref{g1g1T-DQSM},\ref{h1hTperp-DQSM},\ref{h1L-DQSM}) clearly illuminate the role of orbital angular momentum $L^z$ within a scalar diquark spectator. The $(m+x M)^2$  and ${\bf k}^2_\perp$ proportional terms arise from the (diagonal) overlap of  $L^z=l^z=0$  and $L^z=l^z=1$  LFWFs, respectively. In particular the $l^z=1$  LFWF overlap reduces the amount of the polarized $g_1$  PDF (\ref{g1g1T-DQSM}).
We note that the difference in sign of the ${\bf k}^2_\perp$ proportional terms in the unpolarized and polarized uPDFs, given in Eqs.~(\ref{f1-DQSM}) and (\ref{g1g1T-DQSM}), indicates an inconsistency of such models with perturbative QCD. Namely, the leading log behavior for PDFs arises just from these $l^z=1$ overlaps and according to the evolution equations in leading order approximation they have the same sign, which contradicts the scalar diquark model. Similarly, we find also an inconsistency from the evolution equation for the transversity PDF $h_1$. Namely, in any scalar diquark model the transversity $h_1$  saturates the Soffer bound (\ref{constraint-scadiquark-2}) for the unintegrated quantities and, hence, this is also true for the integrated ones,
\begin{eqnarray}
\label{Def-Tra_satured}
h_1(x,\mu^2) = \frac{1}{2} \left[f_1(x,\mu^2) + g_1(x,\mu^2)\right].
\end{eqnarray}
However, this equation is again in conflict with leading order evolution equations, since they are different in the chiral odd and even sectors.
Let us also note that QCD equations-of-motion \cite{ShuVai81} were utilized to relate twist-two with twist-three related PDFs \cite{MulTan95,GoeMetPobPol03}, e.g.,  $g_1$, $h_1$, and  $g_{1\rm T}^\perp$, where the latter is accessible in transversally polarized deep inelastic scattering and results from the overlap of  $l^z=0$ and $l^z=1$ LFWFs. In such kind of dynamic QCD relations quark-gluon-quark correlations enters, too, and one might wonder to which extend such interaction dependent terms are mimicked in a given quark model, e.g., the chiral quark soliton or the bag model \cite{AvaEfrSchYua10}.

If we ignore our critics on the  scalar diquark model rather rely on it, the various uPDFs provide us a model dependent handle on the quark orbital angular momenta. As we realized, the quark orbital angular momenta can be also traced within PDFs.  Furthermore, we might reverse the logic and ask what we can learn about the ${\bf k}_\perp$ dependence.
To realize that even the integrated PDFs remember their ${\bf k}_\perp$ dependence, it is instructive to express them  in terms of the integrated LFWF overlap (\ref{def-TMD-overlap}),
\begin{eqnarray}
\label{def:LFWFoverlap}
{\bf \Phi}(x,\mu^2)=\int\!d^2{\bf k}_{\perp}\, {\bf \Phi}(x,{\bf k}_\perp)\,,
\end{eqnarray}
and the ${\bf k}^2_\perp$ average
\begin{eqnarray}
\label{def:k2perp}
\langle{\bf k}_\perp^2\rangle(x,\mu^2) =
\frac{\int\!d^2{\bf k}_{\perp}\, {\bf k}_{\perp}^2{\bf \Phi}(x,{\bf k}_\perp)}{
\int\!d^2{\bf k}_{\perp}\, {\bf \Phi}(x,{\bf k}_\perp)}\,.
\end{eqnarray}
These equations have to be taken with care. In QCD one should find after integration over ${\bf k}_\perp$ a logarithmical scale dependence, which should show up in the definition (\ref{def:LFWFoverlap}), hence, the ${\bf k}_\perp^2$--average (\ref{def:k2perp}) is expected to be a divergent quantity. Let us ignore here the challenge to understand QCD dynamics, and let us proceed with the scalar diquark model in the common fashion.  Formally, we can express then the unpolarized and polarized  PDFs by means of the integrated LFWF overlap (\ref{def:LFWFoverlap}) and  ${\bf k}_\perp^2$--average (\ref{def:k2perp}):
\begin{eqnarray}
\label{f1-sca}
f_1(x,\mu^2) &\!\!\! = \!\!\!&
\frac{(m+ x M)^2 + \langle{\bf k}_\perp^2\rangle(x,\mu^2)}{M^2}\, {\bf \Phi}(x,\mu^2)\,,\\
\label{g1-sca}
g_1(x,\mu^2) &\!\!\! = \!\!\!&
\frac{(m+ x M)^2 -\langle{\bf k}_\perp^2\rangle(x,\mu^2)}{M^2}\, {\bf \Phi}(x,\mu^2)\,,\\
\label{h1-sca}
h_1(x,\mu^2) &\!\!\! = \!\!\!&
\frac{(m+ x M)^2}{M^2}\, {\bf \Phi}(x,\mu^2)\,,
\end{eqnarray}
respectively. Hence, we read off that in a scalar diquark model the ${\bf k}_\perp^2$--average  ratio
\begin{eqnarray}
\label{kt2-ratio}
\frac{\langle{\bf k}_\perp^2\rangle(x,\mu^2)}{(m+ x M)^2} = \frac{f_1(x,\mu^2)-g_1(x,\mu^2)}{f_1(x,\mu^2)+g_1(x,\mu^2)}
\end{eqnarray}
is given by the ratio of parton densities with helicities opposite  and aligned to the longitudinal proton spin. According to large $x$ counting rules
\cite{Brodsky:1994kg}, this ratio should vanish as $(1-x)^2$ for $x\to 1$.  In the limit $x\to 0$  Regge arguments might tell us that the unpolarized PDF dominates over the unpolarized one and so $\langle{\bf k}_\perp^2\rangle(x,\mu^2)$ approaches the struck quark mass square $m^2$.

From what was said in Sec.~\ref{uPDF-classification-spherical} we can easily extend our scalar diquark model by means of Eq.~(\ref{LFWF-spinor}) and the LFWF ``spinor" (\ref{LFWF-spinor-model}) to a ``spherical" model of rank-four by adding an unpolarized uPDF, which might be ambiguously chosen for $u$ and $d$ quarks.  If we like to adopt SU(4) flavor-spin symmetry to dress our uPDFs with flavor, we can immediately take the formulae (\ref{model-sperical-SU(4)-1}) and (\ref{model-sperical-SU(4)-2}), where then only the unpolarized uPDF in the $u/3+4d/3$ sector is to be chosen. To reach equivalence with
the axial-vector model \cite{JakMulRod97} or the three-quark LFWF model \cite{PasCazBof08} we have to fix the unpolarized uPDF as following
$$
f_1^{u/3+4d/3}(x,{\bf k}_\perp) \stackrel{\rm sph\&SU(4)}{=} 2\, f_1^{2u/3-d/3}(x,{\bf k}_\perp)\,,
$$
where $f_1^{2u/3-d/3}$ is taken from Eq.~(\ref{f1-DQSM}).

\section{Generalized parton distributions}
\label{Sec-GPDs&uPDFs}

Field theoretically GPDs are defined as
off-diagonal matrix elements of two field operators, connected by a gauge link, that lives on
the light cone \cite{Mueller:1998fv,Radyushkin:1996nd,Ji:1996nm}. Generically, in the light-cone gauge
a GPD definition looks like as following
\begin{eqnarray}
\label{F-gen}
F(x,\eta,t,\mu^2)=
\int\frac{d y^-}{8\pi}\;e^{ix P^+y^-/2}\;
\langle P'| {\bf \hat{\phi}}(0)  {\bf \hat{\phi}}(y)
\big|_{y^+=0, {\bf y}_\perp=0}\,|P \rangle,
\end{eqnarray}
where the skewness parameter\footnote{Up to a minus sign we use the variable conventions of
\cite{Mueller:1998fv}.} is defined as
\begin{equation}
\label{eta}
\eta = \frac{P^+-P^{\prime +}}{P^++P^{\prime +}} \ge 0\,.
\end{equation}
As already stated above in Sec.~\ref{Sect-uPDF-overlap}, the operator on the light cone possesses divergencies which are removed within a renormalization procedure and the scale dependence is governed by evolution equations, see Refs.~\cite{Dittes:1988xz,Belitsky:1999hf} and references therein.

GPDs are accessible in hard exclusive electroproduction \cite{Mueller:1998fv,Radyushkin:1996nd,Ji:1996nm} of photons or mesons \cite{Collins:1996fb} and their moments are measurable in QCD Lattice simulations \cite{Hagler:2009ni}.   The support of a GPD might be divided  into three regions
$$ -1\le x \le -\eta\,,\quad -\eta \le x \le \eta\,,\quad \mbox{and}\quad \eta \le x \le 1\,\quad\mbox{where}\quad 0\le \eta.$$
In the {\em outer} region $\eta \le x$ and $x \le -\eta$  a GPD is the probability amplitude for a $s$-channel exchange of a parton
and anti-parton, respectively. We will adopt the common habit and map anti-parton distributions to the region $\eta \le x$.  The
GPD in the {\em central} region $-\eta \le x \le \eta$, interpreted as a mesonic-like $t$-channel exchange, is the dual counterpart of the GPD
in the outer regions \cite{Mueller:2005ed,Kumericki:2008di}, see also explanations given below in Sec.~\ref{Sec-GPDs}, in particular Sec.~\ref{sect-LI},
and in Sec.~\ref{subsec-Dterm}.

The zero-skewness case $\eta=0$ is of specific partonic interest, since it allows for a probabilistic interpretation and allow so to provide a three dimensional picture of the nucleon. In particular in the infinite momentum frame a GPD  in the
impact parameter space \cite{Ralston:1979ys} is the probability for  finding a parton in dependence on the momentum fraction $x$ and the transverse distance of the proton center \cite{Burkardt:2002hr}, see also Ref.~\cite{Ralston:2001xs,Diehl:2002he}.
Since they generalize the common probabilistic interpretation of PDFs to non-forward kinematics,
we will denote them in the following  as non-forward PDFs (nfPDFs).

In Sec.~\ref{Sec-GPD-definitions} we give our twist-two GPD conventions, present their LFWF overlap representations, and discuss the parameterization  of uGPDs.  In Sec.~\ref{sec:GPD-relations} we introduce then in analogy to uPDF models a classification scheme for GPD models. Finally, in Sec.~\ref{sec:GPDs-uGPDs} we utilize  the LFWF overlap representation to have a closer look to the model dependent cross talks of zero-skewness GPDs and uPDFs.

\subsection{General definitions}
\label{Sec-GPD-definitions}

In the following we will employ the standard notation of GPDs, appearing in the form factor decomposition of matrix elements such as generically given in
(\ref{F-gen}), where the bilocal operators read as
$$
{\bf \hat{\phi}}(0)  {\bf \hat{\phi}}(y) \to \hat{\bar{\psi}}(0)\Gamma \hat\psi(y)\quad\mbox{with}\quad \Gamma \in\{\gamma^+,\gamma^+\gamma^5,i\sigma^{+j}
\}\,, \quad \sigma^{+j} =\frac{i}{2}\left[\gamma^+  \gamma^j - \gamma^j \gamma^+ \right].
$$
In the chiral even and parity even [odd] sector we adopt the form factor decomposition of Ji \cite{Ji:1996nm}. Here, we deal with the GPDs $H$ and $E$ [$\widetilde H$ and $\widetilde E$],  where the latter one is related to a target helicity-flip contribution,
\begin{eqnarray}
\label{HE-def}
F^{[\gamma^+]} &\!\!\! = \!\!\!&
{1\over 2\bar P^+}\ {\bar U}(P',{S}') \left[
H(x,\eta,t)\ {\gamma^+}
 +
E(x,\eta,t)\
\frac{\sigma^{+\alpha}\Delta_\alpha}{2 i M}
\right]  U(P,{S})\,,
\\
\label{tHtE-def}
F^{[\gamma^+\gamma^5]} &\!\!\! = \!\!\!&
{1\over 2\bar P^+}\ {\bar U}(P',{S}') \left[
{\widetilde H}(x,\eta,t)\ {\gamma^+\gamma_5}
 +
{\widetilde E}(x,\eta,t)\
\frac{-\Delta^+}{2M}\, {\gamma^5}
\right]  U(P,{S})\,,
\end{eqnarray}
where $\bar P = (P+P^\prime)/2$ is the averaged nucleon momentum.
In the chiral odd sector we adopt the definition of Diehl \cite{Diehl:2001pm}, see also references therein, and we have the four twist-two GPDs
$H_T$, $E_T$, ${\widetilde H}_T$, and ${\widetilde E}_T$, defined as
\begin{eqnarray}
\label{FT-def}
F^{[i\sigma^{+j}]}&\!\!\! =\!\!\! &
{1\over 2\bar P^+}\ {\bar U}(P',{S}') \Bigg[
H_T(x,\eta,t)\ i\sigma^{+j}
+
E_T(x,\eta,t)\ {\Delta^+\gamma^j-\gamma^+ \Delta^j\over 2M}
\\
&&\qquad\qquad\qquad\quad +
{\widetilde H}_T(x,\eta,t)\
{\Delta^+{\bar P^j} -{\bar P^+} \Delta^j\over M^2}
+
{\widetilde E}_T(x,\eta,t)\ {\gamma^+{\bar P^j}-{\bar P^+}\gamma^j\over M}
\Bigg]  U(P,{S})\ .
\nonumber
\end{eqnarray}
All GPDs are real valued functions and combining hermiticity  and time-reversal invariance tells us that they are even in $\eta$, except for the chiral odd function ${\widetilde E}_T(x,\eta,t)$ that is odd in $\eta$:
$$
F(x,\eta,t) = F(x,-\eta,t)\;\;\mbox{for}\; F\in\{H,E,\widetilde H,\widetilde E,H_{\rm T},E_{\rm T},\widetilde H_{\rm T}\}
\quad \mbox{and}\quad \widetilde E_{\rm T}(x,\eta,t) = -\widetilde E_{\rm T}(x,-\eta,t).
$$
Hence, the zero-skewness GPD $\widetilde E_{\rm T}(x,\eta=0,t)$ vanishes, however, the limit $\lim_{\eta\to 0} \widetilde E_{\rm T}(x,\eta,t)/\eta$
will generally exist and pro

Of course, we have might also defined unintegrated GPDs (uGPDs) in terms of quark operators, which make contact to the ``phase space" distributions of Ref.~\cite{Belitsky:2003nz}.  However, this will increase the number of our considered quark GPDs from eight to sixteen \cite{Meissner:2009ww}. Moreover, such uGPDs are hardly to access in experiment. Let us only emphasis here that we might decorate the common eight twist-two GPDs with ${\bf k}_\perp$ dependence, introduce other five uGPDs that generalize the set of $f_{1\rm T}^\perp$, $g_{1\rm T}^\perp$, $h_{1\rm L}^\perp$, $h_{1}^\perp$ and $h_{1 \rm T}^\perp$  uPDFs and  in addition we are left with three uGPDs that die out in the
forward limit and by ${\bf k}_\perp$ integration. We find it rather natural to adopt such a parameterization for uGPDs, e.g., for the unpolarized quark uGPD,
\begin{eqnarray}
\label{uGPD-definition}
F^{[\gamma^+]} &\!\!\! = \!\!\!&
{\bar U}(P',{S}') \left[
H\frac{\gamma^+}{ 2\bar P^+}
 +
E\
\frac{\sigma^{+\alpha}\Delta_\alpha}{4 i M \bar P^+}
+
if_{1\rm T}^\perp\
\frac{ i \sigma^{+j}{\bf k}^j_\perp}{4 M \bar P^+}
+
f_{1,4}\
\frac{{\bf k}^j_\perp i\sigma^{j\alpha}\Delta_\alpha}{2  M}
\right]  U(P,{S})\,,
\end{eqnarray}
we have two ${\bf k}_\perp$ dependent GPDs $H$ and $E$, one skewed and with $t$-dependence decorated Sivers function, and one new function $f_{1,4}$.

As outlined in Appendix \ref{sec-app-LFWF2GPD}, the LFWF overlap representation for all twist-two GPDs can be straightforwardly derived \cite{Diehl:1998kh,Brodsky:2000xy,Diehl:2000xz}. We will focus here on the region $\eta \le x$ in which GPDs are obtained  from the parton number conserved LFWF overlaps.  We find it convenient to group the results in a spin-correlation matrix
\begin{eqnarray}
\label{bfF-LFWFs}
{\bf F}_{(n)}(x,\eta,t|\varphi)=\;
\intsumm \left(
          \begin{array}{cccc}
\psi^{\ast\,\Rightarrow}_{\rightarrow,n}\psi^{\Rightarrow}_{\rightarrow,n} & \psi^{\ast\,\Rightarrow}_{\rightarrow,n} \psi^{\Rightarrow}_{\leftarrow,n}
& \psi^{\ast\,\Rightarrow}_{\rightarrow,n}\psi^{\Leftarrow}_{\rightarrow,n} & \psi^{\ast\,\Rightarrow}_{\rightarrow,n}\psi^{\Leftarrow}_{\leftarrow,n}
\\
\psi^{\ast\,\Rightarrow}_{\leftarrow,n} \psi^{\Rightarrow}_{\rightarrow,n}& \psi^{\ast\,\Rightarrow}_{\leftarrow,n}\psi^{\Rightarrow}_{\leftarrow,n}
& \psi^{\ast\,\Rightarrow}_{\leftarrow,n} \psi^{\Leftarrow}_{\rightarrow,n}& \psi^{\ast\,\Rightarrow}_{\leftarrow,n}\psi^{\Leftarrow}_{\leftarrow,n}
\\
\psi^{\ast\,\Leftarrow}_{\rightarrow,n}  \psi^{\Rightarrow}_{\rightarrow,n}& \psi^{\ast\,\Leftarrow}_{\rightarrow,n} \psi^{\Rightarrow}_{\leftarrow,n}
 & \psi^{\ast\,\Leftarrow}_{\rightarrow,n}\psi^{\Leftarrow}_{\rightarrow,n} & \psi^{\ast\,\Leftarrow}_{\rightarrow,n}\psi^{\Leftarrow}_{\leftarrow,n}
 \\
\psi^{\ast\,\Leftarrow}_{\leftarrow,n} \psi^{\Rightarrow}_{\rightarrow,n} & \psi^{\ast\,\Leftarrow}_{\leftarrow,n} \psi^{\Rightarrow}_{\leftarrow,n}  &
 \psi^{\ast\,\Leftarrow}_{\leftarrow,n} \psi^{\Leftarrow}_{\rightarrow,n} & \psi^{\ast\,\Leftarrow}_{\leftarrow,n}\psi^{\Leftarrow}_{\leftarrow,n} \\
          \end{array}
        \right)\!\!(X_i,{\bf k}^\prime_{\perp i},s_i|X_i,{\bf k}_{\perp i},s_i)\,,\quad
\end{eqnarray}
where $\varphi$ denotes now the polar angle in transverse momentum space
$$\left(\Delta^1,\Delta^2\right) = |{\bf \Delta}_{\perp}| \left(\cos\varphi,\sin\varphi \right) .$$
The phase space integral (\ref{intsum1}), which might be written as
\begin{eqnarray}
\label{intsum11}
\intsumm \cdots \equiv \sum_{s_2,\dots, s_n}
\int\![dX\, d^2{\bf k}_{\perp}]_n \left(\frac{1+\eta}{1-\eta}\right)^{n-2}\,
 \frac{1}{\sqrt{1-\eta^2}}\, \delta\!\left(\frac{x+\eta}{1+\eta}-X_{1}\right)
\cdots\,,
\end{eqnarray}
includes now also the integral over the transverse momenta of the struck quark. The transverse momenta of the outgoing LFWFs are fixed by momentum conservation and read in Leipzig convention as
\begin{equation}
\begin{array}[t]{lll}
X^\prime_{1} = {\displaystyle \frac{x-\eta}{1-\eta}}\, ,\
&{{\bf k}}^\prime_{\perp 1} ={{\bf k}}^{~}_{\perp 1}
- {\displaystyle  \frac{1-x}{1-\eta}}\, {{\bf \Delta}}_\perp
&\mbox{for the struck parton,}
\\[2ex]
X^\prime_i = {\displaystyle \frac{1+\eta}{1-\eta}}\, X_i\, ,\
&{{\bf k}}^\prime_{\perp i} ={{\bf k}}^{~}_{\perp i}
+ {\displaystyle  \frac{1+\eta}{1-\eta}}\, X_i\, {{\bf \Delta}}_\perp
&\mbox{for the spectator $i\in\{2, \cdots, n\}$}\,,
\end{array}
\label{t2-1}
\end{equation}
cf.~Eqs.~(\ref{zeta2eta}--\ref{xeta2XZeta}).

Rather analog to the uPDF spin-density matrix (\ref{tPhi}), the GPD spin-correlation matrix results from summing up the individual LFWF overlaps (\ref{bfF-LFWFs}),
$${\bf F}_{a b}(x,\eta,t|\varphi) = \sum_{n} {\bf F}_{a b}^{(n)}(x,\eta,t|\varphi) \quad\mbox{with}\quad
a=\Lambda^\prime\lambda^\prime,\; b=\Lambda\lambda \in \{\Rightarrow\rightarrow,\Rightarrow\leftarrow,\Leftarrow\rightarrow,\Leftarrow\leftarrow\}\,.
$$
The entries in this matrix might be easily read off from the GPD definitions (\ref{HE-def}--\ref{FT-def}) or the LFWF overlap representations  (\ref{t1},\ref{t1f2},\ref{t155},\ref{t1f255},\ref{t155s2}--\ref{t1f255s1}),
\begin{eqnarray}
\label{bfF}
{\bf F}(x\ge \eta,\eta,t|\varphi) &\!\!\!=\!\!\!& \phantom{\Bigg|}
\\
\!\!\! \!\!\! &\!\!\!=\!\!\!& \left(\!\!\!
\begin{array}{cccc}
\frac{H +\frac{t_0}{4M^2} E +\widetilde H +\frac{t_0}{4M^2} \widetilde{E}}{2}
&
\frac{- |{\bf \bar \Delta}_\perp|\, e^{i\varphi}}{2M}\, \frac{\overline{E}_{\rm T}-\eta \hat{E}_{\rm T}}{2}
&
\frac{ |{\bf \bar \Delta}_\perp|\, e^{-i\varphi}}{2M}\,\frac{E-\eta \widetilde E}{2}
&
\overline{H}_{\rm T}+\frac{t_0 \left(\widetilde H_{\rm T}+ \hat{E}_{\rm T}\right)}{4M^2}
\\
\frac{|{\bf \bar \Delta}_\perp|\, e^{-i\varphi}}{2M}\, \frac{\overline{E}_{\rm T}+\eta \hat{E}_{\rm T}}{2}
&
\frac{H +\frac{t_0}{4M^2} E -\widetilde H -\frac{t_0}{4M^2} \widetilde{E}}{2}
&
\frac{-{\bf \bar \Delta}_\perp^2\,(1-\eta^2)\, e^{-i2\varphi}}{4M^2}  \widetilde H_{\rm T}
&
\frac{|{\bf \bar \Delta}_\perp|\, e^{-i\varphi}}{2M}\, \frac{E+\eta\widetilde E}{2}
\\
\frac{-|{\bf \bar \Delta}_\perp|\, e^{i\varphi}}{2M}\, \frac{E+\eta \widetilde E}{2}
&
\frac{-{\bf \bar \Delta}_\perp^2\,(1-\eta^2)\, e^{i2\varphi}}{4 M^2} \widetilde H_{\rm T}
&
\frac{H +\frac{t_0}{4M^2} E -\widetilde H -\frac{t_0}{4M^2} \widetilde{E}}{2}
&
\frac{-|{\bf \bar \Delta}_\perp|\, e^{i\varphi}}{2M}\, \frac{\overline{E}_{\rm T}+ \eta \hat{E}_{\rm T}}{2}
\\
\overline{H}_{\rm T} +\frac{t_0 \left(\widetilde H_{\rm T}+ \hat{E}_{\rm T}  \right)}{4M^2}
&
\frac{-|{\bf \bar \Delta}_\perp|\, e^{i\varphi}}{2M}\, \frac{E-\eta \widetilde E}{2}
&
\frac{|{\bf \bar \Delta}_\perp|\, e^{-i\varphi}}{2M}\, \frac{\overline{E}_{\rm T}-\eta \hat{E}_{\rm T}}{2}
&
\frac{H +\frac{t_0}{4M^2} E +\widetilde H +\frac{t_0}{4M^2} \widetilde{E}}{2}  \\
\end{array}
\!\!\!
\right)\!,
\nonumber
\end{eqnarray}
where we use the shorthands
$$ {\bf \bar \Delta}_\perp = \frac{{\bf\Delta}_\perp}{1-\eta}= \frac{\sqrt{t_0-t}}{\sqrt{1-\eta^2}}
\quad \mbox{and}\quad
t_0=-\frac{4 M^2\eta^2}{1-\eta^2}$$
that are symmetric functions under $\eta\to -\eta$,
for kinematics see Eqs.~(\ref{a1}--\ref{zeta2eta}).
With our choice of variables $-t_0$ is the exact expression for the minimal
value of the negative square of the momentum transfer  $-t = -\Delta^2$.
Furthermore, we introduced the chiral odd GPD combinations
\begin{eqnarray}
\label{FT-newbasis}
\overline{H}_{\rm T}=H_{\rm T} -\frac{t}{4 M^2} \widetilde H_{\rm T}\,,
\quad
\overline{E}_{\rm T} = E_{\rm T}+2 \widetilde H_{\rm T}-\eta \widetilde{E}_{\rm T}\,,
\quad
\hat{E}_{\rm T} = E_{\rm T} - \frac{1}{\eta} \widetilde{E}_{\rm T}\,.
\end{eqnarray}
The spin-correlation matrix ($\ref{bfF}$) has eight non-vanishing real valued
GPD entries and we can read off the following property, implied by time-reversal and parity invariance,
\begin{eqnarray}
\label{bfF-properties}
{\bf F}^\dagger(x,\eta,t|\varphi) = {\bf F}(x,-\eta,t|\varphi+\pi)\,.
\end{eqnarray}
The trace of the spin correlation matrix, given by
$
2 H(x,\eta,t) - 2\eta^2 E(x,\eta,t)/(1-\eta^2),$
is  for vanishing skewness expressed by the unpolarized GPD $H$, see (\ref{tPhi}).

As alluded above, zero skewness GPDs, called nfPDFs, in the impact space
have their own interest. Such densities can be obtained from a two-dimensional
Fourier transform of certain GPD combinations. In particular, their spin density matrix is obtained from the GPD
spin-correlation matrix (\ref{bfF})  for $\eta=0$ and reads as
\begin{eqnarray}
\label{tbfF}
\widetilde {\bf F}(x,{\bf b}) = \int\!\!\!\int\! \frac{d^2{\bf \Delta}_\perp}{(2\pi)^2}\, e^{-i {\bf b} \cdot {\bf \Delta}_\perp}
{\bf F}(x,{\bf \Delta}_\perp)
\quad\mbox{with}\quad
{\bf F}(x,{\bf \Delta}_\perp)= {\bf F}(x,\eta=0,t=-{\bf \Delta}_\perp^2|\varphi)\,,
\end{eqnarray}
where the two dimensional vector ${\bf b} = b  (\cos\varphi, \sin\varphi)$ with $b=|{\bf b}|$ is the distance from the nucleon center and  $\varphi$ denotes now
the polar angle in the two-dimensional impact space.  From the properties (\ref{bfF-properties}) of the spin-correlation matrix in momentum space
it is easy to realize that the Fourier transformation provides a  hermitian  spin-density matrix in impact parameter space,
\begin{eqnarray}
\widetilde {\bf F}^\dagger(x,{\bf b}) = \widetilde{\bf F}(x,{\bf b})\,.
\end{eqnarray}
Its trace, ${\rm Tr}\, \widetilde{\bf F}(x,{\bf b})= 2 H(x,b)$, is expressed by the unpolarized nfPDF, which depends besides the momentum fraction $x$ only on the distance $b$.
Furthermore, this matrix is also semi-positive definite and it can be used to derive positivity constraints \cite{DieHag05}, even for the general $\eta\neq 0$ case \cite{Pobylitsa:2002iu}.
The explicit expression of the spin-density matrix is straightforwardly obtained by means of Eq.~(\ref{tbfF}):
\begin{eqnarray}
\label{bfF-b}
&&\widetilde {\bf F}(x,{\bf b}) = \left(\!\!\!
\begin{array}{cccc}
\frac{H  +\widetilde H}{2}
&
i\, e^{i\varphi}\, \frac{\overline{E}^\prime_{\rm T}}{2}
&
- i\, e^{-i\varphi}\,\frac{E^\prime}{2}
&
\overline{H}_{\rm T}
\\
- i\, e^{-i\varphi}\, \frac{\overline{E}^\prime_{\rm T}}{2}
&
\frac{H -\widetilde H }{2}
&
e^{-i2\varphi} \widetilde H^{\prime\prime}_{\rm T}
&
-i\, e^{-i\varphi}\, \frac{E^\prime}{2}
\\
i\, e^{i\varphi}\, \frac{E^\prime}{2}
&
 e^{i2\varphi} \widetilde H^{\prime\prime}_{\rm T}
&
\frac{H  -\widetilde H }{2}
&
i\, e^{i\varphi}\, \frac{\overline{E}^\prime_{\rm T}}{2}
\\
\overline{H}_{\rm T}
&
i\, e^{i\varphi}\, \frac{E^\prime}{2}
&
-i\, e^{-i\varphi}\, \frac{\overline{E}^\prime_{\rm T}}{2}
&
\frac{H +\widetilde H }{2}  \\
\end{array}
\!\!\!
\right)\!(x,b)\,,
\end{eqnarray}
where the (dimensionless) derivatives of GPDs are denoted as
\begin{eqnarray}
E^\prime(x,b) &\!\!\!=\!\!\!& \frac{b}{M} \frac{\partial}{\partial b^2} E(x,b)\,,
\qquad\qquad
\overline{E}^\prime_{\rm T}(x,b) = \frac{b}{M} \frac{\partial}{\partial b^2}\left[ E_{\rm T} + 2\widetilde H_{\rm T} \right](x,b)\,,
\\
\widetilde{H}_{\rm T}^{\prime\prime}(x,b) &\!\!\!=\!\!\!& \frac{b^2}{M^2} \frac{\partial}{\partial b^2} \frac{\partial}{\partial b^2}\widetilde{H}_{\rm T}(x,b)
\,,
\quad\;
\overline{H}_{\rm T}(x,b)=H_{\rm T}(x,b)+\widetilde{H}_{\rm T}^{\prime\prime}(x,b)\,,
\nonumber
\end{eqnarray}
and $\varphi$ is now the polar angle in impact parameter space.
Note that we do not introduce new symbols for the GPDs in impact parameter space rather we implicitly distinguish them from those in momentum space by
specifying the argument.
We also emphasize that in the forward case the ``$T$-odd''  GPDs  $\widetilde E$ and $\hat E_{\rm T}$ drop out and so the spin-density matrix is only
parameterized by six real valued GPDs and their first and second order derivatives.
We add that after reordering and redefinition, our spin-density matrix (\ref{bfF-b}) coincides with the result (34) of Ref.~\cite{DieHag05}.

As we saw in Sec.~\ref{Sect-uPDF-overlap} time-reversal and parity invariance allows us to represent uPDFs by different LFWF overlaps. This is also the case for our eight twist-two GPDs.  We relate them here to the LFWF overlaps in such a manner that we can easily realize formal relations  among GPDs and uPDFs. Alternatively, such correspondences between these quantities would  also follow if we introduce uGPD definitions as mentioned above. From the spin-correlation matrix (\ref{bfF}) we can read off for $x\ge \eta$ in the chiral even sector the diagonal $n$-parton LFWF overlaps
\begin{eqnarray}
\label{H-LFWF}
H^{(n)}(x,\eta,t) &\!\!\! =\!\!\! &\; \intsumm
\left[
\psi^{\ast\,\Rightarrow}_{\rightarrow,n}\, \psi^{\Rightarrow}_{\rightarrow,n}+
\psi^{\ast\,\Rightarrow}_{\leftarrow,n}\, \psi^{\Rightarrow}_{\leftarrow,n}
\right]\!
(X^\prime_i,{{\bf k}}^\prime_{\perp i},s_i|X_i, {{\bf k}}_{\perp i},s_i)
-\frac{t_0}{4 M^2} E^{(n)}(x,\eta,t)\,,\\
\label{E-LFWF}
E^{(n)}(x,\eta,t) &\!\!\! =\!\!\! &
{M\over  { \bar \Delta}^{1}}\; \intsumm
\left[
\psi^{\ast\,  \Rightarrow}_{\rightarrow,n} \psi^{ \Leftarrow}_{\rightarrow,n}\,
+
\psi^{\ast\,  \Rightarrow}_{\leftarrow,n} \, \psi^{\Leftarrow}_{\leftarrow,n}
+ \mbox{c.c.}
\right]\!
(X^\prime_i,{{\bf k}}^\prime_{\perp i},s_i|X_i, {{\bf k}}_{\perp i},s_i)\,,
\\
\label{tH-LFWF}
\widetilde H^{(n)}(x,\eta,t) &\!\!\! =\!\!\! & \;\intsumm
\left[
\psi^{\ast\,\Rightarrow}_{\rightarrow,n}\, \psi^{\Rightarrow}_{\rightarrow,n}-
\psi^{\ast\,\Rightarrow}_{\leftarrow,n}\, \psi^{\Rightarrow}_{\leftarrow,n}
\right]\!(X^\prime_i,{{\bf k}}^\prime_{\perp i},s_i|X_i, {{\bf k}}_{\perp i},s_i)-
\frac{t_0}{4M^2} \widetilde E^{(n)}(x,\eta,t)\,,
\qquad
\\
\label{tE-LFWF}
\widetilde E^{(n)}(x,\eta,t) &\!\!\! =\!\!\! &
{M \over  \bar{\Delta}^{1}}\,\frac{1}{\eta}\;\intsumm
\left[
-\psi^{\ast\,  \Rightarrow}_{\rightarrow,n} \psi^{ \Leftarrow}_{\rightarrow,n}\,
+
\psi^{\ast\,  \Rightarrow}_{\leftarrow,n} \, \psi^{\Leftarrow}_{\leftarrow,n}
 +\mbox{c.c.}
\right]\!
(X^\prime_i,{{\bf k}}^\prime_{\perp i},s_i|X_i, {{\bf k}}_{\perp i},s_i)\,.
\end{eqnarray}
The target helicity-conserved unpolarized quark GPD  $H$ and polarized quark GPD  $\widetilde H$ essentially arise
from the sum and difference of diagonal $l^z=0$ and diagonal  $l^z=1$ LFWF overlaps, respectively, where the orbital angular momentum transfer
\begin{eqnarray}
\label{Delta L^z}
\Delta L^z = \Delta l^z = L^{\prime z}-L^{z}= l^{\prime z}-l^{z}
\end{eqnarray}
vanishes.
Directly, from the uPDF overlaps (\ref{OveRep-f1}) and (\ref{OveRep-g1}) one realizes that they generalize $f_1$ and $g_1$ PDFs. However, both of these GPDs contain a $t_0$ proportional $E^{(n)}$ and $\widetilde E^{(n)}$  addenda that presumable will cancel some contributions in the $\Delta L^z=0$ LFWF overlaps.  The target helicity-flip   unpolarized quark GPD  $E$ and polarized quark GPD  $\widetilde E$ are given by  $l_z=0$ and  $l_z=1$ LFWF overlaps and so $|\Delta L^z|=1$. Note that the similarities of $|\Delta L^z|=1$ uPDF and GPD  overlaps (\ref{f_{1T}^perp-overlap-2},\ref{g_{1T}^perp-overlap}) and (\ref{E-LFWF},\ref{tE-LFWF}) suggest that  $E$ and $\widetilde E$ GPDs have a cross talk with $f_{1\rm T}^\perp$ and $g_{1\rm T}^\perp$ uPDFs, respectively. As already said above and explained below in Sec.~\ref{sec:GPD-relations-1}, such a uPDF/GPD relation in the $|\Delta L^z|=1$ sector is presumable lost in QCD.
In the chiral odd sector we write the parton number conserved LFWF overlaps for $x\ge \eta$  in terms of the basis (\ref{FT-newbasis}),
\begin{eqnarray}
\label{oHT-LFWF}
\overline{H}_{{\rm T}}^{(n)}(x,\eta,t)  &\!\!\!=\!\!\! &\frac{1}{2}\;\intsumm
\left[
\psi^{\ast\,\Rightarrow}_{\rightarrow,n} \, \psi^{\Leftarrow}_{\leftarrow,n} +\mbox{c.c.}
\right]\!
 (X^\prime_i,{{\bf k}}^\prime_{\perp i},s_i|X_i, {{\bf k}}_{\perp i},s_i)-
 \frac{t_0}{4M^2}\!\left[\!{\widetilde{H}}^{(n)}_{\rm T}+\hat{E}^{(n)}_{\rm T}\!\right]\!\!(x,\eta,t),\qquad
 \\
\label{oET-LFWF}
 \overline{E}^{(n)}_{\rm T}(x,\eta,t) &\!\!\! =\!\!\! & -\frac{M}{\bar\Delta^1}\;\intsumm
\left[
\psi^{\ast\, \Rightarrow}_{\rightarrow,n}\, \psi^{\Rightarrow}_{\leftarrow,n}
+
\psi^{\ast\, \Leftarrow}_{\rightarrow,n} \, \psi^{\Leftarrow}_{\leftarrow,n}
+\mbox{c.c.}
\right]\!
(X^\prime_i,{{\bf k}}^\prime_{\perp i},s_i|X_i, {{\bf k}}_{\perp i},s_i) \,,
\\
\label{hET-LFWF}
\hat{E}_{{\rm T}}^{(n)}(x,\eta,t)
& \!\!\!=\!\!\! & \frac{M}{\eta \, \bar\Delta^1 }\;\intsumm
\left[
\psi^{\ast\,\Rightarrow}_{\rightarrow,n} \, \psi^{\Rightarrow}_{\leftarrow,n}
-
\psi^{\ast\, \Leftarrow}_{\rightarrow,n} \, \psi^{\Leftarrow}_{\leftarrow,n}
+\mbox{c.c.}
\right]\!
(X^\prime_i,{{\bf k}}^\prime_{\perp i},s_i|X_i, {{\bf k}}_{\perp i},s_i) \,,
\\
\label{tHT-LFWF}
{\widetilde{H}}_{{\rm T}}^{(n)}(x,\eta,t) &\!\!\!=\!\!\! & \frac{M^2} {i(1-\eta^2) \bar{\Delta}^{1} \bar{\Delta}^{2}}\;\intsumm
\left[
\psi^{\ast\,\Leftarrow}_{\rightarrow,n} \, \psi^{\Rightarrow}_{\leftarrow,n}
-
\psi^{\ast\,\Rightarrow}_{\leftarrow,n} \, \psi^{\Leftarrow}_{\rightarrow,n}
\right]\!
(X^\prime_i,{{\bf k}}^\prime_{\perp i},s_i|X_i, {{\bf k}}_{\perp i},s_i) \,.
\end{eqnarray}
All these GPDs probe a longitudinal quark spin transfer (of one unit), where $\overline{H}_{\rm T}$ essentially arise from off-diagonal $l_z=0$  LFWF overlaps, supplemented by $t_0$ proportional correction terms and generalizes the $h_1$ PDF, see Eq.~(\ref{OveRep-h1}). In the $|\Delta L^z|=1$  sector we have $\overline{E}$ and $\hat{E}$ GPDs, accessible with unpolarized and longitudinally polarized target, respectively. Again the similarities with the
uPDF overlap representations  (\ref{h_1^perp-overlap-2}) and (\ref{h1L^perp-overlap}) suggest that
$\overline{E}$ and $\hat{E}$ GPDs are somehow related to $h_1^\perp$  and $h^\perp_{1\rm L}$ uPDFs, respectively.
Finally, we have in the transversity sector the $|\Delta L^z|=2$ GPD ${\widetilde{H}}_{{\rm T}}^{(n)}$ that stems from $l^z=1$ and $l^z=-1$
LFWFs overlaps and might have some connections with the uPDF $h_{1 \rm T}^\perp$, compare Eqs.~(\ref{h_1T^perp-overlap}) and (\ref{tHT-LFWF}).
As for $|\Delta L^z|=1$ sector also for the $|\Delta L^z|=2$ distributions the relation is only suggestive, however, as shown below in Sec.~\ref{sec:GPD2uPDF-Lz2} might not be exist in QCD.

\subsection{GPD model classification and model dependent relations}
\label{sec:GPD-relations}

The classification scheme for uPDF models might be also taken for uGPD ones. However, the number of model constraints among GPDs in momentum space  will in generally be smaller than for uPDFs or in other words an eventual degeneracy of the spin-density matrix for uPDFs is partially removed or even lost in those for GPDs. This is easy to understand, e.g., for the scalar diquark model the uPDFs arise  from the Cartesian product of a LFWF ``spinor", see Eq.~(\ref{bfPhi-rank1}), while the GPD spin-correlation matrix  additionally involves a  ${\bf k}_\perp$-convolution:
\begin{eqnarray}
\label{bfF-rank3}
{\bf F}^{(1)}(x,\eta,t|\varphi) &\!\!\! =\!\!\! & \int\!\!\!\!\!\int\! \frac{d^2{\bf k}_\perp}{1+\eta}\,
 \mbox{\boldmath $\psi^\dagger$}\!
 \left(\!
\frac{x-\eta}{1-\eta},{\bf k}_{\perp 1}- \frac{1-x}{1-\eta} {\bf \Delta}_\perp
\!\right) \otimes
\mbox{\boldmath $\psi$}\!\left(\!\frac{x+\eta}{1+\eta},{\bf k}_{\perp}\!\right),
\;\;
\mbox{\boldmath $\psi$} 
=\left(\!\!
  \begin{array}{c}
    \psi^{\Rightarrow}_{\rightarrow} \\
    \psi^{\Rightarrow}_{\leftarrow} \\
    \psi^{\Leftarrow}_{\rightarrow} \\
    \psi^{\Leftarrow}_{\leftarrow} \\
  \end{array}
\!\!\right).\qquad
\end{eqnarray}
Hence, we expect that for a ``spherical"  model of rank-one and for an ``axial-symmetrical'' one of the first kind with rank-two  the four linear  uPDF constraints, i.e.,  (\ref{constraint-rank1-1}) and the saturated Soffer bound (\ref{Soffer-bound-uPDFs}),  will reduce in the GPD case to three algebraic constraints, where the GPD analogs of the uPDF constraints
\begin{equation}
\label{uPDF-constraint-kperp}
g_1\mp\left[ h_1 + \frac{{\bf k}^2_\perp}{2M^2} h^\perp_{1\rm T}\right] \stackrel{\rm sph}{=}0
\quad\mbox{and}\quad
f_1\mp \left[ h_1 -\frac{{\bf k}^2_\perp}{2M^2} h^\perp_{1\rm T}\right] \stackrel{\rm sph}{=}0
\quad\mbox{or}\quad
f_1-g_1 \pm\frac{{\bf k}^2_\perp}{M^2} h^\perp_{1\rm T} \stackrel{\rm sph}{=}0
\end{equation}
do not hold true anymore, however, their sum, i.e., the GPD analog of a saturated Soffer bound (\ref{Soffer-bound-uPDFs}) exist.
The fifth ``spherical'' uPDF constraint, the quadratic one in Eq.~(\ref{constraint-rank1-2}), is independent on ${\bf k}^2_\perp$, however,
we do not expect that the corresponding GPD analog  holds true, see next section.

Contrarily to the momentum space representation, the spin-density matrices of zero-skewness nfPDFs in impact parameter space (\ref{tbfF}),
might have the same degeneracy as the corresponding uPDFs spin-density matrices.  This is obvious for models that are build within effective two-body LFWFs,
since the convolution integral in momentum space, e.g., Eq.~(\ref{bfF-rank3}), turns then into the product of the Fourier transformed two-body LFWFs
$\widetilde{\psi}^S_s(x,{\bf b})$ in impact parameter space, e.g.,
\begin{eqnarray}
\label{tbfF-rank3}
\widetilde {\bf F}^{(1)}(x,{\bf b}) \propto \mbox{\boldmath $\widetilde \psi^\dagger$}\!\left(\!x,\frac{\bf b}{1-x}\!\right) \otimes
\mbox{\boldmath $\widetilde \psi$}\!\left(\!x,\frac{\bf b}{1-x}\!\right)\,,
\quad
\mbox{\boldmath $\widetilde \psi$}(x,{\bf b})
=\int\!\!\!\!\int\!\frac{d^2{\bf k}_\perp}{(2\pi)^2} e^{-i{\bf b}\cdot {\bf k}_\perp}
\left(\!\!
  \begin{array}{c}
    \psi^{\Rightarrow}_{\rightarrow}(x,{\bf k}_\perp) \\
   \psi^{\Rightarrow}_{\leftarrow}(x,{\bf k}_\perp) \\
    \psi^{\Leftarrow}_{\rightarrow}(x,{\bf k}_\perp) \\
    \psi^{\Leftarrow}_{\leftarrow}(x,{\bf k}_\perp) \\
  \end{array}
\!\!\right).
\end{eqnarray}
Hence, for such nfPDF models in impact parameter space the analogous situation appears as for ``spherical" uPDF models (\ref{bfPhi-rank1}), discussed in
Sec.~\ref{uPDF-classification-spherical}. More generally,  both nfPDF models in impact parameter space and uPDF model in momentum space the $n$-parton
LFWF overlap contain only a $(n-1)$-parton phase space integral rather than a $n$-parton one as it appear for GPDs in momentum space.
Since the impact parameter representation plays a certain role for both the partonic interpretation of GPDs and for the description of partonic processes in hadron-hadron collision, we will classify the nfPDF models in Sec.~\ref{sec:GPDs-classification-nfPDFs}.
Indeed, from this analysis we will gain some new insights even for GPD models in momentum space.

\subsubsection{Algebraic GPD model constraints in momentum space}

After these heuristic thoughts, let us derive algebraic GPD  model constraints, where the classification scheme is adopted
from the uPDF models, given in Sec.~\ref{uPDF-classification}. To do so we employ the eigenvalues of the spin-correlation matrix (\ref{bfF}),
\begin{eqnarray}
\label{bfF-eigenvalues}
&&\!\!\!\!\!{\bf F}^{\rm e.v.}(x\ge \eta,\eta,t) =
\\
&&= \left\{\!\!
\begin{array}{c}
\frac{H-H_{\rm T}-\eta \widetilde E_{\rm T}}{2}+\frac{t\, \widetilde H_{\rm T}}{4 M^2}  + \frac{t_0\, \Delta E}{8M^2}
-
\frac{1}{2}\sqrt{
\left(\widetilde H-H_{\rm T}  + \frac{t_0}{4M^2} \Delta \widetilde E \right)^2 -
\frac{{\bf\bar\Delta}_\perp^2}{4 M^2}\left(\Delta E^2- \eta^2 \Delta \widetilde E^2\right)
}
\\
\frac{H-H_{\rm T}-\eta \widetilde E_{\rm T}}{2}+\frac{t\, \widetilde H_{\rm T}}{4 M^2}  + \frac{t_0\, \Delta E}{8M^2}
+
\frac{1}{2}\sqrt{
\left(\widetilde H- H_{\rm T} + \frac{t_0}{4M^2} \Delta \widetilde E \right)^2 -
\frac{{\bf\bar\Delta}_\perp^2}{4 M^2}\left(\Delta E^2- \eta^2 \Delta \widetilde E^2\right)
}
\\
\frac{H+H_{\rm T}+\eta \widetilde E_{\rm T}}{2}-\frac{t\, \widetilde H_{\rm T}}{4 M^2}  +\frac{t_0\,  \Sigma E}{8M^2}
-
\frac{1}{2}\sqrt{
\left(\widetilde H+ H_{\rm T}  + \frac{t_0}{4M^2} \Sigma \widetilde E \right)^2-
\frac{{\bf\bar\Delta}_\perp^2}{4 M^2}\left(\Sigma E^2- \eta^2 \Sigma \widetilde E^2\right)
}
\\
\frac{H+H_{\rm T}+\eta \widetilde E_{\rm T}}{2}-\frac{t\, \widetilde H_{\rm T}}{4 M^2}  +\frac{t_0\, \Sigma E}{8M^2}
+
\frac{1}{2}\sqrt{
\left(\widetilde H+H_{\rm T} + \frac{t_0}{4M^2} \Sigma \widetilde E \right)^2-
\frac{{\bf\bar\Delta}_\perp^2}{4 M^2}\left(\Sigma E^2- \eta^2 \Sigma \widetilde E^2\right)
}\\
\end{array}
                                   \right\}(x,\eta,t).
\nonumber
\end{eqnarray}
Here, we denote the differences and sums of chiral even and odd quantities as
$$
\left\{{ \Delta E \atop \Sigma E}\right\}  = E \mp \overline{E}_{\rm T}\,,
\quad
\overline{E}_{\rm T} = E_{\rm T} + 2 \widetilde{H}_{\rm T} - \eta \widetilde{E}_{\rm T}\,,
\qquad
\left\{{ \Delta \widetilde E \atop \Sigma \widetilde E}\right\}  = \widetilde E \mp \hat{E}_{\rm T}\,,
\quad
\hat{E}_{\rm T}= E_{\rm T} - \frac{1}{\eta} \widetilde E_{\rm T}\,.
$$
For a ``spherical"  uPDF model of rank-one and an ``axial-symmetrical" uPDF model of the first kind with rank-two  we expect that the corresponding GPD spin-correlation matrix is of rank less than four. Hence, we have at least one vanishing eigenvalue and so one of the following two equations
are satisfied
\begin{eqnarray}
\label{GPD-zeromode1}
&&2\left(H+ \widetilde H-2 H_{\rm T} + \frac{t}{2 M^2} \widetilde H_{\rm T}-\eta \widetilde E_{\rm T}\right)
\left(H-\widetilde H + \frac{t}{2 M^2} \widetilde H_{\rm T}-\eta \widetilde E_{\rm T}\right)=
\\
&&\quad=
\frac{t_0}{M^2}\!\!\left[
\widetilde H - H_{\rm T} + \frac{t}{8M^2}  \Delta\widetilde E
\right]\! \Delta \widetilde E
-
\frac{t_0}{M^2}\!\!\left[
H -H_{\rm T} +\frac{t}{2M^2} \widetilde H_{\rm T}-\eta \widetilde E_{\rm T} +\frac{t+4 \eta^2 M^2 }{4 \eta^2 M^2} \Delta E
\right]\!\Delta E\,,
\nonumber\\
\label{GPD-zeromode2}
&&2\left(H+ \widetilde H+2 H_{\rm T} - \frac{t}{2 M^2} \widetilde H_{\rm T} +\eta \widetilde E_{\rm T}\right)
\left(H-\widetilde H - \frac{t}{2 M^2} \widetilde H_{\rm T}+\eta \widetilde E_{\rm T}\right)=
\\
&&\quad=
\frac{t_0}{M^2}\!\!\left[ \widetilde H + H_{\rm T} + \frac{t}{8M^2}  \Sigma\widetilde E
\right]\! \Sigma \widetilde E
-
\frac{t_0 }{M^2}\!\left[
H +H_{\rm T} -\frac{t}{2M^2} \widetilde H_{\rm T} + \eta \widetilde E_{\rm T} -\frac{t+4 \eta^2 M^2 }{4 \eta^2 M^2} \Sigma E
\right]\!\Sigma E\, .
\nonumber
\end{eqnarray}
As argued above, in the  constraint that holds true each of the three terms separately vanishes,
where the $t$-dependent combination
$$H-\widetilde H \pm \left[\frac{t}{2 M^2} \widetilde{H}_{\rm T} - \eta \widetilde E_{\rm T}\right]\neq 0,$$
as analog of the third relation in Eq.~(\ref{uPDF-constraint-kperp}),
might not be  necessarily equated to zero, however, one of the combinations
$$
H+ \widetilde H\mp2 H_{\rm T} \pm \frac{t}{2 M^2} \widetilde H_{\rm T}\mp \eta \widetilde E_{\rm T} =
H+ \widetilde H\mp (2 \overline H_{\rm T} + \eta \widetilde E_{\rm T})
$$
might vanish.
Consequently, with the requirement that the r.h.s.~of the constraint (\ref{GPD-zeromode1}) [or (\ref{GPD-zeromode2})], i.e.,
$\Delta E =0$ and $\Delta \tilde E =0$ [or $\Sigma E =0$ and $\Sigma \tilde E =0$] we find the following tree linear GPD model constraints
that should be valid for a ``spherical''  model of rank-one or an ``axial-symmetrical'' one of the first kind with rank-two
\begin{eqnarray}
\label{GPD-constraints-linear}
H +\widetilde H \mp \left[2 \overline{H}_{\rm T} + \eta \widetilde E_{\rm T} \right] \stackrel{{\rm sph}^3\atop {\rm ax1}^2}{=} 0\,,\;\;
E \mp \left[E_{\rm T}+ 2 \widetilde{H}_{\rm T} -\eta \widetilde E_{\rm T}\right]\stackrel{{\rm sph}^3\atop {\rm ax1}^2}{=}0\,,
\;\;
\widetilde{E}  \mp  \left[E_{\rm T}-\frac{1}{\eta}\widetilde E_{\rm T} \right]\stackrel{{\rm sph}^3\atop {\rm ax1}^2}{=}0\,.\;\;
\end{eqnarray}

We note that a ``hidden" quadratic relation among transversity GPDs
\begin{eqnarray}
\label{GPD-constraints-quadratic}
4 \overline{H}_{\rm T}(x,\eta,t) \widetilde{H}_{\rm T}(x,\eta,t) +
\left[E_{\rm T}(x,\eta,t)+2 \widetilde{H}_{\rm T}(x,\eta,t)\right]^2 - \left[\widetilde{E}_{\rm T}(x,\eta,t)\right]^2 = 0
\end{eqnarray}
does also not necessarily imply a  degeneration of the eigenvalues. As we will see below in Sec.~\ref{Sec:``Spherical" GPD model constraints}
a quadratic constraint holds true in the DD representation.
However, this relation would necessarily arise in the case that the GPD spin-correlation matrix (\ref{bfF}) of our ``spherical" model possesses three zero modes, i.e.,
the linear combinations
$$
\widetilde H \mp \left[\overline{H}_{\rm T} + \frac{t}{4M^2}\widetilde{H}_{\rm T} \right]
\quad \mbox{and}\quad
H\mp \left[\overline{H}_{\rm T} - \frac{t}{4 M^2} \widetilde H_{\rm T} +\eta \widetilde{E}_{\rm T}\right]
\quad \mbox{or}\quad
H-\widetilde H \pm \left[\frac{t}{2 M^2} \widetilde{H}_{\rm T} - \eta \widetilde E_{\rm T}\right]
$$
as analog of the ``spherical" uPDF model constraints (\ref{uPDF-constraint-kperp}) vanish.
However, as argued above, this is rather unlikely.

In summary, a ``spherical" uPDF model of rank one might imply only one vanishing eigenvalue (\ref{bfF-eigenvalues}) of the GPD spin-correlation matrix (\ref{bfF}).
The same happens for an ``axial-symmetrical"  model of the first kind with a rank-two. In both cases the three
linear constraints (\ref{GPD-constraints-linear}) hold true, where also for the ``spherical" model a possible quadratic relation
(\ref{GPD-constraints-quadratic}) is unlikely to hold.
If these two models have rank-four the third GPD model constraint in Eq.~(\ref{GPD-constraints-linear}), the analog of the saturated Soffer bound (\ref{Soffer-bound-uPDFs}), is not valid and so we are left with the first two  linear constraints in Eq.~(\ref{GPD-constraints-linear}).
For other uPDF models that we have discussed in Sec.~\ref{uPDF-classification}, i.e., the ``axial-symmetrical"   models of the second kind and rank-three ones,   we do not expect that general {\em algebraic} GPD model relations in momentum space representations are valid. However, if the number of  LFWFs utilized in a GPD model is limited and the various couplings are fixed, specific algebraic model relations among GPDs might be found.

\subsubsection{Model constraints for non-forward parton distribution functions}
\label{sec:GPDs-classification-nfPDFs}

In full analogy to our uPDFs model treatment in Sec.~\ref{uPDF-classification}, we can now introduce a classification scheme for nfPDFs models in impact space.  Indeed, comparing the uPDF and GPD spin-density matrices (\ref{tPhi}) and (\ref{bfF-b}), we can read off the replacement rules:
\begin{eqnarray}
\label{GPD02uPDF}
&&\!\!\!\!\!\!  H(x,b) \leftrightarrow f_1(x,{\bf k}_\perp)\,,\;\; \widetilde H(x,b) \leftrightarrow g_1(x,{\bf k}_\perp)\,,\;\;
 \overline{H}_{\rm T}(x,b) \leftrightarrow h_1(x,{\bf k}_\perp) \,,
\\
&&\!\!\!\!\!\!
 E^\prime(x,b) \leftrightarrow -\frac{|{\bf k}_\perp|}{M} f_{1\rm T}^\perp(x,{\bf k}_\perp) \,,
\;\;
\overline{E}^\prime_{\rm T}(x,b) \leftrightarrow -\frac{|{\bf k}_\perp|}{M} h_{1}^\perp(x,{\bf k}_\perp)\,,
\quad
\widetilde{H}_{\rm T}^{\prime\prime}(x,b)  \leftrightarrow \frac{{\bf k}^2_\perp}{2M^2}h_{1\rm T}^\perp(x,{\bf k}_\perp) \,.
\nonumber
\end{eqnarray}
Also note that $
H_{\rm T}(x,b) =  \overline{H}_{\rm T}(x,b) + \widetilde{H}_{\rm T}^{\prime\prime}(x,b) \leftrightarrow h_1(x,{\bf k}_\perp) + \frac{{\bf k}^2_\perp}{2M^2}h_{1\rm T}^\perp(x,{\bf k}_\perp)$.
Setting $g_{1\rm T}^\perp$ and $h_{1\rm L}^\perp$ to zero in the model constraints of Sec.~\ref{uPDF-classification} will then provide us the one-to-one correspondence with the nfPDF constraints, which we present now.
The eigenvalues
\begin{eqnarray}
\label{bfF-eigenvalues-b}
\widetilde {\bf F}^{\rm e.v.} (x,{b})=
 \left\{\!\!
\begin{array}{c}
\frac{H-H_{\rm T} }{2}+ \widetilde{H}_{\rm T}^{\prime\prime}
-
\frac{1}{2}\sqrt{
\left(\widetilde H-H_{\rm T}\right)^2+
\left(E^\prime - \overline{E}^\prime_{\rm T} \right)^2
}
\\
\frac{H-H_{\rm T}}{2}+ \widetilde{H}_{\rm T}^{\prime\prime}
+
\frac{1}{2}\sqrt{
\left(\widetilde H-H_{\rm T}\right)^2+
\left(E^\prime - \overline{E}^\prime_{\rm T} \right)^2
}
\\
\frac{H+H_{\rm T} }{2}- \widetilde{H}_{\rm T}^{\prime\prime}
-
\frac{1}{2}\sqrt{
\left(\widetilde H+H_{\rm T}\right)^2+
\left(E^\prime +\overline{E}^\prime_{\rm T} \right)^2
}
\\
\frac{H+H_{\rm T}}{2}- \widetilde{H}_{\rm T}^{\prime\prime}
+
\frac{1}{2}\sqrt{
\left(\widetilde H+H_{\rm T}\right)^2+
\left(E^\prime +\overline{E}^\prime_{\rm T} \right)^2
}\\
\end{array}
                                   \right\}(x,b)
\end{eqnarray}
of the spin-density matrix (\ref{bfF-b}) might be also obtained by means of the replacement rules (\ref{GPD02uPDF})
from those for uPDFs (\ref{tPhi}). It is also useful to have the formal correspondence to the eigenvalues (\ref{bfF-eigenvalues})
of the GPD spin-correlation matrix in momentum space for the $\eta=0$ case,
\begin{eqnarray}
\label{GPD02GPD0}
&&\!\!\!\!\!\!
H(x,b) \leftrightarrow H(x,0,t)\,, \quad \widetilde H(x,b) \leftrightarrow \widetilde H(x,0,t) \,, \quad
H_{\rm T}(x,b) \leftrightarrow H_{\rm T}(x,0,t)\,,
\\
&&\!\!\!\!\!\!
E^\prime(x,b) \leftrightarrow -\frac{\sqrt{-t}}{2M} E(x,0,t)\,, \quad
\overline{E}_{\rm T}^\prime(x,b) \leftrightarrow -\frac{\sqrt{-t}}{2M} \overline{E}_{\rm T}(x,0,t)
\,, \quad
\widetilde{H}_{\rm T}^{\prime\prime}(x,b) \leftrightarrow \frac{t}{4M^2}\widetilde{H}_{\rm T}(x,0,t)\,,
\nonumber
\end{eqnarray}
where $|{\bf \bar \Delta}_\perp| = -\sqrt{-t}$ and $\eta$ proportional terms such as  $\eta \widetilde E$, $\eta \hat E_{\rm T}$, and $t_0\times \cdots$ drop out.

For a ``spherical" or ``axial-symmetrical" model of the first kind one set of the linear constraints
\begin{eqnarray}
\label{GPD0-constraint1}
\left[\widetilde H\mp H_{\rm T} \right](x,b) \stackrel{\rm sph\atop ax1}{=}0\,,
\;\;
\left[E\mp \overline{E}_{\rm T}\right](x,b)\stackrel{\rm sph \atop ax1}{=}0\,,
\;\;   \lim_{\eta\to 0}\left[
\widetilde E \mp E_{\rm T} \pm \frac{1}{\eta} \widetilde{E}_{\rm T}
\right](x,b)\stackrel{\rm sph \atop ax1}{=}0\;
\end{eqnarray}
is valid.
Here, the last equation does not follow from the degeneracy of the eigenvalues rather it is the analog of the third formula in Eq.~(\ref{GPD-constraints-linear}). We also conclude from
Eq.~(\ref{GPD-constraints-linear}) that the equalities among first derivative quantities, e.g., $E^\prime=\pm \overline{E}^\prime_{\rm T}$,
are valid for the quantities itself. In the case that a ``spherical" or an ``axial-symmetrical" model of the first kind  possesses a rank-one or
-two spin-density matrix, respectively, the three constraints (\ref{GPD0-constraint1})  are supplemented by
\begin{eqnarray}
\label{GPD0-constraint3}
H(x,b)  \mp \left[H_{\rm T}(x,b) -  2\widetilde H_{\rm T}^{\prime\prime}(x,b)\right]   \stackrel{{\rm sph}^3\atop {\rm ax1}^2}{=}0
\quad \mbox{or}\quad
H(x,b) + \widetilde H(x,b) \mp 2\overline{H}_{\rm T}(x,b) \stackrel{{\rm sph}^3\atop{\rm ax1}^2}{=}0 \quad
\end{eqnarray}
with $\overline{H}_{\rm T}(x,b)= H_{\rm T}(x,b)- \widetilde{H}_{\rm T}^{\prime\prime}(x,b)$.
Here, the latter constraint results from the combination of the former with the first model equality of the set (\ref{GPD0-constraint1}). It is the analog of the saturated Soffer bound (\ref{Soffer-bound-uPDFs}) for uPDFs and corresponds to the first algebraic condition in Eq.~(\ref{GPD-constraints-linear}) for a
GPD model in momentum space with rank less than four.
For a ``spherical" nfPDF model the quadratic relation
\begin{eqnarray}
\label{GPD0-constraint2}
 4 \widetilde H^{\prime\prime}_{\rm T}(x,b)  \overline{H}_{\rm T}(x,b)+\left[E_{\rm T}^\prime(x,b)+2\widetilde H_{\rm T}^\prime(x,b)\right]^2
 \stackrel{\rm sph}{=}0
\end{eqnarray}
must hold true in addition to the constraints (\ref{GPD0-constraint1}) and eventually to the rank-one condition (\ref{GPD0-constraint3}).

Analogously to Eq.~(\ref{constraint-axialsym}), for an ``axial-symmetrical" model  of the second kind one of the conditions
\begin{eqnarray}
\label{GPD0-constraint-axialsym}
&&\left[\widetilde{H}(x,b)\right]^2 - \left[H_{\rm T}(x,b)\right]^2
+8\widetilde{H}^{\prime\prime}_{\rm T}(x,b)\left[
H_{\rm T} -\widetilde{H}^{\prime\prime}_{\rm T}\right]\!(x,b)
+\left[E^{\prime}(x,b)\right]^2+\left[\overline{E}^{\prime}_{\rm T}(x,b)\right]^2=
\nonumber\\
&&\qquad\stackrel{\rm ax2}{=}\pm \left[
\sqrt{\left(\widetilde H-H_{\rm T}\right)^2+\left(E^\prime- \overline{E}_{\rm T}^\prime\right)^2}
\sqrt{\left(\widetilde H+H_{\rm T}\right)^2+\left(E^\prime+\overline{E}_{\rm T}^\prime\right)^2}
\right]\!\!(x,b)
\end{eqnarray}
is satisfied. If such a model has rank-two, the two quadratic constraints
\begin{eqnarray}
\label{GPD0-constraint-rank2-2a}
&&\!\!\!\!\!\!
\left[H(x,b)\right]^2-\left[\widetilde{H}(x,b)\right]^2-4\widetilde{H}^{\prime\prime}_{\rm T}(x,b)\left[
H_{\rm T} -\widetilde{H}^{\prime\prime}_{\rm T}\right]\!(x,b)
-
\left[E^\prime(x,b)\right]^2-\left[\overline{E}^\prime_{\rm T}(x,b)\right]^2\stackrel{{\rm ax2}^2}{=}0\,,
\qquad
\\
\label{GPD0-constraint-rank2-2b}
&&\!\!\!\!\!\!
\left[H(x,b)-\widetilde{H}(x,b)\right]H_{\rm T}(x,b)-
2 H(x,b)\widetilde{H}^{\prime\prime}_{\rm T}(x,b) -E^\prime(x,b) \overline{E}^\prime_{\rm T}(x,b)\stackrel{{\rm ax2}^2}{=}0
\end{eqnarray}
are separately valid, cf.~Eqs.~(\ref{constraint-rank2-2a},\ref{constraint-rank2-2b}). Finally, in the case of a rank-three model only one
eigenvalue (\ref{bfF-eigenvalues}) vanishes, i.e., one of the following two quadratic conditions holds true:
\begin{eqnarray}
\left[H+\widetilde{H} \mp H_{\rm T} \pm 2 \widetilde{H}_{\rm T}\right]
\left[H-\widetilde{H} \pm 2 \widetilde{H}^{\prime\prime}_{\rm T}\right](x,b)
-
\left[E^\prime\mp \overline{E}^\prime_{\rm T}\right]^2(x,b) \stackrel{{\rm mod}^3}{=}0\,,
\end{eqnarray}
cf.~Eqs.~(\ref{constraint-rank2-1a},\ref{constraint-rank2-1b}).

Let us finally have a closer look to a ``spherical" model or ``axial-symmetrical" model of the first kind.
The main difference to the GPD model classification in  momentum space is the appearance of the first equality in the model constraints (\ref{GPD0-constraint1}). Utilizing the representation
$$
\widetilde{H}_{\rm T}^{\prime\prime}(x,b)
=
\frac{1}{4 M^2} \frac{\partial^2}{\partial {\bf b}^2}\widetilde{H}_{\rm T}(x,b)  -
\frac{1}{2 M^2\,b^2}\; {\bf b}\cdot \frac{\partial}{\partial {\bf b}}\widetilde{H}_{\rm T}(x,b)
\quad\mbox{and}\quad H_{\rm T}(x,b)  = \overline{H}_{\rm T}(x,b)+ \widetilde{H}_{\rm T}^{\prime\prime}(x,b)\,,
$$
we can convert this model relation into the momentum space and we might also generalize it to the $\eta \neq 0$ case
\begin{eqnarray}
\label{GPD-constraints-linear-2}
\widetilde H(x,\eta,t)  \mp \left[ H_{\rm T}(x,\eta,t)   +
\int_{-\infty}^t\!\frac{ dt^\prime}{4M^2}\, \widetilde{H}_{\rm T}(x,\eta,t^\prime)
\right]
\stackrel{{\rm sph}^3\atop {\rm ax1}^2}{=} 0\,,
\end{eqnarray}
where we assumed that $\widetilde H(x,b) $ and $\overline{H}_{\rm T}(x,b)$ corresponds to
$\widetilde H(x,\eta,t)$ and $\overline{H}_{\rm T}(x,\eta,t)$, respectively,  and the GPDs vanish at $-t\to \infty$.
Certainly, this integral relation differs from the analog algebraic  condition for a ``spherical" model of rank-one in momentum space,
$$
\widetilde H(x,\eta,t) \mp H_{\rm T}(x,\eta,t)\,,$$
which one would read off  with $\Delta E=0 $ or $\Sigma E=0 $ from the eigenvalues (\ref{bfF-eigenvalues})  of the spin-correlation matrix (\ref{bfF}).
In the case the Soffer bound is saturated, we can combine Eq.~(\ref{GPD-constraints-linear-2}) with  Eq.~(\ref{GPD-constraints-linear}) to express the
GPD $H$  by the chiral odd ones,
\begin{eqnarray}
\label{GPD-constraints-linear-2a}
H(x,\eta,t) \mp
\left[
H_{\rm T}(x,\eta,t) -
\int_{-\infty}^t\!\frac{ dt^\prime}{4M^2}\, \widetilde{H}_{\rm T}(x,\eta,t^\prime)+\eta \widetilde E_{\rm T}(x,\eta,t)
\right] \stackrel{ {\rm sph}^3 \atop {\rm ax1}^2}{=} 0.
\end{eqnarray}

\subsection{Non-forward parton densities and GPD/uPDF correspondences}
\label{sec:GPDs-uGPDs}

Both GPDs and uPDFs  allow for the resolution of transverse degrees of freedom and since both of them are embedded in uGPDs one might wonder
whether GPDs and uPDFs can be more directly related to each other. From the Field theoretical definitions such as (\ref{uPDF-gen}) and (\ref{F-gen}), one can formally write down sum rules for twist-two related uPDFs and GPDs,
\begin{eqnarray}
\label{H2q}
H(x,\eta=0,t=0,\mu^2) &\!\!\! = \!\!\!& \intk\, f_1(x,{\bf k}_{\perp})\,,
\\
\label{tH2Dq}
\widetilde H(x,\eta=0,t=0,\mu^2) &\!\!\! = \!\!\!& \intk\, g_1(x,{\bf k}_{\perp})\,,
\\
\label{HT2dq}
H_{\rm T}(x,\eta=0,t=0,\mu^2)  &\!\!\! = \!\!\!& \intk\,  h_1(x,{\bf k}_{\perp})
\,,
\end{eqnarray}
which suggest that GPDs and uPDFs have somehow a cross talk.
There are attempts to find such cross-talks  on a more generic basis \cite{DieHag05} or within models \cite{MeiMetGoe07}. Thereby,  often one employs in such studies the impact parameter space for GPDs and the momentum space for uPDFs.
In our opinion, see below the considerations in Sec.~\ref{Sec-GPDs}, it is more useful to consider both sets of distributions only in
one space rather two different ones.
To get some insights in the expected cross talks we employ the LFWFs overlap representations in momentum space. Thereby, we will restrict ourselves to
nfPDFs (zero-skewness GPDs).

\begin{table}[t]
\begin{center}
\begin{tabular}{|c||c|c||c|c||c|c|c} \hline
$|\Delta L^z|$ & \multicolumn{2}{c||}{parity even}&  \multicolumn{2}{c||}{parity odd}& \multicolumn{2}{c|}{chiral odd} \\
   &  GPD &     uPDF &  GPD &     uPDF &  GPD &  uPDF \\  \hline\hline
  0 & $H$ & $f_1$ & $\widetilde H$ & $g_1$ & $\overline{H}_{\rm T}$ &  $h_1$ \\
  1 & -- &  -- & $\widetilde E $ & $-g_{1{\rm T}}$ & $\hat{E}_{\rm T}$  & $h_{1{\rm L}}^\perp$ \\
  1 & $E$ &  $i f_{1{\rm T}}^\perp$ & -- & -- &$ \overline{E}_{\rm T}$ & $i h_{1}^\perp$ \\
  2 & --  & --  & --  &  --  & ${\widetilde H}_{\rm T}$ & $-h_{1 \rm T}^\perp/2$ \\
  \hline
\end{tabular}
\end{center}
\caption{\label{Tab-DeltaL}\small
Classification of chiral even and odd twist-two GPDs (\ref{HE-def}--\ref{FT-def}) as well as leading--power uPDFs (\ref{Phi^{[gamma^+]},Phi^{[gamma^+gamma^5]}},
\ref{[sigma^{i+}gamma^5]}) with respect to the orbital angular momentum transfer $\Delta L^z$ in the LFWF overlap representation.
}
\end{table}
Comparing the spin-density matrices (\ref{bfF}) and  (\ref{tPhi}) we read off besides the substitution rules ${\bf k}_\perp \to {\bf \Delta}_\perp/2 $
the desired correspondences. In Tab.~\ref{Tab-DeltaL} we group the eight leading twist-two nfPDFs and leading--power uPDFs w.r.t.~the $\Delta L^z$ orbital angular momentum transfer (\ref{Delta L^z}).
Note that $\Delta L^z =0$ LFWF overlap quantities are those that connect directly twist-two distributions, while  $|\Delta L^z| =1$ relate twist-two nucleon helicity-flip nfPDFs with twist-three PDFs. The $|\Delta L^z| =2$ orbital angular momentum transfer can only appear in the chiral odd sector and relates the twist-two nfPDF ${\widetilde H}_{\rm T}$ with the  ``pretzelosity" distribution, related to a twist-four PDF.
To have a closer look to how nfPDFs and uPDFs are related to each other,
we employ the overlap representations of these parton distributions,
where we distinguish between the three cases $|\Delta L^z| \in \{0,1,2\}$.

\subsubsection{Twist-two related uGPDs with $\Delta L^z =0$}
\label{sec:GPD2uPDF-Lz0}

As mentioned above, the three twist-two zero-skewness GPDs and PDFs can be embedded in twist-two related uGPDs:
\begin{eqnarray}
\label{GPD2PDF-twist2}
H(x,0,t)\leftarrow   H(x,0,{\bf \Delta}_\perp,{{\bf k}}_{\perp})\rightarrow f_1(x,{\bf k}_\perp) \,,
&&
\widetilde H(x,0,t) \leftarrow \widetilde H(x,0,{\bf \Delta}_\perp,{{\bf k}}_{\perp}) \rightarrow g_1(x,{\bf k}_\perp)\,,
\nonumber\\
 \overline{H}_{\rm T}(x,0,t) \leftarrow
 \overline{H}_{\rm T}(x,0,{\bf \Delta}_\perp,{{\bf k}}_{\perp})
 \rightarrow h_1(x,{\bf k}_\perp)   \,.\!\!\!\!&&
%
\end{eqnarray}
To find the LFWF overlap contributions for the twist-two related zero-skewness uGPDs,
we simply drop in the LFWFs overlaps  (\ref{H-LFWF}), (\ref{tH-LFWF}) and  (\ref{oHT-LFWF}) the ${\bf k}_\perp$ integration and set $\eta=0$,
\begin{eqnarray}
\label{H^{(n)}}
H^{(n)}(x,\eta=0,{\bf \Delta}_\perp,{{\bf k}}_{\perp}) &\!\!\! = \!\!\!&  \intsum
\left[
\psi^{\ast\,\Rightarrow}_{\rightarrow,n} \psi^{\Rightarrow}_{\rightarrow,n}
+
\psi^{\ast\,\Rightarrow}_{\leftarrow,n} \psi^{\Rightarrow}_{\leftarrow,n}
\right]\left(X_i,{{\bf k}^\prime}_{\perp i},s_i|X_i,{{\bf k}}_{\perp i},s_i\right),
\\
\label{tH^{(n)}}
\widetilde H^{(n)}(x,\eta=0,{\bf \Delta}_\perp,{{\bf k}}_{\perp}) &\!\!\! = \!\!\!& \intsum
\left[
\psi^{\ast\,\Rightarrow}_{\rightarrow,n} \psi^{\Rightarrow}_{\rightarrow,n}
- \psi^{\ast\,\Rightarrow}_{\leftarrow,n} \psi^{\Rightarrow}_{\leftarrow,n}
\right]\left(X_i,{{\bf k}^\prime}_{\perp i},s_i|X_i,{{\bf k}}_{\perp i},s_i\right),
\\
\label{bH^{(n)}_T}
\overline{H}^{(n)}_{\rm T}(x,\eta=0,{\bf \Delta}_\perp,{{\bf k}}_{\perp}) &\!\!\! = \!\!\!&
\frac{1}{2}\;\intsum
\psi^{*\,\Rightarrow}_{\rightarrow,n}(X_i,{{\bf k}}^\prime_{\perp i},s_i) \psi^\Leftarrow_{\leftarrow,n}(X_i, {\bf k}_{\perp i},s_i) +\mbox{c.c.}\,.
\end{eqnarray}
Here, we use the shorthand (\ref{intsum}) for the $(n-1)$ particle phase space integration and the outgoing transverse momenta
(\ref{t2-1}) simplify to
\begin{eqnarray}
\label{t2-1a}
{{\bf k}}^\prime_{\perp 1} ={{\bf k}}^{~}_{\perp 1}
-(1-x)\, {\bf \Delta}_\perp\,, \quad
{{\bf k}}^\prime_{\perp i} ={{\bf k}}^{~}_{\perp i}
+ X_i\, {{\bf \Delta}}_\perp \quad\mbox{for}\quad i\in\{2,\cdots,n\}\,.
\end{eqnarray}
Compared to the incoming transverse momenta, they are shifted by ${\bf \Delta}_\perp$, weighted with a longitudinal momentum fraction.
If we now integrate over ${\bf k}_\perp$ or drop the ${\bf \Delta}_\perp$ dependence, we establish the definitions of nfPDFs or uPDFs,
see definitions (\ref{OveRep-f1},\ref{OveRep-g1},\ref{OveRep-h1}), respectively,
\begin{eqnarray}
\label{F-overlap}
F(x,\eta=0,t) &\!\!\! = \!\!\!&
\sum_{n}
\intk\,  F^{(n)}(x,\eta=0,{\bf \Delta}_\perp,{\bf k}_\perp)\quad\mbox{for}\quad F\in\{H,\widetilde H, H_{\rm T}\}\,,
\\
\label{q-overlap}
q(x,{\bf k}_\perp) &\!\!\! = \!\!\!&
\sum_{n} F^{(n)}(x,\eta=0,{\bf \Delta}_\perp =0,{\bf k}_\perp) \qquad\mbox{for}\quad q\in\{f_1,g_1, h_1\}\,.
\end{eqnarray}
Obviously, from these representations one  easily obtains the sum rules (\ref{GPD2PDF-twist2}).


We realize also from the LFWF overlaps (\ref{H^{(n)}}-\ref{bH^{(n)}_T})
that the $t$-dependency is induced by the ${\bf k}_\perp$-dependency, however, it seems to be impossible to find a model independent
one-to-one map of zero-skewness GPDs and uPDFs. Indeed, LFWFs  overlaps in GPDs, e.g.,
$$
\propto \left[
 e^{-i \varphi}  - (1-x)\frac{|{\bf \Delta}_\perp|}{|{\bf k}_\perp|}  e^{-i \phi}
\right]^{\bar{L}^z}  e^{i \overline{L}^z \varphi} \quad \stackrel{|{\bf \Delta}_\perp| \to 0}{\Longrightarrow}\quad  \propto 1
$$
depend on the partonic overall orbital angular momentum $\bar{L}^z$, see Eq.~(\ref{l^z}), which
imply  specific contributions to the $t$-dependence of GPDs,  while in uPDFs  these phase differences disappear.
Hence, we conclude that in QCD a general map of GPDs to uPDFs or reversely might not exist.

However, let us illustrate that for any model which is based on an effective two-body LFWF that is real valued  and free of nodes we can map a generic scalar ``uPDF" to the generic scalar ``nfPDF".  For doing so we switch to the impact parameter space, where the convolution integral of our ``nfPDF"
\begin{eqnarray}
\label{GPD0-conv}
F(x,\eta=0,t)= \frac{1}{1-x} \intk\,
\phi^\ast\!\left(x,{\bf k}_{\perp}- (1-x){\bf \Delta}_\perp\right)
\phi\!\left(x,{\bf k}_{\perp}\right),
\end{eqnarray}
e.g.,~the scalar version of the spin-correlation function (\ref{bfF-rank3}),
turns into the product of the corresponding LFWFs\footnote{Here $\widetilde F$ denotes the Fourier transform of the generic scalar ``nfPDF"  $F$  and the symbol should not be confused with, e.g., parity odd GPDs. }:
\begin{eqnarray}
\widetilde F(x ,\eta=0,{\bf b}) &\!\!\! =\!\!\! & \int\!\!\!\!\int\!\frac{d^2{\bf \Delta}_\perp}{(2\pi)^2} e^{-i{\bf b}\cdot {\bf \Delta}_\perp} F(x ,\eta=0,t=-{\bf \Delta}^2_\perp)
 \nonumber \\
 &\!\!\! =\!\!\! & \frac{(2\pi)^2}{(1-x)^3} \widetilde{\phi}^\ast(x,{\bf b}/(1-x))\widetilde{\phi}(x,{\bf b}/(1-x))\,,
\end{eqnarray}
where the LFWF in the impact parameter space is the Fourier transform
\begin{eqnarray}
\label{FT-Phi}
\widetilde{\phi}(x,{\bf b}) &\!\!\! =\!\!\! & \int\!\!\!\!\int\!\frac{d^2{\bf k}_\perp}{(2\pi)^2} e^{-i{\bf b}\cdot {\bf k}_\perp} \phi(x,{\bf k}_\perp)
\nonumber\\
&\!\!\! =\!\!\! & \sqrt{1-x}
\int\!\!\!\!\int\!\frac{d^2{\bf k}_\perp}{(2\pi)^2} e^{-i{\bf b}\cdot {\bf k}_\perp} \sqrt{{\bf \Phi}(x,{\bf k}_\perp)}
\,.
\end{eqnarray}
of the square root of the diagonal LFWF overlap, which is the corresponding ``uPDF"
\begin{eqnarray}
\label{q-overlap-1}
{\bf \Phi}(x,{\bf k}_\perp)= \frac{\phi^2(x,{\bf k}_{\perp})}{1-x}.
\end{eqnarray}
Hence, within our assumptions the zero-skewness ``GPD"  in impact parameter space can be obtained from the ``uPDF":
\begin{eqnarray}
\label{q2F-b}
\widetilde{F}(x ,\eta=0,{\bf b}) = \frac{(2\pi)^2}{(1-x)^2}
\left|\int\!\!\!\!\int\!\frac{d^2{\bf k}_\perp}{(2\pi)^2} e^{-i{\bf b}\cdot {\bf k}_\perp/(1-x)} \sqrt{{\bf \Phi}(x,{\bf k}_\perp)}\right|^2\,.
\end{eqnarray}
Mapping back this ``nfPDF" into momentum space,
\begin{eqnarray}
\label{q2F}
F(x ,\eta=0,t=-{\bf \Delta}_\perp^2) =\intk   \sqrt{{\bf \Phi}(x,{\bf k}_\perp-(1-x){\bf\Delta}_\perp)} \sqrt{{\bf \Phi}(x,{\bf k}_\perp)}\,,
\end{eqnarray}
is nothing but taking the square root of the LFWF overlap (\ref{q-overlap-1}) and putting it into the convolution (\ref{GPD0-conv}).
Obviously, if we have overlap contributions from different states, even a superposition of different spectator quark models, the conversion formulae (\ref{q2F-b},\ref{q2F}) are not valid. Hence, even in models the information that is encoded in a uPDF is not sufficient to restore the $t$-dependence of the corresponding nfPDF.

\subsubsection{Mismatches and new model relations in the $|\Delta L^z|=1$ sector}
\label{sec:GPD-relations-1}

We now consider the $|\Delta L^z|=1$ contributions, arising from the off-diagonal LFWF overlaps of $l^z=0$ and $|l^z|=1$ LFWFs. Here, the four  helicity non-conserved twist-two nfPDFs might be somehow related with four  twist-three related uPDFs,
see Tab.~\ref{Tab-DeltaL}.  We already stated  above, see Eq.~(\ref{uGPD-definition}), that these nfPDFs and uPDFs are parameterized by
different sets of uGPDs. The first set of such relations
\begin{eqnarray}
\label{GPD2FPD-twis3-1}
E(x,0,t) \leftarrow E(x,0,{\bf \Delta}_\perp,{{\bf k}}_{\perp})
& \stackrel{?}{\leftrightarrow} &
i f_{1\rm T}^\perp(x,0,{\bf \Delta}_\perp,{{\bf k}}_{\perp})\rightarrow i f_{1\rm T}^\perp(x,{\bf k}_\perp) \,,
\\
\overline{E}_{\rm T}(x,0,t)  \leftarrow \overline{E}_{\rm T}(x,0,{\bf \Delta}_\perp,{{\bf k}}_{\perp})
&\stackrel{?}{\leftrightarrow}&
  i h_{1}^\perp(x,0,{\bf \Delta}_\perp,{{\bf k}}_{\perp}) \rightarrow i h_{1}^\perp(x,{\bf k}_\perp)  \,,
\nonumber
\end{eqnarray}
refers to the correspondence of $T$-even  nfPDF $E$ and the $T$-odd Sivers function $f_{1\rm T}^\perp$, in both  the quarks are unpolarized and the nucleon is polarized,
and to the correspondence of  transverity   $T$-even nfPDF  $\overline{E}_{\rm T}$  and the $T$-odd Boer-Mulders function $ h_{1}^\perp$, where now the quarks are
transversally polarized and the nucleon is unpolarized.
It has been intensively argued in the literature, see, e.g.~Refs.~\cite{Bur02a}, that $T$-even nfPDF  $E$ is related with the $T$-odd Sivers function.
However, in a pure quark model the Sivers function simply vanishes, telling us that the connection to the twist-two  nfPDF $E$ is lost. Including a transverse gauge link
in the quark correlators  one might establish a  model dependent nfPDF/uPDF connection, where it has been suggested to write this as a convolution integral with a so-called lensing function.

The second set of relations
\begin{eqnarray}
\label{GPD2FPD-twis3-2}
\widetilde{E}(x,0,t) \leftarrow\widetilde{E}(x,0,{\bf \Delta}_\perp,{{\bf k}}_{\perp})
& \stackrel{?}{\leftrightarrow} &
-g_{1 \rm T}^\perp(x,0,{\bf \Delta}_\perp,{{\bf k}}_{\perp})  \rightarrow - g_{1 \rm T}^\perp(x,{\bf k}_\perp)\,,
\\
\hat E_{\rm T}(x,0,t) \leftarrow\hat{E}_{\rm T}(x,0,{\bf \Delta}_\perp,{{\bf k}}_{\perp})
& \stackrel{?}{\leftrightarrow} &
h_{1 \rm L}^\perp(x,0,{\bf \Delta}_\perp,{{\bf k}}_{\perp})
\rightarrow h_{1 \rm L}^\perp(x,{\bf k}_\perp)
\nonumber
\,,
\end{eqnarray}
where $\hat E_{\rm T}(x,0,t)= \lim_{\eta\to 0} E_{\rm T}(x,\eta,t)-\widetilde E_{\rm T}(x,\eta,t)/\eta$,
tells us that the nfPDF $\widetilde E$  and $\hat E_{\rm T}$ might be related to the transverse polarized  uPDF $g_{1{\rm T}}$ and
the longitudinal polarized one $h_{1 \rm L}^\perp$ , respectively. Since of the $T$-parity mismatch these nfPDFs drop out in the
spin-density matrix and so a possible relation among them is to our best knowledge not discussed in the literature.

To understand that  nfPDF/uPDF relations (\ref{GPD2FPD-twis3-1},\ref{GPD2FPD-twis3-2}) can be found in model studies, however, might not exist in QCD
we define for $\eta \ge 0$ the ${\bf k}_{\perp}$ unintegrated  $|\Delta L^z|=1$ LFWF overlaps as
\begin{eqnarray}
\label{bbE}
\mathbb{E}^{(n)}(x,\eta,{\bf \Delta}_\perp,{{\bf k}}_{\perp}) 
&\!\!\! =\!\!\! &
\intsum
\left[
\psi^{*\Rightarrow}_{\leftarrow,n} \, \psi^{\Leftarrow}_{\leftarrow,n} +
\psi^{*\,\Rightarrow}_{\rightarrow,n}\, \psi^{\Leftarrow}_{\rightarrow,n}
\right](X^\prime_i,{{\bf k}}^\prime_{\perp i},s_i|X_i, {{\bf k}}_{\perp i},s_i)+\mbox{c.c.}\,,
\\
\widetilde{\mathbb{E}}^{(n)}(x,\eta,{\bf \Delta}_\perp,{{\bf k}}_{\perp}) 
&\!\!\! =\!\!\! &
\intsum
\left[
\psi^{*\Rightarrow}_{\leftarrow,n} \, \psi^{\Leftarrow}_{\leftarrow,n} -
\psi^{*\,\Rightarrow}_{\rightarrow,n}\, \psi^{\Leftarrow}_{\rightarrow,n}
\right](X^\prime_i,{{\bf k}}^\prime_{\perp i},s_i|X_i, {{\bf k}}_{\perp i},s_i)
+\mbox{c.c.}\,,
\\
\overline{\mathbb{E}}^{(n)}(x,\eta,{\bf \Delta}_\perp,{{\bf k}}_{\perp}) 
&\!\!\! =\!\!\! &
\intsum
\left[-
\psi^{\ast\, \Rightarrow}_{\rightarrow,n}\, \psi^{\Rightarrow}_{\leftarrow,n}
-
\psi^{\ast\, \Leftarrow}_{\rightarrow,n} \, \psi^{\Leftarrow}_{\leftarrow,n}
\right](X^\prime_i,{{\bf k}}^\prime_{\perp i},s_i|X_i, {{\bf k}}_{\perp i},s_i)
+\mbox{c.c.}
\\
\label{bbhE}
\hat{\mathbb{E}}^{(n)}(x,\eta,{\bf \Delta}_\perp,{{\bf k}}_{\perp}) 
&\!\!\! =\!\!\! &
\intsum
\left[
\psi^{\ast\, \Rightarrow}_{\rightarrow,n}\, \psi^{\Rightarrow}_{\leftarrow,n}
-
\psi^{\ast\, \Leftarrow}_{\rightarrow,n} \, \psi^{\Leftarrow}_{\leftarrow,n}
\right](X^\prime_i,{{\bf k}}^\prime_{\perp i},s_i|X_i, {{\bf k}}_{\perp i},s_i)
+\mbox{c.c.}
\end{eqnarray}
where now the $(n-1)$ parton phase space integral is defined as
\begin{eqnarray}
\label{intsum1a}
\intsum \cdots \equiv \sum_{s_2,\dots, s_n}
\int\![dX\, d^2{\bf k}_{\perp}]_n \left(\frac{1+\eta}{1-\eta}\right)^{n-2}\,
 \frac{1}{\sqrt{1-\eta^2}}\, \delta\!\left(\frac{x+\eta}{1+\eta}-X_{1}\right) \delta^{(2)}({\bf k}_\perp-{\bf k}_{1 \perp})
\cdots\,,
\end{eqnarray}
see Eq.~(\ref{intsum11}), and the outgoing momenta are specified in Eq.~(\ref{t2-1}).
All these four unintegrated $|\Delta L^z|=1$ LFWF overlaps enter in both nfPDF and uPDF definitions, however, with
different projections on the phase. According to
Eqs.~(\ref{f_{1T}^perp-overlap-2},\ref{E-LFWF}), (\ref{g_{1T}^perp-overlap},\ref{tE-LFWF}), (\ref{h_1T^perp-overlap-2},\ref{oET-LFWF}), and
(\ref{h1L^perp-overlap},\ref{hET-LFWF}) we decompose the zero-skewness uGPDs as
\begin{eqnarray}
\label{uGPDs-E-decomposition}
\mathbb{E}^{(n)}\!(x,0,{\bf \Delta}_\perp,{{\bf k}}_{\perp}) &\!\!\! = \!\!\! &
\frac{2k^2}{M}f_{1\rm T}^{(n)\perp}\!(x,\eta=0,{\bf \Delta}_\perp,{{\bf k}}_{\perp}) +
\frac{\Delta^1}{M}  E^{(n)}\!(x,\eta=0,{\bf \Delta}_\perp,{{\bf k}}_{\perp})\,,
\\
\widetilde{\mathbb{E}}^{(n)}\!(x,\eta,{\bf \Delta}_\perp,{{\bf k}}_{\perp}) &\!\!\! = \!\!\! &
\frac{-2k^1}{M} g_{1\rm T}^{(n)\perp}\!(x,\eta=0,{\bf \Delta}_\perp,{{\bf k}}_{\perp}) +
\frac{\eta \Delta^1}{M}  \widetilde E^{(n)}\!(x,\eta=0,{\bf \Delta}_\perp,{{\bf k}}_{\perp}) +{\cal O}(\eta^2),
\qquad\\
\overline{\mathbb{E}}^{(n)}\!(x,0,{\bf \Delta}_\perp,{{\bf k}}_{\perp}) &\!\!\! = \!\!\! &
\frac{-2k^2}{M}h_{1}^{(n)\perp}\!(x,\eta=0,{\bf \Delta}_\perp,{{\bf k}}_{\perp}) +
\frac{\Delta^1}{M}  \overline{E}_{\rm T}^{(n)}\!(x,\eta=0,{\bf \Delta}_\perp,{{\bf k}}_{\perp})\,,
\\
\label{uGPDs-hE-decomposition}
\hat{\mathbb{E}}^{(n)}\!(x,\eta,{\bf \Delta}_\perp,{{\bf k}}_{\perp}) &\!\!\! = \!\!\! &
\frac{2k^1}{M} h_{1\rm L}^{(n)\perp}\!(x,\eta=0,{\bf \Delta}_\perp,{{\bf k}}_{\perp}) +
\frac{\eta \Delta^1}{M}  \hat E_{\rm T}^{(n)}\!(x,\eta=0,{\bf \Delta}_\perp,{{\bf k}}_{\perp}) +{\cal O}(\eta^2)\,.
\end{eqnarray}
Here specific care has to be taken on the definition of $\eta$ proportional $T$-odd GPD overlaps $\widetilde E^{(n)}$ and $\hat E_{\rm T}^{(n)}$.
Since the two projections are induced by QCD dynamics, which is a priory unknown, we should consider them as independent.
Consequently, even nfPDF/uPDF sum rules such as given in Eqs.~(\ref{H2q}--\ref{HT2dq}) for $\Delta L^z=0$ LFWF overlap
quantities (or twist-two related ones) cannot be derived.  Hence, we can only state that one can
obtain from $|\Delta L^z|=1$ uGPDs zero-skewness twist-two GPDs at t=0 and twist-three related PDFs,
\begin{eqnarray}
F(x,\eta=0,t=0) &\!\!\! =\!\!\! &
\sum_{n}\!\! \int\!\!\!\!\!\int\!d^2{\bf k}_{\perp}\,  F^{(n)}(x,\eta=0,{\bf \Delta}_\perp=0,{{\bf k}}_{\perp})\,, \quad F\in\{E,\widetilde E,\overline{E}_{\rm T}, \hat{E}_{\rm T}\}\,,
\\
f(x) &\!\!\! =\!\!\! &
\sum_{n}\!\! \int\!\!\!\!\!\int\!d^2{\bf k}_{\perp}\,
f^{(n)}(x,\eta=0,{\bf \Delta}_\perp=0,{\bf k}_{\perp})\,, \quad f\in\{f_{1\rm T}^{\perp},g_{1\rm T}^{\perp},h_{1}^{\perp},h_{1\rm L}^{\perp} \}\,,
\qquad
\end{eqnarray}
which are a priory not tied to each other. Of course, in any model the projections are ``fixed" and so one will discover some model dependent relations, where however, in all four nfPDF/uPDF correspondences a mismatch under time inversion appears.

Let us point out that one can also sacrifice a mismatch under parity reflection rather time inversion. Then one might ask for the model correspondences of
\begin{eqnarray}
\label{GPD2FPD-twis3-3}
E(x,0,t) \leftarrow E(x,0,{\bf \Delta}_\perp,{{\bf k}}_{\perp})
& \stackrel{?}{\leftrightarrow} &
g_{1\rm T}^\perp(x,0,{\bf \Delta}_\perp,{{\bf k}}_{\perp})\rightarrow g_{1\rm T}^\perp(x,{\bf k}_\perp) \,,
\\
\overline{E}_{\rm T}(x,0,t)  \leftarrow \overline{E}_{\rm T}(x,0,{\bf \Delta}_\perp,{{\bf k}}_{\perp})
&\stackrel{?}{\leftrightarrow}&
h_{1\rm L}^\perp(x,0,{\bf \Delta}_\perp,{{\bf k}}_{\perp}) \rightarrow h_{1\rm L}^\perp(x,{\bf k}_\perp)  \,.
\nonumber
\end{eqnarray}
Let us suppose that the $T$-odd Sivers and Boer-Mulders functions vanish.
Then the zero-skewness nfPDF $E$  and $\overline{E}_{\rm T}$ as well as the  uPDFs $g_{1\rm T}^\perp$ and $h_{1\rm L}^\perp$ can be obtained from
the following $\Delta L^z=1$ LFWF overlaps, respectively:
\begin{eqnarray}
2\;\intsum \psi^{\ast\,\Rightarrow}_{\leftarrow,n}(X_i,{{\bf k}}^\prime_{\perp i},s_i) \, \psi^{\Leftarrow}_{\leftarrow,n}(X_i, {{\bf k}}_{\perp i},s_i) + \mbox{c.c.}
= \mathbb{E}^{(n)}(x,0,{\bf \Delta}_\perp,{{\bf k}}_{\perp}) + \widetilde{\mathbb{E}}^{(n)}(x,0,{\bf \Delta}_\perp,{{\bf k}}_{\perp})\,,
\\
-2\;\intsum \psi^{\ast\,\Leftarrow}_{\rightarrow,n}(X_i,{{\bf k}}^\prime_{\perp i},s_i) \, \psi^{\Leftarrow}_{\leftarrow,n}(X_i, {{\bf k}}_{\perp i},s_i) + \mbox{c.c.}
= \overline{\mathbb{E}}^{(n)}(x,0,{\bf \Delta}_\perp,{{\bf k}}_{\perp}) + \hat{\mathbb{E}}^{(n)}(x,0,{\bf \Delta}_\perp,{{\bf k}}_{\perp})\,,
\end{eqnarray}
which might be expressed by uGPD combinations (\ref{bbE}--\ref{bbhE}).  Indeed, from the uGPD  parmaterizations  (\ref{uGPDs-E-decomposition}--\ref{uGPDs-hE-decomposition}) we read off that nfPDF $E$ and uPDF $g_{1 \rm T}^\perp$
are two different projections of a uGPD combination
\begin{eqnarray}
\label{E2g1T-E}
E(x,\eta=0,t=0) &\!\!\! =\!\!\! &  \sum_{n}\lim_{\Delta\to 0}\frac{M}{\Delta^1}\intk
\left[
\mathbb{E}^{(n)} + \widetilde{\mathbb{E}}^{(n)}
\right](x,\eta=0,{\bf \Delta}_\perp,{{\bf k}}_{\perp})\,,
\\
\label{E2g1T-g1T}
g_{1 \rm T}^\perp(x)  &\!\!\! =\!\!\! &   -\sum_{n}\!\! \intk  \frac{M}{2k^1}
\left[
\mathbb{E}^{(n)} + \widetilde{\mathbb{E}}^{(n)}
\right](x,\eta=0,{\bf \Delta}_\perp=0,{{\bf k}}_{\perp})\,,
\end{eqnarray}
while for  transversity  nfPDF $\overline{E}$ and uPDF $h_{1 \rm L}^\perp$  we find the analog result
\begin{eqnarray}
\label{oE2h1L-oE}
\overline{E}(x,\eta=0,t=0) &\!\!\! =\!\!\! &  \sum_{n}\lim_{\Delta\to 0}\frac{M}{\Delta^1}\intk
\left[
\overline{\mathbb{E}}^{(n)} + \hat{\mathbb{E}}^{(n)}
\right](x,\eta=0,{\bf \Delta}_\perp,{{\bf k}}_{\perp})\,,
\\
\label{oE2h1L-h1L}
h_{1 \rm T}^\perp(x)  &\!\!\! =\!\!\! &   \sum_{n}\!\! \intk  \frac{M}{2k^1}
\left[
\overline{\mathbb{E}}^{(n)} + \hat{\mathbb{E}}^{(n)}
\right](x,\eta=0,{\bf \Delta}_\perp=0,{{\bf k}}_{\perp})\,.
\end{eqnarray}
Let us have a closer look to the two uGPD projections, e.g., $E$ and $g_{1 \rm T}^\perp$. Angular momentum conservation tells us that for a contribution with given $\bar{L}^z$, see Eq.~(\ref{l^z}) and discussion above, we have
$$
\left[
\mathbb{E}^{(n,\bar{L}^z)} + \widetilde{\mathbb{E}}^{(n,\bar{L}^z)}
\right](x,\eta=0,{\bf \Delta}_\perp,{{\bf k}}_{\perp}) \propto
\left[
|{\bf k}_\perp| e^{-i \varphi}  -(1-x) |{\bf \Delta}_\perp| e^{-i \varphi}
\right]^{\bar{L}^z+1} |{\bf k}_\perp|^{\bar{L}^z} e^{i \overline{L}^z \varphi} +\mbox{c.c.}
$$
As one realizes for $\bar{L}^z\neq 0$ the  nfPDF on the r.h.s.~of Eqs.~(\ref{E2g1T-E},\ref{oE2h1L-oE})
depends on the angular momentum $\bar{L}^z$, even its first Taylor coefficient in the vicinity of ${\bf \Delta}_\perp=0$,
while for the corresponding uPDF  (\ref{E2g1T-g1T},\ref{oE2h1L-h1L}) it does not. Consequently, in QCD we would consider both partonic quantities  as independent.
For quark models with a limited number of LFWFs one can easily establish model relations, e.g., for a two-body  ${\bar{L}^z}=0$ LFWF  model
the following two sum rules show up:
\begin{eqnarray}
\label{GPDuPDF-sumrule-EgT}
E(x,\eta=0,t=0,\mu^2) &\!\!\! \stackrel{{\bar{L}^z}=0}{=} \!\!\! & (1-x)\intk\, g_{1\rm T}^\perp(x,{\bf k}_{\perp})
\quad \mbox{for}\quad f_{1\rm T}^\perp=0\,,
\\
\label{GPDuPDF-sumrule-bETh1L}
\overline{E}_{\rm T}(x,\eta=0,t=0,\mu^2)  &\!\!\! \stackrel{{\bar{L}^z}=0}{=} \!\!\! &
-(1-x)\intk\, h_{1\rm L}^\perp(x,{\bf k}_{\perp})
\quad \mbox{for}\quad h_{1}^\perp=0\,.
\end{eqnarray}
Note that due to the ${\bf k}_{\perp}$-integration  the linear ${\bf k}_{\perp}$-term  in GPD $E$ and $\overline{E}_{\rm T}$ yields
$$
\intk\, \left[{\bf k}_\perp  -(1-x) {\bf \Delta}_\perp\right] \cdots  =   -\frac{1-x}{2} {\bf \Delta}_\perp \intk\,
\cdots
\,,
$$
where the ellipses stay for a LFWF overlap that is  symmetric w.r.t.~the permutation of the two arguments ${\bf k}^{\prime}_\perp = \left[{\bf k}_\perp  -(1-x) {\bf \Delta}_\perp\right]$ and ${\bf k}_\perp$ and  symmetric under the simultaneous reflection ${\bf k}^\prime_\perp \to - {\bf k}^\prime_\perp$
and ${\bf k}_\perp \to - {\bf k}_\perp$. Finally, that means that the integrand has definite symmetry w.r.t.~the shifted integration variable $\overline{\bf k}_\perp= {\bf k}_{\perp} - (1-x) {\bf \Delta}_\perp/2 $.

\subsubsection{Does a ``pretzelosity" sum rule exist?}
\label{sec:GPD2uPDF-Lz2}

It has been somehow speculated in Ref.~\cite{MeiMetGoe07} that a sum rule, such as for unpolarized quantities (\ref{H2q}--\ref{HT2dq}),
might also exist among  the ``pretzelosity" uPDF $h_{1 \rm T}^\perp(x)$, given in Eq.~(\ref{h_1T^perp-overlap}),
and the transversity nfPDF ${\widetilde H}_{\rm T}$,
see Eq.~(\ref{tHT-LFWF}), where both arise from a $|\Delta L^z|=2$ LFWF overlap. To show that such a relation is not
generally justified, let us define the embedding uGPD in terms of LFWF  overlaps
\begin{eqnarray}
\widetilde{\mathbb{H}}_{\rm T}^{(n)}(x,\eta=0,{\bf \Delta}_\perp,{{\bf k}}_{\perp}) &\!\!\! = \!\!\!&  \intsum
\psi^{*\ \Rightarrow}_{\leftarrow,n}(X_i,
     {{\bf k}}^\prime_{\perp i},s_i) \
\psi^{\Leftarrow}_{\rightarrow,n}(X_i, {{\bf k}}_{\perp i},s_i)
-\mbox{c.c.}\,.
\end{eqnarray}
Both quantities of interest are then again obtained by two different projections
\begin{eqnarray}
\label{tH_T^perp-overlap}
{\widetilde H}_{\rm T}(x,\eta=0,t=0) &\!\!\! = \!\!\!&
\sum_{n} \lim_{\Delta\to 0} \frac{M^2}{i \Delta^1 \Delta^2}
\intk\,  \widetilde{\mathbb{H}}_{\rm T}^{(n)}(x,\eta=0,{\bf \Delta}_{\perp},{\bf k}_{\perp})\,,
\\
\label{h_T^perp-overlap}
h_{1 \rm T}^\perp(x) &\!\!\! = \!\!\!&  -
\sum_{n}\intk\, \frac{M^2}{2i k^1 k^2}\,  \widetilde{\mathbb{H}}^{(n)}_{\rm T}(x,\eta=0,{\bf \Delta}_\perp=0,{\bf k}_{\perp})\,,
\end{eqnarray}
see uPDF and GPD definitions (\ref{h_1T^perp-overlap})  and (\ref{tHT-LFWF}), respectively. As discussed in the preceding section, also
here the  projection on the nfPDF depends on the overall partonic angular momenta $\bar{L}^z$  while the projection on uPDF  is free of it.
Introducing the shifted integration variable $\overline{\bf k}_\perp= {\bf k}_{\perp} - (1-x) {\bf \Delta}_\perp/2 $, we find
that $\widetilde{\mathbb{H}}_{{\rm T}}^{(n,{\bar{L}^z}) }(x,\eta=0,{\bf \Delta}_\perp,{\bf k}_{\perp})$ is proportional to
$$
\left[
|\overline{\bf k}_\perp| e^{-i \varphi}  -\frac{1-x}{2} |{\bf \Delta}_\perp| e^{-i \varphi}
\right]^{\bar{L}^z+1}
\left[
|\overline{\bf k}_\perp| e^{i \varphi}  +\frac{1-x}{2} |{\bf \Delta}_\perp| e^{i \varphi}
\right]^{\bar{L}^z-1}  - \mbox{c.c.}\,.
$$
Obviously, also the second Taylor coefficient in the expansion around the vicinity ${\bf \Delta}_\perp =0$ depend on  $\bar{L}^z$ and, thus, in general the nfPDF ${\widetilde H}_{\rm T}$ at $t=0$ differs from the twist-four uPDF $h_{1 \rm T}^\perp$.
Even for the scalar diquark model, i.e.,  $\bar{L}^z=0$, we find that the uGPD is proportional to
$$
\widetilde{\mathbb{H}}_{\rm T}^{(n,{\bar{L}^z}=0)}(x,\eta=0,{\bf \Delta}_\perp,{{\bf k}}_{\perp}) \propto i \left(4 \overline{k}^1 \overline{k}^2-
(1-x)^2 \Delta^1 \Delta^2\right)\,.
$$
Consequently, the $\overline{\bf k}_\perp$-integration in the transversity nfPDF (\ref{tH_T^perp-overlap})
and the ``pretzelosity" distribution (\ref{h_T^perp-overlap}) possesses also for this simple model a different weight. Hence, both distributions
are proportional to each other,
\begin{eqnarray}
\label{sumrule-HTh1T}
\widetilde H_{\rm T}(x,\eta=0,t=0) \stackrel{\rm diquark}{\propto} (1-x)^2\, h_{1 \rm T}^\perp(x)\,.
\end{eqnarray}
where, however, the proportionality factor depends on the functional form of the LFWF. Thus,  even
in a model that contains different ${\bar{L}^z}=0$ states a sum rule that connects the twist-two nfPDF $\widetilde H_{\rm T}$ with the twist-four uPDF $h_{1 \rm T}^\perp(x)$ does not exist in general.

\section{GPD modeling in terms of effective two-body LFWFs}
\label{Sec-GPDs}

Let us now apply the model procedure to GPDs, which are
accessible from hard-exclusive reactions, e.g., electroproduction
of mesons and photon.  Generically, a GPD definition looks like as given in Eq.~(\ref{F-gen})
from which follows that the $x$-moments of the GPD $F$  are polynomials in the
$\eta$ skewness parameter (\ref{eta}).
This polynomiality property is manifestly implemented in the so-called double distribution (DD) representation,
e.g., for the ``quark" part of $F$ GPD
\begin{eqnarray}
F(x,\eta,t) = \int_0^1\! dy\int_{-1+y}^{1-y}\! dz\;  \delta(x-y-z \eta) f(y,z,t)\,,
\end{eqnarray}
which lives in the region $-\eta \le x \le 1$. On the other hand,
describing the initial and final proton states
in terms of the LFWFs (\ref{Def-ProSta}), we can
straightforwardly write down LFWF overlap representations of
GPDs \cite{Diehl:1998kh,Brodsky:2000xy,Diehl:2000xz}, see Sec.~\ref{Sec-GPD-definitions} and Appendix \ref{sec-app-LFWF2GPD},
in which positivity constraints are manifestly implemented, however, not the polynomiality property.
Thereby, the GPD in the region $\eta \le x \le 1$ is described by a LFWF overlap  in which
the parton number is conserved (partonic $s$-channel exchange).
In the region $-\eta \le x \le \eta$ the GPD is given by an  overlap of LFWFs  in which
the parton numbers change from $n+1$ to $n-1$ (mesonic-like $t$-channel exchange) \cite{Brodsky:2000xy,Diehl:2000xz}.

Both the central and outer regions are tied to each other by polynomiality or Poincar\`e covariance,
see e.g., Ref.~\cite{Kumericki:2008di}. Employing this GPD duality, we can simplify the GPD modeling in terms of
LFWFs, in particular, we are interested to obtain the GPD from
 the parton number conserved LFWF overlap, which generically reads for a effective two-body LFWF, cf.~the GPD spin-correlation matrix~(\ref{bfF-rank3}):
 \begin{eqnarray}
\label{F-generic}
F(x\ge\eta,\eta,t)= \frac{1}{1-x} \intk\,
\phi^\ast\!\left(\frac{x-\eta}{1-\eta},{\bf k}_{\perp}- \frac{1-x}{1-\eta} {\bf \Delta}_\perp\right)
\phi\!\left(\frac{x+\eta}{1+\eta},{\bf k}_{\perp}\right).
\end{eqnarray}
Note that the normalization of this scalar LFWF differs from those of the ``spinor" ones (\ref{LFWF-spinor-model}) by a common factor $$1/\sqrt{1-X_1}=\sqrt{(1+\eta)/(1-x)}\,,\quad\mbox{where}\quad X_1= \frac{x+\eta}{1+\eta},$$
which we also took into account in the definition of the unintegrated scalar LFWF overlap for $\eta=0$, see, e.g., Eq.~(\ref{def-TMD-overlap}).
Having such a scalar LFWF at hand, one might transform the overlap representation (\ref{F-generic}) into the DD representation
\begin{eqnarray}
F(x\ge\eta,\eta,t)= \int_{\frac{x-\eta}{1-\eta}}^{\frac{x+\eta}{1+\eta}}\!\frac{dy}{\eta}\;
f(y,(x-y)/\eta,t)\,,
\end{eqnarray}
from which one can read off the  DD $f(y,z,t)$ \cite{Hwang:2007tb}. Hence, one can so obtain the GPD $F$  in the whole region. In the case that such a procedure fails, we might conclude that the employed LFWF does not respect the underlying Lorentz symmetry and should not be used in model estimates.
Of course, there are plenty of LFWF models in the literature that are ruled out by this requirement.

In the following we implement in such a LFWF modeling procedure ${\bf k}_\perp$-dependence by considering unintegrated DDs (uDDs),
which also makes contact to the spectral properties of both uGPDs and phase space distributions \cite{BelJiYua03}.
Neglecting the spin content, in Sec.~\ref{sect-LI}  we spell out the constraints from Poincar\'e covariance for the functional form of
a scalar effective LFWF and show that the DD representation for a GPD can be found from the parton number conserved overlap of a two-body LFWF.
Within these restrictions, in Sec.~\ref{subsect-Regge} we discuss then
the implementation of Regge behavior from the $s$-channel point of view. In Sec.~\ref{subsect-constraints}
we present the DD representation for all twist-two nucleon GPDs in terms of a scalar diquark LFWF.
Finally, in Sec.~\ref{GPDs2uPDFs-diquark}  we discuss the GPD model constraints among the eight twist-two GPDs and the model dependent cross talk of nfPDs and uPDFs.

\subsection{Implementing Lorentz invariance}
\label{sect-LI}

Let us restrict to the region $x \ge \eta$ in  which partons are probed within a $s$-channel exchange.
As it turned out for uPDFs in Sec.~\ref{Sect-uPDF-overlap}, it is rather elegant  to introduce a scalar LFWF overlap in the {\rm outer} region, which we write as\footnote{$\phi$ and consequently  $\bf \Phi$  are considered as functions of the spectator quark mass $\lambda$, i.e., in full notation we would write $\phi(\dots|\lambda^2)$ and ${\bf \Phi}(\dots|\lambda^2)$, see Eq.~(\ref{rho(lambda)}). For shortness this dependence is not indicated in this Section.}
\begin{eqnarray}
\label{GPD-overlap}
{\bf \Phi}(x \ge \eta,\eta,{\bf \Delta}_{\perp},{\bf k}_{\perp}) =
\frac{1}{1-x}
\phi^\ast\!\left(\frac{x-\eta}{1-\eta},{\bf k}_{\perp}- \frac{1-x}{1-\eta} {\bf \Delta}_\perp\right)
\phi\!\left(\frac{x+\eta}{1+\eta},{\bf k}_{\perp}\right),
\end{eqnarray}
which might be also directly viewed in a scalar toy theory as a ``uGPD" definition.
A ``uGPD" definition in terms of a scalar LFWF overlap for antiparticles can be analogously defined with a negative momentum fraction $x \le -\eta$ and for convenience they might be mapped to the   $\eta \le x$  region, decorated with a sign according to its charge conjugation parity.  Positivity
constraints, in its most general form, should be
satisfied in the overlap representations, if they are
not spoiled by a subtraction procedure. For a deeper discussion we refer to the work of P.~Pobylitsa \cite{Pobylitsa:2002iu,Pobylitsa:2002vi}.

As explained above, the residual Lorentz covariance, the so-called polynomiality conditions for GPD form factors (GFFs),
is manifestly implemented in the DD representation and it ties the GPD in the outer and central regions to each other.
Since LF quantization implies also that
Lorentz covariance is not explicit, it is not an entirely trivial
issue to find a LFWF that respect these constraints.
If we represent the LFWF as a Laplace transform,
\begin{eqnarray}
\label{Def-LFWF-Laplace} \phi(X,{\bf k}_\perp) = \int_0^\infty\! d\alpha\; \varphi(X,\alpha)  \exp\left\{-\alpha  \frac{{\bf k}_\perp^2 - X(1-X) M^2}{(1-X) M^2}  \right\} \,,
\end{eqnarray}
the restoration of $t$-dependence can be achieved.
Note that the existence of the integral requires that  $\lim_{\alpha \to \infty} \exp\left\{ \alpha X\right\} \varphi(X,\alpha)$  sufficiently fast vanishes, which is ensured by the stability condition for the spectator system.
This allows us to transform the LFWF overlap (\ref{GPD-overlap})  into a uDD representation, see Appendix \ref{sec-appA},
\begin{eqnarray}
\label{Def-DDunint}
{\bf \Phi}(x,\eta,{\bf
\Delta}_{\perp},{\bf
k}_{\perp}) &\!\!\!=\!\!\!& \int_0^1\! dy\int_{-1+y}^{1-y}\! dz\;  \delta(x-y-z \eta)\,
 {\bf \hat \Phi}\!\left(y,z,t,\overline{\bf k}_\perp\!\right),
\end{eqnarray}
where $\overline{{\bf k}}_\perp = {\bf k}_\perp -(1-y+z){\bf \Delta}_\perp/2$.
We also find that the  uDD is represented as a Laplace transform
\begin{eqnarray}
\label{Def-DD-unint}
{\bf \hat \Phi}(y,z,t,\overline{{\bf k}}_\perp)&\!\!\!=\!\!\!& \frac{1}{2}\int_0^\infty\!dA\,A\; \varphi^\ast \left(\frac{y-\eta(1-z)}{1-\eta}, A \frac{1-y+z}{2}\right) \varphi \left(\frac{y+\eta(1+z)}{1+\eta}, A \frac{1-y-z}{2}\right)
\nonumber \\
&&\times
\exp\left\{A \left[ y(1-y) + \left[(1-y)^2-z^2\right] \frac{t}{4 M^2}-\frac{\overline{{\bf k}}^2_\perp}{M^2}\right]
 \right\}.
\end{eqnarray}
However, it remains the (sufficient) condition that the product of Laplace kernels in (\ref{Def-DD-unint})  must be independent on  $\eta$, i.e.,
\begin{eqnarray}
\label{LFWF-constraint}
\frac{d}{d\eta} \left[\varphi^\ast \left(\frac{y-\eta(1-z)}{1-\eta}, A \frac{1-y+z}{2}\right) \varphi \left(\frac{y+\eta(1+z)}{1+\eta}, A \frac{1-y-z}{2}\right)\right]=0\,.
\end{eqnarray}
A rather simple example which provides a non-trivial solution of Eq.~(\ref{LFWF-constraint}) reads:
\begin{eqnarray}
\label{Ans-LFWF}
\varphi(X,\alpha)= \varphi(\alpha) \exp  \left\{-\alpha\frac{m^2}{M^2}-\alpha  \frac{X}{1-X} \frac{\lambda^2}{M^2} \right\},
\end{eqnarray}
which yields in  Eq.~(\ref{Def-DD-unint}) to the replacement
\begin{eqnarray}
\varphi^\ast(X^\prime,\alpha_2) \varphi(X,\alpha_1) \to \varphi^\ast\!\left(\!A \frac{1-y+z}{2}\!\right) \varphi\!\left(\!A \frac{1-y-z}{2}\!\right) \exp\left\{-A (1-y)\frac{m^2}{M^2}  - A y \frac{\lambda^2}{M^2} \right\}\,.
\end{eqnarray}
Hence, we  find for the uDD (\ref{Def-DD-unint}) a rather general representation
\begin{eqnarray}
\label{Def-DD-unint1}
{\bf \hat \Phi}(y,z,t,\overline{{\bf k}}_\perp)&\!\!\!=\!\!\!& \frac{1}{2}\int_0^\infty\!dA\,A\;
\varphi^\ast\!\left(\!A \frac{1-y+z}{2}\!\right) \varphi\!\left(\!A \frac{1-y-z}{2}\!\right)
\\
&&\times
\exp \left\{-A\left[
(1-y)\frac{m^2}{M^2}+y \frac{\lambda^2}{M^2}-y(1-y)-\left[(1-y)^2-z^2\right] \frac{t}{4 M^2}+\frac{\overline{{\bf k}}^2_\perp}{M^2}
\right] \right\},\nonumber
\end{eqnarray}
which depends besides the set $\{m,\lambda, M\}$ of mass parameters from the reduced LFWF $\varphi(\alpha)$.

A few comments are in order.
\begin{itemize}
\item
In the unintegrated DD (\ref{Def-DD-unint1}) the ${\bf \Delta}_\perp$  dependence
has been absorbed in the $t$ and $\overline{{\bf k}}^2_\perp$ variables, and we see that they are tied to each other.  This allows us to rewrite the $\overline{\bf k}_\perp$  average, needed for the evaluation of
DDs and associated GPDs, as an integral over $t$:
\begin{eqnarray}
\label{Def-overlap-moment1}
{\bf \hat \Phi}(y,z,t)&\!\!\!=\!\!\!&  \intk\; {\bf \hat \Phi}\!\left(y,z,t,\overline{\bf k}\!\right) =
\frac{(1 - y)^2 - z^2}{4} \pi \int_{-\infty}^{t}\! dt^\prime\; {\bf \hat \Phi}\!\left(y,z,t^\prime,\overline{\bf k}_\perp =0\!\right)\,.
\end{eqnarray}
Analogously, the $\overline{\bf k}_\perp^2$ moment of the unintegrated DD (\ref{Def-DD-unint1}), needed below, can be understood as an integral of
the DD (\ref{Def-overlap-moment1})  over $t$:
\begin{eqnarray}
\label{Def-overlap-moment2}
{\bf \hat \Phi}^{(2)}(y,z,t) &\!\!\!=\!\!\!&   \int\!\!\!\!\!\int\! d^2\overline{\bf k}_\perp\,
\frac{\overline{\bf k}_\perp^2}{M^2} \,  {\bf \hat \Phi}\!\left(y,z,t,\overline{\bf k}_\perp\!\right) = \left[(1 - y)^2 - z^2\right] \int_{-\infty}^{t}\! \frac{dt^\prime}{4M^2}\;   {\bf \hat \Phi}(y,z,t^\prime) \,.
\end{eqnarray}

\item
We went here along the line of Ref.~\cite{Hwang:2007tb}, i.e., switching to the DD representation (\ref{Def-DDunint}), we  restore the whole GPD from the parton number conserved LFWF overlap (\ref{GPD-overlap}).
Integrating over ${\bf k}_{\perp}$ in Eq.~(\ref{Def-DDunint}), we find the common form of the DD representation
\begin{eqnarray}
\label{Def-DDcommon}
{\bf \Phi}(x,\eta,t) &\!\!\!=\!\!\!& \int_0^1\! dy\int_{-1+y}^{1-y}\! dz\;  \delta(x-y-z \eta)\, {\bf \hat \Phi}(y,z,t)\,,\quad
{\bf \hat \Phi}(y,z,t)=
 \intk\, {\bf \hat \Phi}\!\left(y,z,t,\overline{\bf k}_\perp\!\right),
 \nonumber\\
\end{eqnarray}
ensuring that GPD polynomiality constraints are satisfied.

\item
The $t$ (or ${\bf k}^2_\perp$) dependency has also a cross talk to the skewness one.  For instance, if one takes a reduced LFWF $\varphi(\alpha)$ that is concentrated in some point $\overline{A}$  one obtains exponential ${\bf k}^2_\perp$  or $t$ dependent quantities and a $\eta$-independent GPD.  On the other hand if one choose a power-like dependence $\varphi(\alpha) \propto \alpha^p$, one will recover the model of Ref.~\cite{Hwang:2007tb}.

\end{itemize}

\subsection{Implementing Regge behavior}
\label{subsect-Regge}

It is well known that spectator quark models entirely fail in the small $x$ region, where they predict that PDFs have a constant behavior rather a $x^{-\alpha}$ Regge behavior, which is with $\alpha\sim 0.5$  phenomenologically established for unpolarized PDFs. The same failure will arise
from our uDD (\ref{Def-DD-unint1}), which approaches a constant in the $y\to 0$ limit.
Viewing the GPD in the central region $-\eta\le x\le \eta $ as a mesonic-like $t$-channel exchange makes directly contact to Regge phenomenology
\cite{Polyakov:2002wz,Mueller:2005ed,Kumericki:2007sa}. Alternatively, we will adopt here the $s$-channel point of view.

Namely, one realizes  in Eq.~(\ref{Def-DD-unint1})  that the variable $y$ and the spectator mass square $\lambda^2$ are conjugate to each other.
 We follow now the proposal \cite{Landshoff:1971xb} that the large spectator mass $\lambda^2$ behavior might be considered as ``dual'' to the Reggeon exchange in the $t$-channel. In this manner, see Eq.~(\ref{Def-ProSta-2}), we might cure the small $x$ failure  by an incoherent sum over the spectator mass $\lambda$:
\begin{eqnarray}
\label{Def-SpeRep_1}
{\bf \hat \Phi}(y,z,t,\overline{\bf k}_\perp)  = \sum_{\;\; \lambda^2}\!\!\!\!\!\!\!\!\int\, \rho(\lambda^2) {\bf \hat \Phi}(y,z,t,\overline{\bf k}_\perp|\lambda^2)\,.
\end{eqnarray}
 The $\lambda^2$ integration is restricted from below by the  cut-off mass $\lambda_c \le \lambda$, which is considered as a parameter that is constrained by the stability criteria for the spectator system, $M-m \le\lambda_c.$ To obtain a $y^{-\alpha}$ behavior, the spectator mass spectral function must behave at large $\lambda^2$ as
\begin{eqnarray}
\lim_{\lambda \to \infty}\rho_\alpha(\lambda,\lambda_c)  \propto \, M^{-2\alpha} \lambda^{2\alpha-2}\,,
\end{eqnarray}
where $\alpha$ is the Regge intercept. For the spectral function we use a simple functional form
\begin{eqnarray}
\label{Def-SpeRep1}
\rho_\alpha(\lambda,\lambda_c)  =
\theta(\lambda^2-\lambda_c^2)  \frac{(\lambda^2 -\lambda_c^2)^{\alpha-1}}{\Gamma(\alpha)  M^{2\alpha} }\,,
\end{eqnarray}
which in the limit $\alpha\to 0$ can be viewed as a delta-function
$\rho_0(\lambda,\lambda_c) =\delta(\lambda^2 -\lambda_c^2). $  Convoluting the unintegrated DD (\ref{Def-DD-unint1})  with this spectral density (\ref{Def-SpeRep}) and renaming the variable $\lambda_c\to\lambda$, we obtain a uDD with the desired small $y$ behavior:
\begin{eqnarray}
\label{Def-DD-unint2}
{\bf \hat \Phi}(y,z,t,\overline{{\bf k}}_\perp)&\!\!\!=\!\!\!& \frac{1}{2 y^{\alpha}} \int_0^\infty\!dA\,A^{1-\alpha}\;
\varphi^\ast\!\left(\!A \frac{1-y+z}{2}\!\right) \varphi\!\left(\!A \frac{1-y-z}{2}\!\right)
\\
&&\times
\exp \left\{-A\left[
(1-y)\frac{m^2}{M^2}+y \frac{\lambda^2}{M^2}-y(1-y)-\left[(1-y)^2-z^2\right] \frac{t}{4 M^2}+\frac{\overline{{\bf k}}^2_\perp}{M^2}
 \right]\right\}, \nonumber
\end{eqnarray}
which we will utilize as a ``master" formula for modeling. Note that up to the factor $y^{-\alpha}A^{-\alpha}$ this uDD has the same functional form as the original one in Eq.~(\ref{Def-DD-unint1}).
One might include some $t$-dependence in $\alpha$ by hand, e.g., following the common wisdom, by taking a linear trajectory $\alpha\to \alpha(t) = \alpha+  \alpha^\prime t$. However, we emphasize that then positivity constraints are not guaranteed by construction. A rather non-trivial $t$-dependence might be introduced by a spectral representation w.r.t.~the struck quark mass $m$, which needs some further investigation. We also point out that it is not allowed in our effective scalar LFWFs (\ref{Def-LF-WF1}) and (\ref{Def-LF-WF1}) to change the relative normalization. Hence,
the implemented Regge behavior appears in all
quantities in a unique manner, which might induce a contradiction with common Regge phenomenology.

\subsection{DD representations and model constraints among DDs}
\label{subsect-constraints}

Employing the standard  GPD definitions in terms of bilocal quark operators and the LFWF overlap representations, listed in Eqs.~(\ref{H-LFWF}--\ref{tHT-LFWF}), it is now straightforward to find DD representations for nucleon GPDs in terms of the ``DD'' (\ref{Def-overlap-moment1}).
Thereby, we employ  first the scalar diquark model, where the spin-couplings (\ref{Def-LF-WF1},\ref{Def-LF-WF2}) are fixed, the effective LFWF (\ref{Def-LFWF-Laplace}) within the Laplace kernel (\ref{Ans-LFWF}). For instance, if we take the GPD $E$ from  the parton number conserved LFWF overlap representation, we find then with one effective spectator ($n=1$) from Eq.~(\ref{E-LFWF}):
\begin{eqnarray}
\label{gpdE-int}
E(x\ge\eta,\eta,t)=  (1-x)\intk\; {\bf \Phi}(x \ge \eta,\eta,{\bf
\Delta}_{\perp},{\bf k}_{\perp})\,,
\end{eqnarray}
where the unintegrated LFWF overlap is defined in Eq.~(\ref{GPD-overlap}).
To convert the GPD $E$  (\ref{gpdE-int}) into the DD representation,   we can now employ the formalism of Sec.~\ref{sect-LI} and we obtain
\begin{eqnarray}
E(x,\eta,t) =
(1-x)  \int_{0}^1\!dy\; \int_{-1+y}^{1-y}\! dz\; \delta(x-y-\eta z )\;
2\left(\frac{m}{M}+y\right)  \intk\; {\bf \hat \Phi}(y,z,t,\overline{\bf k}_\perp)\,,
\end{eqnarray}
where the scalar uDD ${\bf \hat \Phi}(y,z,t,\overline{\bf k}_\perp)$ is given in Eq.~(\ref{Def-DD-unint}) and the factor $2\left(\frac{m}{M}+y\right)$ reminds us that the GPD $E$  stems from an overlap of $L^z=0$ and $L^z=1$ LFWFs. Employing the Regge improved representation (\ref{Def-DD-unint2}) rather the scalar one (\ref{Def-DD-unint}) with fixed diquark mass $\lambda$, we find after ${\bf k}_\perp$-integration the scalar DD in terms of a
reduced LFWF
\begin{eqnarray}
\label{DD}
{\bf \hat \Phi}(y,z,t) &\!\!\!=\!\!\! &
\frac{\pi\,  M^2}{2 y^{\alpha}} \int_0^\infty\!dA\,A^{-\alpha}\;
\varphi^\ast\!\left(\!A \frac{1-y+z}{2}\!\right) \varphi\!\left(\!A \frac{1-y-z}{2}\!\right)
\\
&&\times
\exp \left\{-A\left[
(1-y)\frac{m^2}{M^2}+y \frac{\lambda^2}{M^2}-y(1-y)-\left[(1-y)^2-z^2\right] \frac{t}{4 M^2}
 \right]\right\}. \nonumber
\end{eqnarray}
In the following two sections we give the DD representations  for a scalar diquark and axial-vector diquark model, respectively.

\subsubsection{Scalar diquark model}
\label{sec:GPD-sca}

Analogously, for our model LFWFs all other chiral even (\ref{H-LFWF},\ref{tH-LFWF},\ref{tE-LFWF}) and odd (\ref{oHT-LFWF}--\ref{tHT-LFWF}) twist-two GPD overlap representations can be straightforwardly converted into DD representations, providing us a set of formulae in which both Lorentz covariance and positivity constraints are implemented. Alternatively, one might simply use the definition of the spin-correlation function
(\ref{bfF}) and the overlap representation (\ref{bfF-rank3}) with the scalar diquark LFWF ``spinor" that is borrowed from Yukawa theory, i.e., it is build from the LFWFs (\ref{Def-LF-WF1},\ref{Def-LF-WF2}):
\begin{eqnarray}
\label{LFWF-spinor-S}
\mbox{\boldmath $\psi$}^{\rm S}(X,{\bf k}_\perp) = \frac{1}{M} \left(\!\!
               \begin{array}{c}
                 m +X M \\
                -|{\bf k}_\perp| e^{i \varphi} \\
                 |{\bf k}_\perp| e^{-i \varphi}  \\
                 m +X M \\
               \end{array}\!\!
             \right) \frac{ \overline{\phi}(X,{\bf k}_\perp|\lambda^2)}{\sqrt{1-X}}\,.
\end{eqnarray}
Thereby, we will replace the $\overline{\bf k}_\perp^2$ moments, entering in $H$ and $\widetilde H$ GPDs, by the  integral (\ref{Def-overlap-moment2}) over $t$.  Note, however, that the DD representations are not uniquely defined \cite{BelKirMueSch01,Teryaev:2001qm}; we will show those that naturally occur in the scalar diquark model. We  now provide explicit representations for all eight twist-two GPDs and discuss some particularities and model constraints among them.

In the chiral even sector we find it convenient to write the DD representation in a model dependent manner,
giving up a GPD/DD correspondence  $F\in \{H,E,\widetilde H, \widetilde E\} \leftrightarrow f\in \{h,e,\widetilde h, \widetilde e\}$,
\begin{eqnarray}
\label{Def-SpeRep} \left\{
\begin{array}{c}
H \\
E \\
\widetilde H \\
\widetilde E
\end{array}
\right\}(x,\eta,t)=
\int_{0}^1\!dy\; \int_{-1+y}^{1-y}\! dz\;
\delta(x-y-\eta z )\; \left\{
\begin{array}{c}
h + (x-y) e\\
(1-x) e \\
\widetilde h \\
\widetilde e + (1-y-z/\eta)\, e
\end{array}
\right\}(y,z,t)\,,
\end{eqnarray}
where all DDs are symmetric under $z\to -z$ reflection.
In our  GPD $H$ and $E$ representations (\ref{Def-SpeRep}) a first order polynomial in $x=y+\eta z$ appears.  Thus, the odd $x^n$-moments of these GPDs are even polynomials in $\eta$ of order $n+1$  \cite{Ji:1998pc} and so polynomiality is completed%
\footnote{In the original suggested or common DD representation \cite{Mueller:1998fv,Radyushkin:1997ki}, where  a first order polynomial in $x$ is absent,
all $x^n$ moments are only even polynomials of order $n$. Meanwhile, it is clarified that different DD representations are equivalent, a short
presentation on this subject can be found in Ref.~\cite{Mueller:2014hsa}}.
In the GPD $E$   a $1-x= 1-y-\eta z$ factor appears while the GPD $H$ is decorated with $(x-y)\,e=\eta z\, e$ addenda. Thus, the ``magnetic'' GPD combination $H+E$ is given in terms of the DD $h+(1-y)\,e$ and it has a common DD representation in accordance with the fact that its $x^n$-moments are even polynomials of order $n$. The GPD $\widetilde H$  has a common DD representation, too.
In the representation of GPD $\widetilde E$ appears a  $z/\eta$ proportional term, which is commonly not utilized. For shortness we expressed the addenda in terms of the DD $e$, which is in general not the case. Since the DD $e$  is even in $z$, this term does not contradict polynomiality nor time reversal invariance, i.e., the GPD $\widetilde E$ is symmetric under the exchange $\eta\to -\eta$. The DDs are expressible by the  $\overline{\bf k}_\perp$-moments (\ref{Def-overlap-moment1}) and (\ref{Def-overlap-moment2}) of the Regge improved ``uDD" (\ref{Def-DD-unint2}):
\begin{eqnarray}
\label{Def-SpeFun-h}
h &\!\!\!=\!\!\! &
\left(\frac{m}{M}+y\right)^2 {\bf \hat \Phi}(y,z,t)
+\left[(1 - y)^2 - z^2\right] \left[\frac{t}{4M^2} {\bf \hat \Phi}(y,z,t) + \int_{-\infty}^{t}\! \frac{dt^\prime}{4M^2} {\bf \hat \Phi}(y,z,t^\prime)\right],
\\
\label{Def-SpeFun-e}
e&\!\!\!=\!\!\! &
2\left(\frac{m}{M}+y\right)  {\bf \hat \Phi}(y,z,t)\,,
\\
\label{Def-SpeFun-th}
\widetilde h &\!\!\!=\!\!\! &
\left(\frac{m}{M}+y\right)^2
{\bf \hat \Phi}(y,z,t)
-\left[(1 - y)^2 - z^2\right] \left[\frac{t}{4M^2} {\bf \hat \Phi}(y,z,t) + \int_{-\infty}^{t}\! \frac{dt^\prime}{4M^2} {\bf \hat \Phi}(y,z,t^\prime)\right],
\\
\label{Def-SpeFun-te}
\widetilde e &\!\!\!=\!\!\! &
2\left((1-y)^2-z^2\right){\bf \hat \Phi}(y,z,t)\,.
\end{eqnarray}

As already emphasized for uPDFs in Sec.~\ref{uPDF-scaDiq}, we  see that in a scalar diquark model the dynamical information is contained in one scalar LFWF overlap, i.e., now in  a scalar ``DD''. To make this property explicit,  we expressed the $\overline{\bf k}^2_\perp$-moment  (\ref{Def-overlap-moment2})  by a $t$-integral over this ``DD''.  This $t$-integral arises from the diagonal $L^z=1$ LFWF overlap and, hence, it only enters  in the target helicity conserved DDs $h$ and $\widetilde h$ within different signs. The various prefactors in Eqs.~(\ref{Def-SpeFun-h}--\ref{Def-SpeFun-te}) arise from the overlap of the spin parts, labeled by orbital angular momenta $L^z=0$ and $L^z=\pm 1$.  The
diagonal $L^z=0$ overlap, proportional to $(m+y M)^2/M^2$, enters in both $h$ (\ref{Def-SpeFun-h}) and $\widetilde h$ (\ref{Def-SpeFun-th}) with a positive sign, while the diagonal $L^z=1$ one, proportional to
$(1 - y)^2 - z^2$, contributes to $h$ and $\widetilde h$ with a positive and negative sign, respectively. Moreover, both diagonal overlaps possess different $t$-dependence. The target spin non-conserved DDs $e$ (\ref{Def-SpeFun-e}) and $\widetilde e$ (\ref{Def-SpeFun-te}) arise from the
$L^z=0$ with $L^z=1$ LFWF overlap
respectively, cf.~Eqs.~(\ref{E-LFWF}) and (\ref{tE-LFWF}), and posses a relative simple structure.
Hence, we have three constraints among the four chiral even GPDs, which might be written in the following form:
\begin{eqnarray}
\label{GPDmodel-const1}
h(y,z,t)+\widetilde h(y,z,t) &\!\!\!=\!\!\! & \left(\frac{m}{M}+y\right)  e(y,z,t)\,,
\\
\label{GPDmodel-const2}
h(y,z,t)-\widetilde h(y,z,t)  &\!\!\!=\!\!\! &
 \frac{t}{4 M^2} \widetilde e(y,z,t) + \int_{-\infty}^t\!\frac{dt^\prime}{4 M^2} \widetilde e(y,z,t^\prime) \stackrel{4 M^2 \gg -t} {\approx} \int_{-\infty}^t\!\frac{dt^\prime}{4M^2} \widetilde e(y,z,t^\prime)\,,
\\
\label{GPDmodel-const3}
\widetilde e(y,z,t) &\!\!\!=\!\!\! &  \frac{ (1-y)^2-z^2}{\frac{m}{M}+ y}e(y,z,t)\,. 
\end{eqnarray}

Adopting the convention of Ref.~\cite{Diehl:2001pm},  we find for the four chiral odd twist-two GPDs
\begin{eqnarray}
\label{Def-GPD-FT}
\left\{
\begin{array}{c}
H_{\rm T} \\
E_{\rm T} \\
\widetilde H_{\rm T} \\
\widetilde E_{\rm T}
\end{array}
\right\}(x,\eta,t) =
\int_{0}^1\!dy\!\int_{-1+y}^{1-y}\! dz\;
\delta(x-y-\eta z )\; \left\{
\begin{array}{c}
h_{\rm T}\\
e_{\rm T} \\
{\widetilde h}_{\rm T} \\
{\widetilde e}_{\rm T}
\end{array}
\right\}(y,z,t)\,,
\end{eqnarray}
a common DD representation, where
\begin{eqnarray}
\label{Def-SpeFun-hT}
h_{\rm T} &\!\!\!=\!\!\! &
\left[
\left(\frac{m}{M}+y\right)^2-\left((1-y)^2-z^2\right)\frac{t}{4M^2} \right] {\bf \hat \Phi}(y,z,t)\,,
\\
\label{Def-SpeFun-eT}
e_{\rm T}&\!\!\!=\!\!\! &
2\left[\left(\frac{m}{M}+y\right)(1-y)+(1-y)^2-z^2\right]{\bf \hat \Phi}(y,z,t)\,,
\\
\label{Def-SpeFun-thT}
\widetilde h_{\rm T} &\!\!\!=\!\!\! &
-\left[(1-y)^2-z^2\right] {\bf \hat \Phi}(y,z,t)\,,
\\
\label{Def-SpeFun-teT}
\widetilde e_{\rm T} &\!\!\!=\!\!\! &
2\left(\frac{m}{M}+y\right) z\, {\bf \hat \Phi}(y,z,t)\,.
\end{eqnarray}
Compared to our findings (\ref{Def-SpeRep}--\ref{Def-SpeFun-te}) in the chiral even sector, we see that the chiral odd ones look rather simple. Although it is not immediately obvious from the
overlap representations (\ref{oHT-LFWF}--\ref{tHT-LFWF}) in the GPD basis (\ref{FT-newbasis}),  we can now easily identify the origin of the various DDs. Namely, the factor $(m+y M)^2/M^2$  in $h_{\rm T}$ arises from the diagonal $L^z=0$ LFWF overlap, while the $t$-dependent part comes from an (off-)diagonal $|L^z|=1$  overlap.
It is also clear that ${\widetilde h}_T$ and ${\widetilde e}_T$  entirely arises from a $|L^z|=1$  and a $L^z=0$ with $|L^z|=1$ off-diagonal LFWF overlap,
respectively, where ${e}_T$ contains both of these two overlaps.
Obviously, we can write down various relations among the different DDs
or we might express the chiral odd ones simply by chiral even DDs, e.g.,
\begin{eqnarray}
\label{Con-even2odd}
h_{\rm T}(y,z,t)  &\!\!\!=\!\!\! &   \frac{1}{2}\left[ h  + \widetilde h\right](y,z,t) - \frac{t}{8 M^2} \widetilde e(y,z,t),\quad
\widetilde h_{\rm T}(y,z,t) = -\frac{1}{2} \widetilde e(y,z,t)\,,
\nonumber \\
\label{Con-even2odd-1}
e_{\rm T}(y,z,t) &\!\!\!=\!\!\! & (1-y) e(y,z,t) + \widetilde e(y,z,t)\,, \quad \widetilde e_{\rm T}(y,z,t) = z\,  e(y,z,t)\,.
\end{eqnarray}

Since  $H$, $E$, and $\widetilde E$ GPDs  have very specific DD representations, a few  comments are in order.

\noindent
{\em i.~D-term}\\
\noindent
To complete polynomiality in a common DD representation,
it has been suggested to add (subtract) a so-called $D$-term to the $H$  ($E$) GPD, which entirely lives in the central region \cite{Polyakov:1999gs},
\begin{eqnarray}
\label{Def-SpecRep2}
\left\{{H\atop E}\right\} = \int_0^1\!dy\int_{-1+y}^{1-y}\!dz\, \delta(x-y-z\eta)\left\{ {h^{\rm new}\atop e^{\rm new}}\right\} (y,z,t) \pm \theta(|x| \le |\eta|) \, d(x/\eta,t)\,,
\end{eqnarray}
where $d(1) = d(-1)=0$ vanishes at the boundaries.  Indeed, the form of the DD representation is not unique \cite{Belitsky:2000vk,Teryaev:2001qm,Radyushkin:2011dh,Radyushkin:2013hca} and our DD representations (\ref{Def-SpeRep}--\ref{Def-SpeFun-e}) might be put in the form  (\ref{Def-SpecRep2}). Note that the form (\ref{Def-SpecRep2}) suggest that the $D$-term  is an independent GPD contribution, which in our model is not true (see discussion in Sec.~3.2.2 of Ref.~\cite{Kumericki:2007sa}). We might project on the $D$-term in different ways
\cite{Kumericki:2007sa,Kumericki:2008di}, taking the ``low-energy'' limit $\eta\to\infty$ with fixed $x/\eta$ provides \cite{Belitsky:2000vk}:
\begin{eqnarray}
\label{Cal-Dterm}
d(z,t) = z \int_{0}^{1-|z|}\!dy\;  e(y,z,t)\,,\quad\mbox{with}\quad e(y,z,t)=2\left(\frac{m}{M}+y\right)  {\bf \hat \Phi}(y,z,t)\,.
\end{eqnarray}
Finding the representations for the new DDs $h^{\rm new}$  and $e^{\rm new}$ is a straightforward
task, which has been considered  in Ref.~\cite{Belitsky:2000vk,Teryaev:2001qm,Mueller:2014hsa}.\\

\noindent{\em ii. $\eta\to 0$ limit}\\
\noindent
Mostly the skewness-zero case can be easily taken in Eqs.~(\ref{Def-SpeRep},\ref{Def-GPD-FT}) by setting $\eta=0$.
For instance, the overall normalization condition for the LFWFs (\ref{Def-LF-WF1},\ref{Def-LF-WF2}), fixed in Eq.~(\ref{LFWF-normalization}), might
be directly expressed in terms of the DD $h$:
\begin{eqnarray}
\label{Fix-Nor}
 \int_{0}^1\!dx\;\left\{{1 \atop x}\right\} f_1(x)= \int_{0}^1\!dy\,\int_{-1+y}^{1-y}\!dz\,
\left\{{1 \atop y}\right\}  h(y,z,t=0) = \left\{{n \atop \langle x \rangle}\right\} \,.
\end{eqnarray}

In the skewness-zero case the six GPDs  $F \in \left\{H,\widetilde H, H_{\rm T}, E_{\rm T}, \widetilde H_{\rm T}, \widetilde E_{\rm T}\right\}$ have the standard representation
\begin{eqnarray}
\label{Def-GPD-F_Fwd}
F(x,\eta=0,t)= \int_{-1+x}^{1-x}\!dz\,
f(x,z,t) \quad\mbox{for}\quad
f \in \left\{h,\widetilde h, h_{\rm T}, e_{\rm T}, \widetilde h_{\rm T}, \widetilde e_{\rm T}\right\}\,,
\end{eqnarray}
while for the GPD $E$   a extra factor $(1-x)$ appears
\begin{eqnarray}
\label{Def-SpeRep_FwdE} \
E (x,\eta=0,t) =
(1-x)\int_{-1+x}^{1-x}\! dz\;  e(x,z,t)\,.
\end{eqnarray}
Setting in addition $t=0$, the $H$, $\widetilde H$, and $H_{\rm T}$ GPDs are reduced to the $f_1$, $g_1$, and $h_1$ PDFs, respectively.

More care is required for the $z/\eta$ term in $\widetilde E$. Its $x$-moments can be straightforwardly evaluated from the DD representation (\ref{Def-SpeRep}), e.g.,  the two lowest ones read
\begin{eqnarray}
\label{Def-MomtE}
\left\{ {\widetilde E_0 \atop \widetilde E_1}\right\}(t)\equiv \int_{-\eta}^1\!dx\,\left\{ {1\atop x}\right\}\widetilde E(x,\eta,t)  = \int_{0}^{1}\! dy\;\int_{-1+y}^{1-y}\! dz \left\{{\widetilde e + (1-y)\, e \atop
y \widetilde e + y (1-y)\, e- z^2\, e}
\right\}(y,z,t)\,.
\end{eqnarray}
We might now remember that GPDs are generalized functions in the mathematical sense and
thus we might formally write
$$
 \lim_{\eta\to 0} \tau(x) \frac{z}{\eta}\delta(x-y-z\eta)  =   \tau(x) \lim_{\eta\to 0} \frac{z}{\eta}\delta(x-y)+ \frac{d \tau(x)}{dx}  z^2 \delta(x-y) ,
$$
where $\tau(x)$ is a test function. One immediately finds for our GPD $\widetilde E$  the non-standard representation
\begin{eqnarray}
\label{DDrep1-tEx0}
\widetilde E(x,\eta=0,t) =
\int_{-1+x}^{1-x}\! dz\;
 \left[\widetilde e + (1-x) e\right](x,z,t) -\frac{\overleftarrow{d}}{dx} \int_{-1+x}^{1-x}\! dz\; z^2 e(x,z,t) \,.
\end{eqnarray}
This equation, considered in the sense of a generalized function, is consistent with the Mellin moments evaluated from the DD representation, see, e.g.,  the two lowest ones (\ref{Def-MomtE}). For a test function that sufficiently  vanish on the boundary   $x=0$, e.g.~$x^2$, we can write Eq.~(\ref{DDrep1-tEx0})
after partial integration also as
\begin{eqnarray}
\label{DDrep2-FwdtEx0}
x^2 \widetilde E(x,\eta=0,t) =
x^2 \int_{-1+x}^{1-x}\! dz\;
 \left[\widetilde e + (1-x) e\right](x,z,t) +x^2 \frac{d}{dx} \int_{-1+x}^{1-x}\! dz\; z^2 e(x,z,t) \,,
\end{eqnarray}
where we assumed that $e(x,z,t)$ vanishes at $z=\pm (1-x)$.\\

\noindent{\em iii.  GPDs on the cross-over line}\\
\noindent
The imaginary parts of hard-exclusive amplitudes to leading order accuracy, e.g., for Compton form factors in deeply virtual Compton scattering
\begin{eqnarray}
\label{calF-LO-per}
{\cal F}(\xi,\cdots) &\!\!\!\stackrel{\rm LO}{=}\!\!\!&\!\!\!\!\sum_{q=u,d,\cdots} e_q^2\int_{-1}^1\!\!dx\!\!\left[ \frac{1}{\xi - x - i\epsilon} \mp \frac{1}{\xi + x - i\epsilon} \right]\! F^q(x,\eta=\xi,\cdots)\; \mbox{for} \; F \leftrightarrow {\cal F} \in \left\{ {{\cal H}, {\cal E} \atop \widetilde{\cal H}, \widetilde{\cal E}} \right. ,\qquad\quad
\end{eqnarray}
are given by the corresponding GPDs on the cross-over line  $\eta=x$,
\begin{eqnarray}
\label{F-LO-Im}
\Im{\rm m} {\cal F}(\xi,\cdots) &\!\!\!\stackrel{\rm LO}{=}\!\!\!& \pi \sum_{q=u,d,\cdots} e_q^2  \left[F^q(x=\xi,\eta=\xi,\cdots) \mp F^q(x=-\xi,\eta=\xi,\cdots)\right].
\end{eqnarray}
Here, $e_q$ are the quark charges and according to the discussion at the beginning of Sec.~\ref{Sec-GPDs&uPDFs} the
GPD $F^q$ can be decomposed  in a quark and anti-quark GPD.
One can simplify the DD-representation for quark (or anti-quark) GPDs at the cross-over line, e.g.,
they can be written as an integral over the variable $w =z/(1-y)$.
Let us consider here only  the GPDs  $E$ and $x \widetilde E$ for which  we find from Eq.~(\ref{Def-SpeRep})
the integral representations
\begin{eqnarray}
E(x,x,t) &\!\!\!= \!\!\! &\int_{-1}^{1}\!dw \frac{(1-x)^2}{(1- x\, w)^2 }\; e\left(\frac{x(1-w)}{1- x\, w },(1-x)w,t\right)\,,
\\
x \widetilde{E}(x,x,t) &\!\!\!= \!\!\! & \int_{-1}^{1}\!dw \frac{1-x}{(1- x\, w)^2 }\;
\left[x\,\widetilde e  +\frac{(x - w) (1 - x)}{1 -x\, w} e\right]\left(\frac{x(1-w)}{1- x\, w },(1-x)w,t\right)\,,
\end{eqnarray}
where $e$ and $\widetilde e$ are specified in Eqs.~(\ref{Def-SpeFun-e},\ref{Def-SpeFun-te}).
It is instructive to compare the asymptotically small $x$-behavior of  $E(x,x,t) $  and $x \widetilde{E}(x,x,t)$:
\begin{eqnarray}
\lim_{x\to 0}E(x,x,t) &\!\!\! =\!\!\! &  x^{-\alpha}\int_{-1}^{1}\!dw \; (1-w)^{-\alpha}\, e^{\rm res}(w,t)\,,\\
\lim_{x\to 0} x \widetilde{E}(x,x,t) &\!\!\! =\!\!\! &  - x^{-\alpha}\int_{-1}^{1}\!dw \; w(1-w)^{-\alpha}\, e^{\rm res}(w,t)\,,
\end{eqnarray}
where the residue function $e^{\rm res}(w,t) = \lim_{x\to 0} x^\alpha e^{\rm res}(x,w,t)$. Other twist-two GPDs possess an analogous small-$x$ behavior as
GPD $E$  and we clearly realize that even  GPD $x \widetilde E$ has the same small $x$ asymptotics. Hence, it could provide an important contribution, too, and one might be concerned about the assumption, often employed in GPD phenomenology, that only the pion pole \cite{Mankiewicz:1998kg,Frankfurt:1999fp} is essential in the Compton form factor $\widetilde {\cal E}$, cf.~Ref.~\cite{Vanderhaeghen:1999xj}.\\

\noindent{\em iv. ``dispersion relations''}\\
\noindent
The real parts of hard-exclusive amplitudes to leading order accuracy can be alternatively obtained from a ``dispersion relation" \cite{Teryaev:2005uj,AniTer07} (for a treatment beyond leading order accuracy see also Refs.~\cite{Kumericki:2007sa,Diehl:2007jb}).
The appearance of a subtraction constant in the ``dispersion relation" is tied to the kind of DD representation.
For the  Compton form factors $\cal H$ and $\cal E$, given in Eq.~(\ref{calF-LO-per}), we have the  ``dispersion relation''
 \begin{eqnarray}
 \label{HandE-DR}
\Re{\rm e} \left\{{{ \cal H} \atop {\cal E} }\right\}(\xi,t) \stackrel{\rm LO}{=} {\rm PV}\int_0^1 \frac{2x}{\xi^2-x^2} \left\{{{ H} \atop { E} }\right\}(x,x,t)\pm  {\cal D}(t)\,,
 \quad  {\cal D}(t) = \int_{-1}^{1}\!dz \frac{2 z}{1-z^2} d(z,t)\,,
 \end{eqnarray}
where the subtraction constant $\cal D$ is expressed in terms of the $D$-term (\ref{Cal-Dterm}) and $H$, $E$, and $d$ contain here both
quark and anti-quark GPDs, weighted with the corresponding charge factors.
For $\widetilde H$ and transversity GPDs, which possess the common DD representation, a subtraction constant is absent.
Some attention should be given to the GPD $\widetilde E$. Since of the $x^{-1-\alpha}$ behavior, it is more appropriate to work with an over-subtracted dispersion relation \cite{Bechler:2009me}. The real part of the corresponding Compton form factor reads then as following
 \begin{eqnarray}
 \label{tE-DR}
\Re{\rm e} \widetilde {\cal E}(\xi,t) \stackrel{\rm LO}{=}\frac{1}{\xi} {\rm PV}\int_0^1\!dx\, \frac{2x^2}{\xi^2-x^2}
\widetilde{E}(x,x,t)+\frac{1}{\xi} \widetilde{\cal C}(t)\,, \quad
\widetilde{\cal C}(t) =   \int_{-1}^{1}\!dz \frac{2}{1-z^2} \widetilde{d}(z,t)\,,
 \end{eqnarray}
where the subtraction constant can be evaluated from a $\widetilde{d}$ function that is commonly associated with the pion pole.
As the $d$-term, the $\widetilde d$ one can be obtained from the limit $\eta\to \infty$ with fixed $x/\eta$, where the $z/\eta$-term
in the DD representation (\ref{Def-SpeRep}) drops out:
\begin{eqnarray}
 \label{tE-td}
\widetilde{d}(z,t)
&\!\!\! = \!\!\!&\int_{0}^{1-|z|}\!dy\, \left[e +(1-y)\widetilde{e}\right](y,z,t)
\\
&\!\!\! = \!\!\!&
2 \int_{0}^{1-|z|}\!dy\,\left[\frac{m}{M}+y + (1-y)^3-(1-y) z^2\right]{\bf \hat \Phi}(y,z,t)\,.
\nonumber
\end{eqnarray}
It is obvious  that in our model the subtraction constant inherits the rather
smooth $t$-dependence of GPD $\widetilde E$  and so the steep $t$-dependence of the pion pole will be
generally absent.\\

\subsubsection{Other scalar and pseudo-scalar diquark LFWF models}
\label{sec:GPD-psesca}

As already emphasized above, utilizing an arbitrary LFWF ``spinor" could yield GPDs that do not satisfy the  polynomiality conditions. To find appropriate LFWF ``spinors",  we employ the building blocks
$$
 \frac{m}{M}+ X\,, \quad  \frac{|{\bf k}_\perp|}{M} e^{i\varphi}\,, \quad  \frac{|{\bf k}_\perp|}{M} e^{-i\varphi}
$$
for a $L^z=0$, $L^z=+1$, and $L^z=-1$ coupling, respectively.  We construct from these building blocks all possible LFWF ``spinors" that provide us a set of GPDs, i.e., the polynomiality conditions are satisfied.
We emphasize that all $\Delta L_z=0$ LFWF overlap contributions  in the spin-correlation matrix (\ref{bfF}), i.e.,
the diagonal  entries ${\bf F}_{11}={\bf F}_{44}$, ${\bf F}_{22}={\bf F}_{33}$,  and ${\bf F}_{14}={\bf F}_{41}$,
contain a  $t_0/4M^2= -\eta^2/(1-\eta^2)$ proportional addenda that it expressed in terms $\Delta L^z \neq 0$ GPDs.
Apart from the parity/time-reversal constraints, this $\Delta L_z=0$ and $\Delta L^z  \neq 0$ cross talk provides us the restrictions for the LFWF ``spinors".

For a scalar spectator it turns out that  there exist only one more LFWF ``spinor",
\begin{eqnarray}
\label{LFWF-spinor-A0}
\mbox{\boldmath $\psi$}_{0}^{\rm A}(X,{\bf k}_\perp) = \frac{1}{\sqrt{3}M} \left(\!\!
               \begin{array}{c}
                 m +X M \\
                |{\bf k}_\perp| e^{i \varphi} \\
                 |{\bf k}_\perp| e^{-i \varphi}  \\
                - m -X M \\
               \end{array}\!\!
             \right) \frac{ \overline{\phi}(X,{\bf k}_\perp|\lambda^2)}{\sqrt{1-X}}\,,
\end{eqnarray}
with four non-vanishing entries that meets the GPD requirements. It might be interpreted as the coupling of quarks to an axial-vector diquark spectator with zero spin projection $s_2$ on the $z$-axis. Compared to the scalar diquark LFWF ``spinor" (\ref{LFWF-spinor-S}),
the relative sign of the first/third  and second/forth entries  in the ``spinor" (\ref{LFWF-spinor-A0}) is changed and, moreover,
the normalization is altered by a factor $1/\sqrt{3}$.
The corresponding GPDs are obtained from the scalar diquark ones (\ref{Def-SpeRep}--\ref{Def-SpeFun-te},
\ref{Def-GPD-FT}--\ref{Def-SpeFun-teT}), given in Sec.~\ref{sec:GPD-sca},  by multiplying them with a factor $1/3$
and decorating all chiral odd GPDs with an additional minus sign.

Taking the spin-correlation matrices of the scalar and $s_2=0$ axial-vector diquark spectator model and
forming a positive semi-definite linear combination with them, we can obtain a model in which
the normalization of the chiral odd GPDs can be easily tuned.  Obviously, we can even combine these models in such a manner that the GPDs (of course, also the corresponding uPDFs) in the chiral odd sector vanish. Such a model can be also build from a pair of LFWF ``spinors",
\begin{eqnarray}
\label{LFWF-spinors-Spair}
{\bf\Psi}_{+}^{\rm S}&\!\!\! =\!\!\!&
\frac{1}{ M}\left(\!\!\!\!
\begin{array}{c}
m+X M \\
0 \\
|{\bf k}_\perp| e^{-i \varphi} \\
0  \\
\end{array}\!\!\!\!
\right)\frac{ \overline{\phi}(X,{\bf k}_\perp|\lambda^2)}{\sqrt{1-X}}\,,\;\;\;
{\bf\Psi}_{-}^{\rm S}=
\frac{1}{ M}\left(\!\!\!\!
\begin{array}{c}
0 \\
-|{\bf k}_\perp| e^{i \varphi} \\
0 \\
m+X M  \\
\end{array}\!\!\!\!
\right)\frac{ \overline{\phi}(X,{\bf k}_\perp|\lambda^2)}{\sqrt{1-X}}\,,
\end{eqnarray}
that belong to states in which only struck quarks with spin projection $+1/2$ and $-1/2$, respectively, occur.
Again the chiral even GPDs can be taken from Sec.~\ref{sec:GPD-sca} while the chiral odd ones are simply zero.

Finally, we like to add that one can easily obtain the results for a pseudo-scalar or $s_2=0$ vector diquark model. Essentially, we have to
replace the $L^z=0$ coupling $m/M+X$  by $-m/M+X$ and adopt the relative sign of the first/third and second/forth entries. For a pseudo-scalar target we find then in accordance with an explicit calculation (up to an overall minus sign)
\begin{eqnarray}
\label{LFWF-spinor-PS}
\mbox{\boldmath $\psi$}^{\rm PS}(X,{\bf k}_\perp) = \frac{1}{M} \left(\!\!
               \begin{array}{c}
                 -m +X M \\
                |{\bf k}_\perp| e^{i \varphi} \\
                 |{\bf k}_\perp| e^{-i \varphi}  \\
                 m -X M \\
               \end{array}\!\!
             \right) \frac{ \overline{\phi}(X,{\bf k}_\perp|\lambda^2)}{\sqrt{1-X}}\,,
\end{eqnarray}
and for the $s_2=0$ state of a vector diquark  we have
\begin{eqnarray}
\label{LFWF-spinor-V0}
\mbox{\boldmath $\psi$}_0^{\rm V}(X,{\bf k}_\perp) = \frac{1}{\sqrt{3}M} \left(\!\!
               \begin{array}{c}
                 -m +X M \\
                -|{\bf k}_\perp| e^{i \varphi} \\
                 |{\bf k}_\perp| e^{-i \varphi}  \\
                - m +X M \\
               \end{array}\!\!
             \right) \frac{ \overline{\phi}(X,{\bf k}_\perp|\lambda^2)}{\sqrt{1-X}}\,.
\end{eqnarray}
Since this $\pm m/M+X$ coupling  is the only term in which the quark mass enters linearly in our  GPD models,
we can again obtain the GPD expressions from the scalar diquark ones in Sec.~\ref{sec:GPD-sca}, namely, by the replacement $m\to -m$. Thereby, for pseudo-scalar   diquark GPDs
the overall sign of the chiral odd GPDs changes in addition and the overall normalization of all vector diquark  GPDs is conventionally
modified by a factor $1/3$.
We emphasize that for smaller values of $x$  the  sign of the $L^z=0$ and $|L^z|=1$ coupling, given by the factor $y-m/M$, becomes in such models negative. In particular the GPD $E$  possesses now a node and is for smaller $x$ values naturally negative, see the  DD $e$ expression (\ref{Def-SpeFun-e}).

\subsubsection{A minimal axial-vector diquark model}
\label{sec:GPD-axialvector}

For an (axial-)vector diquark model it is perhaps not an entirely trivial task to find the LFWFs that respect the underlying Lorentz symmetry
and possess the required frame independent form (\ref{Def-ProSta}), i.e., they are in particular independent on the transverse proton
momentum  ${\bf p}_\perp$.
Assuming from the beginning that LFWFs possess the frame independent form (\ref{Def-ProSta}) and constructing them in a frame where the nucleon transverse momentum vanishes might yield erroneous results. It is beyond the scope of this paper to construct
(axial-)vector diquark LFWFs from explicitly  covariant models rather we   propose now a minimal model for an axial-vector diquark spectator which is relatively simple, cf.~Ref.~\cite{Tiburzi:2004mh}.
The guidance for finding such a  model we take from the observation that the axial-vector diquark model of Ref.~\cite{JakMulRod97} is a ``spherical" uPDF model, which spin-density matrix (\ref{model-sperical-SU(4)-1},\ref{model-sperical-SU(4)-2}) can be represented by some unpolarized $f_1$ uPDF,  proportional to  the identity matrix, and a non-trivial part that can be taken from a scalar diquark model.  If we employ the LFWF ``spinors" (\ref{LFWF-spinor-A0}) and
\begin{eqnarray}
\label{LFWF-spinors-AV}
{\bf\Psi}^{\rm A}_{+1}&\!\!\! =\!\!\!&
\frac{\sqrt{2}}{\sqrt{3} M}\left(\!\!\!\!
\begin{array}{c}
0 \\
m+X M \\
0  \\
|{\bf k}_\perp| e^{-i \varphi} \\
\end{array}\!\!\!\!
\right)\frac{ \overline{\phi}(X,{\bf k}_\perp|\lambda^2)}{\sqrt{1-X}}\,,\;\;\;
{\bf\Psi}^{\rm A}_{-1}=
\frac{\sqrt{2}}{\sqrt{3} M}\left(\!\!\!\!
\begin{array}{c}
|{\bf k}_\perp| e^{i \varphi} \\
0 \\
-m-X M  \\
0 \\
\end{array}\!\!\!\!
\right)\frac{ \overline{\phi}(X,{\bf k}_\perp|\lambda^2)}{\sqrt{1-X}}\,,
\end{eqnarray}
where the subscripts $\{0,+1,-1\}$ label the polarization states of the axial-vector diquark
[compare with the generic ``spinors" (\ref{LFWF-spinors-Identity}) ], we can easily verify that these ``spinors" yield by an appropriate choice of the effective LFWF  the leading--power uPDFs of Ref.~\cite{JakMulRod97}. Forming the GPD spin-correlation matrix (\ref{bfF-rank3}) for each  axial-vector diquark
state and summing up the findings, our axial-vector diquark GPD spin-correlation matrix can be straightforwardly obtained:
\begin{eqnarray}
\label{GPD-av-model}
{\bf F}^{\rm av}(x,\eta,t|\varphi)  =\frac{4}{3} {\bf 1\!\!\!1}_{4\times 4}\, H(x,\eta,t) +\frac{4}{3} {\bf e}_{4\times 4}\,  E(x,\eta,t)  -\frac{1}{3} {\bf F}^{\rm sca}(x,\eta,t|\varphi)\,.
\end{eqnarray}
Here, ${\bf 1\!\!\!1}_{4\times 4}$ is again the identity matrix, we introduced the matrix
$$
{\bf e}_{4\times 4} = \frac{{\bf \Delta}_\perp}{(1-\eta)4 M}
\left(
  \begin{array}{cccc}
    0 & 0 & e^{-i \varphi} & 0 \\
    0 & 0 & 0 & e^{-i \varphi} \\
    -e^{i \varphi}  & 0 & 0 & 0 \\
    0 & -e^{i \varphi} & 0 & 0 \\
  \end{array}
\right),
$$
${\bf F}^{\rm sca}$ is the spin-correlation matrix (\ref{bfF}) of the scalar diquark model, where all GPDs are defined by the double distribution
representations (\ref{Def-SpeRep}--\ref{Def-SpeFun-te},\ref{Def-GPD-FT}--\ref{Def-SpeFun-hT}), given in Sec.~\ref{sec:GPD-sca}. Our axial-vector diquark $H$ and $E$ GPDs are the same as in the scalar diquark model. The renaming six GPDs can be also taken from the scalar diquark model, however, they are decorated with an additional $-1/3$ factor. We add that the sum of transverse $s_2=+1$ and $s_2=-1$ LFWF overlaps provide us already a model for chiral even GPDs, where compared to the scalar diquark model the parity odd GPDs $\widetilde H$ and $\widetilde E$  have an additional overall minus sign. Moreover, a vector diquark model for GPDs and uPDFs can be easily found by replacing the struck quark mass $m\to -m$:
\begin{eqnarray}
\label{LFWF-spinors-V}
{\bf\Psi}^{\rm V}_{+1}&\!\!\! =\!\!\!&
\frac{\sqrt{2}}{\sqrt{3} M}\left(\!\!\!\!
\begin{array}{c}
0 \\
-m+X M \\
0  \\
|{\bf k}_\perp| e^{-i \varphi} \\
\end{array}\!\!\!\!
\right)\frac{ \overline{\phi}(X,{\bf k}_\perp|\lambda^2)}{\sqrt{1-X}}\,,\;\;\;
{\bf\Psi}^{\rm V}_{-1}=
\frac{\sqrt{2}}{\sqrt{3} M}\left(\!\!\!\!
\begin{array}{c}
|{\bf k}_\perp| e^{i \varphi} \\
0 \\
m-X M  \\
0 \\
\end{array}\!\!\!\!
\right)\frac{ \overline{\phi}(X,{\bf k}_\perp|\lambda^2)}{\sqrt{1-X}}\,,
\end{eqnarray}
where the $s_2=0$ vector diquark LFWF ``spinor" is given in Eq.~(\ref{LFWF-spinor-V0}).

According to SU(6) symmetry, we might associate the axial-vector diquark GPDs with $d$ quarks, while the $u$-quark GPDs are given as combination of scalar diquark ones $2u/3-d/3$ and axial-vector diquark ones. We might write this model in a compact form as
\begin{eqnarray}
\label{model-sperical-SU(4)-1-GPD}
\widetilde{\bf F}^{u} &\!\!\! \stackrel{\rm sph\&SU(4)}{=} \!\!\!&
\frac{2}{3} {\bf 1\!\!\!1}_{4\times 4}\,  H^{u/6+2d/3}(x,\eta,t) + \frac{2}{3} {\bf e}_{4\times 4} E^{u/6+2d/3}(x,\eta,t) +  \frac{4}{3}\widetilde{\bf F}^{2u/3-d/3}(x,\eta,t)\,,
\\
\label{model-sperical-SU(4)-2-GPD}
\widetilde{\bf F}^{d}&\!\!\! \stackrel{\rm sph\&SU(4)}{=} \!\!\!&
\frac{4}{3} {\bf 1\!\!\!1}_{4\times 4}\,  H^{u/6+2d/3}(x,\eta,t)
+ \frac{4}{3} {\bf e}_{4\times 4} E^{u/6+2d/3}(x,\eta,t) -\frac{1}{3}\widetilde{\bf F}^{2u/3-d/3}(x,\eta,t) \,,
\end{eqnarray}
where we might also impose the conditions $ H^{u/6+2d/3}=H^{2u/3-d/3}$ and $ E^{u/6+2d/3}=E^{2u/3-d/3}$.
We emphasize that in the uPDF analog SU(6) symmetric ``spherical" model  constructions  (\ref{model-sperical-SU(4)-1},\ref{model-sperical-SU(4)-2}) no analog of GPD $E$   appears, where, of course, the nfPDF/uPDF correspondence $H \leftrightarrow  f_1$  holds. Contrarily to a  ``spherical"  uPDF model, we have according to Eq.~(\ref{GPD-av-model}) in the linear combination
$${\bf F}^{\rm av}+ \frac{1}{3} {\bf F}^{\rm sc} -\frac{4}{3} {\bf 1\!\!\!1}_{4\times 4}\,  H(x,\eta,t) = \frac{4}{3} {\bf e}_{4\times 4} \, E(x,\eta,t)$$
polarization effects that arise from the GPD $E$ . This GPD $E$  appearance can be traced back to the restoration of Lorentz symmetry. Namely, in the diagonal entries of the GPD spin-correlation matrix (\ref{bfF}),  which are now identically given by
$$\frac{1}{2}H(x,\eta,t) +\frac{t_0}{8M^2} E(x,\eta,t) =\frac{1}{2}H(x,\eta,t) -\frac{\eta^2}{2(1-\eta^2)} E(x,\eta,t),$$
a $t_0/8M^2$ addenda appears in our model calculation that is absorbed  by the non-vanishing $E$ GPD. If this absorption does not occur, an additional $\eta^2/(1-\eta^2)$  proportional term exist in the GPD $H$  and, thus, polynomiality is broken.

\begin{table}[t]
\begin{center}
\begin{tabular}{|c|c||c||c|c||c|c||c|c|c} \hline
diquark coupling & LFWFs &rel.~presign
&\multicolumn{1}{c||}{parity even}&  \multicolumn{1}{c||}{parity odd}& \multicolumn{1}{c|}{chiral odd} \\
       & Eqs. & of quark&\multicolumn{1}{c||}{(\ref{Def-SpeRep},\ref{Def-SpeFun-h},\ref{Def-SpeFun-e})}&
\multicolumn{1}{c||}{(\ref{Def-SpeRep},\ref{Def-SpeFun-th},\ref{Def-SpeFun-te})}& \multicolumn{1}{c|}{(\ref{Def-GPD-FT}--\ref{Def-SpeFun-teT})} \\
                     &  &  mass term &  norm. &  norm.  &     norm.   \\  \hline\hline
  scalar & (\ref{LFWF-spinor-S}) &  $+$  & +1& $+1$ &  $+1$ \\
  pseudo-scalar &(\ref{LFWF-spinor-PS}) & $-$  & $+1$ & $+1$ & $-1$  \\
  (pseudo-)scalar & (\ref{LFWF-spinors-Spair}) &  $+(-)$  & +1& $+1$ &  $0$ \\
  lon. axial-vector & (\ref{LFWF-spinor-A0}) & $+$  & +1/3 & $+1/3$ &$-1/3$  \\
  tra. axial-vector & (\ref{LFWF-spinors-AV}) & $+$  & +2/3 & $-2/3$ &$0$  \\
  min. axial-vector & (\ref{LFWF-spinor-A0},\ref{LFWF-spinors-AV}) & $+$  & +1& $-1/3$ &$-1/3$  \\
  lon. vector & (\ref{LFWF-spinor-V0}) & $-$   &  +1/3  & $+1/3$ & $+1/3$ \\
  tra. vector & (\ref{LFWF-spinors-V}) & $-$   &  +2/3  & $-2/3$ & $0$ \\
  min. vector & (\ref{LFWF-spinor-V0},\ref{LFWF-spinors-V}) & $-$   &  +1  & $-1/3$ & $+1/3$ \\
  \hline
\end{tabular}
\end{center}
\caption{\label{tab:LFWFs2GPDs}\small
Diquark models built from various LFWF ``spinors". The resulting uPDF and  GPD expressions are taken from the scalar diquark model, given in Eqs.~(\ref{f1-DQSM},\ref{g1g1T-DQSM}--\ref{h1L-DQSM}) and Eqs.~(\ref{Def-SpeRep}--\ref{Def-SpeFun-te},\ref{Def-GPD-FT}--\ref{Def-SpeFun-teT}), respectively, where they
are modified corresponding to the formal (relative) presign setting of the struck quark mass term $m$ in the LFWF ``spinors"   and the overall normalization factors in the parity even,  parity odd, and chiral odd sectors.
}
\end{table}
To set up more flexible  GPD and uPDF models that respect Lorentz symmetry and positivity, one can relax the SU(6) symmetry
conditions and one might write the  spin-correlation and -density matrices for each quark flavor as a linear combination of
those we have discussed, see Tab.~\ref{tab:LFWFs2GPDs}. In each of these six independent sectors (\ref{LFWF-spinor-S},\ref{LFWF-spinor-PS},\ref{LFWF-spinor-A0}, \ref{LFWF-spinors-AV},\ref{LFWF-spinor-V0},\ref{LFWF-spinors-V}) we might even choose a
separate effective two-body LFWF. In this way one can mostly eliminate all model dependent relations.

\subsection{Model dependent relations}
\label{GPDs2uPDFs-diquark}

Our (pseudo-)scalar and longitudinal (axial-)vector LFWF models are ``spherical" GPD models where their spin-correlation matrices (\ref{bfF}) have rank-three and so the model constraints, derived in Sec.~\ref{sec:GPD-relations}, must be valid.
In analogy to the uPDF relations of a ``spherical" model, given in Sec.~\ref{uPDF-scaDiq}, we collect now the ``spherical" GPD  model relations and the GPD/uPDF cross talks for the scalar diquark model.
Some of these relations are also valid for the class of ``spherical"
models of rank-four and so they are supposed to hold true for the bag model \cite{AvaEfrSchYua10}, chiral quark soliton model  \cite{LorPasVan11}, covariant parton model \cite{EfrSchTerZav09}, and the axial-vector diquark model in the version of Ref.~\cite{JakMulRod97}. Apart from the purpose of model estimates such relations might be also directly confronted with phenomenological findings,
which in  principle allow to judge on the quark-diquark coupling without further specification of the effective two-body LFWF.

\subsubsection{``Spherical" GPD model constraints}
\label{Sec:``Spherical" GPD model constraints}
Using the DD results (\ref{Def-SpeRep}--\ref{Def-SpeFun-te},\ref{Def-GPD-FT}--\ref{Def-SpeFun-teT}) of our ``spherical" model, one can conveniently check that  within the upper sign the three linear constraints (\ref{GPD-constraints-linear}), i.e., also the quadratic relation (\ref{GPD-zeromode1}), and the integral equation (\ref{GPD-constraints-linear-2}) hold true.  These four constraints allow us   in analogy to a ``spherical" uPDF model of rank-one, see Sec.~\ref{uPDF-classification-spherical} and Sec.~\ref{uPDF-scaDiq}, to express the chiral even GPDs entirely by the chiral odd ones:
\begin{eqnarray}
\label{GPDmod-con1}
H(x,\eta,t) &\!\!\! \stackrel{{\rm sph}^3}{=} \!\!\! & \pm\left[
\overline{H}_{\rm T}(x,\eta,t) - \frac{t}{4 M^2}\widetilde{H}_{\rm T}(x,\eta,t)  -
\int_{-\infty}^t\!\frac{ dt^\prime}{4M^2}\, \widetilde{H}_{\rm T}(x,\eta,t^\prime)+\eta \widetilde E_{\rm T}(x,\eta,t)\right],
\qquad
\\
\label{GPDmod-con2}
E(x,\eta,t) &\!\!\! \stackrel{{\rm sph}^3}{=} \!\!\! & \pm\left[
E_{\rm T}(x,\eta,t)+ 2 \widetilde{H}_{\rm T}(x,\eta,t) -\eta \widetilde E_{\rm T}(x,\eta,t)
\right],
\\
\label{GPDmod-con3}
\widetilde H(x,\eta,t) &\!\!\! \stackrel{{\rm sph}^3}{=} \!\!\! & \pm\left[
\overline{H}_{\rm T}(x,\eta,t) + \frac{t}{4 M^2}\widetilde{H}_{\rm T}(x,\eta,t)  +
\int_{-\infty}^t\!\frac{ dt^\prime}{4M^2}\, \widetilde{H}_{\rm T}(x,\eta,t^\prime)\right],
\\
\label{GPDmod-con4}
\widetilde{E}(x,\eta,t)&\!\!\! \stackrel{{\rm sph}^3}{=} \!\!\! & \pm\left[
E_{\rm T}(x,\eta,t)-\frac{1}{\eta}\widetilde E_{\rm T}(x,\eta,t)\right],
\end{eqnarray}
where the upper (lower) sign applies for the (pseudo-)scalar  and longitudinal (axial-)vector  diquark models.
Here, the transversity  GPD $\overline{H}_{\rm T}$ , arising from $L^z=0$ LFWF overlaps, is  in our models given as
$$
\overline{H}_{\rm T} \equiv H_{\rm T} - \frac{t}{4M^2} \widetilde{H}_{\rm T} \stackrel{{\rm sph}^3}{=} \pm  \int_{0}^1\!dy\!\int_{-1+y}^{1-y}\! dz\;
\delta(x-y-\eta z ) \left(\pm \frac{m}{M}+y\right)^2  {\bf \hat \Phi}(y,z,t)\,.
$$
Moreover, the analog of the ``hidden'' quadratic constraint (\ref{GPD-constraints-quadratic}), connecting the four chiral odd GPDs, is not satisfied in general, however, it is easily to see that the analog equation is valid in the DD representation.
Hence, it  can be employed to evaluate one transversity DD in terms of the three remaining chiral odd ones. In addition to this five constraints for a ``spherical" GPD model,
further model constraints (\ref{GPDmodel-const1}--\ref{GPDmodel-const3}, \ref{Con-even2odd},\ref{Con-even2odd-1}) can be formulated in terms of DDs.
In analogy to a ``spherical" uPDF model a ``spherical" GPD model of rank-four might be obtained by adding a GPD $H$  addenda, see analog discussion in Sec.~\ref{uPDF-classification-spherical}, and so only the first relation (\ref{GPDmod-con1}) will alter. However, we emphasize that we do not have a LFWF overlap representation for such a model at hand and that the linear combination of ``spherical" scalar and axial-vector diquark  models, see
Eq.~(\ref{GPD-av-model}), will alter the $H$ and $E$ GPD relations (\ref{GPDmod-con1}) and (\ref{GPDmod-con2}) as well. Utilizing the constraints
(\ref{GPDmod-con1}--\ref{GPDmod-con4}) and the prefactors in Tab.~\ref{tab:LFWFs2GPDs}, one might write down the corresponding GPD relations for our minimal version of an axial-vector diquark model.

Let us emphasize once more that  in the scalar diquark model the analogy to the saturation of the Soffer bound (\ref{Def-Tra_satured}) holds.
Combining the model constraints (\ref{GPDmod-con1},\ref{GPDmod-con3}), the transversity GPD $\overline{H}_{\rm T}$ is essentially given by the average of
GPDs $H$ and $\widetilde H$,
\begin{eqnarray}
\label{GPD-Soffer-bound-analog}
\overline{H}_{\rm T}(x,\eta,t) \stackrel{\rm sca}{=} \frac{1}{2} \left[H+ \widetilde H -\eta \widetilde{E}_{\rm T}\right](x,\eta,t)
\quad\Rightarrow \quad
\overline{H}_{\rm T}(x,\eta=0,t) \stackrel{\rm sca}{=} \frac{1}{2} \left[H+ \widetilde H\right](x,0,t)
\,,
\end{eqnarray}
where in the non-forward case also a $\eta$ proportional $|L^z|=1$ with $|L^z|=0$ LFWF overlap, given in terms of the GPD
$\widetilde E_{\rm T}$, appears. Since the corresponding DD (\ref{Def-SpeFun-teT}) is odd in $z$,
this chiral odd GPD  $\widetilde E_{\rm T}$  vanish already by itself in the zero-skewness case.
Thus, we expect that for small or moderate $\eta$ values the $\widetilde E_{\rm T}$ term in Eq.~(\ref{GPD-Soffer-bound-analog}) is negligible.

Note also that the model relation (\ref{GPDmod-con2}) tells us that the chiral odd  GPD $\overline{E}_{\rm T}$  is given by the chiral even $E$ GPD
\begin{eqnarray}
\label{bEx0}
\overline{E}_{\rm T}(x,\eta,t) \equiv
\left[E_{\rm T}+ 2 \widetilde{H}_{\rm T} -\eta \widetilde E_{\rm T}\right](x,\eta,t)
\stackrel{{\rm sph}^3}{=} \pm E(x,\eta,t)\quad \mbox{for} \quad \left\{ {\mbox{scalar} \atop \mbox{lon.~axial-vector}}\right.   \,.
\end{eqnarray}
Since from the DDs (\ref{Def-SpeFun-thT}) and (\ref{Def-SpeFun-teT}) it follows that $-\widetilde{H}_{\rm T}$ and  $\widetilde{E}_{\rm T}$ are positive (negative) in our scalar (axial-vector) diquark model, we can  also read off from the relation (\ref{bEx0}) that the absolute value of
$ E_{\rm T}$  is even larger as the positive $E$, see corresponding  DD (\ref{Def-SpeFun-e}):
\begin{eqnarray}
\label{ETx0}
\pm E_{\rm T}(x,\eta,t)
\stackrel{{\rm sph}^3}{>}   E(x,\eta,t)
\quad \mbox{for} \quad \left\{ {\mbox{scalar} \atop \mbox{lon.~axial-vector}}\right. \,.
\end{eqnarray}
 Analogously, we can derive from the relation (\ref{GPDmod-con4}) a model dependent inequality for the GPD $\widetilde E$,
\begin{eqnarray}
\label{tE>E}
\pm \widetilde E(x,\eta,t)  \stackrel{{\rm sph}^3}{>}  \pm  E_{\rm T}(x,\eta,t) \stackrel{{\rm sph}^3}{>}   E(x,\eta,t)
\quad \mbox{for} \quad \left\{ {\mbox{scalar} \atop \mbox{lon.~axial-vector}}\right.\,.
\end{eqnarray}
We emphasize that these inequalities (\ref{ETx0}) and (\ref{tE>E}) hold only as long as the scalar LFWF overlap ${\bf \hat \Phi}$ is positive definite for a given kinematical point.

\subsubsection{Cross talks of nfPDFs and uPDFs}

In Sec.~\ref{sec:GPDs-uGPDs} we explained that the appearance of orbital angular momentum implies that in QCD a one-to-one map of uPDFs and
nfPDFs does not exist. In our scalar diquark model the quark angular momentum is restricted to $|L^z|\le 1$ and so we can ask again for a model dependent conversion.  Let us for completeness here also provide the conversion for the case that we have only one effective LFWF as it appears in a common scalar diquark model, see Sec.~\ref{sec:GPD2uPDF-Lz0}. The conversion formula (\ref{q2F}) can be easily adopted in a scalar diquark model to the off-diagonal $L^z=0$ LFWF
overlap, i.e., to the transversity  nfPDF $\overline{H}_{\rm T}$  which is given in terms of  uPDF $h_1$
\begin{eqnarray}
\label{FT-h12HT}
\overline{H}_{\rm T}(x ,\eta=0,t) \stackrel{\rm sca}{=} \intkb   \sqrt{h_1(x,\bar{\bf k}_\perp-(1-x){\bf\Delta}_\perp/2)} \sqrt{h_1(x,\bar{\bf k}_\perp+(1-x){\bf\Delta}_\perp/2)}\,,
\end{eqnarray}
where we use here the shifted variable $\bar{\bf k}_\perp= {\bf k}_\perp +(1-x) {\bf\Delta}_\perp/2$.
We might also switch to diagonal $L^z=0$ LFWF overlap, i.e., we replace in (\ref{FT-h12HT}) the nfPDF $\overline{H}_{\rm T}$ by $H+\widetilde{H} $ and the uPDF $h_1$ by $f_1+g_1$.  The
diagonal $L^z=1$ overlap in the $\Delta L^z=0$ sector is given by the difference $H-\widetilde{H} $ and $f_1-g_1$, respectively, which in
the scalar diquark model includes an extra ${\bf k}_\perp^2$ factor.  Hence, we find now from the effective LFWF in impact parameter space,
cf.~Eqs.~(\ref{f1-DQSM},\ref{g1g1T-DQSM},\ref{FT-Phi}), the map
\begin{eqnarray}
\label{FT-f1mg1}
\left[H-\widetilde{H}\right](x ,\eta=0,t) &\!\!\! \stackrel{\rm sca}{=} \!\!\!& \intkb \frac{\bar{\bf k}_\perp^2+\frac{(1-x)^2}{4} {\bf\Delta}_\perp^2 }{
\sqrt{(\bar{\bf k}_\perp-(1-x){\bf\Delta}_\perp/2)^2} \sqrt{(\bar{\bf k}_\perp+(1-x){\bf\Delta}_\perp/2)^2}
}
\\
&&\!\!\!\!\!\!\!\!\!\!\!\!\times
\sqrt{[f_1-g_1](x,\bar{\bf k}_\perp-(1-x){\bf\Delta}_\perp/2)} \sqrt{[f_1-g_1](x,\bar{\bf k}_\perp+(1-x){\bf\Delta}_\perp/2)}
\;. \phantom{\Bigg|}\nonumber
\end{eqnarray}

We also concluded in Sec.~\ref{sec:GPD2uPDF-Lz0} that for a model, which is build from a superposition of effective two-body LFWFs,
the nfPDF/uPDF convolution formulae (\ref{FT-h12HT},\ref{FT-f1mg1}) do not apply. In our Regge improved  two-body LFWF model the $t$-dependence of nfPDF arises from the
asymmetric ${\bf k}_{\perp}$  overlap (\ref{F-generic}) of two effective LFWFs, while the ${\bf k}_{\perp}$ dependency of uPDF is simply given
by the unintegrated diagonal  LFWF overlap. We like now to illustrate how ${\bf k}_\perp$, $t$, and skewness dependence are talking to each other and we like to provide simple conversion formulae, which are only approximately valid. Thereby, the missing information for the desired model dependent nfPDF/uPDF map is apart from the spin-spin coupling contained in the uDD (\ref{Def-DD-unint1}).

A generic scalar ``uPDF'' is generally obtained from the scalar ``uDD'' by integrating out the $z$-dependency,
\begin{eqnarray}
\label{q2F-0}
{\bf \Phi}(x,{\bf k}_\perp) = 2 (1-x) \int_{0}^{1}\!dz\,{\bf \hat \Phi}(x,(1-x)z,0,{\bf k}_\perp)\,,
\end{eqnarray}
where we used that the ``uDD"  for ${\bf \Delta}_\perp=0$ is even in $z$.
Relying on the functional form of the ``uDD" (\ref{Def-DD-unint1}), arising from the implementation of Lorentz invariance,
the scalar ``nfPDF" can be written as
\begin{eqnarray}
F(x,\eta=0,t) = 2\pi  (1-x)\int_{0}^{1}\!dz\, \int_{- (1-x)^2(1-z^2)t/4}^\infty\!d{\bf k}^2_\perp\
{\bf \hat \Phi}(x,(1-x)z,0,{\bf k}_\perp)\,.
\end{eqnarray}
From this equation it also follows that the $t$-slope of the ``nfPDF" provides us some certain
$z$-moment of the ``uDD'':
\begin{eqnarray}
\label{F2q-0}
 \int_{0}^{1}\!dz\, (1-z^2){\bf \hat \Phi}(x,(1-x)z,0,\sqrt{1-z}\,{\bf k}_\perp)= \frac{2}{\pi(1-x)^2}  \frac{d}{dt}F(x,\eta=0,t)\Big|_{t=-4{\bf k}_\perp^2/(1-x)^2}
\end{eqnarray}
If we employ the mean value theorem in Eqs.~(\ref{q2F-0}) and (\ref{F2q-0})  and assume that the two mean values are
approximately given by the common value $\overline{z}$, we can state that the ``nfPDF"  and ``uPDF" can be approximately mapped to each other:
\begin{eqnarray}
\label{q2F-1}
F(x,\eta=0,t) &\!\!\! \approx \!\!\!& \pi \int_{-(1-x)^2 (1-\overline{z}^2)t/4}^\infty\!d{\bf k}^2_\perp\, {\bf \Phi}(x,{\bf k}_\perp)\,,
\\
\label{F2q-1}
 {\bf \Phi}(x,{\bf k}_\perp) &\!\!\! \approx \!\!\!& \frac{4}{\pi(1-x)^2(1-\overline{z}^2)}  \frac{d}{dt}F(x,\eta=0,t)\Big|_{t=-4{\bf k}_\perp^2/(1-x)^2(1-\overline{z}^2)}\,.
\end{eqnarray}
These two conversion formulae become exact if the scalar ``uDD" (\ref{Def-DD-unint1}) is concentrated in some specific $z$ point, e.g., $z=0$.

In the case that our quantities contain a diagonal $|L^z|=1$ overlap, e.g.,
$$
h(y,z,t,\overline{\bf k}_\perp) - \widetilde{h}(y,z,t,\overline{\bf k}_\perp) =
\left(\left[(1-y)^2-z^2\right]\frac{t}{4M^2} +\frac{\overline{\bf k}^2_\perp}{M^2} \right){\bf \hat \Phi}(y,z,t,\overline{\bf k}_\perp),
$$
the cross talk between nfPDF and uPDF is slightly modified. Going along the same line as above we find the approximate conversion formulae
\begin{eqnarray}
\label{q2F-2}
 \left[H-\widetilde H\right](x,\eta=0,t)  &\!\!\! \approx \!\!\!&
 \pi  \int_{- \frac{(1-x)^2(1-\overline{z}^2)t}{4}}^\infty\!d{\bf k}^2_\perp
\left(\!1+(1-x)^2(1-\overline{z}^2)\frac{t}{4{\bf k}^2_\perp} \!\right)
\left[f_1-g_1\right](x,{\bf k}_\perp)\,,
\\
\label{F2q-2}
 \left[f_1-g_1\right](x,{\bf k}_\perp) &\!\!\! \approx \!\!\!& \frac{4^2\, {\bf k}^2_\perp}{\pi(1-x)^4(1-\overline{z}^2)^2}  \frac{d^2}{dt^2}F(x,\eta=0,t)\Big|_{t=-4{\bf k}_\perp^2/(1-x)^2(1-\overline{z}^2)}\,.
\end{eqnarray}

We conclude that in a ``spherical" model of rank-one/three  an approximate (eventually an exact) uPDF/nfPDF mapping  might be employed:
\begin{eqnarray}
\overline{H}_{\rm T}(x,\eta=0,t)
 \quad
 &\stackrel{\mbox{\tiny (\ref{q2F-1},\ref{F2q-1})}}{\approx\longleftrightarrow} \quad
\left(\stackrel{\mbox{\tiny \ref{FT-h12HT})}}{\longleftarrow}\right)&
  \quad  h_1(x,{\bf k}_\perp) \,,
\\
H(x,\eta=0,t)+\widetilde H(x,\eta=0,t)
\quad
&\stackrel{\mbox{\tiny (\ref{q2F-1},\ref{F2q-1})}}{\longleftrightarrow}
\quad \left(\stackrel{\mbox{\tiny(\ref{FT-h12HT})}}{\longleftarrow}\right)&  \quad
f_1(x,{\bf k}_\perp)+g_1(x,{\bf k}_\perp)
\\
 H(x,\eta=0,t)-\widetilde H(x,\eta=0,t)
 \quad
&\stackrel{\mbox{\tiny(\ref{q2F-2},\ref{F2q-2}})}{\longleftrightarrow}
\quad \left(\stackrel{\mbox{\tiny(\ref{FT-f1mg1})}}{\longleftarrow}\right)&
  \quad  f_1(x,{\bf k}_\perp)- g_1(x,{\bf k}_\perp) \,.
\end{eqnarray}
In a ``spherical"  rank-four model, i.e., adding a nfPDF $H$  and the corresponding uPDF $f_1$,
the second and third relations will be in general broken.
In the case that the $f_1$ addenda are (not) decorated with an additional ${\bf k}_\perp^2$ factor, arising from a diagonal $L^z=1$ LFWF overlap, the (second)
third relation holds still true.

\subsubsection{uPDF sum rules}

In Sec.~\ref{sec:GPD-relations-1} we pointed out that also model dependent relations (\ref{GPD2FPD-twis3-3}) among the target helicity flip   $E$ and $\overline{E}_{\rm T}$ nfPDFs and
twist-tree related  $g_{1\rm T}^\perp$ and $h_{1\rm L}^\perp$ uPDFs exist. Using the uPDF and nfPDF representations (\ref{g1g1T-DQSM},\ref{h1L-DQSM}) and (\ref{Def-SpeFun-e},\ref{GPDmod-con2}), respectively, one can easily verify that the relations  (\ref{GPDuPDF-sumrule-EgT}) and (\ref{GPDuPDF-sumrule-bETh1L}) are valid, i.e., they hold true in a scalar diquark model,
\begin{eqnarray}
\label{E2g1T-sum-scalar}
\left\{ E \atop \overline{E}_{\rm T}\right\}
(x,\eta=0,t=0)\stackrel{\rm sca}{=}(1-x)\intk\,
\left\{ g^\perp_{1 {\rm T}} \atop -h^\perp_{1 {\rm L}}\right\} (x,{\bf k}_{\perp}^2)\,.
\end{eqnarray}
Thereby, we employed the fact that the forward GPD is given by  both  the ${\bf k}_\perp$ integral of the uPDF  and
the $z$-integral of the DD at $t=0$,
\begin{eqnarray}
\label{Phi-DD}
{\bf \Phi}(x) = \intk  {\bf \Phi}(x,{\bf k}_\perp)  =\int_{-1+y}^{1-y}\! dz\; {\bf \hat \Phi}(x,z,t=0),
\end{eqnarray}
which simply follows from the DD definition in Sec.~\ref{sect-LI}. From Tab.~\ref{tab:LFWFs2GPDs} we read off that the scalar diquark relation (\ref{E2g1T-sum-scalar}) among $E$ and $g^\perp_{1 {\rm T}}$ is modified in a minimal axial-vector diquark model   by a factor $-3$ while the
$\overline{E}_{\rm T}$ and $h^\perp_{1 {\rm L}}$ still holds:
\begin{eqnarray}
\label{E2g1T-sum-scalar-1}
\left\{ E \atop \overline{E}_{\rm T}\right\}
(x,\eta=0,t=0)\stackrel{\rm av}{=}(1-x)\intk\,
\left\{-3 g^\perp_{1 {\rm T}} \atop -h^\perp_{1 {\rm L}}\right\} (x,{\bf k}_{\perp}^2)\,.
\end{eqnarray}

Let us finally consider the model dependent sum rule (\ref{sumrule-HTh1T}) for the integrated ``pretzelosity'' distribution
\begin{eqnarray}
\label{h^perp_T}
 h^\perp_{\rm T}(x)= \int\!\!\!\int\! d^2{\bf k}_\perp\,  h^\perp_{\rm T}(x,{\bf k}_\perp)\stackrel{\rm sca}{=} -2  {\bf \Phi}(x)\,,
\end{eqnarray}
cf.~uPDF~(\ref{h1hTperp-DQSM}), which provides the  GPD  ${\widetilde H}_{\rm T}$ (\ref{Def-GPD-FT},\ref{Def-SpeFun-thT}) in the forward kinematics.
As explained in Sec.~\ref{sec:GPD2uPDF-Lz2}, in our effective two-body LFWF models  different $\overline{\bf k}_\perp$-weights
enter in the definitions of these functions and determine the proportionality factor in the sum rule (\ref{sumrule-HTh1T}).
In Sec.~\ref{sect-LI} we saw that the $z$-dependency of the DD originates from the functional form of the LFWF and so we might expect that the relation among $h_{\rm T}^\perp$ and $\widetilde H_{\rm T}$ depends finally on the skewness effect. Indeed, plugging the DD representation (\ref{Phi-DD}) into Eq.~(\ref{h^perp_T}), we find the following formula for the ``pretzelosity'' distribution
\begin{eqnarray}
\label{h^perp_T-DD}
h^\perp_{\rm T}(x)\stackrel{\rm sca}{=} -2(1-x)\int_{-1}^1\!dw\,{\bf \hat \Phi}(x,(1-x) w,t=0)\,,
\end{eqnarray}
while the DD $\widetilde h_{\rm T}$ contains an additional factor $(1-y)^2 (1-w^2)$, see Eq.~(\ref{Def-SpeFun-thT}), which tells us that the
GPD $\widetilde H_{\rm T}$ in forward kinematics is given by
\begin{eqnarray}
\widetilde H_{\rm T}(x,\eta=0,t=0)\stackrel{\rm sca}{=} -(1-x)^3 \int_{-1}^1\!dw\, (1-w^2)\, {\bf \hat \Phi}(x,(1-x)w,t=0)
\end{eqnarray}
arises from a different $w$-moment. At $t=0$ the $y$ and $w=z/(1-y)$ dependency factorizes in  ${\bf \hat \Phi}$, see below Eq.~(\ref{Our2RDDA}) in Sec.~\ref{subsec-Dterm},  and so we can also write the model relation among these distributions as
\begin{eqnarray}
\label{H_T/ h^perp_T}
\widetilde H_{\rm T}(x,\eta=0,t=0) \stackrel{\rm sca}{=} \frac{\int_{-1}^1\!dw\, (1-w^2) \Pi(w)}{2\int_{-1}^1\!dw\,  \Pi(w)} (1-x)^2\, h^\perp_{\rm T}(x)\,,
\end{eqnarray}
where the proportionality factor is determined by the specific $w$-moment of a so-called profile function  $\Pi(w)$. The profile function of the model \cite{MeiMetGoe07} is just a constant. For such a choice we find on the r.h.s.~of Eq.~(\ref{H_T/ h^perp_T}) a proportionality factor $1/3$, which confirms  the very specific relation of Ref.~\cite{MeiMetGoe07}, given there in Eq.~(110). We add that the model dependent relation (\ref{H_T/ h^perp_T}) holds  true in all our  diquark models, cf.~Tab.~\ref{tab:LFWFs2GPDs}.

\section{Effective LFWF models versus phenomenology}
\label{Sec-phenomenology}

In the previous sections (\ref{uPDF-scaDiq}) and (\ref{subsect-constraints}) we constructed (u)PDFs and GPDs from  LFWF
``spinors'', giving emphasize to the scalar diquark  ``spinor'' (\ref{LFWF-spinor-S}) [or its components (\ref{Def-LF-WF1},\ref{Def-LF-WF2})].
We like now to confront such  models with experimental/phenomenological results and expectations from Lattice simulations.
In Sec.~\ref{sec:phe-models} we set up two concrete models from a power-likely and exponentially ${\bf k}_\perp$-dependent effective two-body LFWF.
In Sec.~\ref{subsec-Dterm} we provide some general insights into the resulting GPD models.
In Sec.~\ref{sec:pheno-scalar} we compare then the scalar diquark model predictions in the flavor sector $(2 u -d)/3$ with phenomenological as well as Lattice QCD findings and we also provide model predictions for unmeasured partonic quantities. In Sec.~\ref{sec:pheno-axialvector} we have a short look to a minimal axial-vector diquark model, given in terms of the three LFWF ``spinors'' (\ref{LFWF-spinor-A0}, \ref{LFWF-spinors-AV}). It provides as for uPDFs a ``spherical'' SU(6) model of rank-four, see Sec.~\ref{uPDF-classification-spherical}, however,  we will see that its GPD models are ruled out. In Sec.~\ref{sec:GPD-pomeron}  we illustrate that one can even build diquark models that posses a ``pomeron'' like behavior at small $x$.

\subsection{PDF and GPD models from LFWFs}
\label{sec:phe-models}

Choosing an appropriate LFWF ``spinor'', we can consistently build GPDs, uPDFs, and PDFs in terms of a scalar ``DD'' ${\bf \hat \Phi}(y,z,t)$, ``uPDF'' ${\bf \hat \Phi}(x,\overline{\bf k}_\perp)$, and ``PDF'' ${\bf \hat \Phi}(x)$, respectively.  The ``DD'' and ``(u)PDF'' are obtained from a  ``uDD" ${\bf \hat \Phi}(y,z,t,\overline{\bf k}_\perp)$, which is given as overlap of an effective two-body LFWF $\overline{\phi}$. Our representation for this scalar LFWF arises from Eqs.~(\ref{rho(lambda)},\ref{Def-LFWF-Laplace},\ref{Ans-LFWF},\ref{Def-SpeRep1}),
\begin{eqnarray}
\label{LFWF-HM}
\overline{\phi}(X,{\bf k}_\perp|\lambda^2) &\!\!\! = \!\!\! & \theta(\lambda^2-\lambda_c^2)  \sqrt{\frac{(\lambda^2 -\lambda_c^2)^{\alpha-1}}{\Gamma(\alpha)  M^{2\alpha} }} \phi(X,{\bf k}_\perp|\lambda^2)\,,
\\
\label{LFWF_1-HM}
\phi(X,{\bf k}_\perp|\lambda^2) &\!\!\! = \!\!\! & \int_0^\infty\! d\bar{\alpha}\; \varphi(\bar{\alpha})
\exp  \left\{-\bar{\alpha}  \frac{{\bf k}_\perp^2 +m^2(1-X)+X \lambda^2 - X(1-X) M^2}{(1-X) M^2}  \right\} ,
\end{eqnarray}
and it guarantees that both Regge behavior and Lorentz symmetry are implemented. To evaluate the ``uDD", we utilize the Regge improved master formula (\ref{Def-DD-unint2}).  Specifying the reduced Laplace transform $\varphi(\bar{\alpha})$, we utilize
in the following Sec.~\ref{subsec-powerlike} and Sec.~\ref{subsec-exponential}  a power-like and an exponential ${\bf k}_\perp$-dependent LFWF, respectively.

\subsubsection{Power-likely ${\bf k}_\perp$-dependent LFWF}
\label{subsec-powerlike}
The ansatz (\ref{LFWF_1-HM}) within the reduced Laplace transform
\begin{eqnarray}
\label{HMpow-varphi}
\varphi(\overline{\alpha}) = \frac{g\, \bar{\alpha}^{p+\alpha/2}}{M \Gamma(1+p+\alpha/2)} 
\end{eqnarray}
provides a power-likely scalar LFWF
\begin{eqnarray}
\label{HMpow-LFWF}
\phi(X,{\bf k}_\perp|\lambda^2) = \frac{g}{M} \left[\frac{{\bf k}_\perp^2 +m^2(1-X)+X \lambda^2 - X(1-X) M^2}{(1-X) M^2} \right]^{-1-p-\alpha/2},
\end{eqnarray}
which is nothing but the generalized LFWF  from the Yukawa theory ($\alpha=p=0$) where $g$ is the coupling \cite{Brodsky:2000ii}. For $p=1$ and $\alpha=0$ we
have the model that we utilized in Ref.~\cite{Hwang:2007tb}. The $A$ integration in Eq.~(\ref{Def-DD-unint2}) is easily performed and yields the ``uDD''
\begin{eqnarray}
\label{HMpow-Phi_trans}
{\bf \hat \Phi}(y,z,t,\overline{\bf k}_\perp) =  \frac{N (2p+1)}{\pi M^2}
\frac{y^{-\alpha}\, ((1-y)^2-z^2)^{p+\alpha/2}}{\left[(1-y)
\frac{m^2}{M^2}+ y \frac{\lambda^2}{M^2} - y(1-y) -((1-y)^2-z^2)
\frac{t}{4 M^2} + \frac{\overline{\bf k}_\perp^2}{M^2}\right]^{2p+2}}\,,
\end{eqnarray}
where $\overline{{\bf k}}_\perp = {\bf k}_\perp -(1-y+z){\bf \Delta}_\perp/2$ and the normalization is expressed in term of the coupling $g$:
$$N= \frac{\pi g^2\, \Gamma(2p+1)}{2^{2p+1+\alpha} \Gamma^2(p+1+\alpha/2)} .$$
The $\overline{\bf k}_\perp$-integration of this ``uDD'' (\ref{HMpow-Phi_trans}) leads to the ``DD''
\begin{eqnarray}
\label{HMpow-DD}
{\bf \hat \Phi}(y,z,t)=
N \frac{y^{-\alpha}\, ((1-y)^2-z^2)^{p+\alpha/2}}{\left[(1-y)
\frac{m^2}{M^2}+ y \frac{\lambda^2}{M^2} - y(1-y) -((1-y)^2-z^2)
\frac{t}{4 M^2}\right]^{2p+1}}  \,,
\end{eqnarray}
which provides with the representations (\ref{Def-SpeRep}--\ref{Def-SpeFun-te}) and (\ref{Def-GPD-FT}--\ref{Def-SpeFun-teT}) the chiral even and odd GPDs, respectively. The $\overline{\bf k}_{\perp}^2$-moment (\ref{Def-overlap-moment2}) of the ``uDD'' might be alternatively obtained by $t$-integration:
\begin{eqnarray}
\label{HMpow-DD-2}
{\bf \hat \Phi}^{(2)}(y,z,t) = \frac{1}{2 p} \left[(1-y)
\frac{m^2}{M^2}+ y \frac{\lambda^2}{M^2} - y(1-y) -((1-y)^2-z^2)
\frac{t}{4 M^2}\right] {\bf \hat \Phi}(y,z,t)\,.
\end{eqnarray}
Integrating  over $z$, we find from the ``uDD'' (\ref{HMpow-Phi_trans}) at $t=0$ the ``uPDF'',
\begin{eqnarray}
\label{HMpow-uPDF}
{\bf \Phi}(x, {\bf k}_{\perp}^2) = \frac{g^2 \Gamma(2+2 p)}{M^2\Gamma(2+2p+\alpha)}{
x^{-\alpha} (1-x)^{2p+1+\alpha} \over
\left[(1-x)\frac{m^2}{M^2}+x \frac{\lambda^2}{M^2} -x(1-x)+ \frac{{\bf k}_{\perp}^2}{M^2}
\right]^{2p+2}}\,,
\end{eqnarray}
entering in all uPDFs (\ref{f1-DQSM},\ref{g1g1T-DQSM}--\ref{h1L-DQSM}). From the ``DDs'' (\ref{HMpow-DD}) and (\ref{HMpow-DD-2})  we can now obtain the ``PDF'' (\ref{def:LFWFoverlap}) and the corresponding ${\bf k}_{\perp}^2$-moment,
\begin{eqnarray}
\label{HMpow-PDF}
{\bf \Phi}(x) &\!\!\! =\!\!\!&
\frac{\pi g^2 \Gamma(1+2 p)}{\Gamma(2+2p+\alpha)}
{
x^{-\alpha} (1-x)^{2p+1+\alpha} \over
\left[(1-x)\frac{m^2}{M^2}+x \frac{\lambda^2}{M^2} -x(1-x)
\right]^{2p+1}}\,,
\\
\label{HMpow-PDF-2}
{\bf \Phi}^{(2)}(x) &\!\!\! =\!\!\!& 
\frac{\langle {\bf k}^2_\perp \rangle(x)}{M^2} {\bf \Phi}(x)\,,\quad
\langle {\bf k}^2_\perp \rangle(x)  =
 \frac{1}{2p} \left[(1-x) m^2+ x \lambda^2 - x(1-x) M^2\right]\,,
\end{eqnarray}
entering in all PDFs (\ref{f1-sca}--\ref{h1-sca}).
Here we used the definition (\ref{def:k2perp}) to find the expression for $\langle {\bf k}^2_\perp\rangle$.
Alternatively, these ${\bf k}^2_\perp$--integrated results can be also directly derived from the ``uPDF'' (\ref{HMpow-uPDF}).

\subsubsection{Exponentially ${\bf k}_\perp$-dependent LFWF}
\label{subsec-exponential}

Let us now illustrate that one can also utilize LFWFs with exponential ${\bf k}_\perp$-dependence.  To do so we take the reduced Laplace transform
\begin{eqnarray}
\label{HMexp-varphi}
\varphi(\alpha) = \frac{g\,}{M} \delta(\alpha- \bar{A})\,,
\end{eqnarray}
where  Eq.~(\ref{LFWF_1-HM}) provides us the scalar LFWF
\begin{eqnarray}
\label{HMexp-LFWF}
\phi(X,{\bf k}_\perp|\lambda^2) =  \frac{g\,}{M}
\exp  \left\{-\bar{A}  \frac{{\bf k}_\perp^2 +m^2(1-X)+X \lambda^2 - X(1-X) M^2}{(1-X) M^2}  \right\}.
\end{eqnarray}
Hence, the ``uDD'' (\ref{Def-DD-unint2}) is concentrated in $z=0$ and the $A$ integration in this function is constrained by $A= \bar{A}/(1-y)$, which finally yields
\begin{eqnarray}
\label{HMexp-Phi_trans}
{\bf \hat \Phi}(y,z,t,\overline{\bf k}_\perp) &=& \frac{g^2\,}{M^2} \left(\frac{2 \bar{A} y}{1-y}\right)^{-\alpha} \frac{ \delta(z)}{1-y}
\\ &&\times\exp \left\{-\frac{\bar{A}}{1-y}\left[
(1-y)\frac{m^2}{M^2}+y \frac{\lambda^2}{M^2}-y(1-y)-(1-y)^2 \frac{t}{4 M^2}+\frac{\overline{{\bf k}}^2_\perp}{M^2}
 \right]\right\}
 \nonumber
\end{eqnarray}
with $\overline{{\bf k}}_\perp = {\bf k}_\perp -(1-y){\bf \Delta}_\perp/2$.
As we realize from the Radon transforms (\ref{Def-SpeRep}), with our rather extreme ansatz (\ref{HMexp-varphi}) we lose any skewness dependence in GPDs. Obviously, an exponential LFWF that provides a skewness depended DDs can be obtained from a smeared Dirac $\delta$-function.
Hence, polynomiality constraints and  exponential ${\bf k}_\perp$ fall-off do not contradict each other.
Integrating in Eq.~(\ref{HMexp-Phi_trans}) over $\overline{\bf k}_\perp$   provides  immediately  ``DDs'' that have an exponential $t$-dependence,
\begin{eqnarray}
\label{HMexp-DD}
{\bf \hat \Phi}(y,z,t) &\!\!\! = \!\!\!& \frac{\pi g^2\,}{\bar{A}} \left(\frac{2 \bar{A} y}{1-y}\right)^{-\alpha} \delta(z)
\\
\label{HMexp-DD-2}
&&\times\exp \left\{-\frac{\bar{A}}{1-y}\left[
(1-y)\frac{m^2}{M^2}+y \frac{\lambda^2}{M^2}-y(1-y)-(1-y)^2 \frac{t}{4 M^2}
 \right]\right\}\,,
 \nonumber\\
 {\bf \hat \Phi}^{(2)}(y,z,t)&\!\!\! = \!\!\!& \frac{1-y}{\bar{A}} {\bf \hat \Phi}(y,z,t)\,.
\end{eqnarray}
Integrating Eq.~(\ref{HMexp-Phi_trans}) over $z$, we trivially find in the forward case the ``uPDF''
\begin{eqnarray}
\label{HMexp-uPDF}
{\bf \Phi}(x, {\bf k}_{\perp})=\frac{g^2\,}{M^2} \left(\frac{2 \bar{A} x}{1-x}\right)^{-\alpha} \frac{1}{1-x}\exp \left\{-\frac{\bar{A}}{1-x}\left[
(1-x)\frac{m^2}{M^2}+x \frac{\lambda^2}{M^2}-x(1-x)+\frac{\overline{{\bf k}}^2_\perp}{M^2}
 \right]\right\},
\end{eqnarray}
Analogously, from Eqs.~(\ref{HMexp-DD},\ref{HMexp-DD-2}) the ``PDFs'' can be obtained
\begin{eqnarray}
\label{HMexp-PDF}
{\bf \Phi}(x)&\!\!\! = \!\!\!& \frac{\pi g^2\,}{\bar{A}} \left(\frac{2 \bar{A} x}{1-x}\right)^{-\alpha} \exp \left\{-\frac{\bar{A}}{1-x}\left[
(1-x)\frac{m^2}{M^2}+x \frac{\lambda^2}{M^2}-x(1-x)
 \right]\right\}\,,
 \\
 \label{HMexp-PDF-2}
{\bf \Phi}^{(2)}(x)&\!\!\! = \!\!\!& \frac{\langle{\bf k}^2_\perp \rangle(x)}{M^2} {\bf \Phi}(x)\,,
\quad \langle{\bf k}^2_\perp \rangle(x)  = (1-x) \frac{M^2}{\bar{A}}\,.
\end{eqnarray}

As described in Sec.~\ref{subsec-powerlike}, from the given  ``DD" and ``(u)PDFs" expressions one can easily find the corresponding chiral even (\ref{Def-SpeRep}--\ref{Def-SpeFun-te}) and chiral odd (\ref{Def-GPD-FT}--\ref{Def-SpeFun-teT}) twist-two GPDs, leading--power uPDFs (\ref{f1-DQSM},\ref{g1g1T-DQSM}--\ref{h1L-DQSM}), and twist-two PDFs (\ref{f1-sca}--\ref{h1-sca}). According to Tab.~\ref{tab:LFWFs2GPDs} we can then set up scalar, pseudo-scalar, minimal axial-vector, and minimal vector models.

\subsection{Some  GPD model insights}
\label{subsec-Dterm}

\subsubsection{Recovering Radyushkin`s double distribution ansatz}

It is  rather popular to build GPD models by utilizing Radyushkin's DD ansatz (RDDA) \cite{Radyushkin:1997ki} that has been given for $t=0$, e.g.,
utilized in Refs.~\cite{Goeke:2001tz,Belitsky:2001ns}, used in the so-called VGG code \cite{Vanderhaeghen:1998uc}, and in the Goloskokov/Kroll model \cite{Goloskokov:2005sd,Goloskokov:2007nt}.
Since this ansatz was inspired from a triangle diagram, too, it is not surprising  that  our scalar ``DDs'', e.g., Eq.~(\ref{HMpow-DD}), reduces at $t=0$ to RDDA.  This ansatz is written as product of the ``PDF'' (\ref{def:LFWFoverlap}) and a normalized profile function $\Pi$,
\begin{eqnarray}
\label{Our2RDDA}
{\bf \hat\Phi}(y,z,t=0) &\!\!\!=\!\!\!& \frac{{\bf \Phi}(y)}{1-y} \Pi\!\left(\!\frac{z}{1-y}\!\right),
\\
\Pi(w) &\!\!\!=\!\!\!&  \frac{\Gamma(b+3/2)}{\sqrt{\pi}\Gamma(b+1)} (1-w^2)^b\,,
\quad\mbox{where}\quad
\int_{-1}^1\!dw\, \Pi(w) =1.
\nonumber
\end{eqnarray}
Thereby, the additional factor $1/(1-y)$ in RDDA drops out if we change to the new integration variable $w=z/(1-y)$.
In our powerlike LFWF model (\ref{HMpow-DD}) the parameter $b=p+\alpha/2$  is fixed by $p$ and $\alpha$, while the  $\delta(z)$ function in our exponential
LFWF model (\ref{HMexp-DD}) arises now from the limit $b\to \infty$. The ``PDFs'' for these models are given in Eq.~(\ref{HMpow-PDF}) and (\ref{HMexp-PDF}),
respectively.

For $t\neq 0$ the factorization of $y$- and $w$-dependency  is broken for power-like LFWF model, see Eq.~(\ref{HMpow-DD}),
however, we might write the ``DD'' in analogy  to RDDA as
\begin{eqnarray}
\label{hPhi-RDDA}
{\bf \hat\Phi}(y,z,t) &\!\!\!=\!\!\! & \frac{{\bf \Phi}(y)}{1-y} \Pi(z/(1-y)|y,t),\;
\\
\label{hPhi-RDDA-1}
\Pi(w|y,t) &\!\!\!=\!\!\! &
\frac{\Gamma(b+3/2)}{\sqrt{\pi}\Gamma(b+1)}
\frac{(1-w^2)^{b}}{\left[1-(1-w^2)\frac{t}{M^2_{\rm cut}(y)}\right]^{2p+1}}\,,\quad b=p+\alpha/2\,.
\end{eqnarray}
Here, the $t$-dependency is included in the profile function and is it governed by a $y$-dependent cut-off mass
\begin{equation}
\label{hPhi-RDDA-2}
M^2_{\rm cut}(y) =\frac{m^2+ y \lambda^2 - y(1-y)M^2}{(1-y)^2}.
\end{equation}
Note, however, that our DDs $h$ and $\widetilde h$  (\ref{Def-SpeFun-h},\ref{Def-SpeFun-th}) are more intricate and additional
$t$-dependent terms appear. Even for $t=0$,  $h$ and $\widetilde h$  models will consist of two factorizable profile functions, which arise from a diagonal $L^z=0$
and $L^z=1$ LFWF overlaps,  cf.~Eqs.~(\ref{Def-SpeFun-h}) and (\ref{Def-SpeFun-th}).

\subsubsection{How solid are inclusive-exclusive relations?}

Counting rules for the large $-t$ behavior of form factors \cite{MatMurTav73,BroFar75} or the large $x$ behavior of PDFs \cite{BroFar75,Brodsky:1994kg}
are a useful guidance for model builders. They predict a power-like behavior in the corresponding asymptotic, and, thus, our power-like LFWF model, set up in Sec.~\ref{subsec-powerlike},  is more favorable than the exponential LFWF model, given in Sec.~\ref{subsec-exponential}.  The power-like LFWF model with $p=1$ and $\alpha=0$ was studied in Ref.~\cite{Hwang:2007tb}, and we found that the following power behavior for PDF $f_1$  and elastic form factors at large  $x$ and $-t$, respectively,
$$
\lim_{x\to 1} f_1(x) \sim (1-x)^3\,,
\quad \lim_{-t\to \infty} F_1(t) \sim  (-t)^{-2}\,,
\quad \lim_{-t\to \infty} F_2(t)\sim  (-t)^{-3}\,,
$$
generically expected from counting rules \cite{MatMurTav73,BroFar75}, holds only  approximately true for $F_1$ and even fails for $F_2$.  Thus,  the inclusive-exclusive relation
\begin{eqnarray}
\label{relation-exc2inc}
\lim_{x\to 1} f_1(x) \sim  (1-x)^{2n-1} \quad \Longleftrightarrow   \quad\lim_{-t\to \infty} F_1(t) \sim (-t)^{-n}
\end{eqnarray}
of Drell-Yan \cite{DreYan69}, also obtained by  West \cite{Wes70},   is only to some extent respected. It is worthy to mention that Drell and Yan employed  LFWF overlap representations, general arguments, however, also some ``empirical'' findings to get rid of some contributions, see Ref.~\cite{DreYan69},
to conjecture the inclusive-exclusive  relation (\ref{relation-exc2inc}) between unpolarized deep inelastic scattering structure function (taken here as PDF combination) and Dirac form factor.  Utilizing our LFWF models, we will now have a closer look at the inclusive-exclusive  relation, which reveals that the conjecture (\ref{relation-exc2inc}) is indeed  based on specific assumptions.

In our power-like LFWF model  the large $x$ behavior of PDF $f_1$  is inherited from the ``PDF''  (\ref{HMpow-PDF}), see $f_1$ definition (\ref{f1-sca}) and, consequently, it is given by $(1-x)^{2p+1+\alpha}$.
The large $t$-behavior of the Dirac form factor can be calculated from the lowest $x$ moment of   GPD $H$ (\ref{Def-SpeRep},\ref{Def-SpeFun-h}), where we can  set $\eta=0$. As one realizes the leading $-t$ behavior stems from the  diagonal $L^z=1$ LFWF overlaps, and reads in terms of the
ansatz (\ref{hPhi-RDDA}--\ref{hPhi-RDDA-2}) as
\begin{eqnarray}
\label{F1-asy}
F_1(t) \stackrel{-t\gg}{\propto}   \int_{0}^1\!dx\, {\bf \Phi}(x) \int_{-1}^1\!dw\, \frac{(1-w^2)^{p+\alpha/2}}{\left[1-(1-w^2)\frac{t}{M^2_{\rm cut}(x)}\right]^{2p+1}} \left[\cdots  + \cdots (1-w^2)  \frac{t}{M^2_{\rm cut}(x)}  \right]\,,
\end{eqnarray}
where the omissions denote some $x$-dependent terms.
Some care is needed to find the large $-t$ asymptotics.
The integral over $w$ provides us a Hypergeometric functions and their expansion in the vicinity of $-t=\infty$ gives us after integration over $x$ the following asymptotic behavior
$$
\lim_{t\to -\infty} F_1(t) \sim \cdots (-t)^{-p-\alpha/2-1}(1+{\cal O}(1/t)) + \cdots (-t)^{-2p} (1+{\cal O}(1/t)) \,.
$$
The first term on the r.h.s.~reflects the end-point behavior $|w| \sim 1$ of the integrand in Eq.~(\ref{F1-asy}), while the second one stems from the region $|w|<1$.
If we take $1 < \alpha/2+1 < p$, the first term, proportional to $(-t)^{-p-\alpha/2-1}$, is the leading one and we recover with
$$n=p+\alpha/2+1 > 1\quad\mbox{and}\quad 2n-1 = 2p+\alpha+1 >3$$ the inclusive-exclusive  relation (\ref{relation-exc2inc}). Note, however, that this scenario
excludes the generic answer $n=2$, obtained from (dimensional) counting rules.
Choosing the parameter $p < \alpha/2+1$ the second term, proportional to $(-t)^{-2p}$, governs the form factor asymptotics and the inclusive-exclusive  relation
turns with $n=2p$ into
\begin{eqnarray}
\label{relation-exc2inc-2}
\lim_{x\to 1} f_1(x) \sim (1-x)^{n+1+\alpha} \quad\Longleftrightarrow  \quad\lim_{-t\to \infty} F_1(t) \sim (-t)^{-n}\,.
\end{eqnarray}
For $p=\alpha/2+1$ both terms on the r.h.s.~of the asymptotic formulae (\ref{F1-asy}) contribute and the large $(-t)^{-p-\alpha/2-1}$ behavior  might be accompanied  by a logarithmical modification, too, see also the numerical discussion for the $\alpha=0$ case in Ref.~\cite{Hwang:2007tb}.

For the exponential LFWF model the ``PDF'' (\ref{HMexp-PDF}) and the momentum faction integrated  ``DD" (\ref{HMexp-DD}) are exponentially suppressed in the large $x$ region and $-t$ region, respectively.
Apart from  power-like modifications of these exponential suppressions, we can state our exponential LFWF model possesses the following inclusive-exclusive  relation
\begin{eqnarray}
\label{relation-exc2inc-3}
\lim_{x\to 1} f_1(x) \sim  e^{-\bar{A}/(1-x)} \quad\Longleftrightarrow  \quad\lim_{-t\to \infty} F_1(t) \sim  e^{-\bar{A}(1-\bar x)\frac{t}{4 M^2} }  \,,
\end{eqnarray}
where $\bar x$ denotes the mean value.

Let us spell out that  Drell and Yan  took it for granted that the unpolarized deep inelastic scattering structure function possesses a power-like behavior. In the covariant wave function model study of West \cite{Wes70} an exponential scenario was considered, too, however, with a different model assumption,
$$
\lim_{x\to 1} f_1(x) \sim  e^{-a^\prime/\sqrt{1-x}} \quad\Longleftrightarrow  \quad\lim_{-t\to \infty} F_1(t) \sim  e^{-a \sqrt{-t} }\,,
$$
compare with our model findings (\ref{relation-exc2inc-3}).
We are not aware that the power-like relation (\ref{relation-exc2inc-2}), contradicting the inclusive-exclusive  conjecture (\ref{relation-exc2inc}), was mentioned somewhere else. It arises in the DD representation from the dominance of the region $|w| < 1$. A direct link of our DD considerations and those of Drell and Yan  or West might be hardly visible, however, we note that introducing a ultraviolet cut-off, as assumed in Ref.~\cite{DreYan69} to get rid of contributions from the valence region $ 0< c < 1-x< 1 $, will certainly also weaker the $|w| < 1$
contribution by one power $1/(-t)$.

Let us conclude that even in a simple diquark model the intricate interplay of Regge, ultra-violet, large $-t$, and large $x$ behavior shows up.
The underlying Lorentz symmetry, which we implemented in our LFWF  and West in his covariant model
tells us that the functional form of a valence PDF at large $x$ and of the Dirac form factor at large $-t$ should match each other.
We also found  a simple counter example for which the Drell-Yan conjecture (\ref{relation-exc2inc}) does not hold true,
even for a certain power-like LFWF.

\subsection{Models versus phenomenology}
\label{subsec-pheno}

\subsubsection{Scalar diquark sector}
\label{sec:pheno-scalar}

The models presented above are suitable to describe the scalar diquark content of the nucleon, which is according to SU(6) symmetry related to the flavor combination $2 u/3 - d/3$. We will consider the valence quark sector, where we use for PDFs and GPDs the following notation
\begin{eqnarray}
\label{q-fla-sca}
q^{\rm sca} &\!\!\!=\!\!\!&\frac{2}{3} q^{u_{\rm val}} - \frac{1}{3}q^{d{\rm val}}\,,
\quad\;\;\;
q\in \{f_1,g_1,h_1\}\,,
\\
F^{\rm sca}&\!\!\!=\!\!\!&\frac{2}{3} F^{u_{\rm val}} - \frac{1}{3}F^{d{\rm val}}\,,
\quad
F\in\{H,E,\widetilde H,\widetilde E,H_T,E_T,{\widetilde H}_T,{\widetilde E}_T\}\,.
\end{eqnarray}
In the first place we do not consider our model as a constituent quark model, given at an intrinsic low scale of few hundred MeVs or so, rather as an effective one that matches partonic and effective quark degrees of freedom. Hence, we might it even employ at a larger input scale, e.g., $\mu^2=4\, {\rm GeV}^2$. This scale setting, which is
quite convenient to compare with phenomenological and Lattice results, avoids also the problem that perturbative QCD evolution is utilized at very low scales, where the application of perturbation theory might be questionable.

Certainly, various aspects of the scalar diquark model has been widely studied in the literature, however, we are not aware of a critical and consistent analysis versus the full set of available data.
Utilizing the formulae of Sec.~\ref{subsec-powerlike} and  Sec.~\ref{subsec-exponential}, we are now in the position to confront our Regge improved diquark models with  phenomenological findings for PDFs, (generalized) form factors, GPDs, and to some extend with uPDFs.
After fixing the proton mass $M=0.938\, {\rm GeV}$, and the normalization (\ref{Fix-Nor}), i.e., $n=1$, our two models depend on four parameters $\alpha,\lambda,m$,
and $p$ or $\overline{A}$. To fix these parameters, we require a good description of the valence-like unpolarized PDF $f_1^{\rm sca}$.
Parameterizations from global fits incorporate a $\alpha\sim 0.5$ Regge intercept, which is consistent with the intercept of the $\rho/\omega$ trajectory.
We take here Alekhin`s parameterization as a phenomenological reference PDF at $\mu^2=4\, {\rm GeV}^2$, which has build in $\alpha=0.47$. Moreover,  this PDF possesses the averaged momentum fraction:
\begin{eqnarray}
\langle x \rangle^{\rm sca}(\mu^2=4\, {\rm GeV}^2)\equiv\int_0^1\! dx\, x\, f_1^{\rm sca}(x,\mu^2=4\, {\rm GeV}^2) \approx 0.16\;.
\end{eqnarray}
To match the $x$-shape with more precision, we employ also the second $x$-moment,
\begin{eqnarray}
\langle x^2 \rangle^{\rm sca}(\mu^2=4\, {\rm GeV}^2)\equiv\int_0^1\! dx\, x^2\, f_1^{\rm sca}(x,\mu^2=4\, {\rm GeV}^2) \approx 0.052\;,
\end{eqnarray}
and the PDF value at $x=0.25$,
\begin{eqnarray}
f_1^{\rm sca}(x=0.25,\mu^2=4\, {\rm GeV}^2) \approx 1.38\,.
\end{eqnarray}
We also require that the anomalous magnetic moment is well described. It is given by the zeroth moment of $E^{\rm sca}$ GPD,
\begin{eqnarray}
\langle \kappa\rangle^{\rm sca}\equiv\int_0^1\! dx\,  E^{\rm sca}(x,\eta=0,t=0) =
\frac{2}{3}  \kappa^{u} - \frac{1}{3} \kappa^{d}=\kappa^{p} \approx 1.79\,,
\end{eqnarray}
and it can be directly expressed by the proton anomalous magnetic moment. Furthermore, we utilize
the averaged longitudinal quark spin,
\begin{eqnarray}
\langle s\rangle^{\rm sca}(\mu^2)=\frac{1}{2}\int_0^1\! dx\,g_1^{\rm sca}(x,\mu^2)\,,
\end{eqnarray}
which, however, is not well constrained from phenomenology%
\footnote{In deep inelastic scattering (DIS) only the sum of polarized valence and sea quark PDFs can be accessed;
a separation of them is possible by taking into account semi-inclusive DIS data \cite{FloSasStrVog09}.}
nor from Lattice simulations. We might take some value that lies between a fully-broken `valence' ($\approx 0.19$)
and a standard flavor symmetric sea ($\approx 0.37$) scenario \cite{GluReyStrVog00}:
\begin{eqnarray}
\label{<s>^sca-pheno}
0.19 \lesssim \langle s\rangle^{\rm sca} \lesssim  0.37\;,
\end{eqnarray}
while the NLO analysis of Ref.~\cite{FloSasStrVog09} provides a value $\langle s\rangle^{\rm sca}(\mu^2=4\, {\rm GeV}^2)\approx 0.29$.

\begin{figure}[t]
\begin{center}
\includegraphics[width=17.4cm]{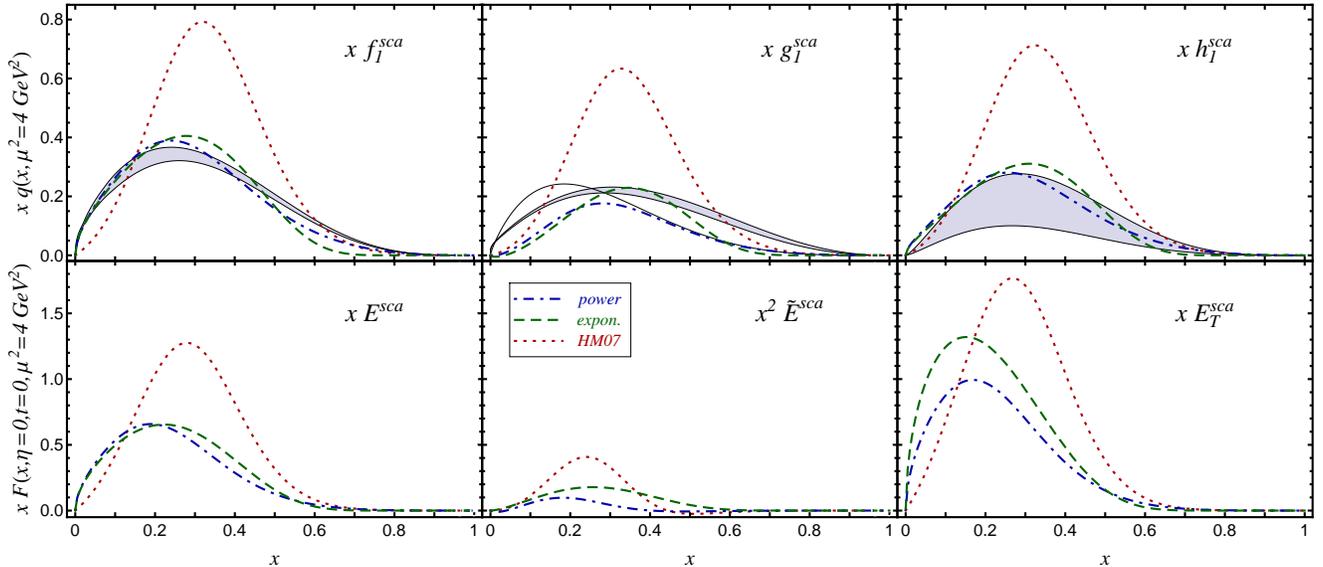}
\end{center}
\vspace{-5mm}
\caption{ \label{Fig-PDFcom} \small
Our Regge improved power-like (dash-dotted), exponential (dashed) LFWFs,  and {\it HM07} (dotted) \cite{Hwang:2007tb} models are
confronted with  unpolarized (left), polarized (middle), and transversity (right) valence quark PDF  combinations (\ref{q-fla-sca}) from leading order fits~\cite{Ale05}, \cite{BluBot02}, and  \cite{Ansetal09}, shown at ${\cal Q}^2_0 = 4\, \GeV^2$  as grayed areas (solid line is from the polarized DIS fit \cite{GehSti95}). In the lower panels the nfPDFs $x E$, $x^2 \widetilde E$, $x E_{\rm T}$ at $t=0$ are presented from left to right.}
\end{figure}

We generated various fitting results, where we observed for fixed  $\langle \kappa\rangle^{\rm sca}=1.79$ an anti-correlation between matching the $x$-shape of unpolarized PDFs and the $\langle s \rangle^{\rm sca} $ value, namely,  a better shape matching requires a lower $\langle s \rangle^{\rm sca} $ value. Here we present a result in which the shape of the unpolarized PDF is fairly matched, see dash-dotted curve in the right upper panel of Fig.~\ref{Fig-PDFcom}:
\begin{eqnarray}
\label{ps[LFWF^pow]}
{\rm LFWF^{\rm pow}}:\quad \alpha=0.47\,,\quad \quad p= 0.455\,,\quad m= 0.415~{\rm GeV} \,,\quad \lambda = 0.858~{\rm GeV}\,,\\
\label{ps[LFWF^exp]}
{\rm LFWF^{\rm exp}}:\quad \alpha=0.47\,,\quad \quad \overline{A}= 4.424\,,\quad m= 0.404~{\rm GeV} \,,\quad \lambda = 0.880~{\rm GeV}\,.
\end{eqnarray}
Mostly independent on the shape of the reduced LFWF our models provide $\langle x \rangle^{\rm sca} \approx 0.16$, $\langle s \rangle^{\rm sca} \approx 0.13$, and $\langle \kappa\rangle^{\rm sca} \approx 1.79$.
Compared to our spectator diquark model \cite{Hwang:2007tb} (dotted curve),
\begin{eqnarray}
\label{ps[HM07a]}
{\it HM07}:\quad \alpha=0\,,\quad \quad p= 1\,,\quad m= 0.45~{\rm GeV} \,,\quad \lambda = 0.75~{\rm GeV}\,,
\end{eqnarray}
which was ``eye fitted" to the $t$-dependence of the Dirac form factor and $\kappa^p$, the $p$ parameter (\ref{ps[LFWF^pow]}) has a rather low value, where the quark and spectator masses are compatible with our new ones (\ref{ps[LFWF^pow]}) and (\ref{ps[LFWF^exp]}).

In the upper panels of Fig.~\ref{Fig-PDFcom} we illustrate that the shape and normalization of the PDFs in the Regge improved models are now much more realistic than in the pure spectator model (dotted), where the exponential LFWF (dashed) is more pronounced in the valence region and exponentially suppressed in the large $x$ region, cf.~Eq.~(\ref{HMexp-PDF}). As alluded above, this is exemplified in the left panel for the unpolarized PDF $f_1^{\rm sca}(x)= H^{\rm sca}(x,\eta=0,t=0)$
where the power-like ansatz (dash-dotted) provides the best matching with a phenomenological PDF (grayed error band).
Our polarized PDF $g_1^{\rm sca}(x)=\widetilde{H}(x,\eta=0,t=0)$ possesses not the steep small-$x$ behavior
that is commonly included in standard PDF parameterizations, cf.~middle panel in the upper row of Fig.~\ref{Fig-PDFcom}. Note that we confronted our valence model with global DIS fits \cite{GehSti95,BluBot02}, which do not allow  discriminating between valence and sea quarks. The global DSSV fit \cite{FloSasStrVog09}, which includes  also  semi-inclusive DIS and polarized proton-proton collision data and so it allows to disentangle the different quark contributions, lies somewhere between the Gehrmann/Stirling (solid curve) and
the Bl\"umlein/Bottcher (grayed area) parameterization.   We emphasize that in our model the Regge behavior is in the parity odd sector inherited from the parity even sector ($\rho/\omega$ exchanges), while on the phenomenological side there is no clear understanding, even not for on-shell high energy scattering processes. In the upper right panel we show that the model prediction for transversity $h_1^{\rm sca}(x)=H_{\rm T}^{\rm sca}(x,\eta=0,t=0)$, saturating the Soffer bound, is compatible with those extracted from semi-inclusive measurements \cite{Ansetal07,Ansetal09}.

In the lower panels of Fig.~\ref{Fig-PDFcom} we  display the resulting model predictions for other nfPDFs at $t=0$ that decouple in the forward limit, i.e., they drop out in the form factor decomposition of matrix elements (\ref{HE-def}--\ref{FT-def}).  The partonic anomalous magnetic moment $\langle\kappa\rangle^{\rm sca}=1.79$ matches  the measurements of the nucleon anomalous magnetic moments and our   nfPDFs $E^{\rm sca}$, having no nodes, is even more sizeable than the unpolarized PDF, shown in the upper row (left).
The  $x^2\, \widetilde E(x,\eta=0, t=0)$  nfPDF (\ref{DDrep2-FwdtEx0}) is shown in the lower middle panel, which has a relative small contribution and a $x^{-\alpha+1}$ behavior at small $x$.  The  chiral odd  nfPDF ${\widetilde H}_{\rm T}$ (not displayed) decouples in
the forward limit, too, however, apart from the sign and a larger suppression at large $x$, see DD (\ref{Def-SpeFun-thT}), its size is compatible
with the unpolarized PDF. Hence, the  nfPDF $E_{\rm T}^{\rm sca}$ at $t=0$, cf.~(\ref{ETx0}), is even larger as the nfPDF $E$  (right).  In the zero-skewness case the chiral odd nfPDF $\overline{E}_{\rm T}^{\rm sca}$  is given by  $E^{\rm sca}$, cf.~the constraint (\ref{bEx0}), and $\widetilde E_{\rm T}^{\rm sca}$ vanishes.

Let us now have a closer look to our model predictions for the quark orbital angular momentum, which plays a central role in the so-called nucleon spin puzzle.  Unfortunately, its definition is controversially discussed, namely, one can define it in terms of an interaction independent operator \cite{Jaffe:1989jz},
which contains a partial derivative, or in a gauge invariant manner by using the covariant derivative \cite{Ji:1996ek}. In a quark model, where the gauge field is absent, both definitions coincide with each other. To calculate the projection of the quark orbital angular momentum on the $z$-axis, one might either use  the definition of the quark angular momentum operator in its overlap representation, which is expressed by the (relative)
orbital angular momentum of the proton \cite{Harindranath:1998ve,Brodsky:2000ii} that is weighted with the momentum fraction $1-x$ of the spectator \cite{Burkardt:2008ua}:
\begin{eqnarray}
\label{L-LFWF}
\langle {\cal L} \rangle^{\rm sca} \stackrel{\rm sca}{=} \int_0^1\!dx\,(1-x) \intk\!\! \int\! d\lambda^2\,|\psi^{\Rightarrow}_{\leftarrow}(x,{\bf k}_\perp,\lambda)|^2
 \stackrel{\rm sca}{=} \int_0^1\!dx\,\frac{1-x}{2}\left[f^{\rm sca}_1(x)-g_1^{\rm sca}(x)\right],
\end{eqnarray}
or Ji`s sum rule \cite{Ji:1996ek},
\begin{eqnarray}
\label{Jisumrule}
\langle L \rangle  \stackrel{\rm EOM}{=} \langle J \rangle  - \langle s \rangle\,,\quad \langle J \rangle = \frac{1}{2} \langle x \rangle + \frac{1}{2} \int_0^1\!dx\,x E(x,\eta=0,t=0)\,.
\end{eqnarray}
Note that equations-of-motion (EOM) play an essential role in the Field theoretical derivation of the angular momentum operator and that
LFWFs enter into both orbital angular momentum definitions (\ref{L-LFWF}) and (\ref{Jisumrule}) in different combinations.
Therefore, in Yukawa theory these both formulae  provide the same value, i.e.,
$\langle L\rangle = \langle {\cal L}\rangle$ \cite{Burkardt:2008ua}%
\footnote{Choosing improper renormalization conditions, one might also render  an inequality in Yukawa theory.}%
. But in a quark model the various LFWFs are not related by equations-of-motion
and so one might find two different values for the quark orbital angular momentum, i.e., $\langle L\rangle \neq \langle {\cal L}\rangle$.

Our spin average for the Regge improved LFWF models,
\begin{eqnarray}
\langle s \rangle^{\rm sca} \approx
\left\{\begin{array}{c}
0.13 \\
0.13  \\
0.36
\end{array}\right\}\quad
\mbox{for}\;\;
\left\{\begin{array}{c}
 {\rm LFWF}^{\rm pow} \\
  {\rm LFWF}^{\rm exp}\\
 \mbox{HM07}
\end{array}\right\},
\end{eqnarray}
is compared to standard PDF parameterizations (\ref{<s>^sca-pheno}) rather low, see also the upper panel in the middle of Fig.~\ref{Fig-PDFcom}.
Hence, in these both models the quark orbital angular momentum, evaluated from Ji`s sum rule (\ref{Jisumrule}),
is a relative small positive value,
\begin{eqnarray}
\label{J^sca-pre1}
\langle L \rangle^{\rm sca} =
\left\{\begin{array}{c}
0.069 \\
0.075  \\
-0.03
\end{array}\right\},\quad
\langle J \rangle^{\rm sca} =
\left\{\begin{array}{c}
0.19  \\
0.20  \\
0.33
\end{array}\right\}\quad
\mbox{for}\;\;
\left\{\begin{array}{c}
 {\rm LFWF}^{\rm pow} \\
{\rm LFWF}^{\rm exp} \\
 \mbox{HM07}
\end{array}\right\},
\end{eqnarray}
while the {\it HM07} model has a large longitudinal spin component and a small negative orbital angular momentum.
Utilizing Ji`s sum rule in lattice simulations, one finds that the orbital angular momentum of sea and valence quarks is
estimated to be $L^{u}\sim -L^d \sim -0.15$ \cite{Hagler:2009ni}, which provides in the scalar sector again a negative quark orbital angular
momentum  of $L^{\rm sca}\sim -0.15$ and a moderate positive angular momentum of $J^{\rm sca}\sim 0.15$. The direct calculation of the quark orbital angular momentum (\ref{L-LFWF}) provides us much larger numbers
\begin{eqnarray}
\label{J^sca-pre2}
\langle {\cal L} \rangle^{\rm sca} =
\left\{\begin{array}{c}
0.33 \\
0.33  \\
0.11
\end{array}\right\},\quad
\langle {\cal J} \rangle^{\rm sca} =
\left\{\begin{array}{c}
0.46  \\
0.46  \\
0.47
\end{array}\right\}\quad
\mbox{for}\;\;
\left\{\begin{array}{c}
 {\rm LFWF}^{\rm pow} \\
{\rm LFWF}^{\rm exp} \\
 \mbox{HM07}
\end{array}\right\},
\end{eqnarray}
where the quark angular momentum $\langle {\cal J} \rangle^{\rm sca} = \langle s \rangle^{\rm sca} + \langle {\cal L} \rangle^{\rm sca}$
has now for all three models (almost) the same large value.  We conclude that a quark model calculation, where usually the dynamics of  LFWFs
is ignored,  can provide  two values, which might largely differ from each other,  for both the orbital and total  quark angular momentum.

\begin{figure}[t]
\begin{center}
\includegraphics[width=15cm]{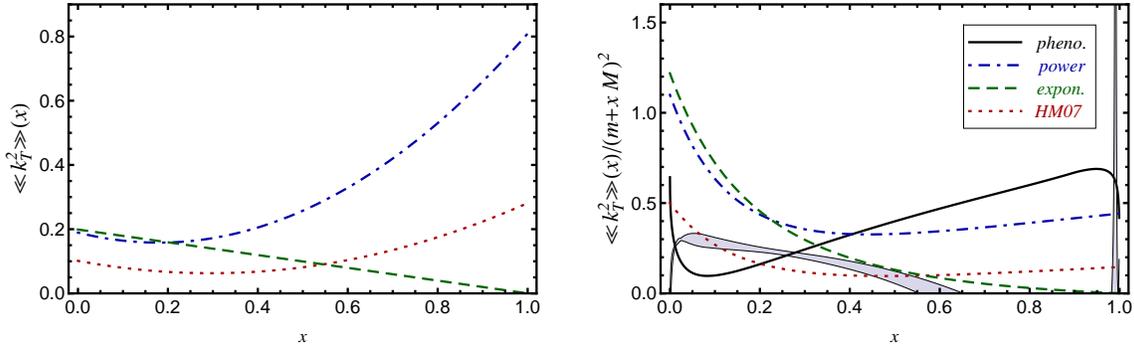}
\end{center}
\vspace{-5mm}
\caption{\small Transverse momentum square average $\langle{\bf k}^2_\perp(x)\rangle$, defined in Eq.~(\ref{def:k2perp}),  (left) and
the ratio $\langle{\bf k}^2_\perp(x)\rangle/(m+x M)^2$ (right) versus the momentum fraction $x$ for a power-like LFWF (dash-dotted), exponential LFWF (dashed), and the {\em HM07} (dotted) model. The grayed area (solid curve) shows the phenomenological
findings, evaluated from Eq.~(\ref{kt2-ratio}), for  Alekhin`s \cite{Ale05} and Bl\"umlein/Bottcher \cite{BluBot02} (Gehrmann/Stirling \cite{GehSti95}) parameterizations.}
\label{Fig-kt2}
\end{figure}
The  $\langle{\bf k}^2_\perp \rangle(x)$ average (\ref{def:k2perp}) possesses in our two LFWF models rather different behavior, see left panel in Fig.~\ref{Fig-kt2}.
As one easily realizes from Eq.~(\ref{HMpow-PDF-2}), for our powerlike LFWF  (dash-dotted and dotted curves)  $\langle{\bf k}^2_\perp \rangle(x) $ takes at the endpoints $x=0$ and $x=1$ the values $m^2/2p$ and $\lambda^2/2p$, respectively, where a non-vanishing minima might exist in the  range $x\in [0,1]$  for $(M-m)^2 <  \lambda^2 <m^2 + M^2$:
$$\langle {\bf k}^2_\perp \rangle_{\rm min}=\frac{1}{2p}\frac{[\lambda^2 -(M-m)^2][(M+m)^2-\lambda^2]}{4 M^2}\,,\quad
x_{\rm min}=\frac{m^2 + M^2 - \lambda^2}{2 M^2}. $$
For the exponential LFWF (dashed) the  ${\bf k}^2_\perp$-average (\ref{HMexp-PDF-2})
takes at $x=0$ the value $M/\overline{A}$, monotonously decreases with growing $x$, and vanishes at $x=0$.
In the {\em HM07} model we find a rather low $\langle {\bf k}^2_\perp \rangle\sim 0.3^2\,{\rm GeV}^2 $
in the small $x$ and valence region, which slightly increases to $0.5^2\,{\rm GeV}^2$ in the large $x$ region. Both of our Regge improved models provides a
larger $\langle {\bf k}^2_\perp \rangle\sim 0.45^2\,{\rm GeV}^2 $ outside the large $x$ region and reaching at $x\to 1$ the values $\sim 0.9^2\,{\rm GeV}^2 $
and zero for the powerlike and exponential LFWF, respectively. In the right panel we show the quantity
$\langle{\bf k}^2_\perp\rangle(x)/(m+x M)^2$, which might be expressed by the PDF ratio (\ref{kt2-ratio}).
The grayed area displays the phenomenological findings, evaluated from Alekhin's \cite{Ale05} and Bl\"umlein/Bottcher \cite{BluBot02} parameterizations. Clearly, for $0.6 \lesssim x$ the polarized PDFs form Bl\"umlein/Bottcher overwhelm the unpolarized PDFs, which yields
a negative $\langle{\bf k}^2_\perp\rangle(x)$ value.  On the other hand, if we take the Gehrmann/Stirling \cite{GehSti95} parameterization we
find a rather reasonable value for the ratio (\ref{kt2-ratio}), shown as solid curve. A similar curve, which is not displayed, we obtain with the DSSV parameterization \cite{FloSasStrVog09}. Surely, the phenomenological uncertainties are so large that a serious confrontation with the scalar diquark picture is impossible.

\begin{figure}[t]
\begin{center}
\includegraphics[width=16cm]{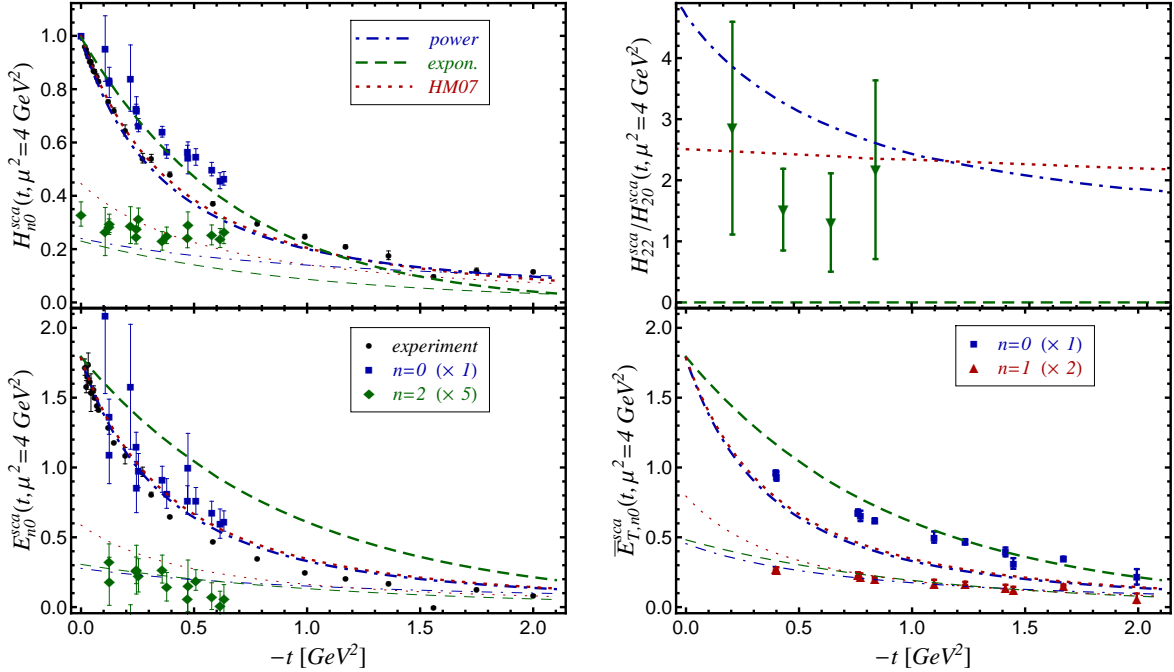}
\end{center}
\vspace{-5mm} \caption{ \label{Fig-FFcom} \small
Our Regge improved power-like (dash-dotted), exponential (dashed) LFWFs,  and {\it HM07} (dotted) \cite{Hwang:2007tb} models are
confronted  versus experimental data  \cite{Hohetal76} and Lattice estimates for the flavor combination $2u^{\rm val}/3-d^{\rm val}/3$.
The  upper [lower] left panel displays the Dirac (Paule) FF combination $F_{1}$ [$F_{2}$] from experiment (circle) \cite{Hohetal76}, $H_{00}$  [$E_{00}$] (squares,thick curves) and  GFF  $H_{20}$  [$H_{20}$] (diamonds, thin curves $\times 5$) from the data set VI
($m_\pi=352.3$ {\rm MeV} ) of Ref.~\cite{Hagetal07}. The  upper right panel displays the ratio $H_{22}/H_{20}$ with
$m_\pi=292.5$ {\rm MeV} \cite{Braetal10}.
In the lower right panel we show the first (squares,thick curves) and second (triangles, thin curves $\times 2$) moments of zero-skewness GPD $\overline{E}_T$ versus Lattice measurement from the heavy pion world $m_\pi=600$ {\rm MeV} \cite{Gocetal06}.}
\end{figure}
Elastic form factors and GFFs are presented in Fig.~\ref{Fig-FFcom}. Here, for our powerlike LFWF model we took into account $\alpha^\prime=0.75\,{\rm GeV}^2$, which compensates the relative small $p$ value. Hence, in this model (dash-dotted) we have almost the same $-t$ dependence as in the ${\it HM07}$ model (dotted),
which slightly under- and overestimate the experimental data (circles) for the Pauli (upper, left) and Dirac (down, left) form factor, respectively. For the exponential LFWF model (dashed) we find that a positive value of $\alpha^\prime$ cannot cure the larger discrepancy to experimental data.
Comparing the thick and thin curves in the upper right and lower panels, it is clearly to realize that the $t$-dependence of GFFs is dying out
with increasing $n$ (thin). This simply reflects the fact that in the zero skewness case the $t$-dependence is accompanied by a factor $(1-x)^2$, cf.~Eqs.~(\ref{HMpow-DD},\ref{HMpow-DD-2}) and (\ref{HMexp-DD-2},\ref{HMexp-DD-2}), or enters via
$x^{-\alpha(t)}$.

Our powerlike LFWF  model is also supported by lattice estimates, which give a positive large ratio\footnote{In lattice nomenclature this ratio is given by $4 A_{3,2}/A_{3,0}$.} (upper, right panel)
$$ H_{2,2}(t)/H_{2,0}(t) = \frac{d^2}{d\eta^2}\int_{-1}^{1}\!dx\, x^2 H(x,\eta,t)\Big|_{\eta=0}\Bigg/  \int_{-1}^{1}\!dx\, x^2 H(x,\eta=0,t),$$
which quantifies the skewness effect.
The sizeable ratio, displayed in the upper right panel by the dash-dotted and dotted curves, are easily understood from the explicit formulae in the DD representation. Roughly spoken, in the DD the moments $H_{2,0}$ and $H_{2,2}$  are multiplied with $y^2$ and $z^2$, respectively, where $y^2$ provides a larger suppression effect. Our exponential LFWF  model possesses a zero-skewness effect and, therefore, this ratio is zero. The GFF $H_{1,2}=4 C_{2}$ has been measured in Lattice calculations, too, and it was found that its value is small in the iso-vector sector \cite{Hagetal07}. Our powerlike LFWF models provide also a small value
\begin{eqnarray}
\label{predictions-C2}
H_{1,2}(t=0)=
\left\{
\begin{array}{c}
  0.36 \\
  0.00 \\
  0.28
\end{array}
\right\}\quad \mbox{for}\;\;
\left\{\begin{array}{c}
 {\rm LFWF}^{\rm pow} \\
{\rm LFWF}^{\rm exp} \\
 \mbox{HM07}
\end{array}\right\},
\end{eqnarray}
qualitative consistent with the statement from the chiral quark soliton model that the valence contributions are positive and (relatively) small \cite{Wak07}.
Transversity GFFs have been also studied in lattice measurements \cite{Gocetal06}.
The $\overline{E}_T =  E_T + 2 \widetilde{H}_T$ GPD, given in our model by $E$ GPD, cf.~Eqs. (\ref{Con-even2odd}), which is qualitative consistent with lattice data from the heavy pion world, see the lower right  panel.

\begin{figure}[t]
\begin{center}
\includegraphics[width=17.4cm]{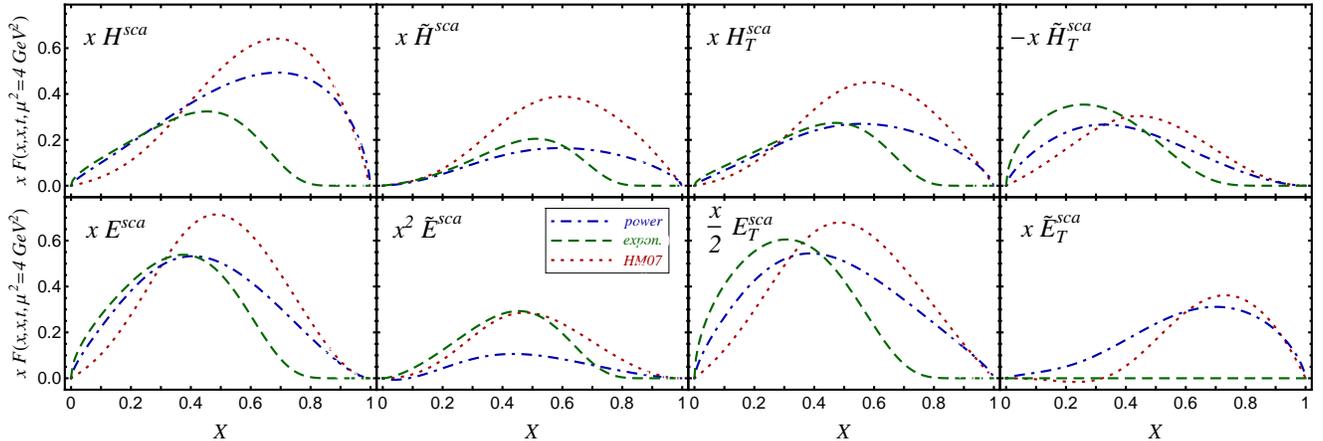}
\end{center}
\vspace{-5mm}
\caption{ \label{Fig-GPDcom} \small
Our Regge improved power-like (dash-dotted), exponential (dashed) LFWFs,  and {\it HM07} (dotted) \cite{Hwang:2007tb} models are
displayed for valence quark  GPDs $F^{\rm sca}=2 F^{u_{\rm val}}/3- F^{d_{\rm val}}/3$ at $\eta =x$ and $t=-0.2\, {\rm GeV}^2$ versus $X=2x/(1+x)$.
Upper [lower] panels from left to right: $x H$ [$x E$] ,  $\widetilde H$ [$x^2 \widetilde E$], $H_{\rm T}$ [$x E_{\rm T}/2$], and
$-\widetilde H_{\rm T}$ [$x {\widetilde E}_{\rm T}$].}
\end{figure}
Let us now consider GPD predictions, relevant for the phenomenology of hard exclusive processes. In a leading order (LO) description of such processes the imaginary part of the amplitude is proportional to the GPDs on the cross-over line $\eta=x$, while the real part can be obtained from a (subtracted)
"dispersion relation", see Eqs.~(\ref{F-LO-Im},\ref{HandE-DR},\ref{tE-DR}). In Fig.~\ref{Fig-GPDcom} we illustrate
that all GPDs might be sizeable, where the pronounced one is the chiral odd GPD $E_T^{\rm sca}$, which we rescaled by a factor $x/2$ rather $x$, while the polarized
GPD ${\widetilde H}^{\rm sca}$ and the chiral odd GPD ${\widetilde E}^{\rm sca}_{\rm T}$ are less pronounced.
We note that transversity contribution are utilized to describe the hard exclusive $\pi^+$ and $\pi^0$  electroproduction
in a handbag model analysis \cite{Ahmad:2008hp,Goloskokov:2009ia,Goloskokov:2011rd}. The GPD  $x {\widetilde E}^{\rm sca}$ is a small quantity,
however, it is clearly different from zero.

\begin{figure}[t]
\begin{center}
\includegraphics[width=17.4cm]{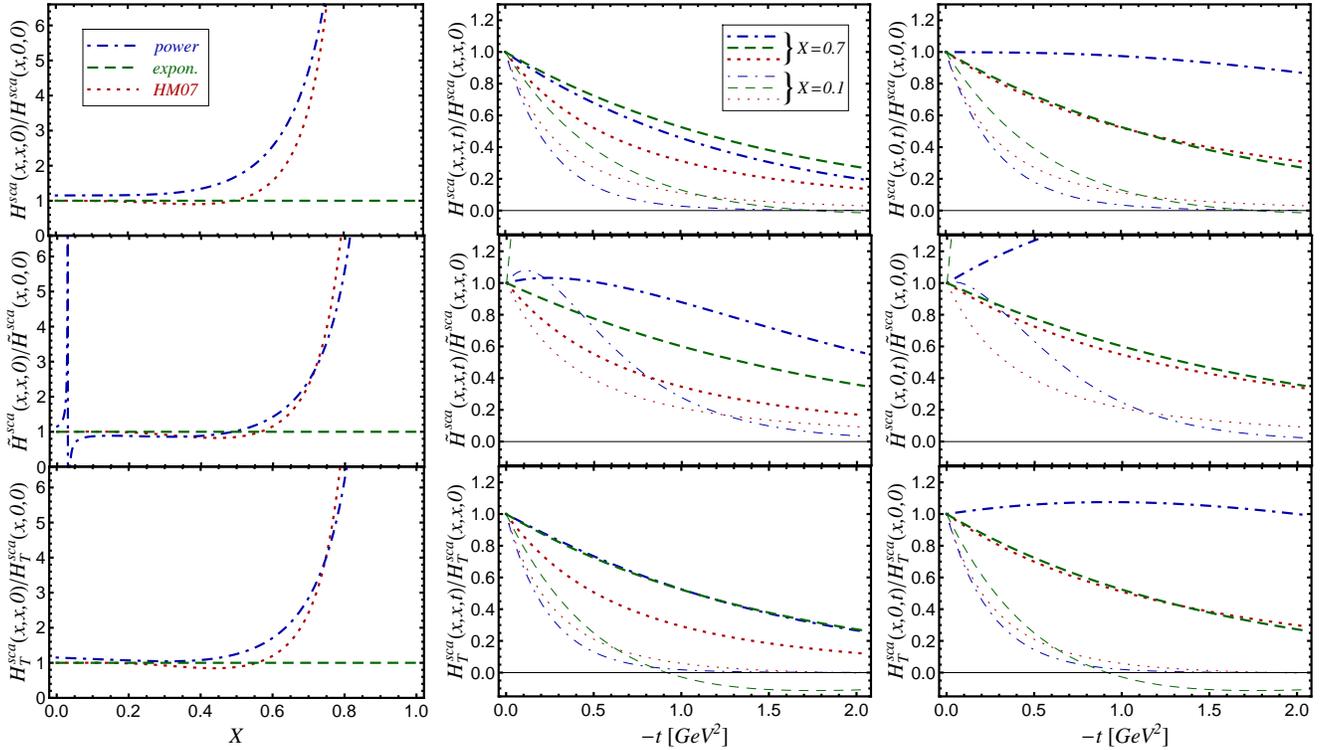}
\end{center}
\vspace{-5mm}
\caption{ \label{Fig-gpd_ratios} \small
Ratios of our Regge improved power-like (dash-dotted), exponential (dashed)  LFWFs,  and {\it HM07} (dotted) \cite{Hwang:2007tb} models are
displayed for valence quark  GPD combinations $F^{\rm sca}=2 F^{u_{\rm val}}/3- F^{d_{\rm val}}/3$ with
$F=\{H^{\rm sca}\ ({\rm top}),{\widetilde H}^{\rm sca}\ ({\rm middle}),  H_T^{\rm sca}\ ({\rm bottom}\}$:
$r$ ratios (\ref{gpd-ratios1}) versus $X=2x/(1+x)$ (left), $F^{\rm sca}(x,\eta=x,t)/F^{\rm sca}(x,\eta=x,t=0)$ (middle)  and  $F^{\rm sca}(x,\eta=0,t)/F^{\rm sca}(x,\eta=0,t=0)$ (right) for $X=0.1$ (thin) and $X=0.7$ (thick) versus $-t$ .}
\end{figure}
A possibility to quantity the skewness effect of $H$, $\widetilde H$, and $H_T$  GPDs is provided by the ratios
\begin{eqnarray}
\label{gpd-ratios1}
r(x) = \lim_{t\to 0}  \frac{H(x,x,t)}{f_1(x)}\,,
\quad
\widetilde r(x) = \lim_{t\to 0}  \frac{\widetilde H(x,x,t)}{g_1(x)}\,,
\quad
r_T(x) = \lim_{t\to 0}  \frac{H_T(x,x,t)}{h_1(x)}\,,
\end{eqnarray}
which are in principle accessible from experimental data.
Phenomenologically, it was found from deeply virtual Compton scattering measurements that at small $x$ the $r$ ratio is compatible with one \cite{Kumericki:2009uq}. As shown in the left panels%
\footnote{To make contact with the kinematics in experimental measurements,
we show here the ratios versus the momentum fraction $X=2x/(1+x)$, which can be equated to the Bjorken variable $x_{\rm Bj}$, rather $x$.}  of Fig.~\ref{Fig-gpd_ratios}, this feature is implemented in all our models. Thereby, the $\widetilde r$ ratio for the Regge improved power-like LFWF model
(dash-dotted) has a singular behavior in the vicinity of $X\sim 0.1$ ($x\sim 0.05$), which is caused by nodes in $\widetilde H(x,x,0)$ and $\widetilde H(x,0,0)$
that are slightly shifted. These nodes (sign changes) are caused by the competition of the diagonal $L^z=0$ and $L^z=1$  LFWF overlaps, see
discussion below Eq.~(\ref{Def-SpeFun-te}). Note that in the limit $x\to 0$ the value of all three ratios of the power-like LFWF models depend on
both the Regge intercept $\alpha$ and the parameter $p$. For the Regge improved model we have values that are slightly larger than one.
For the exponential LFWF model, which has no skewness dependence, the skewness ratios are one
(dashed curves). For the power-like LFWF models the GPDs at the cross-over line have a weaker large-$x$ fall-off than the corresponding PDFs, where such
a behavior is expected from large-$x$ counting rules \cite{Yua03}. Hence, the ratios increase drastically in the large-$x$ region.

In the middle (right) column of  Fig.~\ref{Fig-gpd_ratios} we display the $t$-dependence of GPDs
\begin{eqnarray}
\label{gpd-ratios2and3}
\frac{F(x,\eta=x,t)}{F(x,\eta=x,t=0)}\;({\rm middle})\quad\mbox{and}\quad\frac{F(x,\eta=0,t)}{F(x,\eta=0,t=0)}\;({\rm right})\quad\mbox{for}\quad F\in\{H,\widetilde H, H_T\}\, ,
\end{eqnarray}
normalized to one at $t=0$, at the cross-over ($\eta=0$) line
for fixed Bjorken-like momentum fraction
$X=0.1$ and $X=0.7$ as thin and thick curves, respectively.
For both $H$ (top) and $H_T$ (bottom) GPDs one clearly realizes that the $t$-dependence  flatters out at large $x=X/(2-X)$. Comparing the panels in the middle and right columns, one realizes that for the power-like LFWFs models (dash-dotted, dotted) this effect is more pronounced in the zero-skewness case. Of course,  the
exponential LFWF model (dashed) does not dependent on the skewness parameter and so the $t$-dependence is the same in the panels of the
middle and right columns. Note that in this model $H_T$ becomes negative at large $-t$, e.g., for $x\sim 0.05$ at $-t \gtrsim  1\,{\rm GeV}^2$.
This arises from the $L^z=1$ LFWF which induces a $t$-dependence in the $h_T$ DD, given in Eq.~(\ref{Def-SpeFun-eT}).  For $\widetilde H$ GPD (middle)
the interplay of diagonal $L^z=0$ and $L^z=1$ LFWF overlap induces for the Regge improved models (dash-dotted, dashed) a more intricate pattern
as for $H$ or $H_{\rm T}$  GPD. Here, we observe now that at small $-t$ the ratios (\ref{gpd-ratios2and3}) increase with growing $-t$. Interestingly,
in Ref.~\cite{Guidal:2010de} it was argued that deeply virtual Compton scattering data indicate that GPD $\widetilde H$ has a
weaker $t$-dependence than GPD $H$.

As already pointed out in Ref.~\cite{Hwang:2007tb}, both features that valence-like GPDs are dying out slower as PDFs at large $x$ and that
the $t$-dependence vanishes at large $x$ are in qualitative agreement with hard exclusive $\rho^0$
electroproduction cross section measurements at very large $x_{\rm Bj} \gtrsim 0.6$ \cite{Guidal:2007cw,Morrow:2008ek}.
However, since $-t_{0}$ also increases with growing $x_{\rm Bj}$, the condition $ -t \ll Q^2$ within a rather small value of the photon virtuality
$2\, {\rm GeV}^2 \lesssim Q^2 < 6\, {\rm GeV}^2$  might be not satisfied for the longitudinal part of the $t$-integrated cross section and so it might be doubtful to employ the collinear factorization analysis within the GPD framework to the $t$-integrated cross section itself.

\subsubsection{Comments about the axial-vector sector}
\label{sec:pheno-axialvector}

As we have discussed in Sec.~\ref{uPDF-classification-spherical}, by adding an identity proportional spin density  matrix a scalar diquark model for uPDFs can be easily
extended to a ``spherical" model of rank-four. If we rely on SU(6) flavor-spin symmetry,
then all off-diagonal entries of the uPDF spin-density  matrix in the $u/6+2d/3$ sector have to vanish. Indeed, this is somehow supported by phenomenological
findings.   For the quark spin average we might quote the value as
$$
\langle s^{\rm axi} \rangle = \frac{1}{6} \langle s^{u_{\rm val}} \rangle  + \frac{2}{3}  \langle s^{d_{\rm val}} \rangle   \sim -0.03\,,
$$
where its modulus is indeed  smaller than in the scalar sector, cf.~Eq.~(\ref{<s>^sca-pheno}).  For the lowest moment of the transversity PDF $h_1$, often called ``tensor charge",  we even find from the fit results \cite{Ansetal09} a term that is compatible with zero
$$
\int_0^1\!dx\, \left[\frac{1}{6} h_1^{u_{\rm val}}(x)+\frac{2}{3} h_1^{d_{\rm val}}(x)\right] \approx -0.05^{+0.11}_{-0.07}\,,\quad
\int_0^1\!dx\, h_1^{\rm sca}(x) \approx 0.44^{+0.7}_{-0.16}\,.
$$
Hence, a ``spherical" quark model of rank-four that relies on SU(6) flavor-spin symmetry, such as the three-quark LFWF  \cite{PasCazBof08},
axial-vector diquark \cite{JakMulRod97},
bag \cite{AvaEfrSchYua10}, the chiral quark soliton \cite{LorPasVan11}, and the covariant parton \cite{EfrSchTerZav09} model, might be tuned somehow compatible w.r.t.~to (u)PDF phenomenological findings.

However, for GPDs the situation is more intricate. Taking our minimal axial-vector diquark GPD model, which respect the underlying Lorentz symmetry,  and associating it with $d$-quark, as dictated by  SU(6) symmetry,  leads to a complete
failure w.r.t.~the anomalous magnetic moment of the nucleon. The GPD spin-correlation matrix (\ref{GPD-av-model}) suggest that $\kappa^d$
is positive while its experimental value
$$\kappa^d = F^p_2(t=0)+2 F^n_2(t=0)\approx -2.03$$ is negative. On the other hand it is well known that  SU(6) symmetry, implemented in a constituent quark model, is consistent with the ratio of proton and neutron magnetic moments. Indeed, the three quark LFWF model of Ref.~\cite{BofPas07}
provides a negative $d$-quark  GPD $E$, however, this model has not been developed to the stage where it respect the underlying Lorentz symmetry.
Also the bag model calculation of Ref.~\cite{Ji:1997gm} provides a negative $d$-quark  GPD $E$. Interestingly, the covariant axial-vector diquark model of Ref.~\cite{Tiburzi:2004mh}, where the polarization sum of the axial-vector diquark is taken to be $-g_{\mu\nu}$, provides a $d$-quark  GPD $E$ which possesses a node. According to Eq.~(38) of Ref.~\cite{Tiburzi:2004mh} the $L^z=0$ and $|L^z|=1$ coupling reads in our notation
$$
e^{d} \stackrel{\displaystyle \cite{Tiburzi:2004mh}}{\propto} \frac{8}{9}\left[\frac{2m}{M}(2 x-y)-(1-2 x+y) y\right]\quad{\Rightarrow}\quad
 E^d(x,\eta=0,t)\stackrel{\displaystyle \cite{Tiburzi:2004mh}}{\propto} \frac{8}{9}\left[\frac{2m}{M}-(1-x) \right]x
$$
For zero-skewness this GPD  is positive at large $x$ and might become slightly negative at small $x$, however, the $d$-quark anomalous magnetic moment remains positive.   It is interesting to note that the chiral quark soliton model of Ref.~\cite{Ossetal04} gives for the iso-singlet combination a
qualitative different behavior, namely, its GPD $E$  is negative and positive at large and small $x$, respectively. Obviously, this contradicts both the axial-vector diquark  \cite{Tiburzi:2004mh} and pseudo-scalar diquark model predictions, where the latter is given by
$$
e^{d} \stackrel{\rm PS}{\propto} -(1-x)\left[\frac{m}{M}-y\right]\quad{\Rightarrow}\quad
 E^d(x,\eta=0,t)\stackrel{\rm PS}{\propto} -(1-x)\left[\frac{m}{M}-x\right],
$$
see Sec.~\ref{sec:GPD-psesca}.
We add that a  LFWF model for nfPDFs in the axial-vector sector predicts a negative $E$ for $d$-quarks without nodes \cite{LuSch10},
where, unfortunately, the Gaussian  form of the scalar two-body LFWF does not respect the underlying Lorentz symmetry.   To our best knowledge  it is not investigated whether the GPD results of the equivalent ``spherical" uPDF  models, i.e., the three-quark LFWF  \cite{PasCazBof08},
axial-vector diquark \cite{JakMulRod97},
bag \cite{AvaEfrSchYua10}, the chiral quark soliton \cite{LorPasVan11}, and the covariant parton \cite{EfrSchTerZav09} model are still equivalent
and can be partially represented by a ``spherical"  GPD model, too.  Let us conclude we are not aware on a LFWF model in the axial-vector sector that
has been shown to be consistent with Lorentz symmetry and is in accordance with phenomenological findings.


\subsubsection{Modeling pomeron like behavior}
\label{sec:GPD-pomeron}

We  employ now effective two-body LFWF modeling in the sea quark sector, which we consider here as flavor symmetric. Large $N_c$ counting rules might suggest that polarized $T$-even parton distributions in the iso-singlet sector are relatively suppressed, compared to the iso-vector channel \cite{DiaPetPobPolWei96,PobPol96,Goeke:2001tz,Pobylitsa:2003ty}. While for polarized  PDFs this is consistent with polarized DIS measurements, not much is known about the behavior of  GPD $E$, which might be addressed in deeply virtual Compton scattering off a transversal polarized proton. It is important to note that at large $N_c$ the iso-singlet  GPD $E$ is relatively suppressed to the iso-vector  GPD $E$, however, it is on the same footing as the iso-singlet GPD $H$ \cite{Goeke:2001tz}.  Moreover, GPDs $E$ and $H$  are rather loosely tied to each other by Lorentz invariance and so we might wonder how GPD $E$, associated with  a proton helicity non-conserved ``pomeron'' exchange that plays no role in Regge phenomenology, behaves at small $x$.

According to Table \ref{tab:LFWFs2GPDs}, a diquark model which provides only GPDs $H$ and $E$  can be build from the pair of scalar LFWFs (\ref{LFWF-spinors-Spair}) and the two transverse ones  (\ref{LFWF-spinors-AV}) for a minimal axial-vector diquark coupling,
\begin{eqnarray}
\label{LFWF-overlap-sea}
{\bf F}^{\rm sea} = \sum_{\;\; \lambda^2}\!\!\!\!\!\!\!\!\int\, \rho(\lambda^2) \!\sum_{\sigma=-,+}\int\!\!\!\!\int\!\frac{d^2{\bf k}_\perp}{1+\eta}\!\!\!&&\!\!\!\!\!\!\Bigg[
\frac{1}{2} \mbox{\boldmath $\psi^\dagger$}_\sigma^{{\rm S}}\!
 \left(\!
\frac{x-\eta}{1-\eta},{\bf k}_{\perp 1}- \frac{1-x}{1-\eta} {\bf \Delta}_\perp\Big|\lambda^2
\!\right) \otimes
\mbox{\boldmath $\psi$}_\sigma^{\rm S}\!\left(\!\frac{x+\eta}{1+\eta},{\bf k}_{\perp}\Big|\lambda^2\!\right)
\\
&&\!\!\!\!\!\!\!\!+\frac{3}{4}
\mbox{\boldmath $\psi^\dagger$}_{\sigma 1}^{{\rm A}}\!
 \left(\!
\frac{x-\eta}{1-\eta},{\bf k}_{\perp 1}- \frac{1-x}{1-\eta} {\bf \Delta}_\perp\Big|\lambda^2
\!\right) \otimes
\mbox{\boldmath $\psi$}_{\sigma 1}^{\rm A}\!\left(\!\frac{x+\eta}{1+\eta},{\bf k}_{\perp}\Big|\lambda^2\!\right)
 \Bigg].
\nonumber
\end{eqnarray}
This spin-correlation matrix ${\bf F}^{\rm sea}$ has indeed only GPD $H$ and GPD $E$  entries that are given by the DD representation (\ref{Def-SpeRep}),
\begin{eqnarray}
\label{Def-SpeRep-HE} \left\{\!
\begin{array}{c}
H^{\rm sea} \\
E^{\rm sea}
\end{array}\!
\right\}(x,\eta,t)=
\int_{0}^1\!dy\; \int_{-1+y}^{1-y}\! dz\;
\delta(x-y-\eta z )\; \left\{
\begin{array}{c}
h^{\rm sea} + (x-y) e^{\rm sea}\\
(1-x) e^{\rm sea}
\end{array}
\right\}(y,z,t)\,,
\end{eqnarray}
where the DDs are defined in Eqs.~(\ref{Def-SpeFun-h}, \ref{Def-SpeFun-e}). We recall that because of the explicit $x$ dependence in the DD representation polynomiality is completed in our model and so we can extract by means of Eq.~(\ref{Cal-Dterm}) the $D$-term. Note that the corresponding spin-density matrix for uPDFs is an identity matrix, proportional to
\begin{eqnarray}
f_1^{\rm sea}(x,{\bf k}_\perp) = \left[\left(\frac{m}{M}+x\right)^2+\frac{{\bf k}_\perp^2}{M^2}\right]
{\bf \Phi}(x,{\bf k}_\perp)\,,
\end{eqnarray}
where the ``uPDF'' ${\bf \Phi}$ is specified for the power-like and exponential LFWF model in (\ref{HMpow-uPDF}) and (\ref{HMexp-uPDF}), respectively.

To adjust the parameters of our model,  we take for our convenience now the input scale to be $Q^2_0=2\, {\rm GeV}^2$.
From hard exclusive  electroproduction of mesons and photons it is known
that the Regge slope parameter is smaller than  the soft pomeron one, $\alpha^\prime=0.25$, and we will  simply set it here to zero.
Thus, our GPD models  satisfy positivity constraints.
The effective pomeron intercept $\alpha=1.13$, slightly larger than one, is consistent with global PDF and GPD fit results \cite{Ale05,Kumericki:2009uq}. We set in agreement with  phenomenological findings  $p=3$, i.e., our sea quark PDF vanishes at large $x$
as $(1-x)^{8.13}$ and it is slightly stronger suppressed as $(1-x)^7$, stated by counting rules. The normalization (\ref{Fix-Nor}) is fixed by the averaged sea quark momentum fraction
\begin{eqnarray}
A^{\rm sea} = \int_0^1\!dx\; x\, H^{\rm sea}(x,\eta=0,t=0) = \sum_{q=u,d,s} \int_0^1\! dx\;  2 x\,  \bar{q}(x) = 0.15\,.
\end{eqnarray}
The mass parameters $m$ and $\lambda$ are taken from phenomenological PDF \cite{Ale05} and GPD \cite{Kumericki:2009uq} findings, extracted at leading order, where  for the later we utilized the {\em KM09a} parameterization.  Analogously, we also adjust the exponential LFWF model. Altogether, our sea quark GPD model is for ${\cal Q}^2_0 = 2\, \GeV^2$ fixed by the following parameters
\begin{eqnarray}
\label{gpdH^sea-power}
\mbox{\rm LFWF}^{\rm pow}: && \alpha=1.13\,,\quad p=3\,, \quad m=0.63\, \GeV\,, \quad \lambda=1.00\, \GeV \,,
\\
\label{gpdH^sea-expon}
\mbox{\rm LFWF}^{\rm exp}: && \alpha=1.13\,,\quad A=3\,, \quad m=0.63\, \GeV\,, \quad \lambda=1.45\, \GeV \,.
\end{eqnarray}

\begin{figure}[t]
\begin{center}
\includegraphics[width=17.4cm]{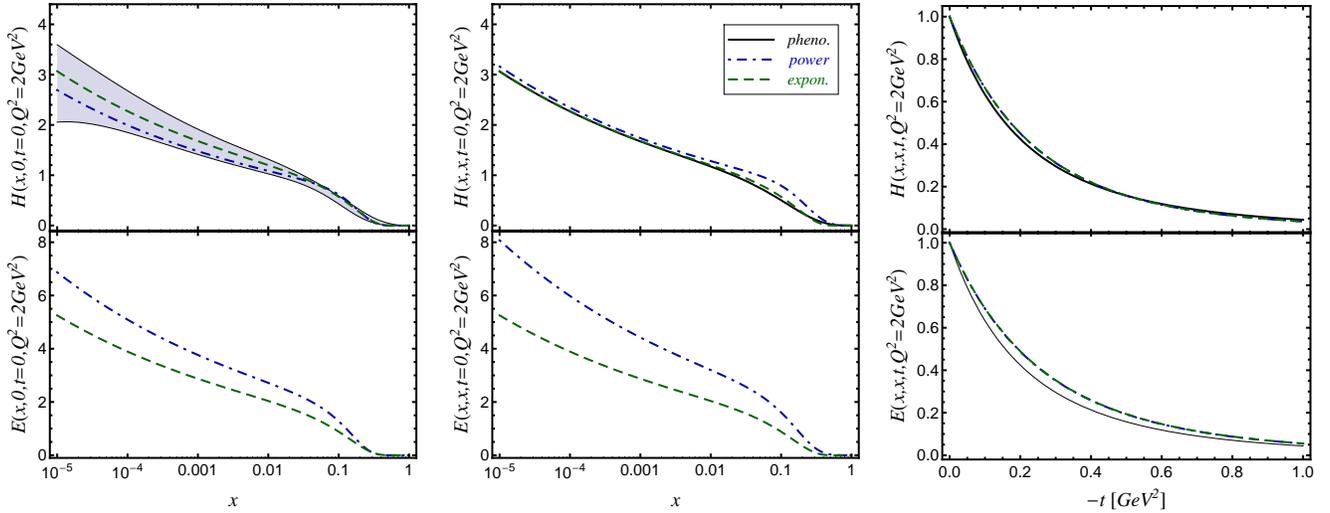}
\end{center}
\vspace{-5mm} \caption{\small Upper panels: the unpolarized sea quark PDF (left) and $x$-dependence for $t=0$ (middle)
and $t$-dependence for $x=10^{-3}$ (right) of GPD $H(x,x,t)$  at the input scale ${\cal Q}^2_0 = 2\,
\GeV^2$, extracted from LO fits \cite{Ale05} and \cite{Kumericki:2009uq}, are shown as grayed area and solid curves.
Our adjusted sea quark GPD models with power-like (\ref{gpdH^sea-power}) and exponential (\ref{gpdH^sea-expon})  LFWF are displayed as dash-dotted and dashed curves, respectively.
In the three lower panels we show from left to right the analog predictions for the GPD $E$.
The thin solid line in the lower third panel shows for comparison the $H$ fit result \cite{Kumericki:2009uq}.
\label{Fig-sea}}
\end{figure}
We illustrate in the upper left and middle panel of Fig.~\ref{Fig-sea} that our both sea quark models can fairly describe the PDF \cite{Ale05} and GPD \cite{Kumericki:2009uq} fit results at ${\cal Q}^2= 2\, \GeV^2$, respectively. Also the $t$-dependence of the power-like
and exponential ansatz, shown in the upper right panel, are hardly to distinguish in the experimentally accessible region $|t| < 1\,{\rm GeV}^2$ from the fit result to the deeply virtual Compton scattering cross section \cite{Kumericki:2009uq}.
The $r$-ratio for the power-like LFWF model is slightly larger than one, however, we expect that both of our GPD models will grow too fast with increasing ${\cal Q}^2$ and  so the $r$-ratio will increase \cite{DieKug07a}, while the experimental data require a constant $r\approx 1$  over a large ${\cal Q}^2$ lever arm \cite{Kumericki:2009uq}.

Our models might now serve to estimate the  $E$ GPDs  and in particular they allow us to calculate the associated $D$-term (\ref{Cal-Dterm}).
In the lower panels of Fig.~\ref{Fig-sea} we display the  $E$ GPDs at $\eta=0$ (left) and $\eta=x$ (middle) versus $x$ at $t=0$.
As we realize with the specific spin-spin coupling (\ref{LFWF-overlap-sea}) the resulting $E$ GPDs  is for small $x$ about twice times bigger than the
corresponding $H$ GPDs,  while the $t$-dependency, shown in the right panel,  differs for $|t| \le 1\, \GeV^2$ only slightly from those of $H$ GPDs.
Note that all these findings respect positivity constraints, e.g.,
$$
-H(x,b) \le \frac{1}{2 M} \frac{\partial}{\partial b} E(x,b) \le  H(x,b),
$$
a special case of those in Refs.~\cite{Pol02,Bur03}  that follow from the positivity of spin-correlation matrix eigenvalues (\ref{bfF-b}) in
impact parameter space.

We can also easily determine the sea quark angular momentum,
\begin{eqnarray}
J^{\rm sea} = \frac{1}{2} \left[A^{\rm sea} + B^{\rm sea}\right]\,, \quad B^{\rm sea}=\int_0^1\!dx\,x\, E^{\rm sea}(x,\eta=0,t=0),
\end{eqnarray}
in terms of the anomalous gravitomagnetic moment $B^{\rm sea}$. Its estimation  from Lattice simulations or
deeply virtual Compton scattering measurements has been not reached at present \cite{Kumericki:2009uq}.
Our models provide a rather large value
\begin{eqnarray}
B^{\rm sea} \approx \left\{ {0.31 \atop 0.23} \right\} \quad \Rightarrow \quad
J^{\rm sea} \approx \left\{ {  0.23 \atop 0.19} \right\}\quad
\mbox{for}\;\;
\left\{
\mbox{\rm LFWF}^{\rm pow} \atop
\mbox{\rm LFWF}^{\rm exp}
\right\} \quad \mbox{at}\quad {\cal Q}^2 = 2\, \GeV^2\,.
\end{eqnarray}
Hence, we find from the angular momentum sum rule and the phenomenological values of $A^{\rm val}$ and $A^{\rm G}$
\begin{eqnarray}
J^{\rm val}  =\frac{1}{2} \left[A^{\rm val}  + B^{\rm val}  \right] \approx 0.22 +  \frac{1}{2}  B^{\rm val}\,,
\quad
J^{\rm G}  =\frac{1}{2} \left[A^{\rm G}  + B^{\rm G}  \right] \approx  0.2 +  \frac{1}{2} B^{\rm G}\,,
\end{eqnarray}
where angular momentum conservation requires that the anomalous gravitomagnetic moment vanishes, i.e.,
\begin{eqnarray}
B^{u^{\rm val}} + B^{d^{\rm val}}  + B^{\rm G} = - B^{\rm sea} \approx  \left\{ {-0.31 \atop -0.23} \right\}
\quad
\mbox{for}\;\;
\left\{
\mbox{\rm LFWF}^{\rm pow} \atop
\mbox{\rm LFWF}^{\rm exp}
\right\}.
\end{eqnarray}
Assuming  that the valence  GPDs $E^{u^{\rm val}}$ and $E^{d^{\rm val}}$ have a similar functional form than GPDs  $H^{u^{\rm val}}$
and $H^{d^{\rm val}}$, one finds%
\footnote{The valence anomalous gravitomagnetic moment
$ B^{ \rm val} =  P^{u^{\rm val}} \kappa^u + P^{d^{\rm val}} \kappa^d \approx 0.836\, P^{u^{\rm val}}  -2.033\, P^{d^{\rm val}}$
might be expressed by the  partonic anomalous magnetic moments
and the Mellin-moment fractions
$
P^i = \int_0^1\!dx\,x\, E^{i}(x,\eta=0,t=0)\Big/\int_0^1\!dx\, E^{i}(x,\eta=0,t=0)\,.
$
If we assume that $ P^{u^{\rm val}}/ P^{d^{\rm val}}$ nearly coincides with the averaged momentum fraction ratio
$ \langle x \rangle^{u^{\rm val}}/ \langle x \rangle^{d^{\rm val}} \approx 2.5$,
we might conclude that $B^{ \rm val}$ is compatible with zero.}
that the  anomalous gravitomagnetic moment in the valence quark sector is compatible with zero.  Hence,
we have in our model a large gluonic anomalous gravitomagnetic moment  $ B^{\rm G} \approx  - B^{\rm sea} \sim -0.28$
and so the gluonic angular momentum $J^{\rm G} \sim 0.05$ is rather small. This
angular momentum scenario is to our best knowledge
not much discussed in the literature \cite{KumMuePas08a,EIC11Die}, which might be related to the fact that sea quarks and gluon degrees of freedom are naturally
ignored by quark model builders and that they are also so far not (directly) accessible from Lattice simulations, see, e.g., Ref.~\cite{Hagler:2009ni}.

\begin{figure}[t]
\begin{center}
\includegraphics[width=17.4cm]{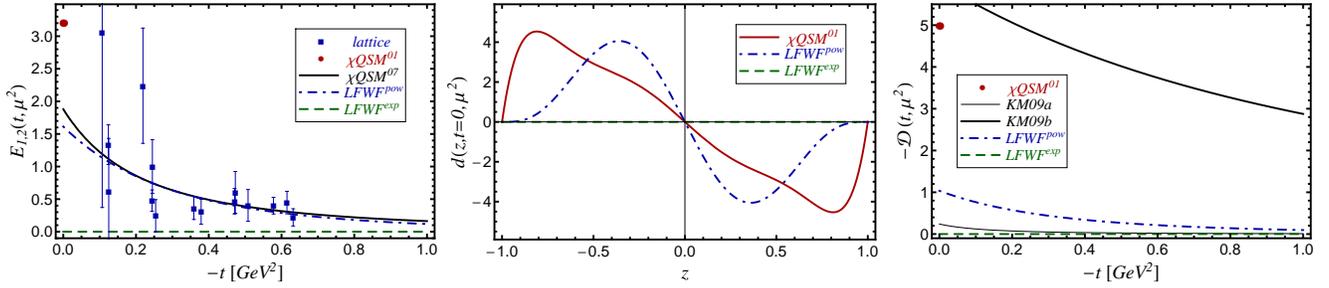}
\end{center}
\vspace{-5mm} \caption{\small The GFF $E^{u+d}_{1,2}$ versus $-t$ (left), $D$-term versus $z$ (middle), and subtraction constant $-{\cal D}$ versus $-t$ (right). The squares are from the data set VI
($m_\pi=352.3$ {\rm MeV} ) of the lattice simulation \cite{Hagetal07}, the solid line  in the left and middle visualizes the chiral quark soliton model results form Ref.~\cite{Goeetal07} and  \cite{Goeke:2001tz}, respectively, and the circles in the first and third panels are the results from \cite{Goeke:2001tz} at $t=0$. The thin ({\em KM09a}) and thick ({\em KM09b}) solid lines in the right panel show the subtraction constant extracted from dispersion relation fits \cite{MueKum09}.
Our adjusted sea quark GPD models with power-like (\ref{gpdH^sea-power}) and exponential (\ref{gpdH^sea-expon})  LFWF are displayed as dash-dotted and dashed curves, respectively.
\label{Fig-D^sea}}
\end{figure}
Let us also consider the first Mellin moments of GPDs $H$ and $E$
that belong to the highest possible power in $\eta^2$ and  so they differ by sign,
\begin{eqnarray}
\label{h_1^{(2)}}
 H_{1,2}(t)= -E_{1,2}(t)\,, \quad
 E_{1,2}(t) = \frac{1}{2} \frac{d^2}{d\eta^2} \int_{-1}^1\! dx\, x\, E(x,\eta,t)\,.
\end{eqnarray}
This GFF appears in the form factor
decomposition (of the quark part) of the energy momentum tensor and should be somehow been governed
by the chiral dynamics \cite{Polyakov:1999gs}.  We recall that the part of the Mellin moments that complete polynomiality is in our model
incorporated in both GPD regions and so the  duality
between the partonic $s$-channel and mesonic like
$t$-channel exchanges is fully incorporated in the model, see,  e.g., in Ref.~\cite{Kumericki:2007sa}. Loosely spoken the
GFF (\ref{h_1^{(2)}}) is associated with the ``pomeron'' exchange, which dictate in particular the sign.
From the point of view of chiral dynamics it appears more natural to interpret such terms
as meson-like $t$-channel exchanges that is essential associated with a $J=0$ $\sigma$ -exchange.
The physical
interpretation of the energy-momentum form factor (\ref{h_1^{(2)}}) has been illuminated in Ref.~\cite{Goeetal07}.
Namely, the absence of a pole at $t=0$ in $ H_{1,2}(t)$,
$$\lim_{t\to 0} t\,  H_{1,2}(t) = 6 M \int\! d^3 {\bf r}\,
p({\bf r}) =0\,,\quad  H_{1,2}(t=0) = M \int\! d^3 {\bf r}\, {\bf r}^2 \,
p({\bf r})\, ,$$ is nothing but the proton stability criteria which  tells us that the net
pressure $\int d^3 {\bf r}\, p({\bf r})$  vanishes. From that it appears to be plausible that $p({\bf r})$ is negative
at large distances $|{\bf r}|$ and so it follows that
the partonic net contribution to the GFF $ H_{1,2}(t)$ $[ E_{1,2}(t)]$
at $t=0$ should be negative [positive]. Our sea quark model, build from a power-like LFWF, is in accordance with this expectation and
we find
\begin{eqnarray}
\label{EMT-For-val}
H_{1,2}^{\rm sea}(t=0) = -E_{1,2}^{\rm sea}(t=0) \approx \left\{ { -1.61 \atop 0 }\right\}\quad \mbox{for}\;\;
\left\{
\mbox{\rm LFWF}^{\rm pow} \atop
\mbox{\rm LFWF}^{\rm exp}
\right\}.
\end{eqnarray}
This  value quantitatively agrees with the
estimate of both the chiral quark soliton model \cite{Goeetal07},  given as $-(4/5) \times 2.35 \approx -1.9$
at some intrinsic scale%
\footnote{Note that a variety of chiral model estimates for this quantity \cite{Goeke:2001tz,SchBofRad02,Goeetal07,Wak07} exist. }
and lattice simulations \cite{Hagetal07}, quoted there as $-4\times 0.421 \approx -1.7$ at ${\cal Q}^2 = 4\,\GeV^2$. In
the left panel of Fig.~\ref{Fig-D^sea} we show the GFF $E_{1,2}$   versus $t$, where its $t$-dependency is compatible with both
the chiral quark soliton model result \cite{Goeetal07} and Lattice simulations \cite{Hagetal07}.

Let us finally have a look to the complete $D$-term (\ref{Cal-Dterm}), which for our power-like LFWF model reads
\begin{eqnarray}
\label{Cal-Dterm-1} d(z,t) = z \int_{(0)}^{1-|z|}\! dy \;   \frac{N
\left(\frac{m}{M} + y\right) y^{-\alpha} ((1-y)^2-z^2)^{p+\alpha/2}}{\left[(1-y)
\frac{m^2}{M^2}+ y \frac{\lambda^2}{M^2} - y(1-y) -((1-y)^2-z^2)
\frac{t}{4 M^2}\right]^{2p+1}}\,,
\end{eqnarray}
and identically vanishes for our exponential one. To evaluate the $y$-integral for $1<\alpha <2$, we employ analytic (or canonical) regularization
\cite{GelShi64}.
 Let us adopt the Gegenbauer expansion of Ref.~\cite{Goeke:2001tz},
\begin{eqnarray}
\label{Cal-Dterm1} d(z,t=0) \approx (1-z^2)\left[-2.02\,
C_1^{3/2}(z) + 1.10\, C_3^{3/2}(z) - 0.26\, C_5^{3/2}(z) +\cdots
\right],
\end{eqnarray}
where the first coefficient is given by  $5 H_{1,2}/4$. Interestingly, in contrast to the chiral quark soliton model result of Ref.~\cite{Goeke:2001tz},
\begin{eqnarray}
\label{Cal-Dterm2} d(z,t=0) \approx (1-z^2)\left[-4.0\,
C_1^{3/2}(z) -1.2 \, C_3^{3/2}(z) - 0.4\, C_5^{3/2}(z) +\cdots
\right],
\end{eqnarray}
our model has a sign alternating series and so our $d$-term is end-point suppressed while the chiral  quark soliton model
result is rather end-point enhanced, cf.~middle panel in Fig.~\ref{Fig-D^sea}.
Note that our first coefficient is closer to the
chiral quark soliton model result of Ref.~\cite{Goeetal07} and that the first three Gegenbauer modes provide a very accurate approximation for our model, visually not distinguishable from the displayed curve.

The subtraction constant in the ``dispersion relation'' (\ref{HandE-DR}) for  Compton form factors $\cal H$ and  $\cal E$, can be  straightforwardly evaluated from the $D$-term and reads within our flavor decomposition as
$$
{\cal D}(t) = \frac{2}{9} \int_{-1}^1\!dz \frac{2 z}{1-z^2}\, d(z,t)\,,
$$
where the additional factor $2/9$ is the average of the three light squared quark charges.  The result is displayed in the right panel of Fig.~\ref{Fig-D^sea} and confronted with phenomenological findings from ``dispersion relation'' fits. Unfortunately, such fits to the present experimental data set provide only a weak constraint on the subtraction constant.

Let us summarize. In the iso-singlet sector the $D$-term has been estimated in the chiral quark soliton model \cite{Goeke:2001tz} and in  Lattice simulations to be negative and sizeable, which is in agreement with the negative subtraction constant found in  ``dispersion relation" fits of deeply virtual Compton scattering data \cite{Kumericki:2009uq}. In our reggezised models a large negative $D$-term might be related to an effective `pomeron' exchange with $\alpha$ slightly larger than one.

\section{Summary and conclusions}

The goal of this article, which  contains various new results,
was to provide a start up for a consistent description of uPDFs and GPDs in terms of LFWFs and to
illuminate the challenges for the modeling procedure and LFWF phenomenology. Thereby, we followed the diquark  picture and so we restricted ourselves to effective two-body LFWFs. The model topics, we have are addressed, are the following:
\begin{itemize}
\item model classification based on the symmetry of the spin-density or -correlation matrices
\item model dependent cross talks of GPDs and uPDFs
\item implementing of the underlying Lorentz symmetry in the effective LFWF approach
\item Regge improvement as a collective phenomena from the $s$-channel point of view
\item emphasizing GPD duality
\end{itemize}

We introduced a new classification scheme for quark models and clearly realized that various uPDF models, including a three quark LFWF model, can be mapped to ``spherical" diquark models, which supports the diquark picture. We also clearly spelled out that although
uPDFs and GPDs can be obtained from the proton LFWFs the cross talk among them might be washed out in QCD. This we illustrated in the axial-vector
channel, where one might even have various LFWF models that provide the same results for uPDFs, however, different predictions for GPDs.  We stress that this sector can be modeled in various manner, that it is not well understood, and models consistent with phenomenology are known to be inconsistent with Lorentz symmetry requirements and reverse. We emphasize that the confrontation of highly symmetric models with phenomenological findings are a useful method to provide insights into the validity of  ``dynamical" QCD symmetries. However, to set up consistent, simple, and more flexible models, the spin-spin coupling of the struck quark with the spectator system has to be studied in more detail.
We only gave here some rather simple examples that are known to provide consistent models.

We also like to emphasize that the quark orbital angular momentum is only a useful partonic quantum number for simple models, however, in QCD a partonic distribution is build from an infinite number of states with various quark orbital angular momenta that contributions are indistinguishable.  On the other hand, it is just the  orbital angular momenta which destroys the cross talk between the three twist-two related uPDFs $f_1$, $g_1$, and $h_1$ with the corresponding zero-skewness GPDs $H$, $\widetilde H$, and ${\overline H}_{\rm T}$. Hence, nucleon tomography with GPDs and uPDFs might provide us two complementary three-dimensional pictures. Their (qualitative) comparison might provide us in the future some handle on the collective phenomena of
orbital angular momentum.
We also demonstrated that in a model one naturally renders two numbers for the quark orbital angular momentum, namely, one from its direct definition and the other one from utilizing Ji`s sum rule. The reason for this ``puzzle'' is obvious, namely, in static LFWF modeling the equations-of-motion are violated or even do not exist.

Let us summarize our GPD modeling attempts. We derived a relatively general representation for GPDs in terms of LFWFs, which satisfies both polynomiality and positivity constraints by construction. The uPDFs can be easily given in terms of LFWF overlaps, too. Here, we restricted ourselves for shortness to the twist-two sector where the extension to the twist-three one is straightforward.  We showed how $t$ and ${\bf k}_\perp^2$ dependencies are tied to each other
in a model dependent manner. We also illustrated that a rather simple LFWF parameterization is able to describe experimentally measured form factors, polarized and unpolarized PDFs from global fits, Lattice estimates, and is also compatible with phenomenological GPD features.

Certainly, LFWF modeling, considered as an attempt to address QCD aspects from the Field theoretical side, is challenging and we consider our work, presented here, as partially simple minded. Various rather old problems were not addressed here. For instance, the ${\bf k}_\perp$ and  $t$ interplay yields to a paradox that the large ${\bf k}_\perp$ behavior predicted by perturbative QCD is rather different from the $t$-behavior from dimensional and perturbative counting rules. Of course, such a paradox challenges for a deeper understanding of QCD dynamics, which might be obtained during exploration of effective LFWFs. Let us list a few of the problems, starting with the simple ones, which must be addressed along this road:
\begin{itemize}
\item Is the implementation of  Regge behavior in highly symmetric quark models realistic?
\item Can one utilize (simple) spin-spin parton spectator couplings to set up flexible models?
\item How $t$-dependence of Regge trajectories can be consistently implemented?
\item How to set up a flexible parameterization of the skewness effect in term of LFWFs?
\item How one can understand the interplay of ${\bf k}_\perp$, $t$, and renormalization scale dependence?
\end{itemize}
Even if the list can easily be extended, we finally conclude that a phenomenological LFWF approach might be a feasible alternative to the unsolved QCD bound state problem.

\vspace{5 mm}

\noindent D.S.H.~thanks the Institute for Theoretical Physics II
at the Ruhr-University Bochum for the hospitality at the stages of this work.
We are thankful for discussions and comments to B.~Pasquini, A.~Prokudin,
A.~Radyushkin, Ch.~Weiss, and F.~Yuan.
This work was supported in part by the BMBF (Federal Ministry for
Education and Research), contract FKZ 06 B0 103,
by the International Cooperation Program of the KICOS (Korea
Foundation for International Cooperation of Science \& Technology),
by Basic Science Research Program through the National
Research Foundation of Korea (NRF) funded by the Ministry of Education,
Science and Technology (2010-0011034).

\appendix
\renewcommand{\theequation}{\Alph{section}.\arabic{equation}}

\section{Unintegrated double distributions}
\label{sec-appA}
\setcounter{equation}{0}

To derive the unintegrated double distribution representation (\ref{Def-DDunint}) we start from the definition (\ref{GPD-overlap}), valid for $x\ge \eta$,
and the Laplace transform (\ref{Def-LFWF-Laplace}), which yields
\begin{eqnarray}
{\bf \Phi}(x,\eta,t,{\bf k}_{\perp}) &\!\!\! = \!\!\!& \frac{1}{1-x} \int_0^\infty\! d\alpha_1 \int_0^\infty \!d\alpha_2\,
\varphi^\ast(X^\prime,\alpha_2) \varphi(X,\alpha_1) \\
 &&\times \exp\left\{
-\alpha_2  \frac{ {\bf k}_\perp^{\prime 2} - X^\prime(1-X^\prime) M^2}{(1-X^\prime) M^2}  -\alpha_1  \frac{{\bf k}_\perp^2 - X(1-X) M^2}{(1-X) M^2}
 \right\}. \nonumber
\end{eqnarray}
Setting $\alpha_i=x_i A$, the scalar LFWF overlap is  given by
\begin{eqnarray}
{\bf \Phi}(x,\eta,{\bf \Delta}_{\perp},{\bf
k}_{\perp}) &\!\!\!=\!\!\!& \int_0^\infty\! dA A \int_0^1\! dx_1\int_0^1\! dx_2\;  \delta(1-x_1 -x_2) \varphi^\ast (X^\prime, A x_2) \varphi (X, A x_1) \\
&&\times
\exp\left\{
-A\left[ x_2 \frac{{\bf k}^{\prime 2}_\perp - X^\prime(1-X^\prime) M^2}{(1-X^\prime) M^2} + x_1 \frac{{\bf k}_\perp^2 - X(1-X) M^2}{(1-X) M^2 }
\right]
 \right\}.
\nonumber
\end{eqnarray}
Next we introduce the DD variables $y$ and $z$ by setting $x_1= \frac{1-y-z}{2(1-y)}$ and $x_2= \frac{1-y+z}{2(1-y)}$ with the constraint $x=y+z \eta$. This leads to the expression
\begin{eqnarray}
{\bf \Phi}(x,\eta,{\bf \Delta}_{\perp},{\bf k}_{\perp})
&\!\!\!=\!\!\!&
\frac{1}{2}\int_0^\infty\! \frac{dA A}{(1-y)^2} \int_0^1\! dy\int_{-1+y}^{1-y}\! dz\;  \delta(x-y-z \eta)
\nonumber\\
&&\times
\varphi^\ast \left(\frac{x-\eta}{1-\eta}, A \frac{1-y+z}{2(1-y)}\right) \varphi \left(\frac{x+\eta}{1+\eta}, A \frac{1-y-z}{2(1-y)}\right)
\\
&&\times
\exp\left\{ \frac{-A}{1-y} \left[
\frac{{\bf \overline{k}}_{\perp}^2}{M^2}  -\left[(1-y)^2-z^2\right] \frac{t}{4 M^2} - y(1-y)
\right] \right\},
\nonumber
\end{eqnarray}
where ${\bf \overline{k}}_{\perp}={\bf k}_\perp -(1 - y+z){\bf \Delta}_\perp/2$. We can now relax the constraint for the momentum fraction $x$ and consider the function as defined in the central region $-\eta \le x\le \eta$, too.
After rescaling the integration variable, $A \to A (1-y)$, we arrive at
the DD representation (\ref{Def-DDunint}), where the DD is given in (\ref{Def-DD-unint}).

\section{LFWF overlap representation for nucleon GPDs}
\label{sec-app-LFWF2GPD}
\setcounter{equation}{0}

We specify the frame by choosing a convenient
parameterization of the light-cone coordinates for the initial and
final proton:
\begin{eqnarray}
P&=&
\left(\ P^+\ ,\ {\bf 0}_\perp\ ,\ {M^2\over P^+}\ \right)\ ,
\label{a1}\\
P'&=&
\left( (1-\zeta)P^+\ ,\ -{\bf \Delta}_\perp\ ,\ {M^2+{\bf
\Delta}_\perp^2 \over (1-\zeta)P^+}\right)\ ,
\end{eqnarray}
where $M$ is the proton mass. We use the component notation $V = (V^+,
{\bf V}_\perp, V^-)$, and our metric is specified by $V^\pm = V^0 \pm
V^z$ and $V^2 = V^+ V^- - {\bf V}_\perp^2$. The four-momentum
transfer from the target is
\begin{eqnarray}
\label{delta}
\Delta&=&P-P'\ =\
\left( \zeta P^+\ ,\ {\bf \Delta}_\perp\ ,\
{t+{\bf \Delta}_\perp^2 \over \zeta P^+}\right)\ ,
\end{eqnarray}
where the Lorentz invariant Mandelstam variable $t \equiv \Delta^2$ and energy-momentum
conservation requires $\Delta^- \equiv P^- - P'^- = (t+{\bf \Delta}_\perp^2)/\zeta P^+$.  The
latter equation connects the transverse
momentum transfer, parameterized in the following as
\begin{eqnarray}
{\bf \Delta}_\perp \equiv (\Delta^1, \Delta^2) = |{\bf \Delta}_\perp|(\cos\varphi,\sin\varphi)\,,
\end{eqnarray}
the skewness variable $\zeta=\Delta^+/P^+$, and the Mandelstam variable $t$ according to
\begin{eqnarray}
\label{na1anew}
t \equiv \Delta^2  &\!\!\! = \!\!\!&
-{\zeta^2M^2+{\bf \Delta}_\perp^2 \over 1-\zeta}
\\
&\!\!\! = \!\!\!& t_0 -\frac{4 (1-\zeta )}{(2-\zeta )^2}\, {\bf \bar\Delta}_\perp^2
\quad\mbox{with}\quad
t_{0} = -{\zeta^2M^2\over 1-\zeta}
\quad\mbox{and}\quad {\bf \bar\Delta}_\perp = \frac{2-\zeta}{2 (1-\zeta )}\, |{\bf \Delta}_\perp|\, (\cos\varphi,\sin\varphi)\,.
\nonumber
\end{eqnarray}
Here, $-t_0$ is the kinematical allowed  minimal value of $-t$ and ${\bf \bar\Delta}_\perp$ is a convenient definition that absorbs
a $\zeta$-dependent  prefactor.

We note that w.r.t.~symmetries, e.g., $s$ and $u$-channel or time reflection, it is more appropriate to use the symmetric skewness variable  $\eta$, defined in Eq.~(\ref{eta}), rather than $\zeta$. Both scaling variables are related to each other by a SL($2|\mathbb{R}$) transformation
\begin{eqnarray}
\label{zeta2eta}
\zeta= \frac{2\eta}{1+\eta} \quad\mbox{or} \quad \eta=\frac{\zeta}{2-\zeta}\,.
\end{eqnarray}
The momentum transfer square (\ref{na1anew}) reads now in an explicit symmetric manner as
\begin{eqnarray}
t= t_0 - (1-\eta^2){\bf \bar \Delta}_\perp^2\,,\quad\mbox{where}\quad
t_0 = -\frac{4 M^2 \eta^2}{1-\eta^2} \quad\mbox{and}\quad
{\bf \bar \Delta}_\perp = \frac{\sqrt{t_0-t}}{\sqrt{1-\eta^2}}\,(\cos\varphi,\sin\varphi)
\end{eqnarray}
are symmetric functions w.r.t.~$\eta\to -\eta$ reflection.

In the following we list the LFWF overlap representations for all twist-two GPD combinations, where the initial and final nucleon state, having
momenta $P$ and $P^\prime$ as well as  spin projections $S$ and $S^\prime$, are described by $n$-parton LFWFs
$$\psi^{S}_{(n)}(X_i, {{\bf k}}_{\perp i},s_i)
\quad \mbox{and}\quad
\psi^{*\, {S^\prime}}_{(n)}(X^\prime_i,{{\bf k}}^\prime_{\perp i},s^\prime_i),$$
respectively.
Here and in the following we use common  LF variables, namely, the momentum fraction $X_i$  and ${\bf k}_{\perp i}$ for the incoming partons. We have to
deal with the parton number conserved and changing LFWF overlaps. In the former case the momenta of the outgoing partons are given by
\begin{equation}
\begin{array}[t]{lll}
X^\prime_{1} = {\displaystyle \frac{X_{1}-\zeta}{1-\zeta}}\, ,\
&{{\bf k}}^\prime_{\perp 1} ={{\bf k}}^{~}_{\perp 1}
- {\displaystyle \frac{1-X_{1}}{1-\zeta}}\, {{\bf \Delta}}_\perp
&\mbox{for the struck parton,}
\\[2ex]
X^\prime_i = {\displaystyle \frac{X_i}{1-\zeta}}\, ,\
&{{\bf k}}^\prime_{\perp i} ={{\bf k}}^{~}_{\perp i}
+ {\displaystyle \frac{X_i}{1-\zeta}}\, {{\bf \Delta}}_\perp
&\mbox{for the spectator $i\in\{2, \cdots, n\}$.}
\end{array}
\label{t2}
\end{equation}
In the latter case the struck partons, appearing in the initial state, are labeled by $i=1$ and $n+1$ and their parton momenta are related by
\begin{eqnarray}
X_{n+1} = \zeta - X_{1}\,, \quad {{\bf k}}_{\perp n+1} =
{{\bf \Delta}}_\perp-{{\bf k}}_{\perp 1} &&  \mbox{for the struck partons,}
\label{t3a}
\end{eqnarray}
where the outgoing momenta of the spectator system ($i=2, \cdots, n$) are given in Eq.~(\ref{t2}).
We add that we alternatively might use the skewness variable $\eta$ together with the momentum fractions
\begin{eqnarray}
\label{xeta2XZeta}
x_i = \frac{2X_i -\zeta}{2-\zeta}
\quad
\mbox{or}
\quad
X_i=\frac{x_i+\eta}{1+\eta}\,.
\end{eqnarray}

For the spinors of a free spin-$1/2$ we adopt the Brodsky--Lepage convention from Ref.~\cite{Lepage:1980fj},
\begin{eqnarray}
U(p,s=1/2)={1\over {\sqrt{2p^+}}}
\left(
\begin{array}{c}
p^++m\\
p^1+i p^2\\
p^+-m\\
p^1+i p^2
\end{array}
\right)
\ ,
\qquad
U(p,s=-1/2)={1\over {\sqrt{2p^+}}}
\left(
\begin{array}{c}
-p^1+i p^2\\
p^++m\\
p^1-i p^2\\
-p^++m
\end{array}
\right)
\ ,
\label{s2}
\end{eqnarray}
which can be represented as linear combination of Bjorken and Drell spinors.
The definition of Dirac matrices can be found in \cite{BjoDreBoo}.

\subsection{Chiral even and parity even GPDs: $H$ and $E$}
\label{sec-app-LFWF2HE}

The  $H$ and $E$ GPDs appear as form factors in the matrix elements
of a  vector operator (\ref{HE-def}),
Using the LFWF (\ref{Def-ProSta}) for the initial and final nucleon state and the one-parton matrix element
\begin{eqnarray}
\lefteqn{
\int\frac{d y^-}{8\pi}\;e^{ix P^+y^-/2}\;
\left< 1;\, X'_{1} P\,'^+, {{\bf p}\,'_{\perp 1}, s'_{1}}
  \left| \bar\psi(0)\, {\gamma^+}\,\psi(y)\, \right|
1;\, X^{~}_{1} P^+, {{\bf p}^{~}_{\perp 1}}, s^{~}_{1}\right>
\Big|_{y^+=0, y_\perp=0}
} \hspace{14em}
\nonumber \\
&=& \sqrt{X^{~}_{1} X'_{1}}\,\sqrt{1-\zeta}\ \delta(X-X_{1})\
    \delta_{s'_{1} s^{~}_{1}}\,,
\end{eqnarray}
where for definiteness we have labeled the struck quark with the index
$i=1$, we can now evaluate the $n \to n$ diagonal LFWF overlap to the matrix element on the l.h.s.~of  Eq.~(\ref{HE-def}).
Evaluating the form factor decomposition on the r.h.s.~of Eq.~(\ref{HE-def}) in terms of helicity amplitudes, conveniently written as
\begin{eqnarray}
{1 \over 2\bar P^+}\
{\bar U}(P',{S}'){\gamma^+}\, U(P,S) &\!\!\! =\!\!\!&
{2\sqrt{1-\zeta} \over 2-\zeta}\, \delta_{S ,\, {S}'}\ ,
\label{com1z}\\
{1 \over 2\bar P^+}\
{\bar U}(P',{S}'){\sigma^{+\alpha}\Delta_\alpha\over 2iM} U(P,S)
&\!\!\! =\!\!\!& {2\sqrt{1-\zeta} \over 2-\zeta}\,\left[
 \frac{t_{0}}{4 M^2}\, \delta_{S ,\, {S}'}
 - { |{\bf \bar \Delta}_\perp| \,e^{2i S\varphi} \over 2M } \, 2S\ \delta_{S ,\, -{S}'}
 \right],
\nonumber
\end{eqnarray}
we find from the target spin conserved ($S^\prime=S=1/2$) and flip ($S^\prime=-S=1/2$) cases the following linear combinations
in the outer GPD region $\eta\le x\le 1$:
\begin{eqnarray}
\label{t1}
 H(x,\eta,t)+ \frac{t_{0}}{4M^2}E(x,\eta,t)
=
\sum_{n,s_1}\!\!\int\!\!\!\!\int\!\!d^2{\bf k}_\perp\;\intsum
\psi^{\ast\,\Rightarrow}_{(n)}(X^\prime_i,{{\bf k}}^\prime_{\perp i},s_i)\,
\psi^{\Rightarrow}_{(n)}(X_i, {{\bf k}}_{\perp i},s_i)
\end{eqnarray}
and
\begin{eqnarray}
\label{t1f2}
{|{\bf \bar \Delta}_\perp| \,e^{-i \varphi}\over 2M}\,
E(x,\eta,t)=
\sum_{n,s_1}\!\!\int\!\!\!\!\int\!\!d^2{\bf k}_\perp\;\intsum
\psi^{\ast\,  \Rightarrow}_{(n)}(X^\prime_i,{{\bf k}}^\prime_{\perp i},s_i) \,
\psi^{\Leftarrow}_{(n)}(X_i, {{\bf k}}_{\perp i},s_i)\,,
\end{eqnarray}
respectively.
Here, ${\bf \bar\Delta}_\perp$ is given in Eq.~(\ref{na1anew}) and the integral measure is defined  as
\begin{eqnarray}
\label{intsum1}
\intsum \cdots \equiv \sum_{s_2,\dots, s_n}
\int\!\frac{[dX\, d^2{\bf k}_{\perp}]_n}{\sqrt{1-\zeta}^{n-2}}\,
\frac{2-\zeta}{2\sqrt{1-\zeta}}\, \delta(X-X_{1}) \delta^{(2)}({\bf k}_{\perp}-{\bf k}_{\perp 1})
\cdots\,,
\end{eqnarray}
see also Eq.~(\ref{[dXd2k]}),
and the arguments of the final-state wavefunctions are given in  Eq.~(\ref{t2}).
In Eqs.~(\ref{t1}) and
(\ref{t1f2}) one has to sum over all possible combinations of
spins $s^{~}_i$ and over all parton numbers $n$ in the Fock
states. We also imply a sum over all possible ways of numbering the
partons in the $n$-particle Fock state so that the struck quark has
the index $i=1$.
Note a flip of the target spin projection leaves Eq.~(\ref{t1}) invariant, however, it will change the
phase $\varphi\to -\varphi$ and the sign on the l.h.s.~of Eq.~(\ref{t1f2}).

Analogous formulae hold in the domain $-1 < x <-\eta$, where the
struck parton in the target is an antiquark instead of a quark. Some
care has to be taken regarding overall signs arising because fermion
fields anticommute. For details we refer to
\cite{Diehl:1998kh,Brodsky:2000xy,Diehl:2000xz}.

For the $n+1 \to n-1$ off-diagonal term, let us
consider the case where quark $1$ and antiquark $n+1$ of the initial
wavefunction annihilate into the current leaving $n-1$ spectators.
Then $x_{n+1} = \zeta - x_{1}$ and ${{\bf k}}_{\perp n+1} =
{\vec{\Delta}}_\perp-{{\bf k}}_{\perp 1}$.   The current matrix element  is
\begin{eqnarray}
\lefteqn{
\int\frac{d y^-}{8\pi}\;e^{i X P^+y^-/2}\;
\left< 0 \left| \bar\psi(0)\, {\gamma^+}\,\psi(y)\, \right|
2;\ x^{~}_{1} P^+, x^{~}_{n+1} P^+,\;
    {{\bf p}^{~}_{\perp 1}}, {{\bf p}^{~}_{\perp n+1}},\; s_1, s_{n+1}
\right> \Big|_{y^+=0, y_\perp=0}
} \hspace{21em}
\\
&=& \sqrt{X_{1} X_{n+1} \phantom{1}\!\!}\, \delta(X-X_{1})\
    \delta_{s_1\, -s_{n+1}}\ ,
\nonumber
\end{eqnarray}
and we thus obtain the formulae for the off-diagonal
contributions to $H$ and $E$
in the domain $|x| \le \eta$:
\begin{eqnarray}
\lefteqn{
 H(x,\eta,t)+ \frac{t_{0}}{4 M^2} E(x,\eta,t)
} \hspace{5em}
\label{t3} \\
 &=& \sum_{n,s_1}\!\!\int\!\!\!\!\int\!\!d^2{\bf k}_\perp\;\intsum \sqrt{1-\zeta}\,
\psi^{\ast\,\Rightarrow}_{(n-1)}(X^\prime_i,
  {{\bf k}}^\prime_{\perp i},s_i) \
\psi^{\Rightarrow}_{(n+1)}(X_i, {{\bf k}}_{\perp i},s_i) \
\delta_{s_{1}\, -s_{n+1}} \ ,
\nonumber
\\
\lefteqn{
{|{\bf \bar \Delta}_\perp| \,e^{-i \varphi}\over 2M} \, E(x,\eta,t)
\rule{0pt}{4.5ex} } \hspace{5em}
\label{t3f2} \\
 &=& \sum_{n,s_1}\!\!\int\!\!\!\!\int\!\!d^2{\bf k}_\perp\;\intsum \sqrt{1-\zeta}\,
\psi^{{\ast\,\Rightarrow}}_{(n-1)}(X^\prime_i,
  {{\bf k}}^\prime_{\perp i},s_i) \
\psi^{\Leftarrow}_{(n+1)}(X_i, {{\bf k}}_{\perp i},s_i) \
\delta_{s_{1}\, -s_{n+1}} \ ,
\nonumber
\end{eqnarray}
where the momenta are specified in Eq.~(\ref{t3a}).

A few comments are in order. Comparing Eqs.~(\ref{t1},\ref{t1f2}) and  Eqs.~(\ref{t3},\ref{t3f2}), we realize that the parton number off-diagonal expression can be obtained from the diagonal one by the substitution:
\begin{eqnarray}
\label{dia2offdia}
\psi^{*\,\Rightarrow}_{(n)}(X^\prime_i,{{\bf k}}^\prime_{\perp i},s_i) \
\psi^{S}_{(n)}(X_i, {{\bf k}}_{\perp i},s_i)
\;\Rightarrow\;
\sqrt{1-\zeta}\;
\psi^{{\ast\,\Rightarrow}}_{(n-1)}(X^\prime_i,
  {{\bf k}}^\prime_{\perp i},\lambda^{~}_i) \
\psi^{S}_{(n+1)}(X_i, {{\bf k}}_{\perp i},s_i) \
\delta_{s_{1}\, -s_{n+1}}\,,
\nonumber\\
\end{eqnarray}
where the momenta are specified in Eqs.~(\ref{t2},\ref{t3a}).
The powers of $\sqrt{1-\zeta}$ in (\ref{t1}), (\ref{t1f2}) and
(\ref{t3}), (\ref{t3f2}) have their origin in the integration measures
in the Fock state decomposition (\ref{Def-ProSta}) for the outgoing
proton. The fractions $X_i'$ appearing there refer to the light-cone
momentum $P'^+ = (1-\zeta)\, P^+$, whereas the fractions $X_i$ in the
incoming proton wavefunction refer to $P^+$. Transforming all
fractions so that they refer to $P^+$ as in our final formulae thus
gives factors of $\sqrt{1-\zeta}$. Different powers appear in the
$n\to n$ and $n+1 \to n-1$ overlaps because of the different parton
numbers in the final state wavefunctions.

\subsection{Chiral even and parity odd GPDs: ${\widetilde{H}}$ and ${\widetilde{E}}$ GPDs}

The   ${\widetilde H}$ and  ${\widetilde E}$ GPDs  are defined through matrix elements
of the  axial-vector operator (\ref{tHtE-def}).
The evaluation of the LFWF overlap is analogously performed as in Sec.~\ref{sec-app-LFWF2HE}.
For the $n \to n$ diagonal term, the relevant one-parton matrix element at quark level is
\begin{eqnarray}
\lefteqn{
\int\frac{d y^-}{8\pi}\;e^{ix P^+y^-/2}\;
\left< 1;\, X'_{1} P\,'^+, {{\bf p}\,'_{\perp 1}, s'_{1}}
  \left| \bar\psi(0)\, {\gamma^+}\gamma_5\,\psi(y)\, \right|
1;\, X^{~}_{1} P^+, {{\bf p}^{~}_{\perp 1}}, s^{~}_{1}\right>
\Big|_{y^+=0, y_\perp=0}
} \hspace{14em}
\\ \nonumber
&=& \sqrt{X^{~}_{1} X'_{1}}\,\sqrt{1-\zeta}\ \delta(X-X_{1})\
  2s_1\  \delta_{s'_{1} s^{~}_{1}}
\end{eqnarray}
and the helicity amplitudes of the form factors read:
\begin{eqnarray}
{1 \over 2\bar P^+}\
{\bar U}(P',{S}'){\gamma^+}\gamma^5\, U(P,S)&\!\!\! =\!\!\!&
{2\sqrt{1-\zeta} \over 2-\zeta}\,  2S\,\delta_{S ,\, {S}'}\ ,
\label{com1zaxial}\\
{1 \over 2\bar P^+}\
{\bar U}(P',{S}'){-\Delta^+ \over 2M}\,
\gamma^5 U(P,S)
&\!\!\! =\!\!\!& {2\sqrt{1-\zeta} \over 2-\zeta}\,
\Big[\frac{t_{0}}{4M^2}\,2 S\, \delta_{S ,\, {S}'}
-{\zeta\,|{\bf \bar \Delta}_\perp| \,  e^{i 2S\varphi} \over (2-\zeta) 2M }\,
\delta_{S ,\, -{S}'}\Big]
\ .
\nonumber
\end{eqnarray}
Considering the target spin conserved ($S^\prime=S=1/2$) and flip ($S^\prime=-S=1/2$) cases,
we obtain for the $n\to n$ diagonal overlap contributions to the linear combinations of ${\widetilde{H}}$ and
${\widetilde{E}}$ GPDs in the domain $\eta\le x\le 1$:
\begin{eqnarray}
\label{t155}
 \widetilde H(x,\eta,t)+\frac{t_{0}}{4M^2} \widetilde E(x,\eta,t)
=
\sum_{n,s_1}2s_1\!\!\int\!\!\!\!\int\!\!d^2{\bf k}_\perp\;\intsum
\psi^{\ast\,\Rightarrow }_{(n)}(X^\prime_i,
  {{\bf k}}^\prime_{\perp i},s_i) \
\psi^{ \Rightarrow }_{(n)}(X_i, {{\bf k}}_{\perp i},s_i)
\end{eqnarray}
and
\begin{eqnarray}
\label{t1f255}
 {|{\bf \bar \Delta}_\perp| \, e^{-i \varphi} \over 2M }\,
(- \eta){\widetilde{E}}(x,\eta,t)
=
\sum_{n,s_1}2s_1\!\!\int\!\!\!\!\int\!\!d^2{\bf k}_\perp\;\intsum
\psi^{\ast\,\Rightarrow}_{(n)}(X^\prime_i,{{\bf k}}^\prime_{\perp i},s_i) \
\psi^{\Leftarrow}_{(n)}(X_i, {{\bf k}}_{\perp i},s_i)\, .
\end{eqnarray}
Flipping the target spin changes the sign on the l.h.s.~of Eq.~(\ref{t155})  and the sign of the phase in Eq.~(\ref{t1f255})

For the $n+1 \to n-1$ off-diagonal term the relevant $2\to 0$  partonic matrix element is
\begin{eqnarray}
\lefteqn{
\int\frac{d y^-}{8\pi}\;e^{ix P^+y^-/2}\;
\left< 0 \left| \bar\psi(0)\, {\gamma^+}\gamma_5\,\psi(y)\, \right|
2;\ X_{1} P^+, X_{n+1} P^+,\;
    {{\bf p}^{~}_{\perp 1}}, {{\bf p}^{~}_{\perp n+1}},\;
    s_{1}, s_{n+1}
\right> \Big|_{y^+=0, y_\perp=0}
} \hspace{21em}
\\
&=& \sqrt{X_{1} X_{n+1} \phantom{1}\!\!}\, \delta(X-X_{1})\
  2s_1\  \delta_{s_{1}\, -s_{n+1}}\ .
\nonumber
\end{eqnarray}
We thus obtain that the formulae for the off-diagonal
contributions to ${\widetilde{H}}$ and ${\widetilde{E}}$ GPDs
in the domain $|x| \le \eta$ are obtained from Eqs.~(\ref{t155}, \ref{t1f255}) by the substitution (\ref{dia2offdia}).

\subsection{Chiral odd GPDs: $H_T$, $E_T$, ${\widetilde{H}}_T$ and ${\widetilde{E}}_T$}

The $H_T$, $E_T$, ${\widetilde H}_T$, and ${\widetilde E}_T$  GPDs  are defined trough the matrix elements of the tensor operator (\ref{FT-def}).
The evaluation of the LFWF overlap is analogously performed as in Sec.~\ref{sec-app-LFWF2HE}.
For the $n \to n$ diagonal term, the relevant one-parton matrix element at quark level are
\begin{eqnarray}
\lefteqn{
\int\frac{d y^-}{8\pi}\;e^{ix P^+y^-/2}\;
\big< 1;\, X'_{1} P\,'^+, {{\bf p}\,'_{\perp 1}, s'_{1}}
\, \big| \bar\psi(0)\left\{{i\sigma^{+1}\atop i\sigma^{+2} } \right\}\psi(y)\, \big|
1;\, X^{~}_{1} P^+, {{\bf p}^{~}_{\perp 1}}, s^{~}_{1}\big>
\Big|_{y^+=0, y_\perp=0}
} \hspace{14em}
\\ \nonumber
&=& \sqrt{X^{~}_{1} X'_{1}}\,\sqrt{1-\zeta}\ \delta(X-X_{1})
\left\{{-2s_1\atop -i} \right\}  \delta_{-s'_{1}\, s^{~}_{1}}\ .
\end{eqnarray}
The helicity amplitudes of the bilinear spinors, which we need in the calculation, are the vector one in Eq.~(\ref{com1z}) and
\begin{eqnarray}
{1 \over 2M}\
{\bar U}(P',{S}')  U(P,S) &\!\!\!=\!\!\!&
\frac{2-\zeta }{2 \sqrt{1-\zeta }}\, \delta_{S ,\, {S}'}
+{2\sqrt{1-\zeta} \over 2-\zeta}\,
{|{\bf \bar \Delta}_\perp| \, e^{i2S\varphi}  \over 2 M } \, 2S\ \delta_{S ,\, -{S}'}\,,
\\
{1 \over 2M}\
{\bar U}(P',{S}')\left\{{\gamma^{1}\atop \gamma^{2} } \right\} U(P,S) &\!\!\!=\!\!\!&
{2\sqrt{1-\zeta} \over 2-\zeta}\,
\left\{{-2S\atop -i} \right\}
\left[
{|{\bf \bar \Delta}_\perp|\, e^{-i2S\varphi} \over 2M } \, 2S\, \delta_{S ,\, {S}'} - \frac{(2-\zeta)\zeta }{4 (1-\zeta)} \,\delta_{S ,\, -{S}'}
\right],
\label{tt1001}
\nonumber\\
\label{t1000}
{1 \over 2\bar P^+}\
{\bar U}(P',{S}')\left\{{i\sigma^{+1}\atop i\sigma^{+2} } \right\} U(P,S) &\!\!\!=\!\!\!&
{2\sqrt{1-\zeta} \over 2-\zeta}\,
\left\{{-2S\atop -i} \right\}
\delta_{S ,\, -{S}'}\,.\nonumber
\end{eqnarray}
The $n\to n$ diagonal LFWF overlap contributions to
$H_T$, $E_T$, ${\widetilde{H}}_T$, and ${\widetilde{E}}_T$ GPD combinations
in the domain $\eta\le x\le 1$ read for the transverse component $i=1$ and $i=2$ as
\begin{eqnarray}
\label{t155s2}
&&{|{\bf \bar \Delta}_\perp|\, e^{-i\varphi}\over 4M} \left[
2\, \widetilde{H}_{T}(x,\eta,t)
+\left(1+\eta\right) E_{T}(x,\eta,t)
-\left(1+\eta\right){\widetilde{E}}_{T}(x,\eta,t)
\right]\\
&&\qquad =
\sum_{n,s_1}\frac{1+2s_1}{2}\!\!\int\!\!\!\!\int\!\!d^2{\bf k}_\perp\;\intsum
\psi^{\ast\,\Rightarrow}_{(n)}(X^\prime_i,{{\bf k}}^\prime_{\perp i},-s_{1},s_{i=2,\cdots ,n}) \,
\psi^{\Rightarrow}_{(n)}(X_i, {{\bf k}}_{\perp i},s_i)\,,
\nonumber
\end{eqnarray}
\begin{eqnarray}
\label{t155s3}
&&
{-|{\bf \bar \Delta}_\perp|\, e^{i\varphi}\over 4M}\left[
2\, \widetilde{H}_{T}(x,\eta,t)
+\left(1-\eta\right) E_{T}(x,\eta,t)
+\left(1-\eta\right){\widetilde{E}}_{T}(x,\eta,t)
\right]
\\
&&\qquad =
\sum_{n,s_1}\frac{1-2s_1}{2}\!\!\int\!\!\!\!\int\!\!d^2{\bf k}_\perp\;\intsum
\psi^{\ast\,\Rightarrow}_{(n)}(X^\prime_i,{{\bf k}}^\prime_{\perp i},-s_{1},s_{i=2,\cdots ,n}) \,
\psi^{\Rightarrow}_{(n)}(X_i, {{\bf k}}_{\perp i},s_i)\,,
\nonumber
\end{eqnarray}
and
\begin{eqnarray}
\label{t1f255s}
&&
 H_{T}(x,\eta,t) + {t_0-t\over 4M^2}\, {\widetilde{H}}_{T}(x,\eta,t) +\frac{t_{0}}{4M^2}
 \left[E_{T} - \frac{1}{\eta}\, {\widetilde{E}}_{T}\right](x,\eta,t)
\\
&&\qquad =
\sum_{n,s_1}\frac{1-2s_1}{2}\!\!\int\!\!\!\!\int\!\!d^2{\bf k}_\perp\;\intsum
\psi^{\ast\,\Rightarrow}_{(n)}(X^\prime_i,{{\bf k}}^\prime_{\perp i},-s_1,s_{i=2,\cdots ,n}) \,
\psi^{\Leftarrow}_{(n)}(X_i, {{\bf k}}_{\perp i},s_i) \,,
\nonumber
\end{eqnarray}
\begin{eqnarray}
\label{t1f255s1}
&&
{-{\bf \bar \Delta}_\perp^2 (1-\eta^2)\over 4M^2} \, e^{-i2\varphi}
{\widetilde{H}}_{T}(x,\eta,t)
\\
&&\qquad =
\sum_{n,s_1}\frac{1+2s_1}{2}\!\!\int\!\!\!\!\int\!\!d^2{\bf k}_\perp\;\intsum
\psi^{\ast\,\Rightarrow}_{(n)}(X^\prime_i,{{\bf k}}^\prime_{\perp i},-s_1,s_{i=2,\cdots ,n}) \,
\psi^{\Leftarrow}_{(n)}(X_i, {{\bf k}}_{\perp i},s_i) \,,
\nonumber
\end{eqnarray}
respectively.

From the comparison of the diagonal one-parton matrix elements (\ref{t1000}) with
\begin{eqnarray}
\lefteqn{
\int\frac{d y^-}{8\pi}\;e^{iX P^+y^-/2}\;
\big< 0 \, \big| \bar\psi(0)\left\{{i\sigma^{+1}\atop i\sigma^{+2} } \right\}\psi(y)\, \big|
2;\ X_{1} P^+, X_{n+1} P^+,\;
    {{\bf p}^{~}_{\perp 1}}, {{\bf p}^{~}_{\perp n+1}},\;
    s_{1}, s_{n+1}
\big> \Big|_{y^+=0, y_\perp=0}
} \hspace{21em}
\nonumber\\
&=& \sqrt{X_{1} X_{n+1} \phantom{1}\!\!}\, \delta(X-X_{1})\left\{{-2s_1\atop -i} \right\}  \delta_{s_{1}\, s_{n+1}}\ ,
\nonumber
\end{eqnarray}
we read off that the formulae for the off-diagonal contributions to the chiral odd GPD combinations
in the central region $|x| \le \eta$ are obtained from Eqs.~(\ref{t155s2}
--\ref{t1f255s1}) by the substitution:
\begin{eqnarray}
\label{dia2offdia_1}
&&\psi^{\ast\,\Rightarrow}_{(n)}(X^\prime_i,{{\bf k}}^\prime_{\perp i},-s_{1},s_{i=2,\cdots ,n}) \
\psi^{S}_{(n)}(X_i, {{\bf k}}_{\perp i},s_i)
\\
&& \hspace{2cm}
\;\Rightarrow\;
\sqrt{1-\zeta}\;
\psi^{{\ast\,\Rightarrow}}_{(n-1)}(X^\prime_i,{{\bf k}}^\prime_{\perp i},s_i)\ \psi^{S}_{(n+1)}(X_i, {{\bf k}}_{\perp i},s_i) \
\delta_{s_{1}\, s_{n+1}}\,,
\nonumber
\end{eqnarray}
where the momenta are specified in Eqs.~(\ref{t2},\ref{t3a}).


\end{document}